\documentclass[dsc, eng]{ita}    
\usepackage{ae}
\usepackage{graphicx}
\usepackage{epsfig}
\usepackage{amsmath}
\usepackage{amssymb} 
\usepackage{multirow}
\usepackage{float}
\usepackage{makeglo}

\usepackage{subfigure}
\usepackage{array}
\usepackage{lineno}

\usepackage{tablefootnote}
\usepackage{threeparttable}

\usepackage{comment}

\newcommand{\argmin}{\mathop{\mathrm{argmin}}\limits}

\course{Electronics and Computing Engineering}
\area{Informatics}

\authorgender{masc}                     
\author{Gabriel de Souza Pereira}{Moreira}
\itaauthoraddress{Rua Fauze Dimas Lumumba Gon\c{c}alves, 307, Jd. Santa In\^es I}{12248-460}{S\~ao Jos\'e dos Campos, S\~ao Paulo}

\title{CHAMELEON: A Deep Learning Meta-Architecture for News Recommender Systems}

\advisorgender{masc}                    
\advisor{Prof.~Dr.}{Adilson Marques da Cunha}

\bossgender{masc}												
\boss{Prof.~Dr.}{Pedro Teixeira Lacava}

\kwcip{Recommender Systems}
\kwcip{Deep Learning}
\kwcip{Recurrent Neural Networks}

\examiner{Prof. Dr.}{Paulo Marcelo Tasinaffo}{Presidente}{ITA}
\examiner{Prof. Dr.}{Luiz Alberto Vieira Dias}{Membro interno}{ITA}
\examiner{Prof. Dr.}{Leandro Balby Marinho}{Membro externo}{UFCG}
\examiner{Prof. Dr.}{Rodrygo Santos}{Membro externo}{UFMG}

\date{09}{Dezembro}{2019}

\makeglossary
\frontmatter

\begin{document}
\maketitle

\begin{itadedication}
I dedicate this work to my dear wife Luiza, for years of support and encouragement throughout my career, to my lovely children Melissa and Samuel, for understanding their busy father, to my parents Francisco and Ros{\^a}ngela for the education and motivation they provided me while growing and above all to my God for sustaining and inspiring me.
\end{itadedication}

\begin{itathanks}
First at all, I would like to thank God for providing me health, intelligence, and inspiration to accomplish this research.

Then, I would like to thank my wife Luiza, who encouraged me all the time, dedicating great love, patience, and support so that I could carry out this work.

My children Melissa and Samuel, for all their love, inspiration and comprehension for their busy dad. In special, Melissa deserves the credit for the name \emph{CHAMELEON} in the title of this research.

My parents Rosangela and Francisco, for the provided education and constant motivation that allowed me to get up to here.

My parents-in-law, for all the support to my children and for their prayers.

My sister Liliane, for being a partner of all my challenges and achievements.

Prof. Adilson Marques da Cunha, for dedicated guidance since my Masters, which I take not only for my thesis, but for life.

Prof. Dietmar Jannach, which was my mentor at the \emph{ACM RecSys'18 Doctoral Symposium} and a great research partner during the final phase of this work.



The Brazilian Aeronautics Institute of Technology (\emph{Instituto Tecnol\'ogico de Aeron\'autica} - ITA), for providing me with an excellent learning environment and professional development.

The Globo.com, for providing context on their technical challenges on news recommendation and for sharing an exclusive large and high-quality dataset for this research.

The company CI\&T, for allowing me to carry out this research in parallel with my responsibilities and to internalize the learned knowledge into my company's projects.

Finally, all family and friends who encouraged and supported me in this journey.
\end{itathanks}

\thispagestyle{empty}
\ifhyperref\pdfbookmark[0]{\nameepigraphe}{epigrafe}\fi
\begin{flushright}
\begin{spacing}{1}
\mbox{}\vfill
{\sffamily\itshape
``For the Lord gives wisdom;\\
 from his mouth come knowledge and understanding.''\\}
--- \textsc{Proverbs 2:6}
\end{spacing}
\end{flushright}

\begin{abstract}
\noindent
Sistemas de Recomenda\c{c}\~{a}o (SR) t\^{e}m sido cada vez mais populares para auxiliar os usu\'{a}rios em suas escolhas, melhorando seu engajamento e satisfa\c{c}\~{a}o com os servi\c{c}os online. Os SR se tornaram um t\'{o}pico de pesquisa popular e, desde 2016, m\'{e}todos e t\'{e}cnicas de \emph{Deep Learning} t\^{e}m sido explorados nesta \'{a}rea.

Os SR de not\'{\i}cias s\~{a}o projetados para personalizar a experi\^{e}ncia dos usu\'{a}rios e ajud\'{a}-los a descobrir artigos relevantes em um amplo e din\^{a}mico espa\c{c}o de busca. Contudo, este \'{e} um cen\'{a}rio desafiador para recomenda\c{c}\~{o}es. Grandes portais publicam centenas de not\'{\i}cias por dia, implicando que \'{e} necess\'{a}rio lidar com o crescimento acelerado do n\'{u}mero de itens, que se tornam rapidamente obsoletos e irrelevantes para a maioria dos usu\'{a}rios. Pode ser observado no comportamento dos leitores de not\'{\i}cia que suas prefer\^{e}ncias s\~{a}o mais inst\'{a}veis do que as de usu\'{a}rios de outros dom\'{\i}nios onde sistemas de recomenda\c{c}\~{a}o s\~{a}o tradicionalmente aplicados, como entretenimento (e.g., filmes, seriados, m\'{u}sicas). Adicionalmente, o dom\'{\i}nio de not\'{\i}cias possui um alto n\'{\i}vel de esparsidade, pois a maioria dos usu\'{a}rios \'{e} an\^{o}nima, sem comportamento passado observado. 

A principal contribui\c{c}\~{a}o desta pesquisa \'{e} a meta-arquitetura de \emph{Deep Learning} CAMALE\~{A}O, em ingl\^es  \emph{CHAMELEON}, projetada para endere\c{c}ar os desafios espec\'{\i}ficos de recomenda\c{c}\~{a}o de not\'{\i}cias. Ela consiste de uma arquitetura de refer\^{e}ncia modular, que pode ser instanciada usando diferentes componentes arquiteturais de redes neurais. 

Como informa\c{c}\~{o}es sobre intera\c{c}\~{o}es passadas de usu\'{a}rios s\~{a}o escassas no dom\'{\i}nio de not\'{\i}cias, o contexto do usu\'{a}rio pode ser utilizado para lidar com o problema de \emph{cold-start} de usu\'{a}rios. O conte\'{u}do textual do artigo tamb\'{e}m \'{e} importante para endere\c{c}ar o problema de \emph{cold-start} de itens. Adicionalmente, o decaimento temporal da relev\^{a}ncia dos itens (artigos) \'{e} bastante acelerado no dom\'{i}nio de not\'{i}cias. Finalmente, acontecimentos externos importantes podem atrair temporariamente o interesse p\'{u}blico global, um fen\^{o}meno geralmente conhecido como \emph{concept drift} em aprendizado de m\'{a}quina. Todas estas caracter\'{i}sticas s\~{a}o explicitamente modeladas nesta pesquisa por um sistema de recomenda\c{c}\~{a}o h\'{i}brido baseado em sess\~{o}es, utilizando Redes Neurais Recorrentes. 

A tarefa de recomenda\c{c}\~{a}o endere\c{c}ada neste trabalho consiste na predi\c{c}\~{a}o do pr\'{o}ximo clique de um usu\'{a}rio, utilizando apenas informa\c{c}\~{a}o dispon\'{i}vel em sua sess\~{a}o atual. Um m\'{e}todo \'{e} proposto para uma real\'{\i}stica avalia\c{c}\~{a}o  
\emph{offline} temporal de tal tarefa, reproduzindo o fluxo de intera\c{c}\~{o}es de usu\'{a}rios e de novos artigos sendo continuamente publicados em um portal de not\'{\i}cias. 

Experimentos executados nesta pesquisa, utilizando duas grandes bases de dados, mostram a efetividade do \emph{CHAMELEON} na recomenda\c{c}\~{a}o de not\'{\i}cias em muitos aspectos de qualidade, como acur\'{a}cia, cobertura de cat\'{a}logo de itens, capacidade de recomendar itens n\~{a}o-populares e redu\c{c}\~{a}o do problema de \emph{cold-start} de itens, quando comparado com outros algoritmos tradicionais e com o estado-da-arte para recomenda\c{c}\~{a}o baseada em sess\~{a}o. 
\end{abstract}

\begin{englishabstract}
\noindent
Recommender Systems (RS) have been increasingly popular in assisting users with their choices, thus enhancing their engagement and overall satisfaction with online services. RS have became a popular research topic and, since 2016, Deep Learning methods and techniques have been increasingly explored in this area. News RS are aimed to personalize users experiences and help them discover relevant articles from a large and dynamic search space. Therefore, it is a challenging scenario for recommendations. Large publishers release hundreds of news daily, implying that they must deal with fast-growing numbers of items that get quickly outdated and irrelevant to most readers. News readers exhibit more unstable consumption behavior than users in other domains such as entertainment. External events, like breaking news, affect readers interests. In addition, the news domain experiences extreme levels of sparsity, as most users are anonymous, with no past behavior tracked. The main contribution of this research was named \emph{CHAMELEON}, a Deep Learning meta-architecture designed to tackle the specific challenges of news recommendation. It consists of a modular reference architecture which can be instantiated using different neural building blocks. As information about users' past interactions is scarce in the news domain, the user context can be leveraged to deal with the \emph{user cold-start problem}. Articles' content is also important to tackle the \emph{item cold-start problem}. Additionally, the temporal decay of items (articles) relevance is very accelerated in the news domain. Furthermore, external breaking events may temporally attract global readership attention, a phenomenon generally known as \emph{concept drift} in machine learning. All those characteristics are explicitly modeled on this research by a contextual hybrid session-based recommendation approach using Recurrent Neural Networks. The task addressed by this research is session-based news recommendation, i.e., next-click prediction using only information available in the current user session. A method is proposed for a realistic temporal offline evaluation of such task, replaying the stream of user clicks and fresh articles being continuously published in a news portal. Experiments performed with two large datasets have shown the effectiveness of the \emph{CHAMELEON} for news recommendation on many quality factors such as accuracy, item coverage, novelty, and reduced item cold-start problem, when compared to other traditional and state-of-the-art session-based recommendation algorithms. 

\end{englishabstract}

\listoffigures 

\listoftables 

\listofabbreviations
\begin{longtable}{ll}
ACE & Article Content Embedding \\
ACM & Association for Computing Machinery \\
ACR & Article Content Representation module (CHAMELEON)\\
BERT & Bidirectional Encoder Representations from Transformers\\
BFR@n & Batches before First Recommendation metric\\
BP & Back-propagation\\
BPR & Bayesian Personalized Ranking \\
BPR-MF & Bayesian Personalized Ranking Matrix Factorization \\
CAR & Contextual Article Representation sub-module (CHAMELEON)\\
CARS & Context-Aware Recommender Systems\\
CB & Content-Based baseline algorithm\\
CBF & Content-Based Filtering\\
CHAMELEON & Contextual Hybrid session-bAsed MEta-architecture\\             & applying deep LEarning On News recommender systems\\
CET & Content Embeddings Training sub-module (CHAMELEON)\\
CBOW & Continuous Bag Of Words\\
CF & Collaborative Filtering\\
CNN & Convolutional Neural Network\\
CO & Co-Occurence baseline algorithm\\
COV@n & Item Coverage metric\\
CTR & Click-Through Rate\\
DAE & Deep Auto Encoder\\
DBN & Deep Belief Network\\
DBM & Deep Boltzmann Machine\\
DeepJoNN & Deep Joint Network for session-based News recommendations\\
DFFNN & Deep Feed-Forward Neural Network\\
DKN & Deep Knowledge-aware network for News recommendation\\
DL & Deep Learning\\
DNN & Deep Neural Network\\
DRL & Deep Reinforcement Learning\\
DSSM & Deep Structured Semantic Model\\
EILD-R & Expected Intra-List Diversity with Rank-sensitivity metric\\
EILD-RR & Expected Intra-List Diversity with Rank- and Relevance-sensitivity metric\\
ESI-R@n & Expected Self-Information with Rank-sensitivity metric\\
ESI-RR@n & Expected Self-Information with Rank- and Relevance-sensitivit metric\\
HR@n & Hit Rate metric\\
HRNN & Hierarchical RNN\\
IC & Input Configuration\\
FF & Feed-Forward\\
FM & Factorization Machines\\
GNN & Graph Neural Networks\\
GRU & Gated Recurrent Unit\\
GRU4Rec & Gated Recurrent Unit network for Recommendation\\
GPU & Graphics Processing Units\\
IR & Information Retrieval\\
KNN & K-Nearest Neighbors\\
LDA & Latent Dirichlet Allocation\\
LSA & Latent Semantic Analysis\\
MF & Matrix Factorization\\
LSTM & Long Short Term Memory\\
MAP & Mean Average Precision\\
MLP & Multi-Layer Perceptron\\
MRR@n & Mean Reciprocal Ranking metric\\
MSE & Mean Squared Error\\
MSI & Mean Self-Information\\
MTL & Multi-Task Learning\\
MV-DNN & Multi-View Deep Neural Network\\
NARM & Neural Attentive Recommendation Machine\\
NDCG & Normalized Discounted Cumulative Gain\\ NAR & Next-Article Recommendation module (CHAMELEON)\\
NAS & Neural Architecture Search \\
NER & Named Entity Recognition\\
NLP & Natural Language Processing\\
PCA & Principal Component Analysis\\
PMF & Probabilistic Matrix Factorization\\
QRNN & Quasi-Recurrent Neural Networks\\
RA-DSSM & Recurrent Attention DSSM\\
RBM & Restricted Boltzmann Machine\\
ReLU & Rectified Linear Unit\\
RL & Reinforcement Learning\\
RMSE & Root Mean Squared Error\\
RNN & Recurrent Neural Network\\
RQ & Research Question\\
RRQ & Research Requirement\\
RR & Recommendations Ranking sub-module (CHAMELEON)\\
RS & Recommender Systems\\
SDA-GRU & Sequence Denoising GRU Autoencoder\\
SGD & Stochastic Gradient Descent\\
SER & SEssion Representation sub-module (CHAMELEON)\\
SR & Sequential Rules baseline algorithm\\
SR-GNN & Session-based Recommendation using Graph Neural Networks\\
STAMP & Short-Term Attention/Memory Priority Model\\
TDSSM & Temporal DSSM\\
TF-IDF & Term Frequency-Inverse Document Frequency\\
TFR & Textual Features Representation sub-module (CHAMELEON)\\
t-SNE & t-Distributed Stochastic Neighbor Embedding\\
UGRNN & Update Gate RNN\\
URL & Uniform Resource Locator (a.k.a web address)\\
V-SkNN & Vector Multiplication Session-Based kNN\\
W2V*TF-IDF & \emph{TF-IDF} weighted \emph{word2vec}\\
\end{longtable}

\listofsymbols
\begin{longtable}{ll}
$ \mathbb{A} $ & A set\\
$ \mathbb{R} $ & The set of real numbers\\
$ a \in \mathbb{R}^d $ & A vector $a$ with dimension $d$ \\
$ a_i $ & Element $ i $ of vector $ a $, with indexing starting at 1\\
$ A_{i,j} $ & Element $ i,j $ of matrix $ A $\\
$ A_{i,:} $ & Row $ i $ of matrix $ A $\\
$ A_{:,i} $ & Column $ i $ of matrix $ A $\\
$ A^\intercal $ & Transpose of matrix $ A $\\
$ a \oplus b $ & Concatenation of vectors $ a $ and $ b $\\
$ a \cdot b $ & Dot product between vectors $ a $ and $ b $\\
$ A \odot B $ & Element-wise (Hadamard) product of $ A $ and $ B$\\
$ f(x;\theta) $ & A function of $x$ parametrized by $ \theta $.\\
$ \lVert a \rVert $ & Euclidean ($ L^2 $) norm of a vector $a$\\
$ P(a) $ & Probability distribution over a discrete variable\\
$ \text{log}(x) $ & Natural logarithm of $ x $\\
$ x^{(i)} $ & The $i$-th example (input) from a dataset\\
$ y^{(i)} $ & The target or label associated with $ x^{(i)} $, for supervised learning\\

\end{longtable}


\tableofcontents

\mainmatter

\chapter{Introduction}
\label{sec:chapter_1}

Recommender Systems (RS) have been increasingly popular in assisting users with their choices and helping them to filter through large information and product spaces, enhancing their engagement and satisfaction with online services \cite{jawaheer2014modeling}.

Typically, RS focus on two tasks: (1) \emph{predict task}, i.e., given a user and an item, what is the user likely preference for the item; and (2) \emph{recommendation task}, i.e., given a user, produce the best ranked list of n-items for user needs \cite{isinkaye2015recommendation}. 

RS became a topic of interest among machine learning, information retrieval, and human-computer interaction researchers \cite{Ekstrand2011}. There has been a vast amount of research in this field, mostly focusing on designing new algorithms for recommendations \cite{Shani2010}.

Research on recommender algorithms garnered significant attention since 2006, when Netflix launched the \emph{Netflix Prize} to improve the state of movie recommendation. The objective of this competition was to build a recommender algorithm that could beat their internal \emph{CineMatch} algorithm in offline tests by 10\% of accuracy increase \cite{Bennett2007, Gomez-Uribe:2015:NRS:2869770.2843948}. That prize motivated high activity, both in academia and amongst hobbyists. The \$1 M prize demonstrates the value that vendors place on accurate recommendations.

Perhaps the most widely-known application of recommender system technologies is Amazon.com, which has evolved its system during the last two decades \cite{smith2017two}. They have increased sales volume by providing product recommendations based on purchase and browsing history, in addition to contextual information (e.g., the item a user is currently viewing).

Supporting discovery in information spaces of such magnitude is a significant challenge, as the sizes of these decision domains are often very large: Netflix had over 17,000 movies in its selection \cite{Bennett2007}, and Amazon.com had over 410,000 titles in its Kindle store alone \cite{Amazon2010}. 

Recommender systems have been researched and applied in online services from different domains, like e-commerce (e.g., Amazon \cite{DBLP:conf/icis/LeeH14}), music \cite{Bu2010,van2013deep,wang2014} (e.g., Spotify, Pandora, Last.fm), videos (e.g. YouTube \cite{davidson2010youtube}, Netflix \cite{Gomez-Uribe:2015:NRS:2869770.2843948}), people \cite{Badenes2014} (e.g., Facebook), jobs \cite{Bastian2014} (e.g., LinkedIn \cite{kenthapadi2017personalized}, Xing \cite{mishra2016bottom}), and research papers \cite{wang2011collaborative, Beel2013b} (e.g., Docear \cite{beel2013introducing}), among others.

Since 2016, it has steeply increased the RS research exploring Deep Learning \cite{hinton2006,hinton2006b,bengio2007,bengio2009} architectures and methods \cite{hidasi2017dlrs}, which have already been successful in other complex domains, such as Computer Vision, Natural Language Processing (NLP), machine translation, and speech recognition.

\section{Contextualization}

The consumption of online news has increased rapidly, in contrast with the decline of traditional newspapers. By 2012, the percentage of users visiting news portals already represented the major portion of overall Web traffic \cite{trevisiol2014cold}.

News recommender systems are aimed to personalize users experiences and help them discover relevant articles from a large and dynamic search space. Popular news portals such as Google News \cite{das2007}, Yahoo! News \cite{trevisiol2014cold}, The New York Times \cite{spangher2015}, Washington Post \cite{graff2015, bilton2016}, among others, have gained increasing attentions from a massive amount of online news readers. 

As an example, the Washington Post reported that when they started sending out personalized newsletters, they experienced that click-through rates for the personalized newsletters were three times the average and the overall open rate consisted of the double of the average for the Post's newsletters \cite{bilton2016, gulla2017adressa}.

Online news recommendations have been addressed by researchers in the last years, using different families of recommendation methods: Content-Based Filtering \cite{li2011scene, capelle2012semantics, ren2013concert, ilievski2013personalized, mohallick2017exploring}, Collaborative Filtering \cite{das2007} \cite{diez2016}, and Hybrid approaches \cite{chu2009personalized, liu2010personalized}  \cite{li2011scene, rao2013personalized, lin2014personalized, li2014modeling, trevisiol2014cold, epure2017recommending}. 

For some domains like news and e-commerce, most users are not logged-in and their short-term reading interests must be estimated from a few interactions within its session. In this scenario, a session can be seen as a sequence of user interactions that takes place within a given time frame \cite{quadrana2017personalizing}.

Session-based recommendation is the task of recommending relevant items given an ongoing user session \cite{QuadranaetalCSUR2018}.  This type of task used to be underappreciated in recommender systems, due to the usual  sparsity of training data \cite{zhang2019deep}. Although, in recent years, it has been observed and increased interest in session-based recommendation \cite{QuadranaetalCSUR2018}. 

Recurrent Neural Networks (RNN) possess several properties that make them attractive for sequence modeling. In particular, they are capable of incorporating input from past consumption events, allowing to derive a wide range of sequence-to-sequence mappings \cite{donkers2017sequential}.

The seminal work in the usage of RNNs for session-based recommendation was the \emph{GRU4Rec} architecture \cite{hidasi2016}, based in a type of RNN known as Gated Recurrent Unit (GRU) \cite{cho2014properties, chung2014empirical}. Since \emph{GRU4Rec}, a research line emerged with subsequent works on the usage of RNNs for session-based recommendation \cite{hidasi2016parallel,hidasi2018recurrent,wu2016recurrent,liu2016context,smirnova2017contextual}. RNNs have also been explored for session-aware recommendation \cite{donkers2017sequential,quadrana2017personalizing,ruocco2017inter, skrede2017inter}, a task where the user can be identified and his previous behaviour is available, possibly helping to model his preferences \cite{quadrana2017personalizing}.

However, as shown in \cite{jannach2017recurrent,ludewig2018evaluation,LudewigMauro2019}, approaches using neural networks that rely only on information about logged interactions (non-hybrid), have certain limitations and they can, depending on the experimental setting, be outperformed by much simpler approaches for session-based recommendations, e.g.,  nearest-neighbor techniques.

\section{Motivation}

News portals are a challenging scenario for recommendations \cite{Zheng:2018:DDR:3178876.3185994} for a number of reasons. Large publishers release thousands of news daily, implying that they must deal with fast-growing numbers of items that get quickly outdated and irrelevant to most readers. This scenario results in an extreme level of the \emph{item cold-start problem}, where it gets hard for the RS to learn quickly who to recommend fresh articles before they get irrelevant, in a couple of hours.

Additionally, the news domain suffers from an extreme \emph{user cold-start problem}, as most users are anonymous, and it gets hard to learn their preferences with almost no past behavior tracked \cite{diez2016}. In such scarcity of information about the user, understanding his context might be helpful to provide meaningful recommendations.

Furthermore, news readers may exhibit more unstable consumption behavior than users in other domains such as entertainment, as their interests usually change over time. External events reported in breaking news may also temporally attract global attention \cite{epure2017recommending}, a phenomenon generally known as \emph{concept drift} in machine learning \cite{vzliobaite2016overview}.

This research was motivated by the potential, envisioned by this author, of leveraging Deep Learning architectures and methods to tackle the specific challenges of news recommendation. 

In special, to deal with the \emph{user cold-start problem}, it could be leveraged contextual information from the user session. The sequence of items previously clicked in a session could be processed by a Recurrent Neural Networks (RNN). Additional contextual information, such as user location, device and session could be easily incorporated by neural networks.

To deal with the \emph{item cold-start problem}, a typical solution is a hybrid recommendation approach, that leverages not only users behaviour, but also the items content. Deep Learning techniques have been effective in extracting relevant features from unstructured data, such as text \cite{bansal2016ask}, music \cite{van2013deep} \cite{wang2014}, and images \cite{mcauley2015image, he2016deep}.

Finally, to deal with the \emph{concept drift}, it could be devised a protocol in which the neural networks are trained incrementally over the continuous stream of users interactions and published articles. Additionally, neural networks could be used to learn a non-linear function to model the temporal decay of news relevance.

During this research, it was identified a research gap, as only six works proposing deep neural architectures for news recommendation had been published\footnote{To the best of the knowledge from this research so far, the only works presenting a deep learning architecture for news recommendation were \cite{song2016multi}, \cite{kumar2017, kumar2017word}, \cite{park2017deep}, \cite{okura2017embedding}, \cite{zhang2018deep}, and \cite{wang2018dkn}.}, and all of those architectures addressed only one or two of the aforementioned challenges. Furthermore, those works evaluated only one aspect of recommendation quality -- accuracy -- and used unrealistic protocols for offline evaluation, which will hardly reproduce the online performance of those methods \footnote{A comparison between the \emph{CHAMELEON} meta-architecture and those other neural architectures for news recommendation is presented in Section~\ref{sec:comparison-works}.}.

\section{Objective}
\label{sec:obj}


The main objective of this research is to propose a Deep Learning meta-architecture -- the \emph{CHAMELEON} --  designed to tackle the specific challenges of news recommendation and improve the recommendation quality of news portals.

For the purpose of this research, a meta-architecture is a reference architecture that collects together decisions relating to an architecture strategy \cite{Malan2004MetaArchitecture}. A meta-architecture might be instantiated as different architectures with similar characteristics that fulfill a common task, in this case, news recommendations. 

The \emph{CHAMELEON} meta-architecture is structured to support changes, operating at the level of inputs, outputs, modules, sub-modules, and their interactions. Modules and sub-modules can be instantiated by different architectures, as they evolve. Such modular structure also make their components evaluation straightforward. 

The \emph{CHAMELEON} acronym stands for \emph{Contextual Hybrid session-bAsed MEta-architecture applying deep LEarning On News recommender systems}. 

\section{Research Questions}
\label{sec:rq}

This thesis addresses the following Research Questions (RQ):

\begin{itemize}
\item \emph{RQ1} - How does a contextual and hybrid RS based on the proposed neural meta-architecture perform in the news domain, in terms of recommendation quality factors (accuracy, item coverage, novelty, and diversity), compared to other traditional and state-of-the-art approaches for session-based recommendation?
\item \emph{RQ2} - What is the effect on news recommendation quality factors of leveraging different types of information in a neural-based contextual hybrid RS?
\item \emph{RQ3} - What is the effect on news recommendation quality of using different textual representations, produced by statistical NLP and Deep NLP techniques?
\item \emph{RQ4} - Is a hybrid RS based in the proposed meta-architecture able to reduce the problem of \emph{item cold-start} in the news domain, compared to other existing approaches for session-based recommendation?
\item \emph{RQ5} - Is it possible for a neural-based RS to effectively balance the trade-off between the recommendation quality factors of \emph{accuracy} and \emph{novelty}?
\end{itemize}

To the best of our knowledge, those  research questions are explored for the first time in this research, as the other six works proposing neural-based approaches for news recommendation\footnote{The existing works on neural-based news recommender systems, to the best of our knowledge, are \cite{song2016multi}, \cite{kumar2017, kumar2017word}, \cite{park2017deep}, \cite{okura2017embedding}, \cite{zhang2018deep}, and \cite{wang2018dkn}.} are only optimized and evaluated for recommendation accuracy, ignoring other recommendation quality attributes very important for news recommendation, such as novelty, diversity and robustness against the \emph{item cold-start problem}. Their evaluation protocols are not realistic, for not considering the temporal dynamics of news publishing and readership. Furthermore, those works do not investigate the individual importance of different types of information or different representations of the articles' textual content.

\section{Research Scope}


The recommendation task addressed in this work is the next-item prediction for user sessions \cite{smirnova2017contextual}, i.e., "what is the next most likely article a user might read in a session?". Supervised machine learning is employed for this task, as models are trained using past user interactions on news portals to predict future article reads (labels).

As information about users' past interactions is scarce in the extreme cold-start scenario of news recommendation and users have a constant drift on their interests. This work focuses in session-based recommendation, the most typical scenario in news portals, mostly accessed by anonymous users. The usage of session-aware recommendation algorithms in the news domain is out of the scope of this research for the aforementioned reason.

Deep neural architectures, like RNNs and Convolutional Neural Networks (CNNs), are investigated, assessed, compared, and combined to provide guidelines for instantiation, implementation, and evaluation of the proposed Deep Learning Meta-Architecture for News Recommendations: \emph{CHAMELEON}. 

Users' context and interactions and also items' static and dynamic information are combined into a contextual and hybrid recommendation approach based on neural networks. RNNs are used to model the sequence of clicks within users sessions, as additional contextual information.

The content of news articles is leveraged in a hybrid recommendation approach to counter the \emph{item cold-start problem}. The textual representation is learned using feature extractors based on CNNs and RNNs.

The dynamics of news readership, like the temporal decay of articles relevance, is learned by the neural networks. They are trained incrementally on the clicks stream, adapting to changes in global interests and temporal seasonality.

A number of experiments address the Research Questions, stated in Section~\ref{sec:rq}, to evaluate the quality of recommendations provided by the instantiations of the proposed meta-architecture and compared to traditional and state-of-the-art algorithms for session-based recommendation. The investigated recommendation quality factors are: (a) accuracy, (b) item coverage, (c) novelty, (d) diversity, and (e) robustness against the item cold-start problem.

The experiments are performed on two large news portals datasets and follow a proposed temporal offline evaluation protocol to provide a more realistic evaluation, which emulates a continuous stream of user clicks, while fresh articles are published.

Regarding to other recommendation techniques, like collaborative filtering based on nearest neighbours or matrix factorization, although relevant, are out of the scope of this research work for, historically, not being successful for recommendations on the news domain.


\section{Presentation Order}
In Chapter~2, it is presented the background concepts on Recommender Systems and a survey on News Recommender Systems.
Chapter~3 presents a survey of related works on Deep Learning architectures applied for Recommender Systems. 
In Chapter~4, the proposed Deep Learning Meta-Architecture for News Recommender Systems is presented. 
In Chapter~5, it is proposed some instantiations of \emph{CHAMELEON} and the experimental design is presented.
In Chapter~6, it is developed the analysis of the main experimental results and the discussion of Research Questions. Finally, in Chapter~7, the main conclusions, recommendations, and suggestions for future work are presented.

\chapter{News Recommender Systems}
\label{sec:chapter_2}

This chapter describes, in the first section, the necessary background of Recommender Systems methods and evaluation approaches. The second section covers related works on News Recommender Systems.

\section{Recommender Systems Background}
\label{sec:recsys_background}

Historically, people have relied on recommendations and mentions from their peers or the advice of experts to support decisions and discover new material. These methods of recommending new things have their limits, particularly for information discovery. There may be an independent film or book that a person would enjoy, but no one in their circle of acquaintances has heard of it yet. Computer-based RS provide the opportunity to expand the set of people from whom users can obtain recommendations. They also enable the mining of users history for patterns, potentially providing a more finely-tuned selection experience \cite{Ekstrand2011}.

Historical user activity is key for building user profiles capable to predict user behavior and affinities in many web applications such as targeting of online advertising, content personalization, social recommendations, and web search \cite{ahmed2011}. 

\begin{figure}[h]
	\centering
	\includegraphics[width=12cm]{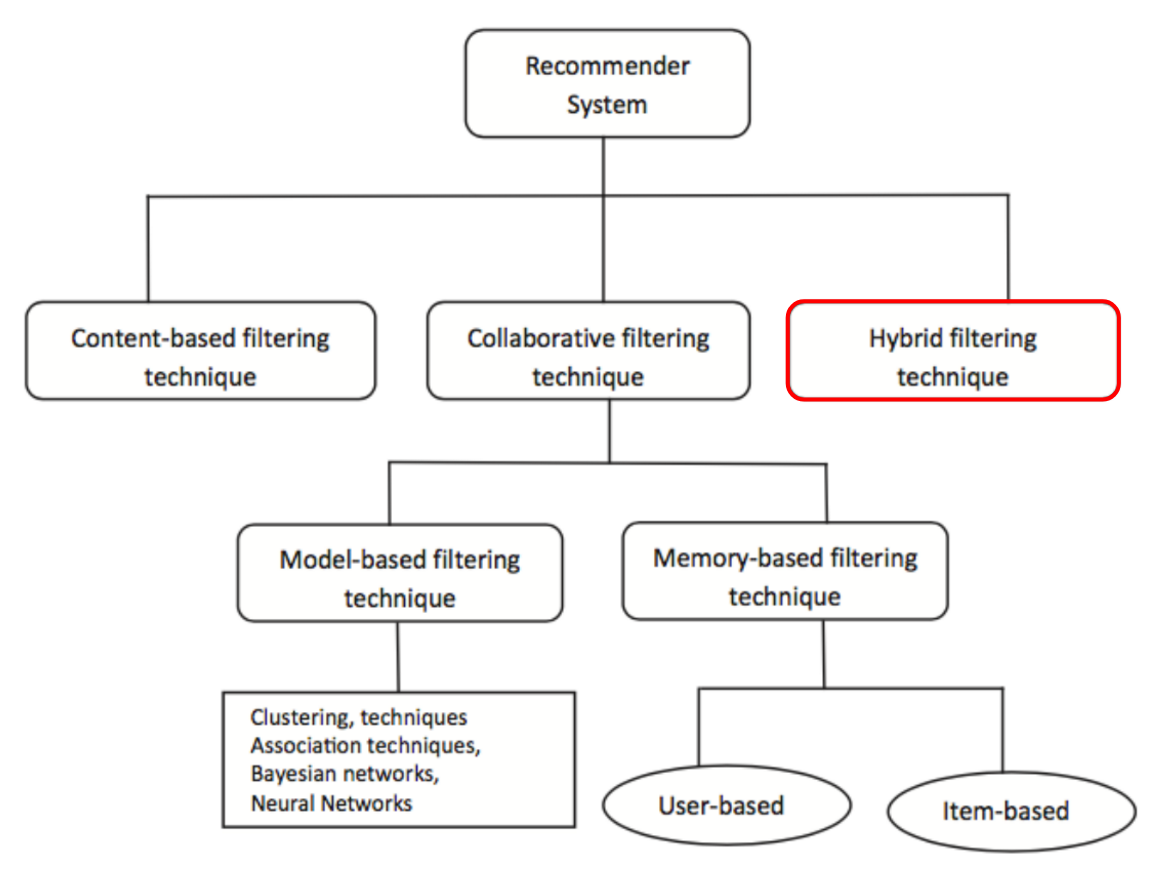}
	\caption{Recommender System methods taxonomy, adapted from \cite{isinkaye2015recommendation}}
	\label{figure:rs_taxonomy}
\end{figure}

The main families of recommendation methods are Collaborative Filtering (CF), Content-Based Filtering (CBF), and Hybrid Filtering, as shown in Figure \ref{figure:rs_taxonomy}, which are briefly described in the next sub-sections. 

This research is specially focused in Hybrid Filtering, which combines item attributes (CBF) and past user behaviours (CF) to provide recommendations, specifically in cold-start scenarios like in news portals.

\subsection{Collaborative Filtering}

Collaborative Filtering (CF) is a popular and well-known family of methods for RS. The main idea behind CF methods is that users that agree about the relevance of many items will probably have similar preferences on other items \cite{Ekstrand2011}. 

Users may explicitly input their preferences through system interfaces, like \emph{movie star ratings} on Netflix or \emph{likes} on Facebook. However, explicit feedback is not always available. Thus, RS may infer user preferences from  more abundant implicit feedback, which indirectly reflects users preferences based on user behavior. Types of implicit feedback include clicks, page views, browsing history, purchase history, search patterns, or even mouse movements. For example, a user that purchased many books by the same author probably likes that author \cite{hu2008}.

Implicit and explicit feedback provide key information for modeling users' preferences on items and for personalizing recommendations \cite{jawaheer2014modeling}. 


The main methods for CF are memory-based (e.g., Nearest Neighbors) and model-based (e.g., Clustering, Association Rules, Matrix Factorization, Neural Networks) \cite{jannach2010}. Matrix Factorization (MF) is a collection of Linear algebra methods that factorizes a matrix into a product of matrices. Traditional CF techniques are based on Matrix Factorization of the User-Item matrix \cite{koren2009}.

The main challenges of CF methods are related to the intensively researched cold-start problem. Generally, lots of items available in the system have just a few ratings per user leading to a sparse user-item matrix. In such scenario, it is difficult to match similar users by usage of their ratings in common items. The item cold-start problem occurs when an item is not rated by a substantial number of users to be eligible for recommendation, while the user cold-start problem happens when the system has not enough users ratings to learn their preferences and make accurate recommendations \cite{Felicio2017}.

In the next sub-sections the main approaches for CF are briefly described: (a) memory-based and (b) model-based CF \cite{zhang2014collaborative}.

\subsubsection{Memory-based CF}
Memory-based CF was one of the first computer-based recommender methods \cite{breese1998empirical}. It tries to find users that are similar to the active user and uses their preferences to predict ratings for the active user \cite{breese1998empirical}. Ratings or interactions are used to calculate the similarity and weight between users or items.

There are several advantages of memory-based CF. First of all, since we only need to calculate similarity, it is easy to implement. Second, memory-based CF systems are scalable to large size data. Third, most of memory-based systems are online learning models. Thus, new arrival of data can be easily handled. At last, the recommendation results can be understood and can provide feedback to explain why to recommend these items \cite{zheng2016}. 

However, several limitations are also existing in memory-based CF techniques. It is very slow as it uses the entire database every time it makes a prediction. The recommendation results are unreliable and not accurate, when data is sparse and common rated items are very few \cite{thorat2015survey, zheng2016}.

Neighbor-based CF \cite{herlocker1999algorithmic} -- a type of memory-based recommender algorithm -- was the seminal work of CF methods and involves two steps: similarity calculation and prediction. In the similarity calculation step, similarity metrics (e.g., Pearson or Cosine similarity) can be measured between users (User-kNN) or items (Item-kNN). To generate recommendations for a given user, a set of similar users (neighbors) is selected and the most preferred items from those users are recommended.

\subsubsection{Model-based CF}

Model-based CF are based on machine learning or data mining models and may leverage complex rating patterns in training data. Model-based CF algorithms are developed to counter the shortcomings of memory-based CF models \cite{zheng2016}.

There are some advantage of model-based CF algorithms, as they can improve prediction performance and are usually scalable for being a compressed version of the users behaviour patterns. Therefore, those algorithms also suffer from the sparsity problem so that it is unable to generate reasonable recommendations for those users who provide few ratings \cite{betru2017}.

Models like Naive Bayes  \cite{miyahara2000collaborative} and Bayesian Networks \cite{shen2003discriminative} were explored for recommendations in the early days. Clustering algorithms were also used as CF models, when user ratings were categorical.

Matrix Factorization (MF) \cite{koren2009} is the most successful collaborative filtering approach. It finds common factors in a joint latent space of user and item features, that may be the underlying reasons behind users interactions. For example, in a movie recommender system, these factors can be genre, actors, director, the special effects, or even the city were the movie was shot. In the end, MF finds a feature set for each user and item, as the result of the factorization process. After trained, the user preference for an item can be estimated by the scalar product between the factor vectors of those user and item \cite{zheng2016}.

MF proceeds by decomposing the original rating matrix $ R  \in \mathbb{R}^{m \times n} $ consisting of ratings by $ m $ users for $ n $ items into two low-rank matrices $ U \in \mathbb{R}^{m \times d} $ and $ V \in \mathbb{R}^{n \times d} $  consisting of the user and item features respectively of rank $ d $, like in Equation \ref{eq:mf_pred}.

\begin{equation} \label{eq:mf_pred}
R  \approx U \times V
\end{equation}

The system learns the latent factors by minimizing the objective function shown as follows

\begin{equation} \label{eq:mf_loss}
	\argmin_{U,V} l(R,U,V) + \beta(\lVert U \rVert + \lVert V \rVert),
\end{equation}

where $ l(R,U,V) $ is the loss function of predicting rating using the latent factors $ U $ and $ V $ and the last term regularizes factors weights, controlled by $\beta$ importance, to avoid overfitting.

The \emph{Weighted Matrix Factorization (WMF)} version \cite{hu2008collaborative} has also being extensively used. Other classical MF method is the \emph{Probabilistic Matrix Factorization (PMF)} \cite{mnih2008probabilistic}. The underlying assumption behind this method is that the prior probability distribution of latent factors and the probability of observed ratings given the latent factors follows a Gaussian distribution. 

Many algorithms have been developed to enhance the performance of \emph{Probabilistic Matrix Factorization (PMF)}, by designing of the Bayesian versions, like \emph{Bayesian Personalized Ranking Matrix Factorization (BPR-MF)} \cite{rendle2009bpr}, \emph{Bayesian PMF} \cite{salakhutdinov2008bayesian}, \emph{Sparse Covariance
Matrix Factorization (SCMF)} \cite{shi2013scmf}, and \cite{xu2013fast} or incorporating side information, such as social relationships \cite{adams2010incorporating, ma2011recommender, zhao2014leveraging}.

Although efficient, matrix factorization methods suffer from the cold-start problem. Another problem often presented in many real world applications is data sparsity or reduced coverage \cite{li2015deep}. Incorporating side information has shown promising performance in collaborative filtering in such scenarios, like in \cite{porteous2010bayesian} \cite{kim2014scalable} \cite{park2013hierarchical}. But Matrix Factorization are still shallow models unable to capture complex rating of existing patterns in user interactions.

More recently, neural networks have being used for model-based CF, like in \cite{salakhutdinov2007} \cite{wang2015}, and \cite{wu2016}. As neural networks for CF are key in this research, they are explored in more detail on Chapter 3.

\subsection{Content-Based Filtering}

Collaborative Filtering models suffer from the sparsity problem. For those users with no previous interactions, model-based CF is unable to generate reasonable recommendations \cite{schein2002}. 

Content-Based Filtering (CBF) is another family of methods which matches users and items by means of their attributes, ignoring other users behaviors. Content information is proved to be effective to reduce the \emph{item cold-start problem} \cite{schein2002}. This problem takes place when an item have being involved in very few interactions, so that there is not enough behavioral information to recommend an item \cite{Burke2007}.

Content-Based Filtering (CBF) methods filters items based on the similarity of the contents the user is interested in. These approaches utilize different resources, such as item information or user profiles, to learn latent factors of users or items. In this manner, even for a user which provided little interactions, his preference can still somehow be inferred. For example, a brand new user in restaurants' recommendation system might get relevant recommendations only considering the city he lives and setting in his profile (e.g., vegetarian).

Another advantage of Content-Based Filtering (CBF) methods is that they provide transparency on how the recommender system works and recommendations are easily explainable \cite{li2014modeling}. Thus, the user might understand recommendations provided. 

A series of user studies conducted by Pu and Chen \cite{pu2006, pu2007} have indicated that explanation interfaces could effectively help build users' trust in the system. Other researchers also state that explanation interfaces can cultivate user trust \cite{sinha2002}, promote loyalty, increase satisfaction, and make it easier for users to find what they want \cite{muhammad2015}.


However, one important limitation of content based recommender system is that it recommends the same types of items -- which is known as the \emph{bubble effect} -- and because of that it suffers from an overspecialization problem.

A CBF popular method for text-based recommendations comes from Information Retrieval. \emph{Term Frequency-Inverse Document Frequency (TF-IDF)} \cite{ramos2003using} is a statistical method able to identify the most relevant words of a document, after being trained in a related corpus of documents. It creates a sparse vector representation of the documents, which can be used to compute the similarity between articles (e.g., cosine similarity). When applied for recommendation, user profiles vectors are aggregated by the weighted average of item's \emph{TF-IDF} vectors and the most similar non-interacted items are recommended to users.


Topic Modeling, a common technique in text mining, have also being explored for CBF, matching items with the user topics of interest, automatically inferred based on his interactions on other items. Topic models provide an interpretable low-dimensional representation of documents \cite{chang2009reading}. Topic models are unsupervised, which means that they do not require human labeling and are able to cluster words that co-occur in different documents. 

In one of our previous work \cite{moreira2016recommender}, \emph{TF-IDF} and Topic Modeling are used based on \emph{Non-negative Matrix Factorization (NMF)} for content and people recommendations in an enterprise collaboration platform named Smart Canvas\texttrademark. The discovered topics were also used to provide to the users explanation for the recommendations.

Topic modeling techniques have evolved from probabilistic \emph{Latent Semantic Analysis (LSA)} to a \emph{Bayesian version -- Latent Dirichlet Allocation (LDA)}\cite{blei2003latent}. 
The \emph{LDA} \cite{blei2003latent} is a graphical model which reveals how the words of each document are assumed to come from a mixture of topics. The topic proportions are document-specific, but the set of topics is shared by the corpus. For example, \emph{LDA} can capture that one article might be about biology and statistics, while another might be about biology and physics \cite{wang2011collaborative}.

The \emph{LDA} has been used to describe users as a mixture of topics and to assume that each of their actions is motivated by choosing a topic of interest first and subsequently words that might happen in a relevant document for the user. \emph{LDA} has being used model users interests along time \cite{ahmed2011} or to model user context  \cite{yu2012towards}. In tag-based recommendation systems, \emph{LDA} is also used to find the latent relation between keywords of item description and item tags created by users, such that the items can be recommended based on the tags \cite{krestel2009latent} \cite{song2011automatic}. 

In \cite{Tamir2016}, authors recommend textual content employing three basic steps: (1) Content understanding, (2) Profile aggregation, and (3) Matching. In the Content understanding step, \emph{LDA} topic modeling technique is used for latent topics discovery on content corpus. In the Profile aggregation step, user's interests on topics are inferred based on their interactions with contents. And in the Matching step, a relevant content is recommended for each user topics of interest.
  
\subsection{Hybrid Recommender Systems}

Hybrid recommender systems \cite{Burke2007} emerged as various recommender strategies have matured, combining two or more algorithms into composite systems, that ideally build on the strengths of their component algorithms to achieve some synergy. One example could be, due to the inability of collaborative filtering to recommend new items, combining it with content-based filtering to minimize the cold-start problem \cite{Burke2007}. 

A survey conducted in \cite{Burke2002} identified seven different strategies for hybrid RS: Weighted, Switching, Mixed, Feature Combination, Feature Augmentation, Cascade, and Meta-level. In \cite{Burke2007}, six of those hybrid types were explored, combining four basic recommendation methods: content-based, collaborative-filtering, heuristic collaborative, and knowledge-based, generating 41 possible hybrid combinations. The author highlighted the importance of examining the design goals for a hybridized system (overall accuracy, cold-start performance, etc.), evaluating the relative performance of each component of the hybrid system under those conditions.

One of the pioneer works that have explored the hybrid combination of textual information reviews with users rating was \cite{jakob2009beyond}. It found that movie reviews are usually related to different aspects, such as price, service, positive or negative feelings, and those aspects that can be exploited for rating prediction.

Some popular hybrid approaches are combining MF on user-item interactions matrix and probabilistic topic models (e.g., \emph{LDA}) on side textual item information, like in \cite{wang2011collaborative} for scientific articles and also in \cite{nikolenko2015svd} for sponsored content.

\subsection{Recommender Systems Evaluation}
With a large variety of algorithms available, selecting the most appropriate algorithm for a system is non-trivial \cite{Beel2013a}. Typically, such decisions are driven by an evaluation process \cite{Herlocker2004}, based on experiments comparing the performance of a number of candidate recommenders. Such evaluations are typically performed by applying some evaluation metrics that can assess different factors of recommendation quality \cite{Shani2010}.

There are three main RS evaluation methods: (1) user studies, (2) online evaluations, and (3) offline evaluations \cite{Ricci2011}.

In user studies or virtual lab studies \cite{Ekstrand2011}, a controlled experiment is conducted with volunteer users, which explicitly rate recommendations generated by different algorithms. The algorithm with the highest average rating is considered the best approach for a given context \cite{Ricci2011}. User studies typically ask their participants to quantify their overall satisfaction with the recommendations, providing a qualitative feedback. Therefore, this approach is rarely used for recommender system evaluations \cite{Beel2013b}, due to its costs and to possible biases in the experiment.

In online evaluations or user trials \cite{Ekstrand2011}, recommendations are shown to real users of the system during their session \cite{Ricci2011}. Users are not asked explicitly to rate recommendations but the recommender system observes how often a user accepts a recommendation. To compare two algorithms, recommendations using each algorithm are randomly assigned to different user sessions and the general acceptance of the algorithms recommendation is compared (A/B testing) \cite{Beel2013a}.

Offline evaluations use offline datasets of users interactions and possibly items metadata, from which some user interactions has been removed. Subsequently, the recommender algorithms are analyzed on their ability to infer users preferences and recommend meaningful items, which are assessed based on the hidden interactions \cite{Beel2013b}. This kind of evaluation can run multiple times at a low cost, and usually deliver the results in some hours. Typically, offline evaluations are meant to identify the most promising recommendation approaches \cite{Knijnenburg2012}, before involving users \cite{Ekstrand2011}.

\subsubsection{Evaluation Metrics}
\label{sec:recsys_eval_metrics}

Most results in RS are evaluated in offline setting by accuracy metrics, which measure the RS ability to recommend items that users had actually interacted \cite{Shani2010}. One popular metric in the past was \emph{Root Mean Squared Error (RMSE)}, used for example to evaluate ratings predictions (e.g., in a 1-5 scale) in the \emph{Netflix Prize} \cite{Bennett2007}. 

Recent research has focused in Top-N accuracy metrics, in which it is measured the relevance of the top-N recommendations from a ranked list. Many of those metrics were borrowed from Information Retrieval (IR), like accuracy metrics (e.g., Precision@N, Recall@N), and ranking metrics (e.g., \emph{Normalized Discounted Cumulative Gain (NDCG)} \cite{jarvelin2002cumulated}, \emph{Mean Average Precision (MAP)}, \emph{Mean Reciprocal Ranking (MRR)}).

Besides accuracy, coverage is another important evaluation factor. Users coverage measures the percentage of users for which a RS is able to provide recommendations, whilst item coverage measures the percentage of items that are ever recommended \cite{maksai2015predicting}. \emph{Gini Index} and \emph{Shannon's Entropy} have also been used as coverage metrics \cite{Shani2010}.

Diversity is an important evaluation factor in many recommendation scenarios, like entertainment and news. If the RS only suggest similar items to the user, this may lead to poor user experiences. The most common metrics for diversity are based on dissimilarity between recommended items \cite{zhang2002novelty, ziegler2005improving, rodriguez2012multiple, li2014modeling}.

Other relevant evaluation factors, specially in the news domain, are novelty and serendipity \cite{iaquinta2008introducing} \cite{murakami2007metrics}.

Novelty occurs when the system suggests to the user an unknown item, that he might have autonomously discovered. \cite{garcin2014offline} define novelty as the fraction of recommended articles that are not among the most popular
items. A variation of such a metric, called \emph{surprisal}, is a weighted sum of negative log frequencies of the items in the recommendation list \cite{zhou2010solving}.

Serendipity is the quality of being both unexpected and
useful \cite{maksai2015predicting}. A serendipitous recommendation helps the user to find a surprisingly interesting item that he could not have otherwise discovered (or it would have been really hard to discover) \cite{iaquinta2008introducing}. Serendipity metrics were proposed by \cite{murakami2007metrics} and  \cite{ge2010beyond} for music and TV show recommendation.

The most popular metric for online evaluation of news recommendation is the Click-Through Rate (CTR), consisting of the ratio between the number of clicked items and the number of recommended items. The CTR is a de-facto standard in the industry, because it is often correlated to the revenues generated by the news website, from either advertisements (ads) displayed on the website or paid articles (or sometimes both) \cite{garcin2014offline}. The CTR is usually very low (e.g., 0.5\%), due to the fact that most users only read a low fraction of all recommended stories in a news portal \cite{lommatzsch2014real}. \cite{garcin2014offline} suggests that the \emph{Success@k} (\emph{Precision@k}) metric would be better than the CTR to compare offline and online evaluations.

A similar metric is Online Accuracy. While CTR consider only clicks on recommended items, online accuracy considers a recommendation as accurate if the user browsed to the recommended page in the future, even if he did not clicked immediately in the provided recommendation \cite{maksai2015predicting}.

\subsubsection{The Limitation in RS Evaluation}

Offline evaluation assesses whether the RS can recommend items that users had actually clicked in the recorded history. Therefore, it is not possible to affirm that non-clicked items, which users may not be aware of, were not relevant to the user \cite{pu2011user}. Such limitation are addressed by live experiments in online evaluations, which are usually more costly for requiring a production recommender system in place, prepared to run A/B testing with a considerable number of users.

Logged users interactions are influenced by the page layout, by the order in which the items were presented (position bias), and also by the recommender algorithms that provided recommendations. In that setting, offline evaluation would be biased to give higher accuracy to algorithms that mimics recommenders that were in place when logs were recorded.

The best scenario for logged interactions database is when users are provided with only random recommendations, thus eliminating the bias from item positions and recommender algorithms. Although, in this setting, when no recommender systems are in place, items on the news portal front page usually attract the most clicks, creating bias towards popularity-based algorithms. In the live evaluation, the popularity-based strategy is clearly not the most interesting, because users do not want to read articles they have already seen on the front page \cite{garcin2014offline}.

Evaluation with historical datasets is reproducible and convenient to test many methods, but it has a variety of shortcomings and may not generalize to real-world performance \cite{kirshenbaum2012live}. A known challenge in recommender systems evaluation is that optimizing recommendation algorithms using offline evaluation metrics do not always correlate with online evaluation metrics and user satisfaction.

In one of our previous work \cite{moreira2015comparing}, a comparison was conducted between offline and online accuracy evaluation of different algorithms and settings in a real-world content recommender system - Smart Canvas\texttrademark. The experiments have shown that, in general, offline accuracy and online CTR did not correlate well, like also observed in \cite{Beel2013a} and \cite{garcin2014offline}. An interesting finding was that by filtering only recommendations of non-popular long-tail articles, which are usually more interesting for users, it was observed better alignment between offline and online accuracies.

In \cite{maksai2015predicting}, they investigated predicting the online performance of news recommendation algorithms by a regression model using offline metrics. Their results confirmed that there is more to online performance than just offline Accuracy. Other metrics, such as Coverage or Serendipity, play important roles in predicting or optimizing online metrics such as CTR. Their model can then be applied to trade-off curves for each algorithm constructed from offline data to select the optimal algorithm and parameters.


\section{News Recommender Systems}

The problem of filtering and recommending news stories has been investigated for more than 20 years now, such as in the early work of \cite{konstan1997}, and it continues to be researched in the last years \cite{diez2016, lommatzsch2017incorporating, mohallick2017exploring, epure2017recommending, ozgobek2014survey}. 


\cite{karimi2018news} provides a comprehensive survey on news recommendation methods, by analyzing 112 papers with new algorithms proposals, published from 2005 to 2016. The most popular approaches were Content-Based Filtering (59 papers) and Hybrid Approaches (45 papers). Collaborative Filtering-based methods for news recommendation were proposed only by 19 papers, although CF is the method of choice in most of other recommender systems domains.

With the large volume of events being reported everyday, an important issue of online news reading services is how to provide personalized news recommendation, helping readers to find interesting stories that maximally match their reading interests \cite{lin2014personalized}.

Differently from other domains (i.e., products or movies), news recommender systems have some peculiar characteristics. The set of items (usually news stories) universe is dynamic, where each item is expected to have a short shelf life. News stories undergoes a constant churn with new stories added every minute and old ones getting dropped \cite{das2007}. The user-item matrix is very sparse, as most online users are anonymous and actually read few stories from the entire repository  \cite{lin2014personalized}. A high volume of news articles overload the web within limited time span. This requires more computation resources for generating fresh personalized news recommendation \cite{mohallick2017exploring}.

Temporal modeling of users' preferences is important in many recommendation scenarios, specially for news RS \cite{li2014modeling, liu2010personalized}. \cite{billsus1999hybrid} was one of the first to find that there were two types of user interest in news reading: short-term and long-term. The short-term interest usually is related to hot news events or user context-specific interests thus changing quickly. In contrast, long term interest often reflects more stable user preferences \cite{liu2010personalized, song2016multi}.

The user context is also important for relevant recommendations to alleviate the cold-start and data sparsity issues in the existing systems. People might be interested in different topics depending on the day of the week, or even the hour \cite{mohallick2017exploring}. 

In \cite{said2013month}, they have analyzed and visualized the weekly and hourly impressions and also the CTR in the \emph{Plista} news recommendation system \cite{kille2013plista}. They found out that traditional news sources are mainly consumed during the first half of the day whereas topic-focused news receive the bulk of their interaction latter in the day. Readers of traditional news are more likely to interact with recommendations (higher CTR) than readers of topic-focused domains like motorcycles, technology, and business.

Popularity is an important readership factor, as breaking news or popular topics usually attract high attention of readers. The content popularity over time is a crucial ingredient in content management, since the commercial value of most content is varying or decaying temporally \cite{chu2009personalized}.

In \cite{lommatzsch2017incorporating} they observed that popular, medium popular, and unpopular items have different, but typical life cycles in terms of number of interactions. They proposed a simple weighted moving average approach to predict the item popularity in a near future (e.g., in the next 15 minutes), so that the trending item popularity information could be used in a recommendation model. Therefore, their proposed approach was not evaluated in the paper.

Recency or freshness is also very important because, specially in the news domain, information value decays rapidly over time \cite{lommatzsch2017incorporating}. In \cite{trevisiol2014cold}, they reported that in Yahoo! News,  80\% of visits are received within the first 30 hours after the article publication and before the first 20\% of the overall article lifespan. In \cite{gulla2016intricacies}, they conducted a log analysis on four different newspapers in Norway, and observed that readers interests on news story usually decays exponentially within 2-3 days after published, but at very different decay rates, depending on the publisher and on the article category.

According to \cite{karimi2018news}, the recency of an article can be considered in the recommendation process at three stages: pre-filtering (removing outdated news before the ranking) \cite{das2007, desarkar2014diversification, saranya2012personalized}, recency modeling (including recency into the recommender algorithms) \cite{bielikova2012effective, pon2007tracking, yeung2010context}, and post-filtering (re-ranking relevant items according to their freshness) \cite{li2011scene, medo2009adaptive, zheng2013penetrate}. The influence of articles recency in its relevance may depend on the article topic, exceptional events, specific contextual conditions or users preferences \cite{karimi2018news}.

On typical landing pages of news portals, popularity and recency are typically the most important ranking criterion. Those properties usually form the basis of hard-to-beat non-personalized baseline recommenders, which can work for most users, generally interested in fresh and popular stories, including first-timers \cite{doychev2014analysis, karimi2018news}.

In news RS, user feedback is implicit, often described by binary ratings indicating whether a user has clicked on a news story. While clicks can be used to capture positive user interest, they don't say anything about a user's negative interest, differently from explicit feedback, like ratings \cite{das2007} \cite{lin2014personalized}.

User engagement in news websites can be evaluated by different metrics such as dwell time (time the user spends reading articles) \cite{dallmann2017improving}, CTR, session duration, page views, among others. The authors of \cite{iqbal2016measuring} propose focus ratio and active ratio, as effective metrics for tracking user engagement. Focus ratio is the difference between the time a web page has been loaded in a browser and the time that page was actually visible in the active tab. On the other hand, active ratio is the difference between the time a web page is visible in the active tab and the user is actually considered viewing or interacting with that page. Such metrics could hopefully be used as a richer implicit feedback than clicks to model user interest in a news story.

\cite{mohallick2017exploring} conducts a relevant review about privacy issues in news recommendation domain. The user demographic information (e.g., name, age, gender, occupation, and relationship status), contextual information like time and location, and his logged behavior in terms of page access patterns, may reveal sensible personal preferences or interests. The paper presents a set of privacy protection techniques, like anonymization, perturbation, and cryptographic procedures.

\subsection{Content-Based Filtering for News RS}

The unstructured format of the news stories makes the recommendation process difficult to analyze and might result in unreliable recommendation. News recommender systems are mostly text-centric as the news domain is rich in text and unstructured in nature \cite{mohallick2017exploring}.

CBF methods have being used for news recommendations, as their textual content is a rich source of information. However, in some scenarios, simply representing user's profile information by a bag of words is insufficient to capture the exact reading interest of the user \cite{li2011scene}.

In recommendation research, news content are usually represented by using vector space model (e.g., \emph{TF-IDF}) \cite{billsus1999personal, capelle2012semantics, ren2013concert} or topic distributions obtained by language models (e.g., \emph{LDA}). Specific similarity measurements were adopted to evaluate the relatedness between news articles \cite{li2011scene}. Interestingly, a user study with volunteers showed up that users' declared interest in news topics did not strongly predict their actual interest in specific news items \cite{sela2015personalizing}.

Typically, news articles describe the occurrence of a specific event. Named entities include when, where, what happened, who are involved, and so on. News readers might have special preference on some particular named entities contained in news articles. Thus, \emph{Named Entity Recognition (NER)} techniques have been largely used in personalized news recommendation \cite{li2011scene}.

Among the first systems, the \emph{News Dude} \cite{billsus1999personal} was a personal news recommender agent that utilizes \emph{TF-IDF} combined with the \emph{K-Nearest Neighbor algorithm} to recommend news items to individual users. \emph{Newsjunkie} \cite{gabrilovich2004newsjunkie} was other system that recommended news stories by formal measures of information novelty and shows how techniques can be used to custom-tailor news feeds based on user's reading history. \emph{YourNews} \cite{ahn2007open} aimed to increase the transparency of adapted news delivery by allowing users to adapt user profiles. \cite{lee2007moners} present a mobile web news recommendation system, which incorporates news article attributes and user preferences with regard to categories into the modeling process.

\cite{ilievski2013personalized} proposed a probabilistic model using a set of a hand-crafted hierarchical taxonomy to represent news articles content like genre, location, keywords, and publisher.

An important limitation in CBF systems is that it tends to create stationary user profiles. That happens because recommendations tend to be always similar to the ones the user has previously seen, creating a \emph{bubble filter}. Lack of news diversity may lead to poor user experiences \cite{li2014modeling}. When visiting a news website, the user is looking for new information, information that he did not know before, that may even surprise him \cite{liu2010personalized}.

A typical approach to tackle this issue of CBF is to model news diversity, as dissimilarity between news. \cite{zhang2008avoiding} recommend news articles with diverse topics in a single recommendation session. In \cite{li2011scene}, they explicitly consider the diversity among news items in the recommended list, modeled as a budgeted maximum coverage problem.

\subsection{Collaborative Filtering for News RS}
There were few works using collaborative filtering (content agnostic) for news recommendations, i.e., ignoring articles content and using only users behaviour.

Google News presented their CF method in \cite{das2007}. They used three approaches, combined using a linear model: collaborative filtering, by using \emph{MinHash} clustering; \emph{Probabilistic Latent Semantic Indexing (PLSI)}; and co-visitation counts.

In \cite{diez2016}, they use matrix factorization to learn mapping of users and items into a common Euclidean space, where the similarities can be computed in a linear geometric context. The purpose was to suggest for each reader an ordered list of news that other readers with similar trajectories have seen in the past.

\subsection{Hybrid RS for News RS}
Hybrid RS have being popular for news recommendations, in which the inability of collaborative filtering to recommend news items is commonly alleviated by combining it with content-based filtering, like in \cite{chu2009personalized} and \cite{rao2013personalized}. 

In a subsequent work from Google News researchers \cite{liu2010personalized}, a Bayesian hybrid framework was introduced, combining CF models presented in \cite{das2007} with a probabilistic CBF model, which uses news categories and takes into account global news trends. By analyzing Google News large-scale logs to measure the stability of users' news interests, they found out that their interests do vary over time but follow the aggregate local news interest trends.

Recommender Systems often face the exploration/exploitation problem, where two competing goals must be balanced: exploiting user previous choices to provide accurate recommendations, and exploring his other possible interests, not specifically related to his previous choices, to reduce overspecialization. In \cite{li2010contextual}, Yahoo! researchers model news recommendations as a contextual bandit problem, a principled approach in which a learning algorithm sequentially selects articles to serve users based on contextual information of the user and articles, while simultaneously adapting its article-selection strategy based on user-click feedback to maximize total user clicks in the long run.

\cite{chu2009personalized} proposed a hybrid feature-based machine learning approach for personalized news recommendation to handle the cold-start problem. They build content profiles combining static and dynamic characteristics (e.g., popularity and freshness), which are updated in real time. User profiles of users included demographic information and a summary of user activities within Yahoo! properties. Based on all features in user and content profiles, they used bi-linear regression models to provide personalized recommendations of new items for both existing and new users.

The \emph{SCENE} \cite{li2011scene} method performs a two-stage clustering on both news attributes (e.g., textual content, named entities) along with their special properties (popularity and recency), and user access patterns. \emph{SCENE} is composed of three major components: Newly-Published News Articles Clustering, User Profile Construction, and Personalized News Items Recommendation. For personalization, user's profile is constructed in three different dimensions - news topic distribution, similar access patterns, and news entity preference.

In \cite{kirshenbaum2012live}, a live comparison of 20 different methods for personalized article recommendation was performed at \emph{forbes.com}. The winning method was a hybrid of item-item collaborative filtering and a content-based \emph{TF-IDF} method, including Wikipedia-based features and a Bayesian score remapping technique.

\cite{lin2014personalized} proposed \emph{PRemiSE}, a hybrid system that embeds content-based (named entities from news stories), collaborative filtering, and social networking approaches into a unified probabilistic framework, to produce predictions that balance the content of news, the reading preferences of users, and recommendations from "experts". The social networks and experts are virtual and inferred from user's access logs, by using the information diffusion theory.

In \cite{li2014modeling}, long-term user history was leveraged to provide coarse grain news recommendation for certain news groups. The short-term history of the user was then used to recommend specific news articles within the selected groups. They select news items from the user-item affinity graph, by using an absorbing random walk model to increase the diversity of the recommended news list.

In \cite{lommatzsch2014real}, they describe how they won the \emph{Plista} news recommendations contest, which allowed research teams to compete by providing the best recommendations for real users (online evaluation). In that work, they implemented a context-aware delegation strategy, which selects a recommender algorithm (e.g., User-based CF, Item-based CF, CBF, Popular) based on the request context (e.g., day of the week, hour, current page popularity). They found out that there is not a single optimal algorithm - the accuracy is strongly dependent on the context and on the publisher.

\cite{garcin2014offline} employs Context Trees for news recommendations. A context-tree recommender system builds a hierarchy of contexts, arranged in a tree such that a child node completely contains the context of its parents. A context can be the sequence of stories read by a user, the sequence of topics, or topic distributions. This class of recommender systems adapts its model to current trends and reader preferences.

In \cite{trevisiol2014cold}, Yahoo! researchers have reported an interesting strategy to deal with cold-start news recommendation. They found out that the referrer URL (from which the user is coming from) has a big predictive potential for next click prediction. They define a special case of \emph{BrowseGraph} model - the \emph{ReferrerGraph}, that consists of a sub-graph built from the browsing sessions with homogeneous referrer URL.

The New York Times started with CBF based on articles tags, but soon observed limitations in the user profiling and recommendation accuracy. After some prototypes and online evaluation, they could improve recommendation effectiveness by usage of Collaborative Topic Modeling \cite{wang2011collaborative}, which initializes the item representation based on a mixture of its topics (\emph{LDA}) and updates it based on audience reading patterns.
\cite{spangher2015}

In \cite{caldarelli2016signal}, they describe a preliminary research on news recommendations based on a signal processing technique (e.g., the discrete wavelet transform). Users profiles are modeled as bag-of-signal, which represent the informative entities and the time use patterns, and allow to compute the similarity between users.

In \cite{epure2017recommending}, they use Markov processes over news categories to model users' short-term, medium-term, and long-term interests. They found out that, within sessions, users are likely to read news within the same category, although users' interest proportion on the categories change every 1 to 4 months.

To the best of the knowledge from this research so far, the only works presenting a deep learning architecture for news recommendation were \cite{song2016multi}, \cite{kumar2017, kumar2017word}, \cite{park2017deep}, \cite{okura2017embedding}, \cite{zhang2018deep}, and \cite{wang2018dkn}. Those works are described with more detail Chapter~\ref{sec:chapter_3} and compared to our proposed meta-architecture in Section~\ref{sec:comparison-works}.

\chapter{Deep Learning for Recommender Systems}
\label{sec:chapter_3}

This chapter presents the related work on the application of Deep Learning architectures and methods for Recommender Systems. A background on Deep Learning methods and techniques is provided in Appendix~\ref{ape:dl-background}.

Deep Learning research  \cite{hinton2006,hinton2006b,bengio2007,bengio2009} has yielded relevant advances in computer vision, audio, speech recognition, and natural language processing. However, applications of deep learning in recommender systems have not been extensively explored yet \cite{zheng2016}. The uptake of deep learning by the recommender systems community was relatively slow, as the topic became popular only in 2016, with the first Deep Learning for Recommender Systems workshop at \emph{ACM RecSys 2016} \cite{hidasi2017dlrs}.

In many recommender systems, scalability and handling dynamic pools of items and users are considered as critical needs.
One of the advantages of Neural Networks for recommender systems is that they scale to the size of a training set (by training using mini-batches) and also support online learning, as updating latent factors of items or users can get performed independently from historical data \cite{zheng2017joint}.

In the survey conducted in \cite{zhang2019deep}, the strengths of Deep Learning-based recommender systems are summarized as follows: (a) \emph{Nonlinear Transformation} -- the ability to model nonlinearity in data, differently from linear techniques such as matrix factorization, factorization machines, and sparse linear models; (b) \emph{Representation Learning} -- reduces the efforts in hand-crafting feature design and enables models to include
heterogeneous content information such as text, images, audio, and even video; (c) \emph{Sequence Modeling}  -- as RNNs and CNNs can be effectively model  sequential data, such as session clicks; and (d) \emph{Flexibility} -- neural networks can be modularized and combined to formulate powerful hybrid recommendation models, such as the architectures proposed in our research.

\cite{Hidasi2017summerschool} divided the main approaches using Deep Learning in Recommender Systems as: (a) \emph{Item Embeddings and 2vec Models}, (b) \emph{Deep Collaborative Filtering}, (c) \emph{Deep Session-Aware and Session-Based RS}, and (d) \emph{Deep Feature Extraction from Heterogeneous Data}, and their combinations. 

In this Chapter, we briefly touch (a) and (b), focusing our review on works on (c) and (d), which are more relevant for this research.

\section{Item Embeddings and 2vec Models}

Different approaches to learn embeddings (also known as distributed representations, or latent features) for items have being proposed in recommender systems research, like Matrix Factorization (MF) methods \cite{grbovic2015commerce}. 

Recent methods focus on item embeddings without user identification and can be used as the basis of more advanced methods or as item-to-item recommenders. Most of the models use some variation of \emph{Word2Vec} \cite{mikolov2013distributed} - originally devised for word embeddings on event data.

The \emph{Prod2Vec} \cite{grbovic2015commerce} is an extension of the Word2Vec algorithm to product shopping sequences. It learns item embeddings for products, trying to predict the other products users have also bough in e-commerce portals. As a result, the \emph{Prod2Vec} can be seen as a matrix factorization technique on the product co-occurence matrix. A similar proposal was presented as the \emph{Item2Vec} \cite{barkan2016item2vec}.

The \emph{Meta-Prod2Vec} \cite{vasile2016meta} has improved upon the \emph{Prod2Vec}, by using the product metadata side information to regularize the final product embeddings. 

The \emph{Content2Vec} jointly embed all product information into a product vector such that the inner-product of any two product vectors is proportional to the probability that the two products will be bought by the same user \cite{nedelec2017specializing}. It has used a multi-model hybrid RS approach, with the \emph{Prod2Vec} for CF, the \emph{AlexNet} CNN architecture for image representation and the \emph{Word2Vec} + the \emph{TextCNN} for text representation.

\section{Deep Collaborative Filtering}

A natural application of deep models is the Collaborative Filtering (CF). Leveraging the versatility of deep models, multiple types of interaction and context are often integrated \cite{hidasi2017dlrs}.

Early pioneer work on CF based on using neural networks was done in \cite{salakhutdinov2007}. They introduced a two-layer RBM to model ratings. An RBM model was trained for each user, and visible \emph{softmax} units corresponds with the items rated by the user. If two users rated the same item, their RBMs shared weights connected to the corresponding visible unit \cite{zheng2016}, with better for computational tractability.

Autoencoders and denoising autoencoders \cite{vincent2008extracting}, with intentionally corrupted input, have being used as the central component of their CF recommender systems \cite{chen2012, sedhain2015, strub2015}. 

The idea of user-based \emph{AutoRec} \cite{sedhain2015} was to learn hidden structures that can reconstruct user's ratings given her historical ratings as inputs using traditional autoencoders. In terms of user personalization, such approach shares a similar spirit as the item-item model \cite{sarwar2001, ning2011} that represent a user as his rated item features. 

In \cite{wang2015}, it was introduced a technique named \emph{Collaborative Deep Learning (CDL)}, which utilizes ratings and also item review texts, in order to address the cold-start problem. They integrate a bayesian \emph{Stack Denoise Auto Encoder (SDAE)} \cite{vincent2010} and \emph{Collaborative Topic Regression (CTR)} \cite{wang2011collaborative}. It learns latent factors of items from review texts and draw a latent user vector from a Gaussian distribution \cite{zheng2016}.

The \emph{CDL} uses tags and metadata instead of the item ID. It was the first deep model to learn from review texts for recommender systems \cite{betru2017}. However, they did not modeled reviews by users, in which users preferences could be better inferred \cite{zheng2016}. In addition, their approach was only suitable for one-class collaborative filtering problems \cite{pan2008}. Other limitation was text representation using bag-of-words, which does not provide semantic similarity, and word order is lost \cite{zheng2016} \cite{betru2017}.

In \cite{li2015deep}, authors proposed a general deep architecture named \emph{Deep Collaborative Filtering (DCF)}, which integrates matrix factorization and deep feature learning. It models the mappings between the latent factors used in CF and the latent layers in deep models. They also present a practical instantiation of the proposed architecture, by utilizing the probabilistic matrix factorization and \emph{Marginalized Denoising Autoencoders (mDA)} \cite{chen2012}, with high scalability and low computational cost.

In \cite{wu2016}, a method for top-N recommendation named \emph{Collaborative Denoising AutoEncoder (CDAE)} is presented. It is probably the first work to utilize, for recommender systems, the idea of \emph{Denoising AutoEncoders (DAE)} \cite{vincent2008extracting}. The DAE extends the classical autoencoder by training to reconstruct each data point from its (partially) corrupted version. The goal of DAE is to force the hidden layer to discover more robust features and to prevent it from simply learning the identity function. The model learns latent representations of corrupted user-item preferences that can best reconstruct the full input, recovering co-preference patterns. 

\cite{wang2016collaborative} developed the \emph{Collaborative Recurrent AutoEncoder (CRAE)}, which can model the generation of item sequences while extracting the implicit relationship between items (and users). It used an encoder-decoder architecture, corrupting input items texts (e.g., movie plot, review) with a \emph{<BLANK>} token. The synergy between denoising and CF has enabled the \emph{CRAE} to make accurate recommendations while learning to fill in the blanks in sequences. The \emph{CRAE} was the first model to bridge the gap between RNN and CF, with respect to hybrid methods for RS.

\emph{Multi-VAE} and \emph{Multi-DAE} \cite{liang2018variational} have proposed variants of the variational autoencoder for recommendation with implicit data and they have shown better performance than \emph{CDAE}. The authors also have introduced a principled Bayesian inference approach for parameter estimation and have shown favorable results over the commonly used likelihood functions.

\cite{he2017} authors devised a general framework named \emph{Neural Collaborative Filtering (NCF)}. They modeled user-item interactions in three different instantiations: the \emph{Generalized Matrix Factorization (GMF)}; the \emph{Multi-Layer Perceptron (MLP)}; and their combination as \emph{Neural Matrix Factorization (NeuMF)}. Their experiments have shown that by using deeper layers of neural networks offered better recommendation performance.

Negative sampling approaches can be used to reduce the number of training unobserved instances.
Follow-up work \cite{niu2018neural, song2018neural} proposed using pairwise ranking loss to enhance performance. \cite{lian2017cccfnet} and \cite{wang2017item} extended the \emph{NCF} model to cross-domain recommendations. It was shown in \cite{xue2017deep} and \cite{zhang2018neurec} that the one-hot identifier can be replaced with columns or rows of the interaction matrix to retain user/item interaction patterns \cite{zhang2019deep}.

\section{Deep Sequence-Based, Session-Based, and Session-Aware RS}

Since historical behaviors from different time periods have different effects on users' behaviors, the importance of sequential information in recommender systems has been gradually recognized by researchers. Methods based on Markov assumption, including \emph{Factorizing Personalized Markov Chain (FPMC)} \cite{rendle2010factorizing} and \emph{Hierarchical Representation Model (HRM)} \cite{wang2015learning}, have been widely used for sequential prediction. However, a major problem of these methods is that they independently combine several components. To solve this deficiency, the RNN has been employed to model global sequential dependency among all possible components \cite{liu2016context}.

Much of the research recommender systems has focused on models that work when a user identifier is available and a clear user profile can be built. Classic recommendation algorithms require many user interactions to start providing useful recommendations. Therefore, in many websites like e-commerce, news and media websites, users are not required to login for browsing. 

In order to provide better recommendations, those websites try to employ techniques like cookies and browser fingerprinting to recognize the user. But those approaches are not reliable for long-term users tracking, as users may use multiple devices, clean their cookies regularly, or even set a browser setting to prevent cookies creation.

The Session-Based RS is a type of a Context-Aware RS (CARS), which leverages information available in the current session for recommendations, ignoring user profiles. Session-based implementations generally use simple item-to-item recommendations, looking for items similar to the ones the user has consumed in that session, or using transition probabilities based in the last click. Publishers observe relatively short sessions, with fewer than ten clicks on average \cite{epure2017recommending}.

In Session-Aware RS, additional information about the user session is incorporated in the prediction process \cite{twardowski2016modelling}. In this setting, users identifiers are present. Thus, it is possible to propagate information from the previous user session to the next, thus improving the recommendation accuracy.  

This research is focused in Session-Based RS. This is a challenging problem, specially when recommended items are ephemeral, with its short life-cycle too short or its availability too dynamic, allowing to identify it only by unique id, like in news recommendations and online auctions for ads.

The Recurrent Neural Networks (RNN) and the Convolutional Neural Networks (CNN) are gaining attention as suitable for session-based/aware recommendations, since the seminal work of \cite{hidasi2016}. Relevant works in this recent research line are described in more detail in the next subsections.

\subsection{The \emph{GRU4Rec}}
\label{sec:gru4rec}
The \emph{GRU4Rec} \cite{hidasi2016} has represented the seminal work in the usage of Recurrent Neural Network (RNN) for session-based recommendations. For better modeling of longer sessions, they have used Gated Recurrent Unit (GRU) \cite{cho2014properties, chung2014empirical}, which deals with the vanishing gradient problem, common in standard RNN.

\begin{figure}[h]
	\centering
	\includegraphics[height=7cm]{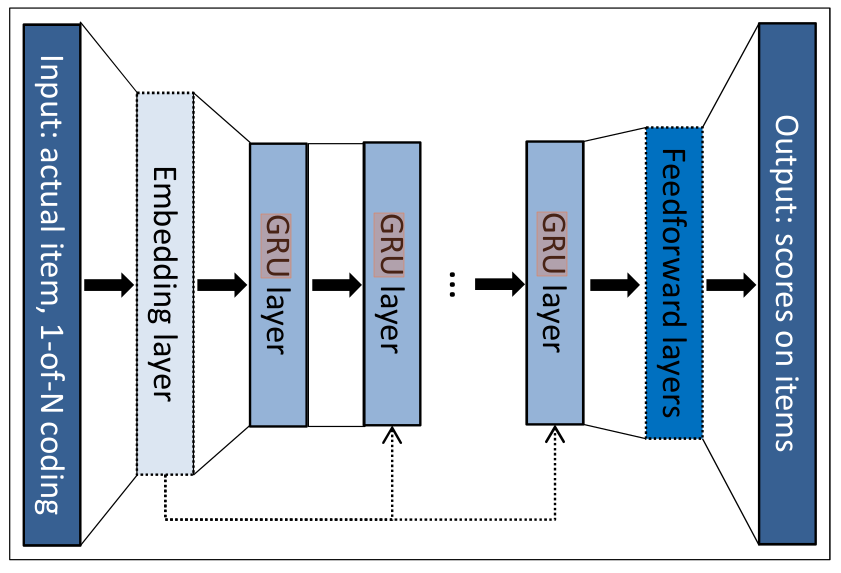}
	\caption{The GRU4Rec general architecture, from \cite{hidasi2016}}
	\label{figure:gru4rec}
\end{figure}

The input vector is a one-hot encoded representation of the item, with the vector length equal to the total number of items. The output is the predicted score for a fixed number of items, i.e. the likelihood of being the next in the session, as illustrated in Figure \ref{figure:gru4rec}.

\begin{figure}[h]
	\centering
	\includegraphics[height=7cm]{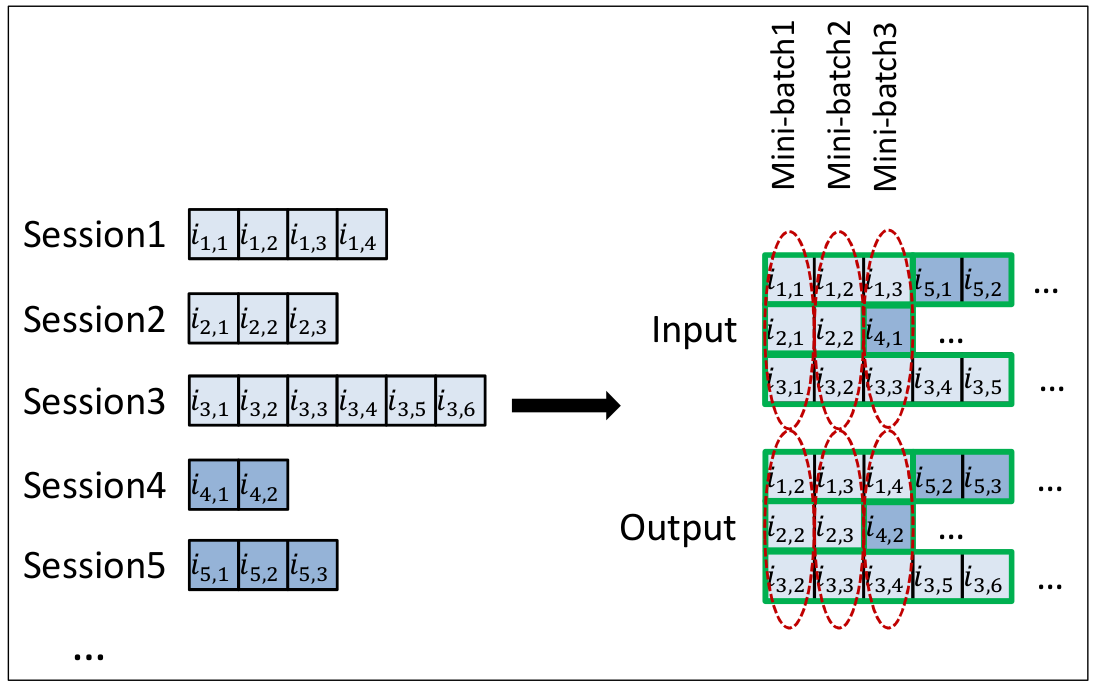}
	\caption{The GRU4Rec - Session-parallel mini-batch creation, from  \cite{hidasi2016}}
	\label{figure:gru4rec-sessionparallel}
\end{figure}

They have used an efficient approach of session-parallel mini-batches, by first creating an order for the sessions and then using the first event of the first X sessions to form the input of the first mini-batch (which desired output is the second events of active sessions). The second mini-batch is formed from the second events and so on. If any of the sessions end, the next available session is put in its place, as shown in Figure \ref{figure:gru4rec-sessionparallel}. Sessions are assumed to be independent, thus they reset the appropriate hidden state, when this switch occurs.

\begin{figure}[h]
	\centering
	\includegraphics[height=5cm]{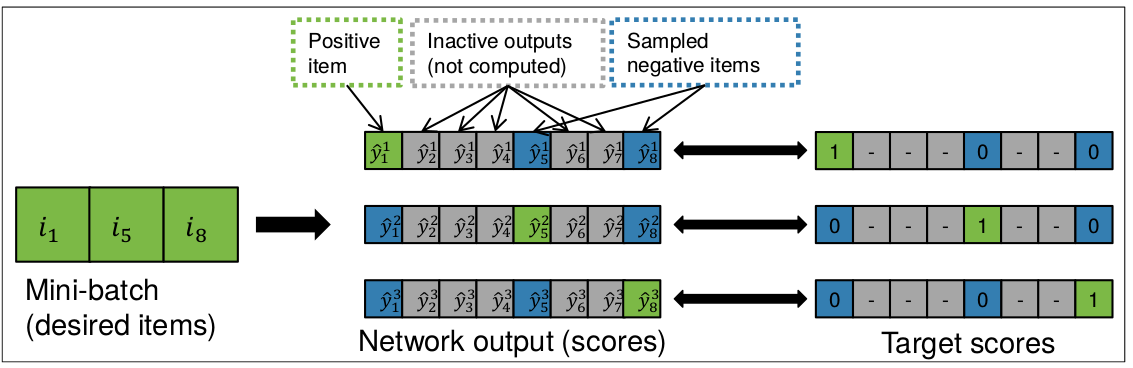}
	\caption{The \emph{GRU4Rec} - Mini-batch sampling, from \cite{hidasi2016}}
	\label{figure:gru4rec2-minibatchsampling}
\end{figure}

The number of items may be in the order of hundreds of thousands or even few millions, making it hard to scale by calculating a score for each item in each step. They have used negative sampling to compute scores for some negative samples, besides the desired output (other items in the same mini-batch), and modify the weights, so that the desired output is highly ranked, as shown in Figure \ref{figure:gru4rec2-minibatchsampling}.

The training procedure uses mini-batch negative sampling, from other items in the same batch that users have not interacted with. It is basically a form of popularity-based sampling, since the training iterates through all events, thus the probability of an item acting as a negative sample is proportional to its support. This popularity bias in the sampling process makes sense, under the hypothesis that popular items missed by the user are more likely to express a dislike than other unknown items, which the user could be interested if he is aware of them.

They evaluated their approach on two datasets: click-streams of an e-commerce and video watching in a Youtube-like service. The evaluation was done by providing the session events from test set one-by-one, and also by checking the rank of the item of the next session event, compared to other negative examples (the most popular 30k items not clicked by the user). The hidden state of the GRU is reset to zero after a session finishes. At the end, the authors have compared the two pairwise ranking losses: the \emph{Bayesian Personalized Ranking (BPR)} and their own \emph{TOP1} ranking loss.

The Recall@20 and the \emph{Mean Reciprocal Rank (MRR@20)} metrics were computed and their results were 20 - 30\% better than baseline approaches like \emph{Item-KNN} and \emph{BPR-MF}.

In a subsequent paper in this research line \cite{hidasi2018recurrent}, the authors designed a better loss function and used additional samples besides the ones in mini-batch. They observed that limiting negative samples from the same mini-batch does not provide flexibility, as its always popularity-based sampling. So, they proposed sampling more items outside the mini-batch, shared with all mini-batch examples. A hyperparameter controls whether the external sampling will use uniform or popularity-based sampling.

They also proposed a family of Ranking-max loss function, in which, instead of averaging the parwise ranking loss between the target item and all sampled items, the loss is computed only by comparing with the most relevant sample score. This approach solved vanishing of gradients as the number of samples increased.

Therefore, despite the success of RNNs in session-based recommendation, \cite{jannach2017recurrent} indicated that some trivial methods (e.g., simple neighbourhood approach) could achieve the same or even better accuracy results as GRU4Rec while being computationally much more efficient.

The GRU4Rec method is one of the baselines used in our experiments, as described in Section~\ref{sec:eval_methodology}.

\subsection{The Multi-modal Data for Session-based Recommendations (\emph{p-RNNs})}
In \cite{hidasi2016parallel}, they extended the GRU4Rec to be a hybrid session-based architecture, where side features (image and textual description) of the item are also used besides the item ID. 

They have proposed multi-modal architecture named \emph{p-RNNs}. It was not end-to-end, in the sense that image features were independently extracted by using transfer learning (last average pooling layer) from a pre-trained \emph{GoogLeNet} \cite{szegedy2015going} architecture implementation. 

The textual representation was not directly learned by a neural network, instead they simply have used \emph{TF-IDF} vectorization \cite{salton1988term}. They reported that in their experiments, word embeddings \cite{mikolov2013distributed} and Language Modeling with RNNs \cite{mikolov2010recurrent} did not perform well, maybe due to the noisier descriptions in the used dataset. 

This complex architecture was composed by different subnets, as each information source (IDs, images, textual descriptions) had their own RNN. All subnets were combined in different configurations and trained with distinct strategies (simultaneous, alternating, residual, and interleaving). 

The p-RNN architecture has presented significant performance improvements over feature-less session models, while all session-based models outperform the item-to-item type of baseline.

\subsection{The \emph{Contextual Recurrent Neural Networks (CRNN)}}

\begin{figure}[h]
	\centering
	\includegraphics[height=7cm]{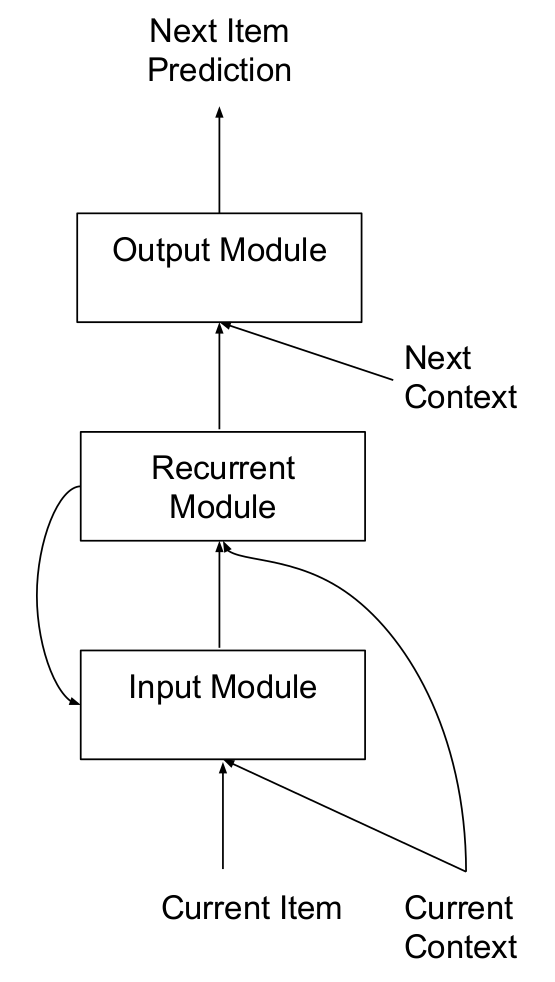}
	\caption{The \emph{CRNN} architecture, from \cite{smirnova2017contextual}}
	\label{figure:crnn}
\end{figure}

In \cite{smirnova2017contextual}, a family of \emph{Contextual Recurrent Neural Networks (CRNN)} was proposed, by using RNN cells to learn context-aware transitions for session-based recommendation. Their experiments on e-commerce datasets have considered contextual information, like type of the interaction (eg. view, add-to-basked, sale), time gaps between events, and time of day of interaction. In their architecture, the current context is used in the input module and the next context on the output module, as shown in Figure \ref{figure:crnn}.

They have experimented feeding the context into a GRU, by using two strategies: the Context-dependent input/output representation and the Context-dependent hidden dynamics. The latter strategy was the most successful in experiments against sequential and non-sequential baselines.

\subsection{The \emph{DLRec}}

\begin{figure}[h]
	\centering
	\includegraphics[height=5cm]{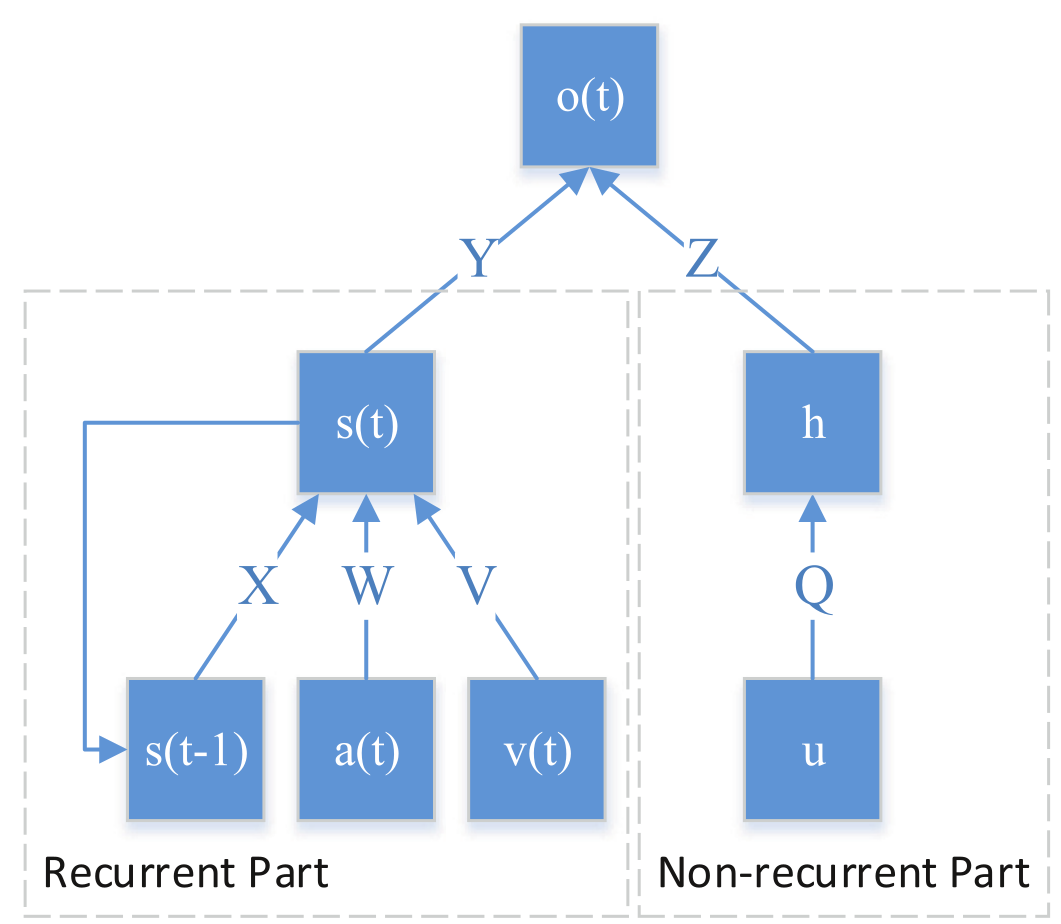}
	\caption{The \emph{DLRec} architecture for recommendation, from \cite{wu2016recurrent}}
	\label{figure:dlrec}
\end{figure}

\cite{wu2016recurrent} proposed a sequence-based recommendation neural model named \emph{DLRec}, to predict the probability that the user will access an item given the time heterogeneous feedback of this user. Their proposed architecture, shown in Figure \ref{figure:dlrec}, contains recurrent and non-recurrent subnets. In the non-recurrent part, the user one-hot vector \textit{u} is mapped to a feed-forward layer. In the recurrent, the input layer consists of vectors \textit{u}, \textit{v(t)}, \textit{a(t)}, and \textit{s(t-1)}, representing the current user, item, feedback activity, and the last hidden layer state, respectively. In the last layer, the two subnets are combined in a softmax layer, with a probability distribution over all items. In this setting, user characteristics are only considered independently from sequence properties. 

Their experiments were performed on media and e-commerce datasets, and \emph{DLRec} was able to provide more accurate recommendations than other non-sequential recommenders, such as \emph{BPR-MF} \cite{rendle2009bpr} and \emph{CLiMF} \cite{shi2012climf}.

\subsection{The Recurrent Recommender Networks (RRN)}
In \cite{wu2017recurrent}, their authors stated that many of the current state-of-the-art approaches and RS do not adequately model the temporal and causal aspects inherent in data, and user profiles and item attributes are generally considered static. 

In their approach, the user and item models are trained separately, and outputs of both networks are subsequently coupled with further auxiliary parameters, by capturing stationary concepts, in order to predict user ratings. That architecture requires learning two RNNs such that user and item properties can yet again only loosely be intertwined.

That model was shown to be able to capture temporal patterns in rating data, outperforming previous works in terms of rating prediction accuracy (\emph{RMSE}) on movies datasets. 

\subsection{The \emph{Context-Aware Recurrent Neural Network (CA-RNN)}}
\begin{figure}[h]
	\centering
	\includegraphics[height=5cm]{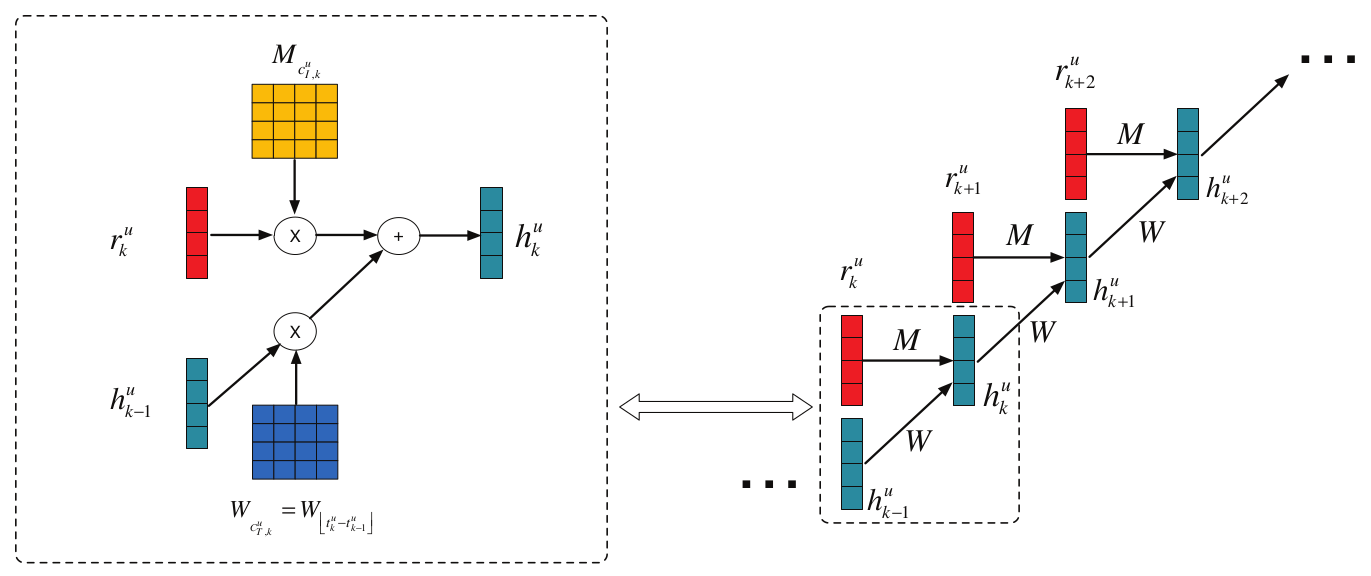}
	\caption{The \emph{CA-RNN} architecture, from \cite{liu2016context}}
	\label{figure:ca-rnn}
\end{figure}

\cite{liu2016context} have proposed the \emph{Context-Aware Recurrent Neural Networks (CA-RNN)}, shown in Figure \ref{figure:ca-rnn}, which employs adaptive context-specific input matrices and adaptive context-specific transition matrices.

The adaptive context-specific input matrices capture external situations where user behaviors happen such as time, location, weather, and so on. These matrices capture how much the length of time intervals between adjacent behaviors in historical sequences affect the transition of global sequential features.

\begin{figure}[h]
	\centering
	\includegraphics[height=5cm]{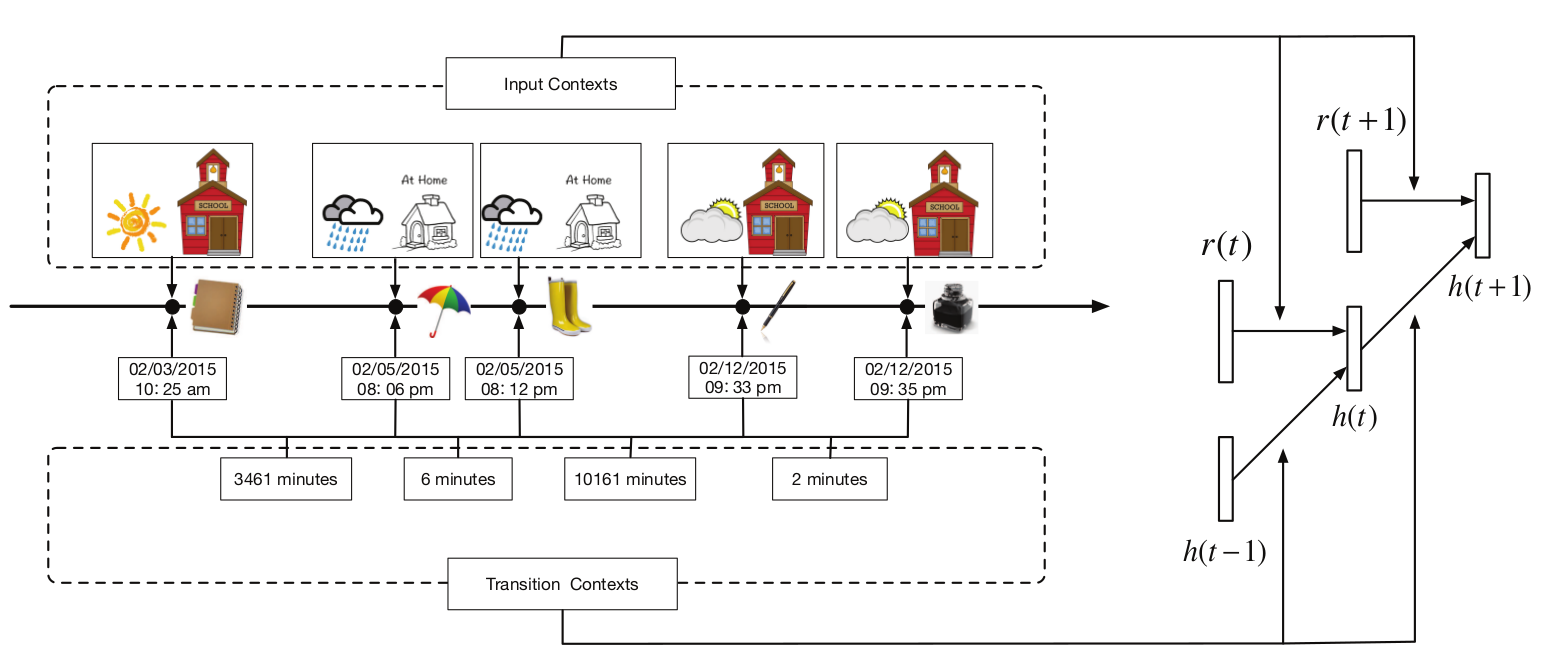}
	\caption{An example of input and transition contexts in a user purchasing sequence, from \cite{liu2016context}}
	\label{figure:ca-rnn_contextexample}
\end{figure}

In Figure \ref{figure:ca-rnn_contextexample}, authors show an example of a purchasing sequence of an user. Input contexts are external contexts like location, time (weekdays or weekends, morning or evening), or  weather (sunny or rainy). Transition contexts are contexts of transitions between two adjacent input elements in historical sequences, i.e. time intervals.

The \emph{Bayesian Personalized Ranking (BPR)} \cite{rendle2009bpr} and the \emph{Back Propagation Through Time (BPTT)} \cite{rumelhart1988learning} are used for the learning of the CA-RNN. Their experimental results on an e-commerce dataset and on a movies dataset have shown that the CA-RNN outperformed other competitive sequential and context-aware models.


\subsection{The User-based RNN}
\label{sec:userbasedRNN}
\begin{figure}[h]
	\centering
	\includegraphics[height=5cm]{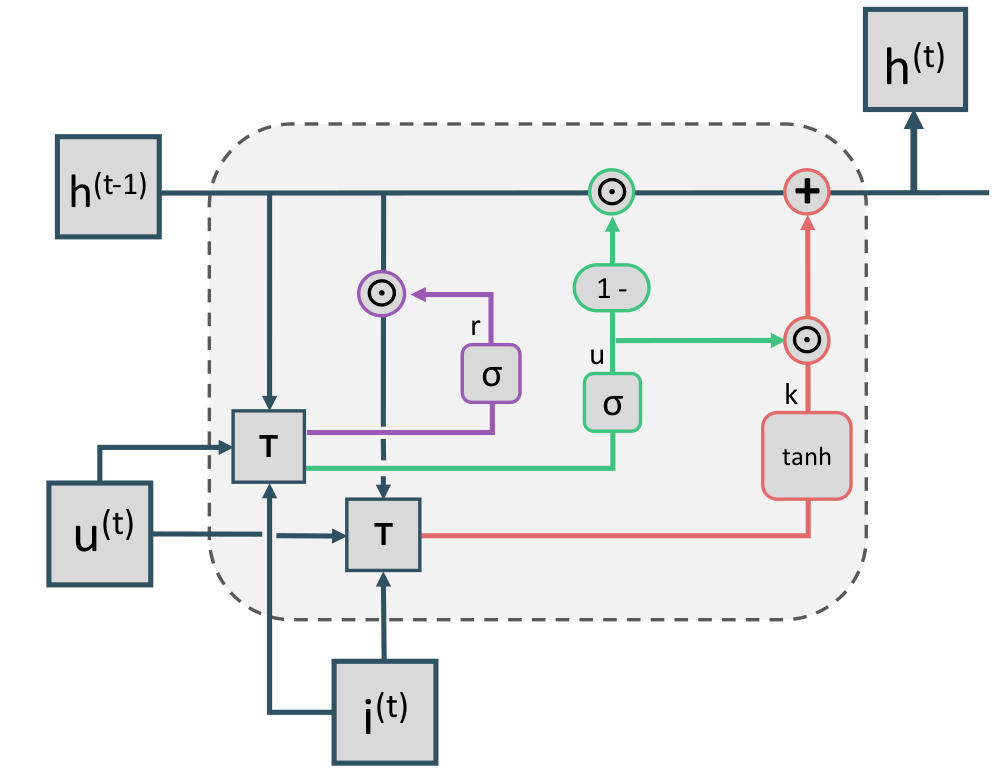}
	\caption{The linear user-based GRU cell, with the addition of user representation, from \cite{donkers2017sequential}}
	\label{figure:sequential_user_rnn}
\end{figure}

In \cite{donkers2017sequential}, they adapted GRU to deeply integrate user embeddings into the update gating process, as shown in Figure \ref{figure:sequential_user_rnn}. These user-based GRUs are designed and optimized for the purpose of generating personalized next item recommendations in sequence-based RS. They have created two strategies for this integration: the \emph{Rectified Linear User Integration} and the \emph{Attentional User Integration}.

Their experiments were performed in movies and music listening datasets. RNN models with user information have presented higher accuracy and the \emph{Attentional User-based GRU} has performed the best in most cases.


\subsection{The \emph{Hierarchical Recurrent Neural Networks (HRNN)}}
\label{sec:HRNN}

\begin{figure}[h]
	\centering
	\includegraphics[height=5cm]{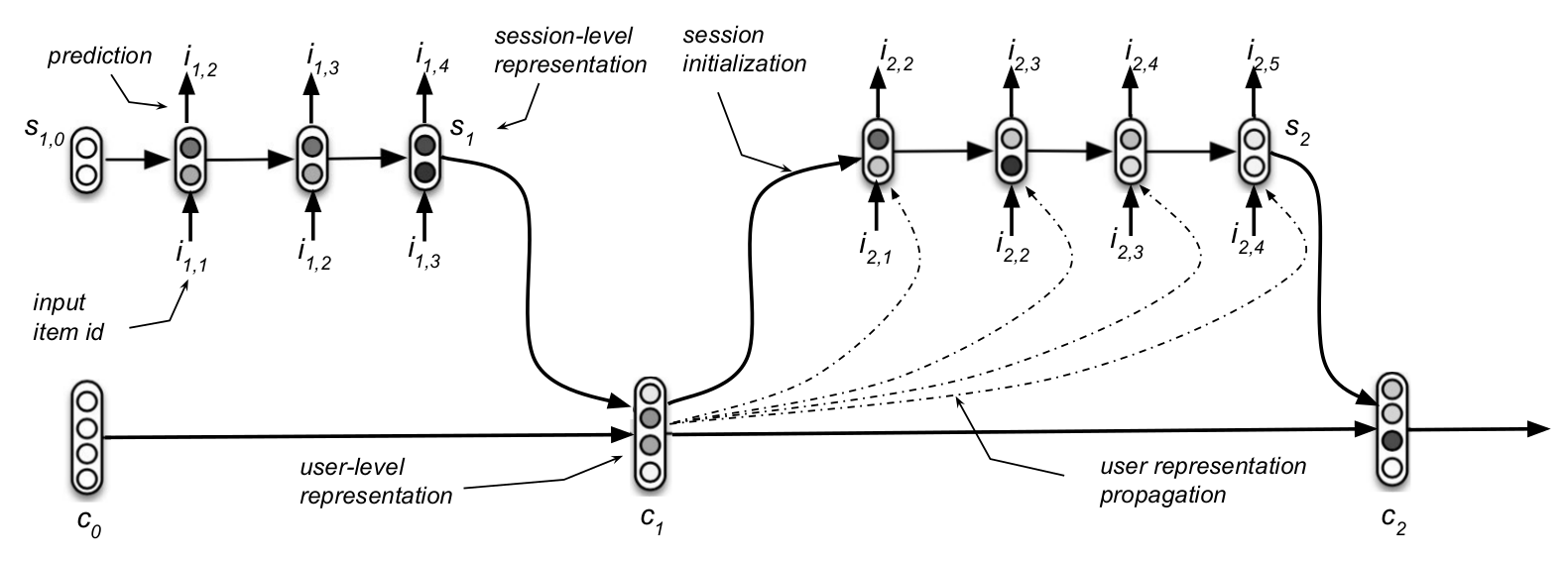}
	\caption{The \emph{Hierarchical RNN} architecture for personalized session-based recommendation, from \cite{quadrana2017personalizing}}
	\label{figure:hrnn}
\end{figure}

\cite{quadrana2017personalizing} also proposed a way to personalize RNN models with users' cross-session information and to provide session-aware recommendations. Their architecture, shown in Figure \ref{figure:hrnn}, is based on a \emph{Hierarchical RNN (HRNN)}, where the hidden state of a lower-level RNN at the end of one user session is passed as an input to a higher-level RNN, which aims at predicting a good initialization (i.e., context vector) for the hidden state of the lower RNN for the next session of the user.

In their experiments, the proposed \emph{HRNN} architecture outperformed state-of-the-art session-based RNNs and other basic personalization strategies for session-aware recommendation on two datasets from jobs and video streaming domains.


\subsection{The \emph{Inter-Intra Session RNN (II-RNN)}}
\label{sec:ii-rnn}
In \cite{ruocco2017inter}, it was proposed the II-RNN to leverage information from previous user sessions to the current one. Their work is similar to \cite{quadrana2017personalizing}, although both researches were independently developed in parallel. In the II-RNN, the second RNN learns from recent sessions and predicts user's interests in the current session. By feeding this information to the original RNN, it is able to improve its recommendations.

Their proposed model especially improved recommendations at the start of sessions and was able to deal with the cold-start problem in session-aware RS.


\subsection{The \emph{3D CNN} Recommendation Architecture}

In \cite{tuan20173d}, the authors describe a method that combines session clicks and content features such as item descriptions and item categories to generate recommendations. 
That was a pioneering work in RS research, by using 3-dimensional CNNs with character-level encoding from all input data.

The \emph{3D CNN} recommendation architecture provides a way to capture spatio-temporal patterns and it was originally introduced for video data \cite{ji20133d, tran2015learning}.

The main difference between 3D and 2D CNN is that in 3D CNN the convolution and pooling are performed in all three dimensions (i.e. spatio-temporally), while in 2D CNNs they are performed
spatially in two dimensions, even if the input is 3-dimensional.

Character-level networks allow modeling different data types, by using their raw textual representation, thus reducing feature engineering effort. In the proposed model, each input feature is represented as its own alphanumeric format, without the need of categorical embeddings.

\begin{figure}[h]
	\centering
	\includegraphics[height=7cm]{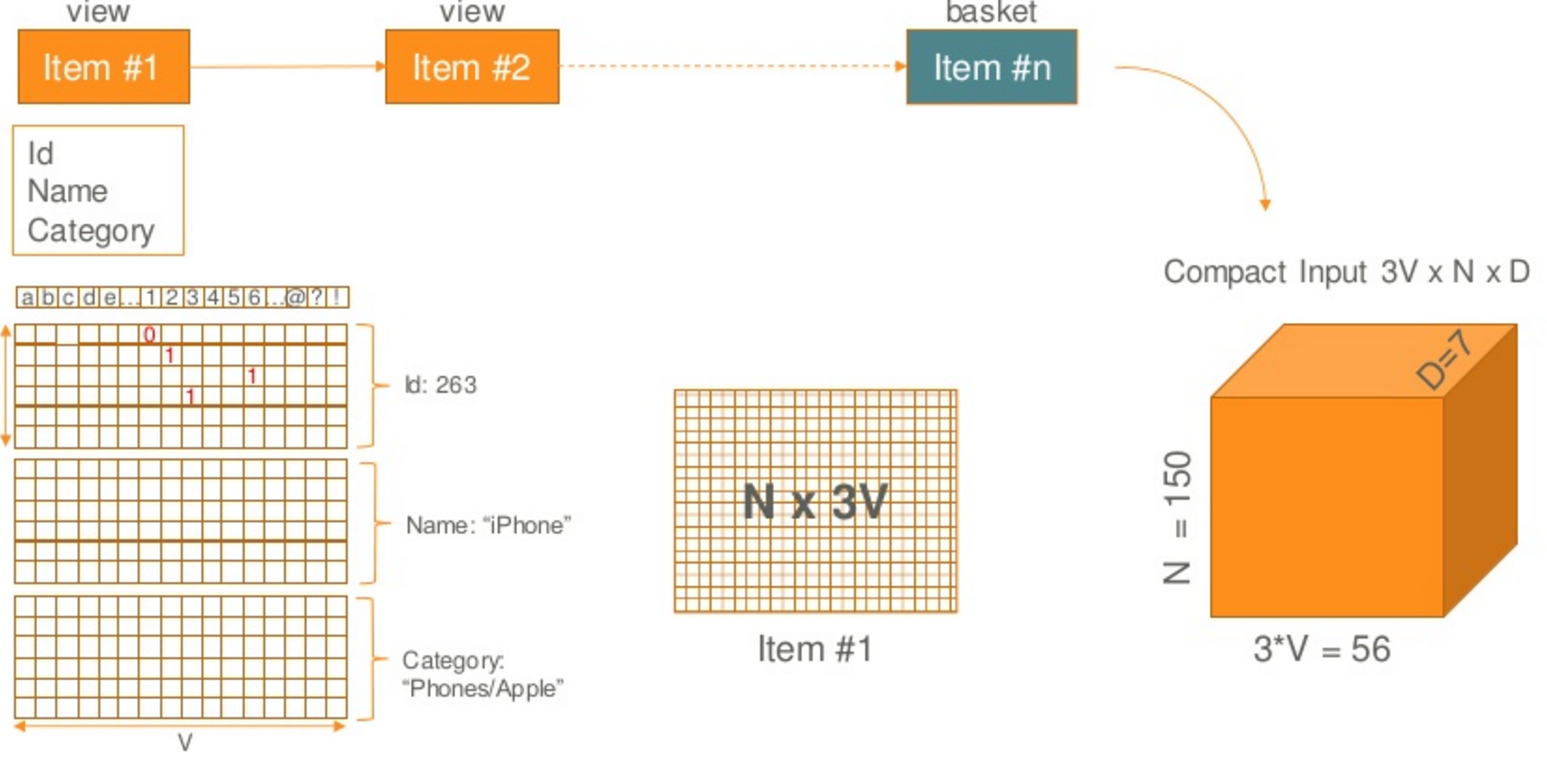}
	\caption{The 3D-CNN architecture, from \cite{tuan20173d}}
	\label{figure:3dcnn-example}
\end{figure}

The input 3D tensor, illustrated in Figure \ref{figure:3dcnn-example}, represents the one-hot encoded characters of each of the features in the first two dimensions, and the third dimension represents the time (item views in the same session). As the input tensor is fixed, it was necessary to truncate the length for textual features and the maximum number of clicks in the session.

They have applied the proposed method to predict add-to-cart events in e-commerce websites. On two real datasets, their method outperformed several baselines and a state-of-the-art method based on recurrent neural networks.

\subsection{The \emph{Deep Joint Network for Session-based News Recommendations (Deep JoNN)}}
\label{sec:deep_jonn}

Inspired by the 3D-CNN architecture from \cite{tuan20173d}, \cite{zhang2018deep} proposed a neural architecture for session-based news recommendation that uses a character-level convolutional network to process articles' input features: (a) item ID, (b) keywords and entities, and (c) category.

\begin{figure}[h]
	\centering
	\includegraphics[height=12cm]{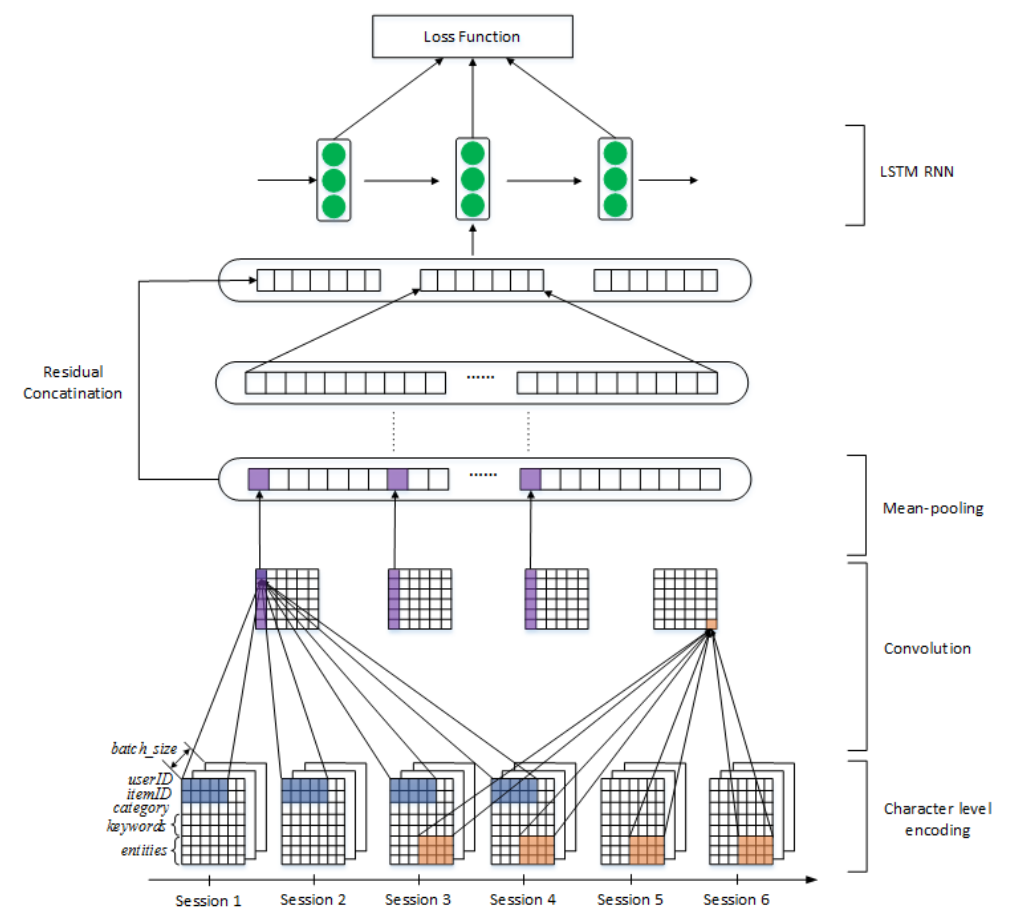}
	\caption{The \emph{DeepJoNN} architecture, from \cite{zhang2018deep}}
	\label{fig:deepjonn-example}
\end{figure}

But differently from \cite{tuan20173d}, instead of processing the sequence of sessions by a dimension of the CNN, they are processed by a Recurrent Neural Network, as shown in Figure~\ref{fig:deepjonn-example}.

They have also proposed a time-decay function to penalize the scores of older news articles, presented in Equation~\ref{eq:deepjonn-timedecay},

\begin{equation} \label{eq:deepjonn-timedecay}
R_{decay} = e^{-\lambda \cdot (t-t_0)},
\end{equation}

where $t$ and $t_0$ represent the
click time and the publication time
of the news, respectively. The decay rates are multiplied by the output values from LSTM RNN layer to form the final outputs.

In their evaluation protocol, they have used the last click of sessions as the test set. Negative items were randomly sampled for training, independently of their publishing time.

They have measured accuracy metrics (\emph{Recall@20, MRR@20}) and compared \emph{DeepJoNN} against some baselines: POP, Item-KNN \cite{hidasi2016}, BPR-MF \cite{rendle2009bpr}, and HRNN \cite{quadrana2017personalizing}.

A discussion on the similarities and differences between \emph{DeepJoNN} and  \emph{CHAMELEON} are presented in Section~\ref{sec:comparison-works}.

\subsection{The \emph{Session-Based Recommendation with Graph Neural Networks (SR-GNN)}}
\label{sec:sr-gnn}

\begin{figure}[h]
	\centering
	\includegraphics[width=16cm]{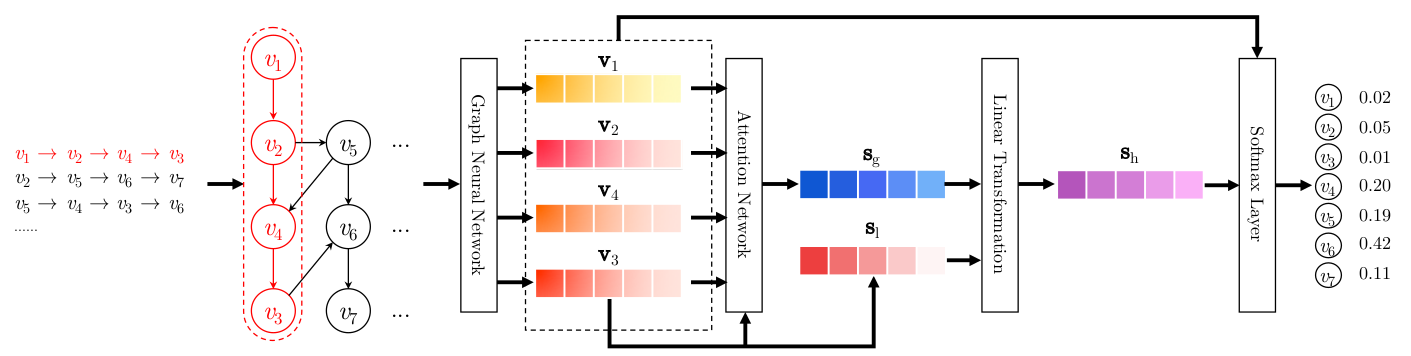}
	\caption{The workflow of the proposed \emph{SR-GNN} method, from \cite{wu2019session}}
	\label{figure:sr-gnn}
\end{figure}

Graph Neural Networks (GNN) are designed for generating representations for graphs \cite{scarselli2008graph, li2015gated}. It was recently proposed a method for \emph{Session-based Recommendation using Graph Neural Networks (SR-GNN)} \cite{wu2019session}, which is illustrated in \ref{figure:sr-gnn}.

In the first step of their method, session sequences are modeled as graph-structured data. Then, their proposed GNN captures transitions of items and generates item embedding vectors correspondingly. Each session is represented as the composition of the global preference and the current interest of that session using an attention network. Finally, it is predicted the probability of each item that will appear to be the next-click one for each session.

The authors of \emph{SR-GNN} argue that their proposed model is more capable to capture complex transitions of items than RNN-based methods for session-based recommendation.

Their experiments were conducted on two e-commerce datasets and the \emph{SR-GNN} indeed outperformed other neural architectures for session-based recommendation methods, such as \textit{GRU4Rec} \cite{hidasi2016}, \textit{NARM} \cite{Li2017narm} and \textit{STAMP} \cite{Liu2018stamp}.

The state-of-the-art \emph{SR-GNN} method is one of the baselines used in our experiments, as described in Section~\ref{sec:eval_methodology}.

\subsection{Other Session-based Approaches}
In \cite{tan2016improved}, they employed data augmentation strategies to generate additional sessions by taking every possible sequence, by starting from the beginning of a session and also randomly removing some items from these sequences (embedding dropout). Naturally, those additional sessions made training time much longer, as pointed by \cite{hidasi2018recurrent}. They also have tried to employ \emph{Student-Teacher training}, by using items sequence clicked by users after a target item as privileged information \cite{vapnik2009new}, by using generalized distillation \cite{lopez2015unifying}, during the training time, even though this "future" information was not available in testing time. Future items in the sequence were trained by a Teacher network, and a Student network was used as input to the previous items of the target item, having Teacher's output as a soft label. Therefore, this model did not performed well, in terms of training time and accuracy (only a slight improvement for small datasets).

In \cite{twardowski2016modelling}, it was also used RNNs in a session-based recommendation system. His network architecture input receives item metadata and contextual information from events in the session. Textual information is represented as: bag-of-words vectors (loosing words ordering and text semantics); categorical information, as one-hot vector; and numeric attributes,  scaled to [0, 1]. Embedding layers convert this input representations into lower-dimensional continuous spaces. Event embeddings are passed in time sequence to an RNN, which hidden state is further merged with item embeddings in a feed-forward network. Their experiments on products and ads datasets have presented superior performance, if compared to some baselines as: a popularity model and a content-based model.

In \cite{Li2017narm}, it was proposed the \emph{Neural Attentive Recommendation Machine (NARM)}, a hybrid neural encoder-decoder
architecture leveraging GRU layers and attention mechanisms to model user's sequential behavior and capture user's main purpose in the current session, which are combined as a unified session representation later \cite{Li2017narm}. The recommendation scores is computed for each candidate item with a bi-linear matching scheme based on this unified session representation. \emph{NARM} trains jointly item and session representations, as well as their matches. Their experiments were performed on e-commerce datasets, and \emph{NARM} was able to outperform \emph{GRU4Rec}, as well as other KNN-based baselines.

In \cite{Liu2018stamp}, the \emph{Short-Term Attention/Memory Priority Model (STAMP)} was proposed. Such model is able to capture users' general interests from the long-term memory of a session context, whilst taking into account users' current interests from the short-term memory of the last-clicks. According to the authors, their idea is similar to \emph{NARM} \cite{Li2017narm}, although \emph{STAMP} explicitly emphasizes the current interest reflected by the last click to capture the hybrid features of current and general interests from previous clicks, while \emph{NARM} only captures the general interests. Furthermore, differently than  \emph{NARM} which uses GRU, the \emph{STAMP} is based on MLP with attention mechanisms. According to their experiments in e-commerce datasets, it resulted in higher accuracy and faster training time than \emph{NARM}.

\section{Deep Feature Extraction from Heterogenous Data}

Traditional CBF and Hybrid RS generally represent item's textual information as bags-of-words \cite{melville2002content} \cite{agarwal2009} or TF-IDF encodings, or rely on topic modeling \cite{wang2011collaborative} \cite{gopalan2014content}. A potential drawback of these approaches is that they do not take word orders and the such as surrounding words and word orders \cite{kim2016convolutional}. Furthermore, the extracted textual features may not be necessarily relevant for recommendation \cite{bansal2016ask}.

With deep learning, it became easier to reliably extract useful features directly from content and use them for recommendations. Depending on the domain, the content and its processing can greatly differ \cite{hidasi2017dlrs}. 

For recommendations based on images \cite{mcauley2015image} \cite{he2016deep}, generally Convolutional Neural Networks (CNNs) are used. For textual items, common approaches are weighted words embeddings, paragraph vectors, CNNs, and RNNs \cite{bansal2016ask}.

Deep architectures for hybrid RS leverage interaction data (Collaborative Filtering) with items metadata (Content-Based Filtering) generally involves the following three main steps:
\begin{enumerate}
	\item \textbf{Initializing} - to obtain item representations based on metadata and use them as initial item features;
	\item \textbf{Regularizing} - to compare metadata representations with interactions' representations (which should be close) adding a regularizing term to loss off this difference; and
	\item \textbf{Joining} - to generate an item feature vector, by concatenating metadata embedding (fixed by item) with learned users' interactions representations.
\end{enumerate}

\subsection{Hybrid RS Leveraging Textual Reviews}

The next subsections cover hybrid approaches for item recommendations, by using their side textual information like movies, products, and restaurants' reviews.

\subsubsection{The \emph{Convolutional Matrix Factorization (ConvMF)}}

\begin{figure}[h]
	\centering
	\includegraphics[width=15cm]{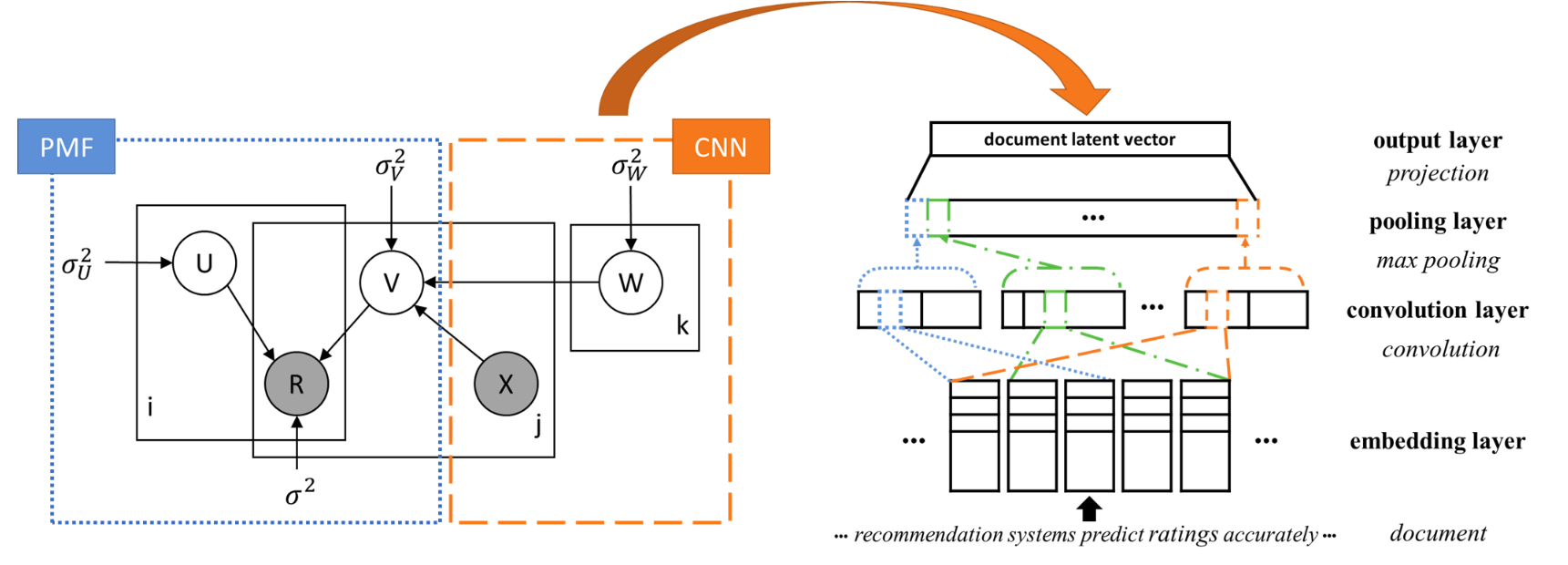}
	\caption{The \emph{ConvMF} architecture, from  \cite{kim2016convolutional}}
	\label{figure:convmf}
\end{figure}

In \cite{kim2016convolutional}, the authors have proposed the \emph{Convolutional Matrix Factorization (ConvMF)} shown in Figure \ref{figure:convmf}, by integrating the CNN with the \emph{Probabilistic Matrix Factorization (PMF)}. Consequently, the ConvMF captures content information from documents and further enhances the rating prediction accuracy. 

The integrated CNN component uses pre-trained word embeddings and is able to capture subtle contextual difference of a word in a document. Their experiments have shown that the ConvMF had a superior accuracy (lower \emph{RMSE}) over \emph{PMF} and \emph{CDL} approaches.

\subsubsection{The \emph{Dual Attention-based Model (D-Attn)}}

In \cite{seo2017interpretable}, authors also proposed to model user preferences and item properties, by using CNNs, because of its ability to extract complex features. They trained user and item networks jointly, enabling the interaction between users and items similarly, as matrix factorization.

But, differently from previous works, like \cite{kim2016convolutional}, they have used dual local and global attention mechanisms. The local attention selects informative keywords from a local window, before the words are fed into the convolutional layer, providing insight on user's preferences or items' properties. The global attention helps CNNs to focus on the semantic meaning of the whole review text. 

Thus, combined local and global attentions have enabled an interpretable and improved representation of users and items.

\begin{figure}[h]
	\centering
	\includegraphics[width=15cm]{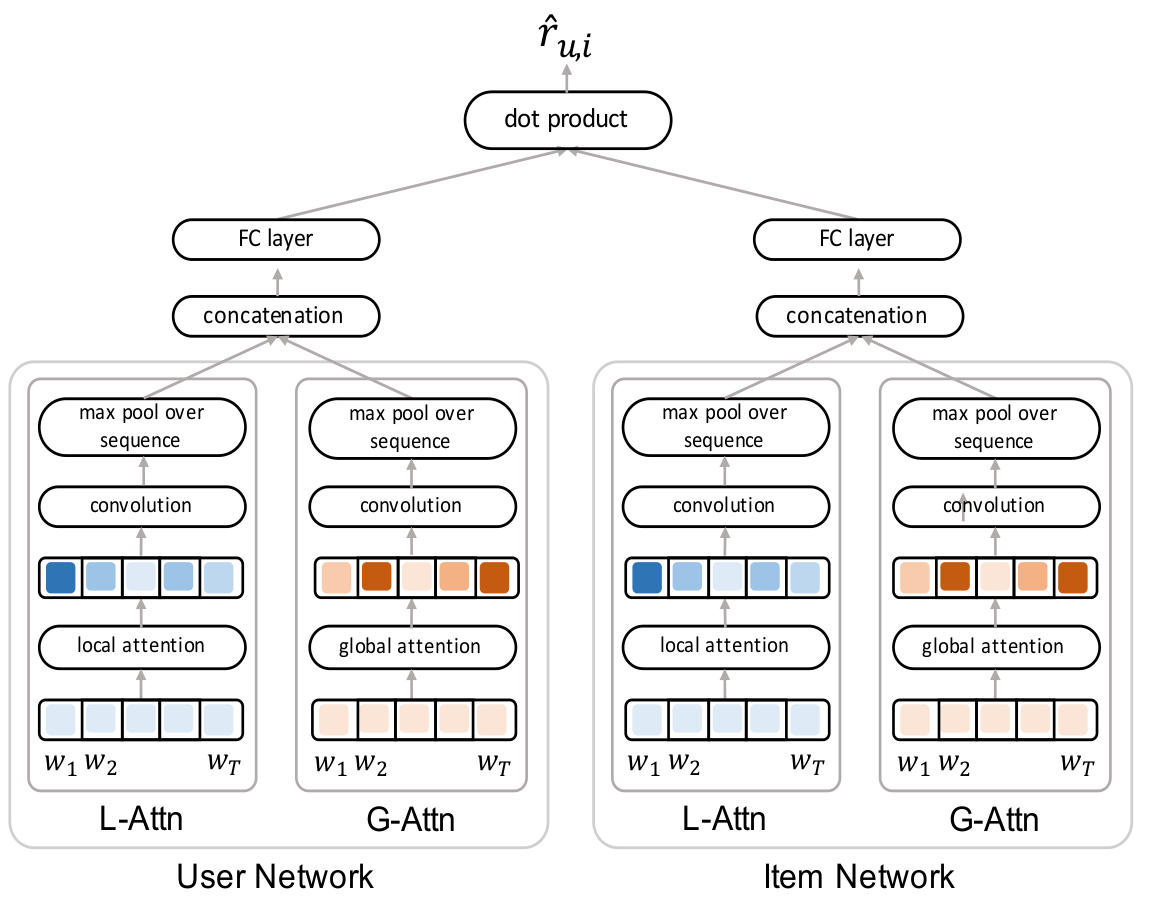}
	\caption{The \emph{D-Attn} architecture, from \cite{seo2017interpretable}}
	\label{figure:dattn}
\end{figure}

Their D-Attn architecture, shown in Figure \ref{figure:dattn}, outperformed \emph{ConvMF+}, matrix factorization, and topical (\emph{HFT}) model on popular review datasets (Yelp and Amazon).

\subsubsection{The Deep Cooperative Neural Networks (DeepCoNN)}
\begin{figure}[h]
	\centering
	\includegraphics[height=10cm]{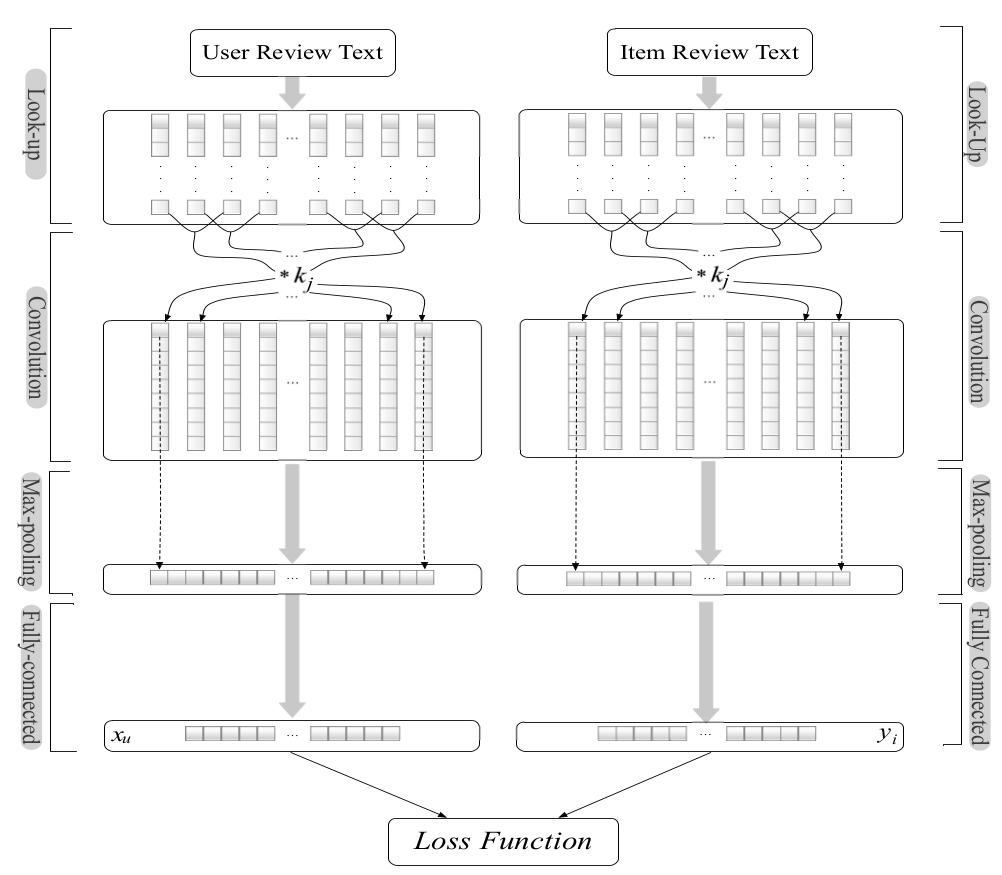}
	\caption{The architecture of \emph{DeepCoNN}, from \cite{zheng2017joint}}
	\label{figure:deepconn}
\end{figure}

In \cite{zheng2017joint}, the authors have proposed a model named Deep Cooperative Neural Networks (DeepCoNN), which consists of two parallel CNNs coupled in the last layers, as presented in Figure \ref{figure:deepconn}. 

One of the networks focuses on learning user behaviors, exploiting reviews written by user, and the other one learns item properties from reviews written for the item. 

A shared layer is introduced on top, to couple these two networks together. The shared layer enables latent factors learned for users and items to interact with each other, in a manner similar to Factorization Machine (FM) techniques. 

To capture the semantic meaning existing in review texts, the \emph{DeepCoNN} represents review texts, by using pre-trained word-embedding techniques \cite{turian2009} \cite{mikolov2010} \cite{mikolov2013} to extract semantic information from reviews. 

Their experimental results demonstrated that the \emph{DeepCoNN} outperformed a set of baseline recommender systems on a variety of datasets.


One drawback of the \emph{DeepCoNN} is that, similarly to MF, it cannot deal with those users or items without ratings. The authors recommend the incorporation of user demographic information such as ages or genders, to enable reasonable recommendations to newly joining users \cite{zheng2016}.


\subsubsection{The \emph{Transformational Neural Network (TransNet)}}

\cite{catherine2017transnets} have proposed an extension of \emph{DeepCoNN} architecture named \emph{Transformational Neural Networks (TransNet)}. 

In their experiments, they have observed that much of the predictive value from the \emph{DeepCoNN} came from reviews of the target user for the target item in training set and that can be considered a type of data leak. Such evaluation setting was unrealistic, as user reviews on unseen items would not be available in operational system (test time).

The \emph{TransNet} extended the \emph{DeepCoNN} model, by introducing an additional latent layer, representing an approximation of the review, and corresponding to the user-target item pair. This layer is used in training time as a regularization, to be similar to the latent representation of the actual review written by the target user for the target item.

\begin{figure}[h]
	\centering
	\includegraphics[height=10cm]{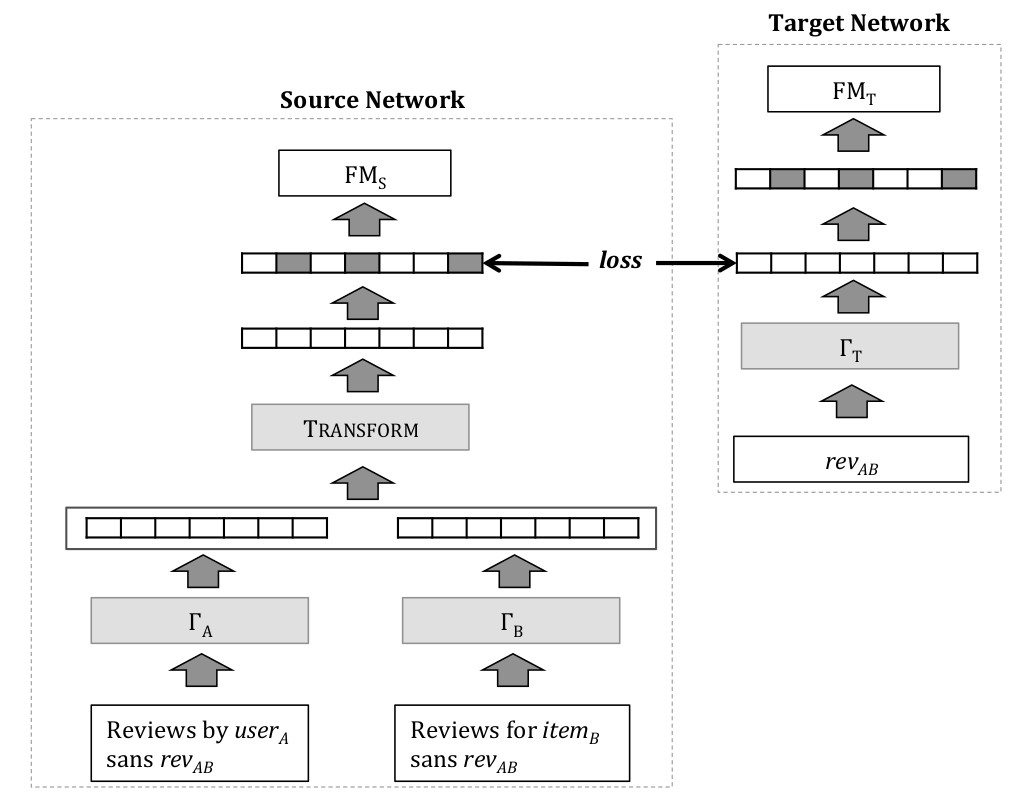}
	\caption{The \emph{TransNet} architecture, from \cite{catherine2017transnets}}
	\label{figure:transnet}
\end{figure}

The \emph{TransNet}, as shown in Figure \ref{figure:transnet}, consists of two networks: a Target Network, that processes the target review $rev_{AB}$; and a Source Network, that processes the texts of the (user, item) pair that does not include the joint review  $rev_{AB}$ .

The \emph{TransNet} also uses only the text of reviews and is user/item identity-agnostic - the user and the item are fully represented, by using review texts, and their identities are not used in the model. They also have proposed the Extended \emph{TransNet} (\emph{TransNet-Ext}) architecture, which uses identities of users and items to learn a latent representation, similarly to Matrix Factorization.




\subsection{Text Recommendations Using Deep Learning}
The next sections present research on textual items of RS like blogs, social media posts, research papers, and news articles using Deep Learning.

\subsubsection{Ask the GRU: Multi-task Learning for Deep Text Recommendations}
In \cite{bansal2016ask}, authors present a method leveraging Recurrent Neural Networks (RNNs) \cite{werbos1990backpropagation} to represent text items for collaborative filtering. 


As RNNs are high capacity models, thus prone to overfitting, authors used multi-task learning for regularization. They have used a simple side task for predicting item metadata such as genres or item tags. In such approach, the network producing vector representations for items directly from their text content is shared for both tag predictions and recommendation tasks. This allows to make predictions even in cold-start conditions, while providing regularization for the recommendation model \cite{bansal2016ask}.

\begin{figure}[h]
	\centering
	\includegraphics[width=15cm]{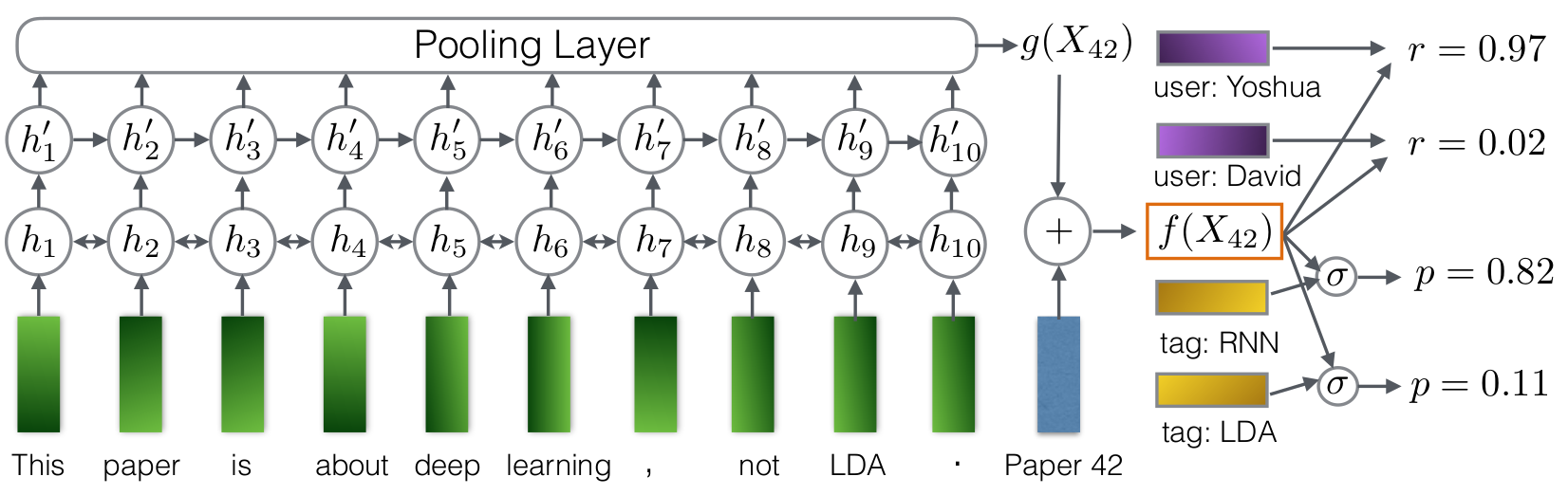}
	\caption{The Architecture for text item recommendation, proposed from \cite{bansal2016ask}}
	\label{figure:askgru}
\end{figure}

The proposed architecture is presented in Figure \ref{figure:askgru}. Items were represented by an item-specific embedding combined with a two-layer GRU (RNN) - a bidirectional layer followed by an unidirectional layer, which hidden states of the sequence were polled to form its representation. Users and tags are also represented by embeddings, which are combined with the item representation to do tag prediction and recommendation (multi-task learning). The model was end-to-end, which means that all parameters were trained simultaneously.

They evaluated their model on the task of scientific paper recommendation, by using two publicly available datasets, where items are associated with text abstracts \cite{wang2011collaborative} \cite{wang2015}.

The authors found that the RNN-based models yielded up to 34\% relative-improvement in \emph{Recall@50} for cold-start recommendation over Collaborative Topic Regression (CTR) approach of \cite{wang2011collaborative} and a word-embedding based model \cite{weston2011wsabie}, while giving competitive performance for warm-start recommendation. Finally, they pointed out that multi-task learning improves the performance of all models significantly, including baselines.


\subsubsection{Quote Recommendation in Dialogue Using Deep Neural Network}
In \cite{lee2016quote}, the authors have introduced the task of recommending quotes which are suitable for given dialogue context.

\begin{figure}[h]
	\centering
	\includegraphics[height=10cm]{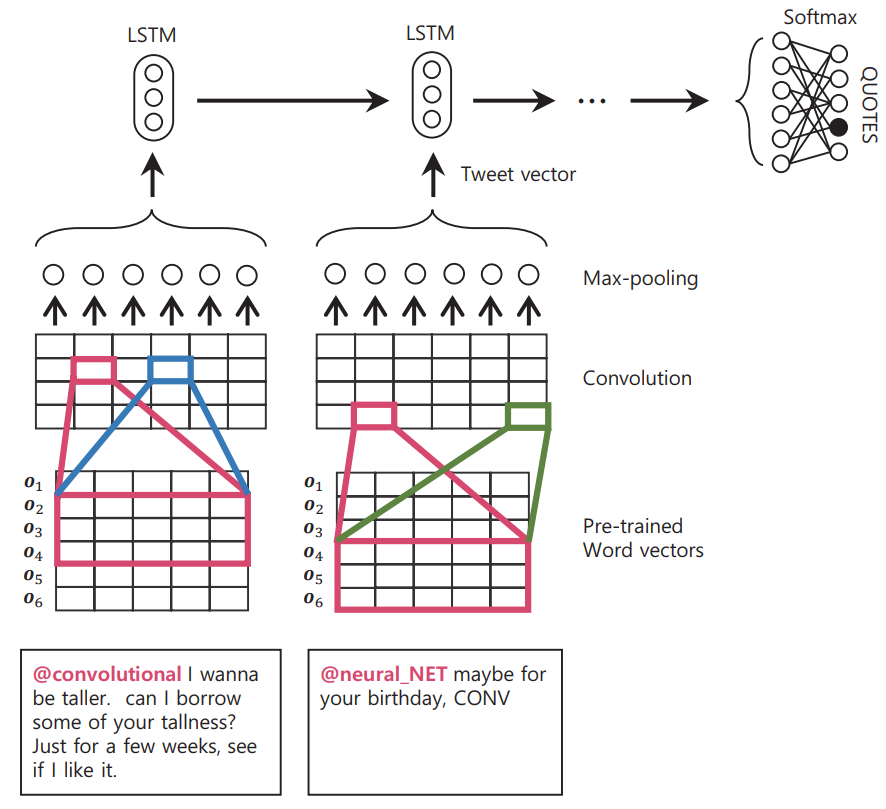}
	\caption{The Architecture of a Deep architecture for quote recommendation, from \cite{lee2016quote}}
	\label{figure:quotes_deep_rec}
\end{figure}

Their architecture, shown in Figure \ref{figure:quotes_deep_rec}, have modeled semantic representation of each utterance and constructed a sequence model for the dialog thread.

The CNN maps tweets in the thread to their distributional vectors. And then, the sequence of tweet distributional vectors are fed to the RNN so as to compute the relevance of target quotes to the given tweet dialogue.

To evaluate the model, they have collected a large set of Twitter dialogues with quote occurrences, in order to evaluate the proposed recommender system. Their experimental results have shown their approach outperforming not only other state-of-the-art algorithms in quote recommendation task, but also other neural network based methods built for similar tasks. 

\subsubsection{News Recommendations with DNN}
\label{sec:news_rec_dnn_park2017}

In \cite{park2017deep}, researchers from \emph{NAVER} news portal proposed a hybrid neural architecture for news recommendation based on RNNs. Their article embeddings were trained from bag-of-words representation from queries, titles, and contents of
the news articles. 

Trials with different losses functions were reported in that work, without (cosine similarity, MSE) and with negative samples (\emph{BPR} and \emph{TOP1} \cite{hidasi2016}).

They have also proposed a personalization re-ranking approach, in which the users content preference is a weighted average of the categories of articles the user has read recently. For articles without an editorial-defined category, they used a CNN-based text classifier to predict the article categories.

They trained two RNN models, session-based and history-based
RNN models, the latter one considering all clicks of a user.

Surprisingly, the accuracy of their session-based RNN model was much higher than for the history RNN model -- which already included the current session clicks. That result reinforces how determinant can be the short-term interests to model next-click predictions in session-based recommendation scenarios.

The neural architecture for news recommendation proposed by \cite{park2017deep} have some similarities and many differences to our research, which are discussed next in Section~\ref{sec:comparison-works}.

\subsubsection{\emph{Deep Knowledge-aware Network for News Recommendation} (DKN)}
\label{sec:dkn}

\begin{figure}[h]
	\centering
	\includegraphics[height=9cm]{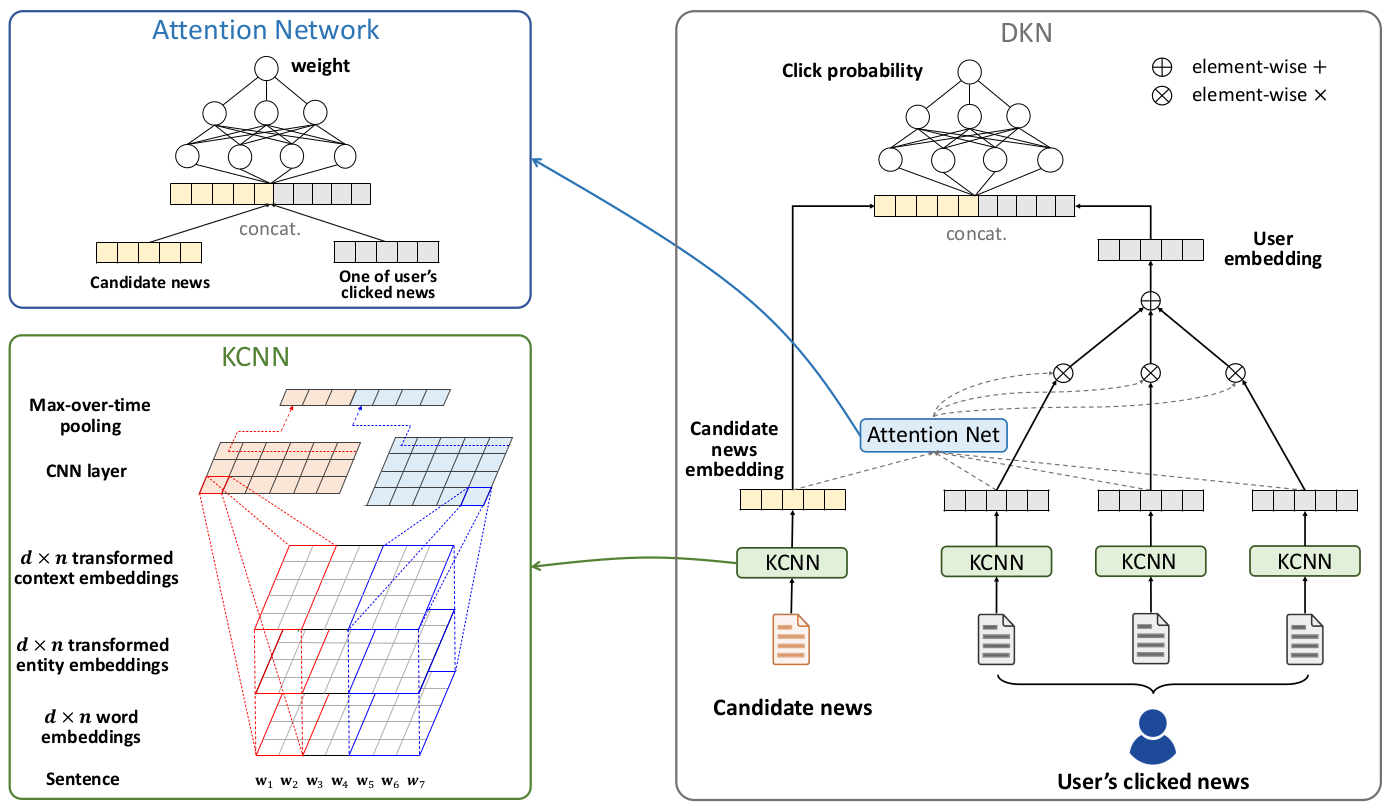}
	\caption{The \emph{DKN} framework, from \cite{wang2018dkn}}
	\label{figure:dkn}
\end{figure}

In \cite{wang2018dkn}, it was proposed the \emph{Deep Knowledge-Aware Network (DKN)}, that incorporates knowledge graph representation for news recommendation, which is presented in Figure~\ref{figure:dkn}.

\emph{DKN} is a content-based model for Click-Through Rate (CTR) prediction, which takes one piece of candidate news and one user's click history as input, and outputs the probability of the user clicking the news.

Specifically, for a piece of input news, their method enrich the news title information by associating each word in the news content with a relevant entity in a knowledge graph and expand with its immediate neighbors to provide complementary information.

The key component of \emph{DKN} is a multi-channel and word-entity-aligned knowledge-aware convolutional neural network that fuses semantic-level and knowledge-level representations of news. It treats words and entities as multiple channels, and explicitly keeps their alignment relationship during convolution. In addition, to address users' diverse interests, they also use an attention mechanism to dynamically aggregate a user's history with respect to current candidate news.

Their experiments showed that the proposed method was able to provide more accurate recommendations than other Deep Learning architectures, such as \emph{Wide \& Deep} \cite{cheng2016wide}, \emph{DeepFM} \cite{guo2017deepfm}, and \emph{DMF} \cite{xue2017deep}.

Some similarities and differences between \emph{DKN} and \emph{CHAMELEON} are discussed in Section~\ref{sec:comparison-works}.

\section{Multi-View Learning for Recommender Systems}

Many domains have multiple data source modalities like visual, audio, textual and structured data. Multi-view or multi-modal learning aims to learn how to model latent factors for each view and jointly optimizes all the functions to improve the generalization performance \cite{zhao2017multi}. Multi-view learning algorithms can be classified into three groups: co-training, multiple kernel learning, and subspace learning. 

Subspace learning-based approaches are of special interest in this research. Such approaches aim to obtain a latent subspace shared by multiple views by assuming that the input views are generated from this latent subspace \cite{xu2013survey}. 

The \emph{Deep Structured Semantic Model (DSSM)} \cite{huang2013learning}, described with more detail in Appendix~\ref{sec:DSSM}, is a deep neural network for learning semantic representations of entities in a common continuous semantic space and measuring their semantic similarities. It projects user queries and documents features to a shared latent space, where the similarity between the clicked documents retrieved by a query is maximized. 

The \emph{DSSM} was further extended to RS by \emph{MV-DNN} \cite{elkahky2015multi}, \emph{TDSSM} \cite{song2016multi}, and \emph{RA-DSSM} \cite{kumar2017} architectures, which are discussed in the next sub-sections.

\subsection{The \emph{Multi-View Deep Neural Network (MV-DNN)}}
\label{sec:MV-DNN}
\cite{elkahky2015multi} proposed a Deep Learning architecture to map users and items to a latent space, where the similarity between users and their preferred items is maximized. This architecture was named \emph{Multi-View Deep Neural Network (MV-DNN)}. In the literature, multi-view learning is a well-studied area which learns from data that do not share common feature space \cite{sun2013survey}.

\begin{figure}[h]
	\centering
	\includegraphics[height=7cm]{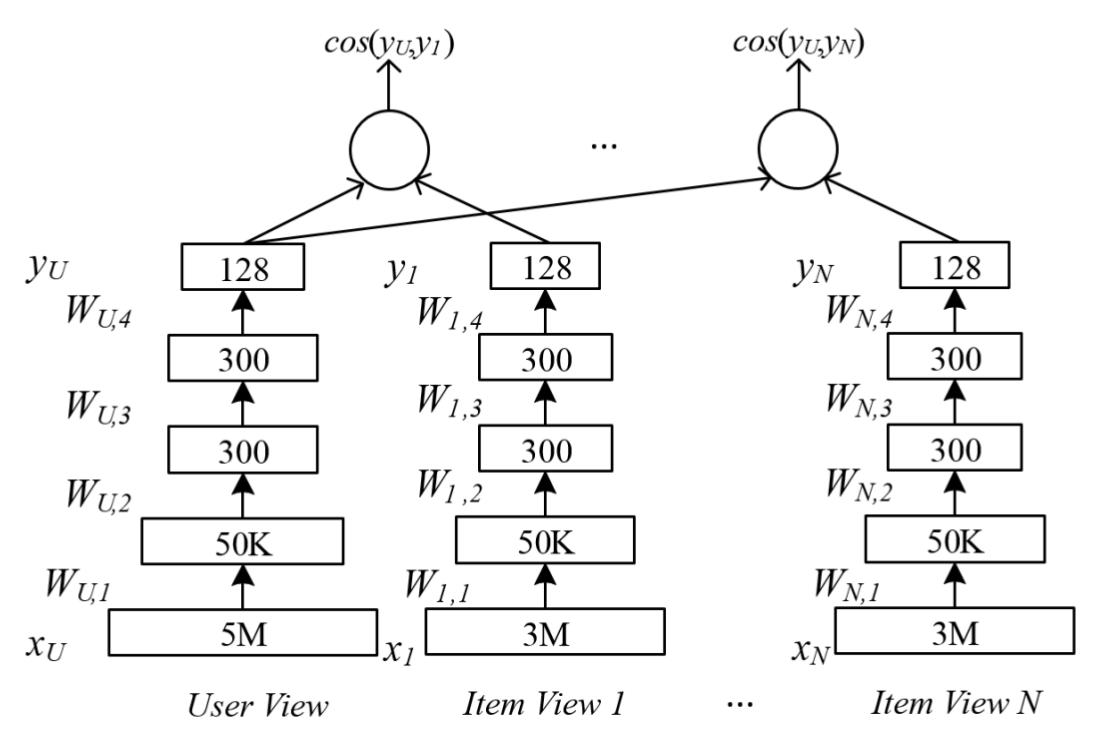}
	\caption{The Multi-View Deep Neural Network, from \cite{elkahky2015multi}}
	\label{figure:mv_dnn}
\end{figure}

The \emph{MV-DNN}, shown in Figure \ref{figure:mv_dnn}, is an extension of the \emph{DSSM} \cite{huang2013learning}. It was designed for cross-domain recommendation, treating users as the pivot view and each domain (supposing Z domains) as an auxiliary view, leading to Z+1 feed forward networks in total. The \emph{MV-DNN} projects users and items, each of which is represented by a rich feature set, through non-linear transformation layer(s) to a compact shared latent semantic space where it is maximized the similarity between the mapping of the user and mappings of items liked by the user. 

The user feature representation has allowed the model to learn relevant user behavior patterns and gives useful recommendations for users who do not have any interaction with the service, given that they have adequate search and browsing history. The combination of different domains into a single model for learning helps improved the recommendation quality across all the domains, as well as having a more compact and a semantically richer user latent feature vector. 

The feature extraction process was not automatically learned by Deep Learning. Textual features were handcrafted, by using \emph{TF-IDF} or letter-\emph{trigrams} and categorical features were one-hot encoded (binary features), leading to highly-dimensional feature vectors (3.5 million features for user vectors and between 50 and 100 thousand features for each item domain features). Because of the high dimension space, some techniques like \emph{K-means} and \emph{Local Sensitive Hashing (LSH)} were used for dimensionality reduction, before the fully connected layers of the neural network.

At that time, authors claimed that their approach was significantly better than the state-of-the-art algorithms (up to 49\% enhancement on existing users and 115\% enhancement on new users). 

In addition, their experiments on a publicly open data set also indicate the superiority of their method in comparison with transitional generative topic models, for modeling cross-domain recommender systems. Experimental results have also shown that combining features from all domains produces much better performance than building separate models for each domain.

\cite{zheng2017joint} critique the aspect that \emph{MV-DNN} outputs were coupled with a cosine similarity objective function to produce latent factors with high similarity. In this way, user and item factors were not learned explicitly in relation to the rating information, with no guarantee that the learned factors could help the recommendation task.

\subsection{The \emph{Temporal DSSM (TDSSM)}}
\label{sec:TDSSM}

\cite{song2016multi} pointed out that in the \emph{MV-DNN} architecture (and \emph{DSSM} as well), both the user view and item view were static, in the sense that the input features represented a holistic view of users interests, during a certain period of time. The model thus lacked the ability of responding promptly to some scenarios where freshness and temporal dynamics of items are sometimes more important than the content relevance itself, like news recommendation.

\begin{figure}[h]
	\centering
	\includegraphics[height=6cm]{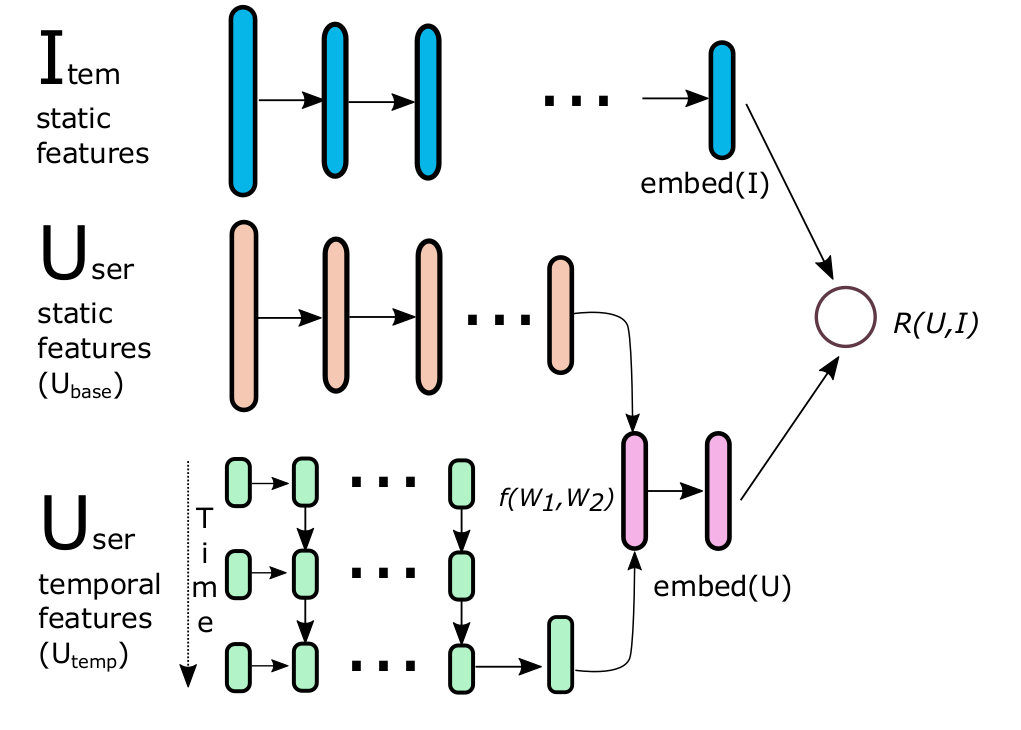}
	\caption{The \emph{Temporal DSSM (TDSSM)} architecture, \cite{song2016multi}}
	\label{figure:tdssm}
\end{figure}

The authors have introduced the \emph{TDSSM} architecture, shown in Figure \ref{figure:tdssm}, where user features are modeled by the combination of a static and a time dependent part \cite{song2016multi}. The key difference to \emph{MV-DNN} is the usage of an RNN to model user interests at different time spots. 

In their experiment, they have used previous two-weeks clicks as short-term history, with each day as a time spot. Therefore, the length of the LSTM for the \emph{TDSSDM} was 14 steps. For the \emph{Multi-Rate TDSSM (MR-TDSSM)}, they use two LSTMs in different rates, where the fast-rate LSTM used daily signals and the slow-rate LSTM leveraged weekly signals.

To speedup training and reduce the number of parameters to learn, they have used the original \emph{DSSM} to pre-train user and item embeddings, which were the inputs to the \emph{MR-TDSSM} model.

Their results were superior than matrix factorization and popularity baselines.

\subsubsection{The Embedding-based News Recommendation}
\label{sec:embeddings_based}

In \cite{okura2017embedding}, it was proposed an embedding-based method
for news recommendation. Their method has three steps: (i) start with distributed representations of articles content (bag-of-words) based on a variant of a denoising autoencoder, (ii) generate user representations by using a recurrent neural network (RNN) with browsing histories as input sequences, and (iii) match and list articles for users based on inner-product operations by taking system performance into consideration.

\begin{figure}[h]
	\centering
	\includegraphics[height=4cm]{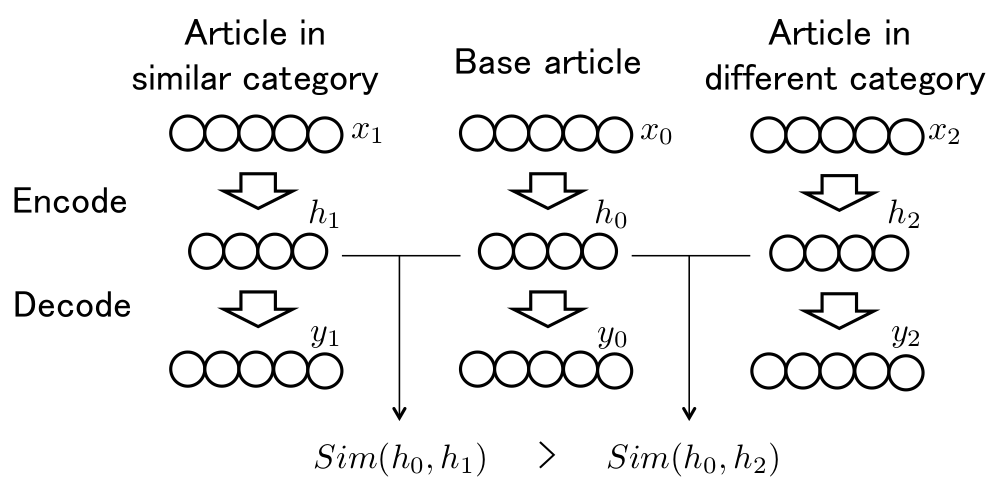}
	\caption{Encoder for triplets of articles, from \cite{okura2017embedding}}
	\label{figure:okura2017embedding}
\end{figure}

For step (i), they proposed an interesting approach to generate article embeddings unsupervisedly, adapting the regular loss function (reconstruction error) used for denoising autoencoder to include a regularization that ensures that the categorical similarities among the embeddings of the original articles are preserved, as shown in Figure~\ref{figure:okura2017embedding}.

For step (2), they have used a loss function based on triplets with the user representation, and a pair of article representations (one positive sample and one negative sample). It is important to note that, instead of using randomly sampling articles as negative instances, they have used as negative samples the non-clicked items recommended for users during their sessions, as they had such data available.

In step (3), the recommendations relevance is obtained by a simple inner-product relation between an article and a user representation (learned by an RNN).

They report their offline and online experiments on Yahoo! Japan's homepage. In the offline evaluation, authors compared different baselines with RNNs-based techniques to learn user representations, and the GRU provided the most accurate recommendations.

In the online evaluation, their proposed model provided an improvement of 23\% for CTR and 10\% for total duration, when compared with their baseline method in production.

\subsection{The \emph{Recurrent Attention DSSM (RA-DSSM)}}
\label{sec:kumar2017}

In \cite{kumar2017user}, it is proposed a neural architecture based on \emph{DSSM} for news recommendation. That model provides sequence-based (not session-based) recommendation, considering the sequence of all available clicks for given user, ignoring the temporal characteristics of news domain. 

Initially, they learned \emph{doc2vec} embeddings to represent the articles content. Then, they proposed simple approaches to build users profiles based on weighted averages of those \emph{doc2vec} embeddings, where the latter clicks are more influential to the user profile. The network is trained by using the loss function from \emph{DSSM} \cite{huang2013learning}. 

\begin{figure}[h]
	\centering
	\includegraphics[height=8cm]{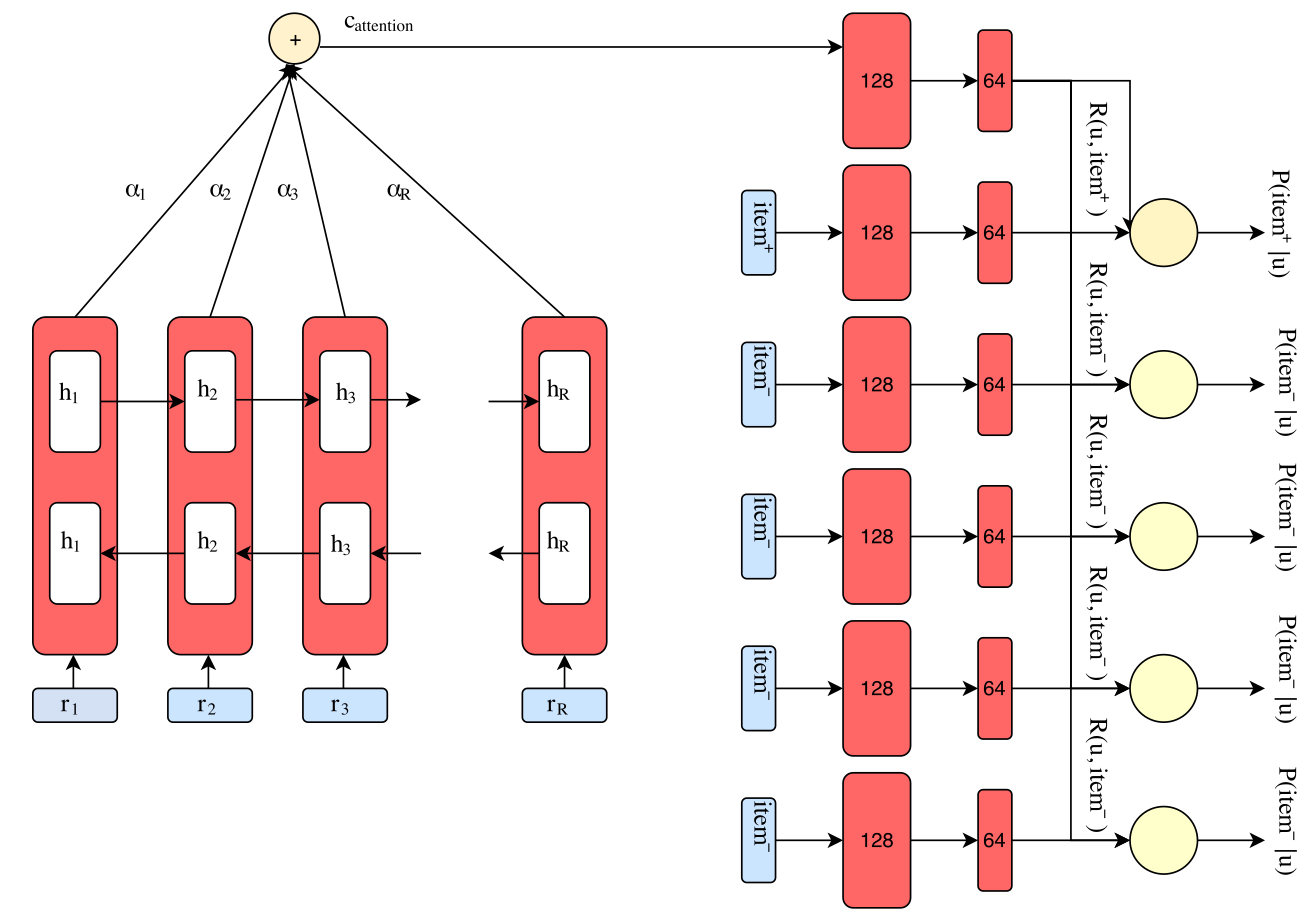}
	\caption{The \emph{Recurrent Attention DSSM} architecture, from \cite{kumar2017}}
	\label{figure:ra-dssm}
\end{figure}

In \cite{kumar2017}, the same authors proposed an RNN-based model to build user profiles, named \emph{Recurrent Attention DSSM (RA-DSSM)}, as shown in Figure \ref{figure:ra-dssm}. 

For learning a non-linear mapping for users, they use an attention-based recurrent layer in combination with fully connected layers. The sequence in which the articles are read by the user encapsulates information about  long-term user interests, which are processed by an LSTM \cite{hochreiter1997long,sutskever2014sequence}. They have used bidirectional LSTMs \cite{schuster1997bidirectional}, to capture both static and dynamic interests, which the user has developed over time. A neural attention mechanism \cite{bahdanau2014neural} was also incorporated to learn the extent of each user interests.

For learning mappings for items, only fully connected layers are used, based on the textual content of news articles. The articles content was represented as \emph{doc2vec} embeddings \cite{le2014distributed} with 300 dimensions.

In order to capture the similarity between users and items, they were projected to the same latent space, by adapting the \emph{Deep Structured Semantic Model (DSSM)} \cite{huang2013learning} presented in Section \ref{sec:DSSM}. For each positive sample, it was used 4 negative instances randomly sampled (i.e., articles not interacted by the user), independently of the articles publishing date. 

Finally, a ranking based loss function was used to learn the parameters of the network. To recommend news articles to the users, they have used the computed inner product between user and item latent vectors.

In their offline experiments in a real-word dataset published by the \emph{CLEF-NewsREEL 2017} competition \cite{kille2013plista, kille2017clef, lommatzsch2017}, they could get a 4.7\% improvement (Hit Ratio@10) over other methods such as \emph{BPR}, \emph{eALS}, and \emph{NeuMF}. Their model also has shown some effectiveness in handling user cold-start and item cold-start problems. 

Therefore, the evaluation protocols used by \cite{kumar2017user} and \cite{kumar2017} were not realistic for  real news portals for some reasons. First at all, their experiments reserved only the latter user click for the test set, and trained upon the all the previous user clicks. This protocol did not emulate session-based recommendations, in which it would not be able to train the model for every new click of a user session to provide next-click recommendation. Another issue is that their protocol ignores the temporal aspects on news domain, training and evaluating on sequences of user clicks from different days, with very different time intervals among the clicks, and providing negative samples that may be not reasonable, e.g., sampling a very old articles or articles published in a far future.

The research group have also proposed a 3D CNN \cite{kumar2017word} to capture temporal changes in users interests and to provide better recommendations, combining users interactions with the content of the articles. 

Finally, in a subsequent article \cite{khattar2018neural}, the news recommendation is treated as a binary classification problem. Similar heuristics of \cite{kumar2017user} are used to build users profiles. Instead of using a similarity-based loss function, the user profile and article representations are element-wise multiplied and passed through a number of feed forward layers to output the predicted score for a given user and item pair.

The same research group have proposed different neural architectures to tackle sequence-based news recommendation \cite{kumar2017user, kumar2017word, kumar2017, khattar2018neural}. All their experiments use the same dataset and evaluation protocol, but they did not report a comparison between their proposed architectures, unfortunately. 

As the \emph{RA-DSSM} \cite{kumar2017} have some similarities to the proposed neural architectures in this research, a comparison is developed in Section~\ref{sec:comparison-works}

\chapter{CHAMELEON - A Deep Learning Meta-Architecture for News Recommender Systems}
\label{sec:chapter_4}

This chapter describes the proposed Deep Learning Meta-Architecture for News Recommender Systems -- the \emph{CHAMELEON} \cite{moreira2018chameleon}. 

When building a news recommender system, one has several design choices regarding the types of data that are used, the chosen algorithms, and the specific network architecture when relying on deep learning approaches. \emph{CHAMELEON} provides an architectural abstraction (a ``meta-architecture''), which contains a number of general building blocks for news recommenders, which can be instantiated in various ways, depending on some particularities of the given problem setting.

"Meta" means the concept applied to itself. A meta-architecture can be defined as an architecture for creating architectures \cite{Hakimi2013metaarchitecture}. When working at the meta level, we're not thinking about the things themselves -- but in how those things may change overtime and how to support that change, by evolving the architecture \cite{Bloomberg2014archictecture}. The fundamental reason to work at the meta level is to create an abstraction that indicates which components might change.

For the purpose of this research, a meta-architecture is a reference architecture that collects together decisions relating to an architecture strategy \cite{Malan2004MetaArchitecture}. In a sense, a meta-architecture can be also be seen as a way to reduce the search space for Neural Architecture Search (NAS), a technique for automating the design of artificial neural networks.

A Meta-Architecture might be instantiated as different architectures with similar characteristics that fulfill a common task, in this case, news recommendations. 

Meta-architectures using deep learning have been proposed for different domains. For example, in \cite{fuentes2017robust} and \cite{niitani2017chainercv}, \emph{Faster Region-based Convolutional Neural Network (Faster R-CNN)}, \emph{Region-based Fully Convolutional Network (R-FCN)}, and \emph{Single Shot Multibox Detector (SSD)} were defined as Deep Meta-Architectures for object detection in computer vision. Those meta-architectures abstract the Feature Extractor module, whose components can be instantiated as more concrete CNN architectures, like \emph{VGG-16} \cite{simonyan2014very}, \emph{Inception-v3} \cite{szegedy2016rethinking}, or \emph{Resnet} \cite{he2016deep}. 

\section{The Problem Context}

In this Section, it is characterized the context of the news recommendation problem, in terms of influencing factors and challenges.

\subsection{Factors Affecting News Relevance}

\begin{figure}[h]
	\centering
	\includegraphics[width=15cm]{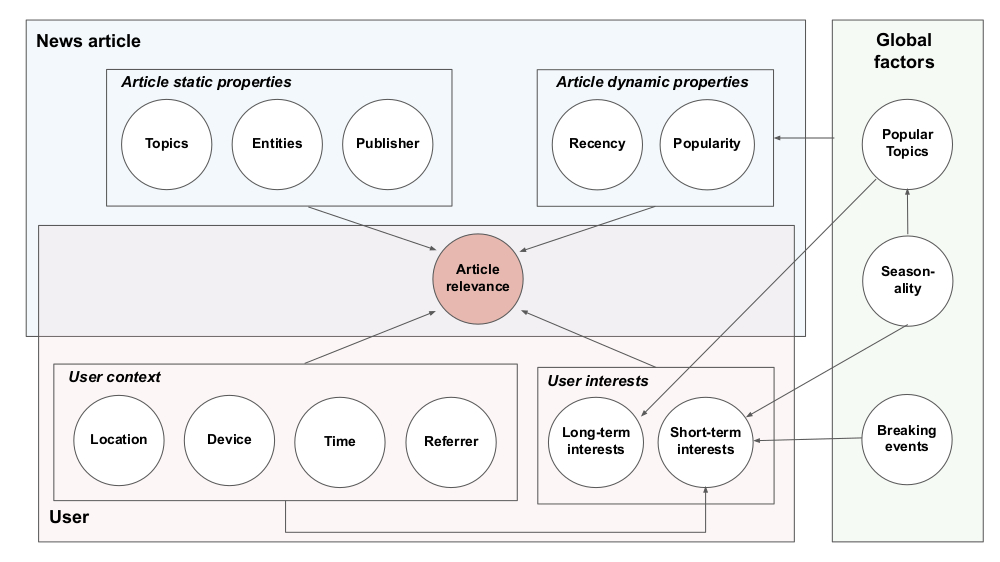}
	\caption{The Conceptual Model of factors affecting news relevance for users}
	\label{figure:relevance-conceptual-model}
\end{figure}

Many factors may influence the relevance of news article for users, as can be seen in the proposed Conceptual Model presented in Figure \ref{figure:relevance-conceptual-model}.

With respect to \emph{article-related} factors, we distinguish between \textbf{static} and \textbf{dynamic properties}. Static properties refer to article's content (text), its title, \textbf{topics}, mentioned \textbf{entities} (e.g., places and people) or other metadata \cite{li2011scene, said2013month}. 
The reputation of the \textbf{publisher} can also add trust to an article \cite{lommatzsch2014real, gulla2016intricacies}.
Some news-related aspects can also dynamically change, in particular its \textbf{popularity}  \cite{chu2009personalized, lommatzsch2017incorporating} and \textbf{recency} \cite{trevisiol2014cold, gulla2016intricacies}. On landing pages of news portals, those two properties are typically the most important ranking criteria and in comparative evaluations, recommending recently popular items often shows to be a comparably well-performing strategy \cite{karimi2018news}.

When considering \emph{user-related} factors, we distinguish between users' \textbf{short-term and long-term interests} and contextual factors. Regarding the context, their \textbf{location}  \cite{fortuna2010real, montes2013towards, tavakolifard2013tailored}, their \textbf{device} \cite{lee2007moners}, and the current \textbf{time} \cite{mohallick2017exploring, montes2013towards} can influence users' short term interests, and thus the relevance of a news article \cite{mohallick2017exploring, said2013month}. In addition, the \textbf{referrer} URL can contain helpful information about a user's navigation and reading context \cite{trevisiol2014cold}.

Considering user's long-term interests can also be helpful, as some user preferences might be stable over extended periods of time \cite{diez2016}. Such interests may be specific personal preferences (e.g., chess playing) or influenced by popular global topics (e.g., on technology). In this work, we address only short-term user preferences, since we focus on scenarios where most users are anonymous. In general, however, as shown in \cite{quadrana2017personalizing}, it is possible to merge long-term and short-term interests by combining different RNNs when modeling user preferences.

Finally, there are \textbf{global factors} that can affect the general popularity of an item, and thus, its relevance for a larger user community. Such global factors include, for example, \textbf{breaking news} regarding natural disasters or celebrity news. Some \textbf{topics} are generally \textbf{popular} for many users (e.g., sports events like Olympic Games); and some follow some \textbf{seasonality} (e.g., political elections), which also influences the relevance of individual articles at a given point in time \cite{chu2009personalized, gulla2016intricacies, lommatzsch2017incorporating}.

The proposed Meta-Architecture models all these factors either explicitly or implicitly to improve recommendations accuracy, depending on current context of users and articles. 

\subsection{News RS Challenges}

The news domain is specially challenging for Recommender Systems \cite{karimi2018news,ozgobek2014survey}. Those challenges are summarized as follows:

The news domain has, however, a number of characteristics that makes the recommendation task particularly difficult \cite{karimi2018news,ozgobek2014survey}, among them the following:

\begin{itemize}
    \item \emph{Extreme user cold-start} - On many news sites, the users are anonymous or not logged in. News portals have often very little or no information about an individual user past behavior \cite{li2011scene, diez2016, lin2014personalized};
    \item \emph{Accelerated decay of item relevance} - The relevance of an article can decrease very quickly after publication and can also be immediately outdated when new information about an ongoing development is available. Considering the recency of items is therefore very important to achieve high recommendation quality, as each item is expected to have a short shelf life \cite{das2007, ozgobek2014survey};
    \item \emph{Fast growing number of items} - Hundreds of new stories are added daily in news portals \cite{spangher2015}. This intensifies the item cold-start problem. However, fresh items have to be considered for recommendation, even if not too many interactions are recorded for them \cite{diez2016}. Scalability problems may arise as well, in particular for news aggregators, due to the high volume of new articles being published \cite{karimi2018news, mohallick2017exploring, ozgobek2014survey}; and
    
    \item \emph{Users preferences shift} - The preferences of individual users are often not as stable as in other domains like entertainment \cite{diez2016}. Moreover, short-term interests of users can also be highly determined by their contextual situation \cite{diez2016, campos2014time, kille2013plista, ma2016user} or by exceptional situations like breaking news  \cite{epure2017recommending}.
\end{itemize}

\section{Requirements}
\label{sec:requirements}
To conceptualize the \emph{CHAMELEON} Meta-Architecture, some requirements were first devised, based on challenges of news recommender systems and also on the capabilities provided by Deep Learning. The \emph{CHAMELEON} Meta-Architecture should be able:

\begin{itemize}
	\item \emph{RR1} - to provide personalized news recommendations in extreme cold-start scenarios, as most news are fresh and most users cannot be identified;
    \item \emph{RR2} - to automatically learn news representations from textual content and news metadata, minimizing the need of manual feature engineering;
	\item \emph{RR3} - to leverage the user session information, as the sequence of interacted news may indicate user's short-term preferences for session-based recommendations;
	\item \emph{RR4} - to leverage users' contextual information, as a rich data source in such information scarcity about the user;
	\item \emph{RR5} - to model explicitly contextual news properties (popularity and recency), as those are important factors on news interest life cycle; 
	\item \emph{RR6} - to support an increasing number of new items and users by incremental model retraining (online learning), without the need to retrain on the whole historical dataset; and
	\item \emph{RR7} - to provide a modular structure for news recommendation, allowing its modules to be instantiated by different and increasingly advanced neural network architectures and methods.
\end{itemize}

\section{The Proposed Meta-Architecture}

For the purpose of this research, the \emph{CHAMELEON} acronym stands for \emph{Contextual Hybrid session-bAsed MEta-architecture applying deep LEarning On News recommender systems}. 

A chameleon was chosen to represent this research for having some commonalities with the proposed Meta-Architecture. First of all, that reptile has a unique ability to adapt his skin coloration, depending upon his context. The proposed Meta-Architecture provides contextual recommendations for the users, depending on their location, time, and used device. 
Secondly, chameleons have the most distinctive eyes of any reptile. Each eye can pivot and focus independently, allowing the chameleon to observe two different objects simultaneously. The proposed Meta-Architecture have two independent and complementary modules: one focused in news articles representation learning and the other targeting news recommendations. 

Finally, chameleons are accurate hunters, as their elastic tongue precisely targets their prey. The proposed Meta-Architecture aims to provide accurate recommendations for news portals' readers.

\begin{figure}[h]
	\centering
	\includegraphics[width=17cm]{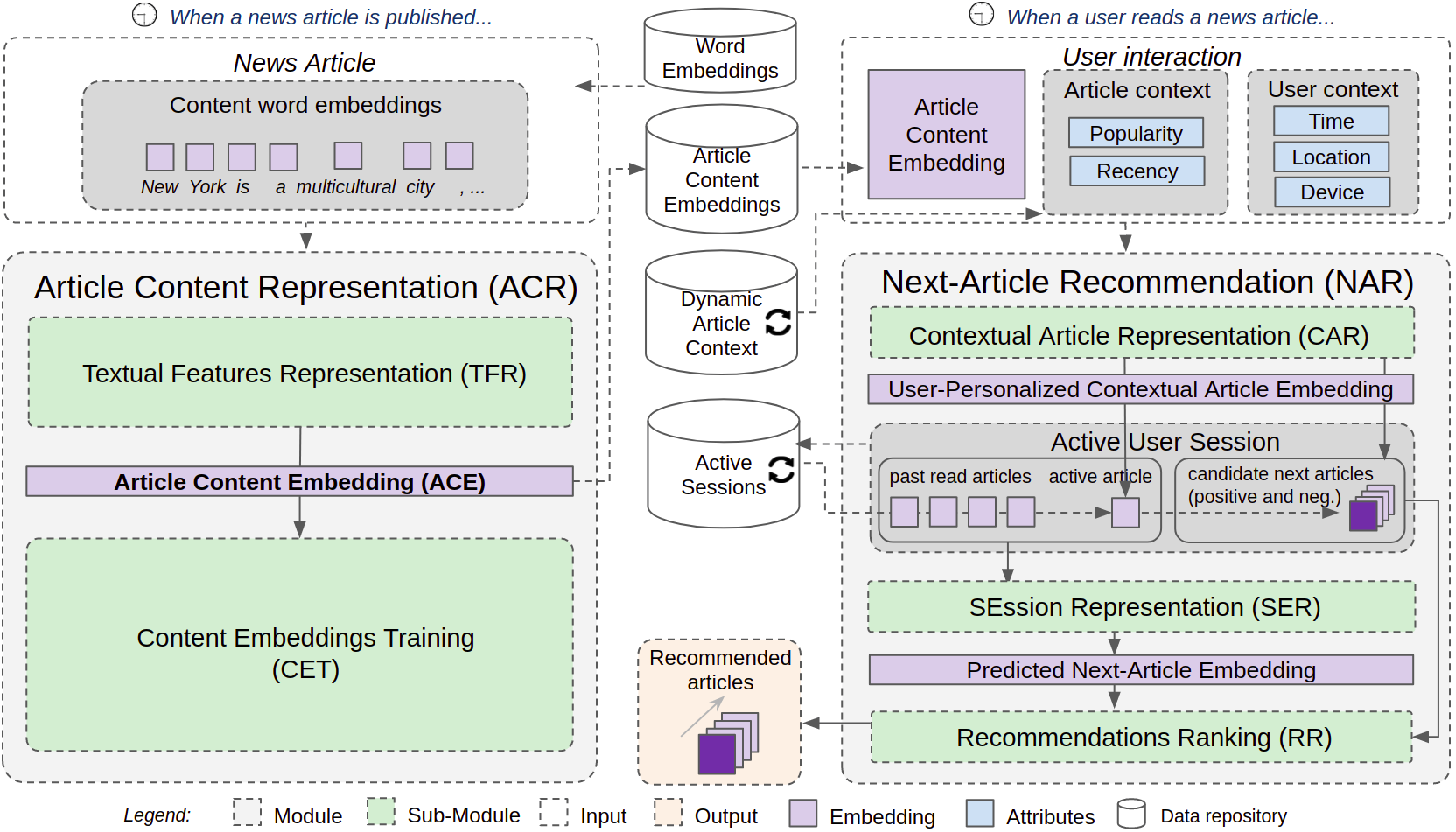}	\caption{CHAMELEON - The Deep Learning Meta-Architecture for News Recommender Systems}
	\label{figure:CHAMELEON}
\end{figure}

The \emph{CHAMELEON} Meta-Architecture was conceptualized based on requirements outlined in Section \ref{sec:requirements}. 

It was structured to support changes, operating at the level of inputs, outputs, modules, sub-modules, and their interactions. Modules and sub-modules can be instantiated by different architectures, as they evolve. Such modular structure also make their components evaluation straightforward.

The main building blocks of \emph{CHAMELEON} are presented in Figure \ref{figure:CHAMELEON}. 
It is based on neural networks, which naturally support incremental online learning from mini-batches (\emph{RR6}).

The \emph{CHAMELEON} is composed of two modules (\emph{RR7}) with independent life cycles for training and inference:

\begin{itemize}
  \item The \emph{Article Content Representation} (\emph{ACR}) module used to learn a distributed representation (an embedding) of articles' content; and
  \item The \emph{Next-Article Recommendation} (\emph{NAR}) module responsible to generate next-article recommendations for ongoing user sessions.
\end{itemize}

They are described in detail in the next sections.

\subsection{The \emph{Article Content Representation (ACR)} Module}
The ACR module is responsible for learning distributed representations (embeddings) from news' contents. 

The input for the ACR module is the article textual content, represented as a sequence of word embeddings. Pre-training word embeddings in a larger text corpus of the target language (e.g., Wikipedia) is a common practice in Deep NLP, by using methods like \emph{Word2Vec} and \emph{GloVe}, as described in Section \ref{sec:deep-nlp}.

The \emph{ACR module} learns an \emph{Article Content Embedding} for each article independently from the recorded user sessions. This is done for scalability reasons, because training user interactions and articles in a joint process would be computationally very expensive, given the typically large amount of daily user interactions in a typical news portal. Instead, the internal model is trained for a side classification task (e.g. predicting target metadata attributes of an article).

The \emph{ACR module} is composed of two sub-modules: \emph{Textual Features Representation (TFR)} and \emph{Content Embeddings Training (CET)}.

\subsubsection{The \emph{Textual Features Representation (TFR)} Sub-module}
\label{sec:news_embeddings_training}
The \emph{TFR} sub-module is responsible to learn relevant features directly from the article textual content (\emph{RR2}). It might be instantiated as a Deep \emph{NLP} architecture, like \emph{CNN} \cite{lee2016quote, catherine2017transnets, seo2017interpretable},  \emph{RNN}, and variations \cite{bansal2016ask}, or \emph{Quasi-Recurrent Neural Networks (QRNNs)} \cite{bradbury2016quasi}. 

The \emph{TFR} sub-module process article's textual content using one of those \emph{Deep NLP} architectures, which are followed by sequence of Fully Connected (\emph{FC}) layers to output the \emph{Article Content Embedding}.

\subsubsection{The \emph{Content Embeddings Training (CET)} Sub-module}

The \emph{Content Embeddings Training (CET)} sub-module is responsible to train the \emph{Article Content Embeddings (ACE)} for a side task. 

It can be instantiated as a \emph{supervised} or \emph{unsupervised} learning model. In the \emph{supervised approach}, the side task is to predict articles' metadata attributes, such as articles categories. In the \emph{unsupervised approach}, the task is to reconstruct the original article text from the learned \emph{ACE}, as an sequence autoencoder \cite{dai2015semi, srivastava2015unsupervised}.

In Annex~\ref{sec:acr-training}, it is described the training methods of the \emph{supervised} and the \emph{unsupervised} instantiations of the \emph{ACR} module.

\subsection{The \emph{Next-Article Recommendation (NAR)} Module}

The \emph{Next-Article Recommendation (NAR)} module is responsible to provide news articles recommendations for active user sessions. It is designed as a hybrid recommender system, considering both the recorded user interactions and the content (\emph{RR2}) of the news articles. 

It is also a context-aware recommender system, in that it leverages information about the user context (\emph{RR4}), e.g., location, device, previous clicks in the session, and article's context --- popularity and recency (\emph{RR5}) -- which quickly decay over time. 

Generally, considering these additional factors can be crucial for the effectiveness of the session-based recommendations, in particular as previous work has shown that RNNs without side information are often not much better than relatively simpler algorithms \cite{jannach2017recurrent, ludewig2018evaluation}. 

The \emph{NAR} module is based on RNNs, which are suitable to work on sequences. In this case, the task is next-click prediction, i.e. to predict the next article a user might be interested in his session (\emph{RR3}).

The \emph{NAR} module is composed by three sub-modules: \emph{Contextual Article Representation (CAR)}, \emph{SEssion Representation (SER)}, and \emph{Recommendations Ranking (RR)}. These sub-modules are described in the following sections.

\subsubsection{The \emph{Contextual Article Representation (CAR)} Sub-module}

The \emph{CAR} sub-module is responsible to combine the inputs for the \emph{NAR} module: (1) the pre-trained Article Content Embedding; (2) the contextual properties of the article (popularity and recency); and (3) the user context (e.g., time, location, device, and referrer). Such contextual information is valuable in the extreme cold-start scenario on news recommendation domain.

This sub-module can be implemented as a sequence of feed-forward \emph{Fully Connected (FC)} layers or by using Factorization Machines (FM) to learn a representation based on its input feature interactions. 

The \emph{CAR} sub-module outputs the  \emph{User-Personalized Contextual Article Embedding}, whose representations might differ for the same article, depending on the user context and on the current article context (popularity and recency).

\subsubsection{The \emph{SEssion Representation (SER)} Sub-module}

The \emph{SEssion Representation (SER)} sub-module is responsible to model users' short-term  preferences, based on the sequence of interactions (reading news) in user's active session. 

The proposed instantiations of the \emph{SER sub-module} are RNN-based, which naturally deals with sequential data. It can be instantiated by different types of RNN cells, such as LSTM \cite{hochreiter1997long}, GRU \cite{chung2014empirical}, or \emph{Update Gate RNN (UGRNN)} \cite{collins2016capacity}.

The usage of RNNs for session-based recommendations was originally proposed by \cite{hidasi2016} and derived work \cite{hidasi2018recurrent, quadrana2017personalizing, donkers2017sequential} has usually trusted on item IDs to model users preferences in their sessions, which knowledgeably intensifies the item cold-start problem.

The input for the \emph{SER} sub-module is the sequence of news readings by the user in a session, represented as \emph{the User-Personalized Contextual Article Embeddings}. Such approach was designed to leverage item's content and context, and be able to recommend fresh articles.


The \emph{SER} sub-module outputs a \emph{Predicted Next-Article Embedding} -- the expected representation of a news content the user would like to read next in the active session.

\subsubsection{The \emph{Recommendations Ranking (RR)} Sub-module}

The \emph{RR} sub-module is responsible to recommend articles for a user session. 

Most deep learning architectures proposed for session-based RS are modeled as a classification problem, in which each available item has an output neuron, i.e., the neural network outputs a vector whose dimension is the number of available items, like \cite{hidasi2016, hidasi2018recurrent}. Such approach may work for domains were the items number is more stable, like movies and books. Although, in the dynamic scenario of news recommendations, where thousands of news stories are added and removed daily, such approach would require to include additional output neurons for newly published articles, and eventually a full retrain of the network.

For this reason, instead of using a \emph{softmax cross-entropy loss}, the \emph{NAR} module optimization goal is to maximize the similarity between the \emph{Predicted Next-Article Embedding} and the \emph{User-Personalized Contextual Article Embedding} corresponding to the next article actually read by the user in his session (positive sample), whilst minimizing its similarity with negative samples (articles not read by the user during the session).


With this strategy, a newly published article might be immediately recommended (\emph{RR6}), as soon as its \textit{Article Content Embedding} is trained and added to a repository. 

One inspiration for this approach came from the \emph{DSSM} \cite{huang2013learning} and its derived works for RS, like the \emph{MV-DNN} \cite{elkahky2015multi}, the \emph{TDSSM} \cite{song2016multi}, and the \emph{RA-DSSM} \cite{kumar2017}, which uses a ranking loss based on embeddings similarity, presented in Sections \ref{sec:DSSM}, \ref{sec:MV-DNN}, \ref{sec:TDSSM}, and \ref{sec:kumar2017}, respectively.

The aforementioned objective is related to recommendation accuracy, i.e. to predict the next article that the user will click next. As described in Section~\ref{sec:recsys_eval_metrics}, it is known for many years that prediction accuracy is not the only factor that determines the success of a recommender. Other quality factors discussed in the literature are, e.g., novelty, catalog coverage, and diversity \cite{di2017adaptive}. Those other quality factors often imposes a trade-off with recommendation accuracy, because the objectives are contradictory \cite{karimi2018news, castells2015novelty}.

For example, in the context of news recommendation, the aspect of novelty is particularly relevant to avoid a ``rich-get-richer'' phenomenon where a small set of already popular articles get further promoted through recommendations and less popular or more recent items rarely make it into a recommendation list. Improving novelty means to recommend less popular and more long-tailed articles, which will also  increase the item coverage, but will inevitably reduce the recommender accuracy \cite{karimi2018news}.

In this work, it is proposed a new approach to balance the trade-off between two conflicting quality factors -- accuracy and novelty -- of a session-based news recommender system. The formal optimization objectives of the \emph{NAR} module is presented in the next sections.

\subsubsection{Optimizing for Recommendation Accuracy}
\label{sec:acc_loss}

Formally, the method for optimizing prediction accuracy is described as follows. The inputs for the \emph{NAR} module are represented by ``$i$'' as the article ID, ``$uc$'' as the user context, ``$ax$'' as the article context, and ``$ac$'' as the article textual content. Based on those inputs, we define ``$cae =\Psi(i, ac, ax, uc)$'' as the \emph{User-Personalized Contextual Article Embedding}, where $ \Psi(\cdot) $ represents a sequence of fully-connected layers with non-linear activation functions to combine the inputs for the RNN.

The symbol $s$ stands for the user session (sequence of articles previously read, represented by their $ cae $ vectors), and ``$ nae=\Gamma (s) $'' denotes the \emph{Predicted Next-Article Embedding}, where $ \Gamma(\cdot) $ is the output embedding predicted by the RNN as the next article.

In Equation~\ref{eq:relevance}, the function $R$ describes the relevance of an item $ i $ for a given user session $ s $ as the similarity between the $nae$ vector predicted as the next-article for the session and the $cae$ vectors from the recommendable articles.

\begin{equation} \label{eq:relevance}
R(s, i) = \text{sim}(nae, cae)
\end{equation}

The $\text{sim}(\cdot) $ function can be the \emph{cosine similarity} (Equation~\ref{eq:cosine_sim}), originally proposed by \cite{huang2013learning} and experimented in \cite{moreira2018news}, or a learnable similarity function, like in our subsequent work \cite{moreira2019contextual}.

\begin{equation} \label{eq:cosine_sim}
sim(\theta) = \frac{a \cdot b}{\lVert a \rVert \lVert b \rVert }
\end{equation}




The ultimate task of the \emph{NAR} module is to produce a ranked list of items (top-n recommendation) that we assume the user will read next\footnote{This corresponds to a typical next-click prediction problem.}. Using $i \in \mathbb{D} $ to denote the set of all items that can be recommended, we can define a ranking-based loss function for a problem setting as follows. The goal of the learning task is to maximize the similarity between the predicted next article embedding ($nae$) for the session and the $cae$ vector of the next-read article (positive sample, denoted as $ i^+ $), while minimizing the pairwise similarity between the $nae$ and the and $cae$ vectors of the negative samples $i^- \in \mathbb{D}^-$. i.e., those that were not read by the user in this session.
Since $\mathbb{D}$ can be large in the news domain, we approximate it through a set $\mathbb{D}'$, which is the union of the unit set of the read articles (positive sample) $ \{i^+\} $ and a set with random negative samples from $\mathbb{D}^-$.

As proposed in \cite{huang2013learning}, we compute the posterior probability of an article being the next one given an active user session with a \emph{softmax} function over the relevance scores, as shown in Equation \ref{eq:prob_click},

\begin{equation} \label{eq:prob_click}
P(i \mid s, \mathbb{D}') = \frac{ \text{exp}(\gamma R(i, s))}{\sum_{\forall i' \in \mathbb{D}'}{ \text{exp}(\gamma R(i', s)})}
\end{equation}

where $\gamma$ is a smoothing factor (usually referred to as \emph{temperature}) for the \emph{softmax} function, which can be trained on a held-out dataset or which can be empirically set.

Using these definitions, the model parameters $\theta$ in the \emph{NAR} module are estimated to maximize the accuracy of the recommendations, i.e, the likelihood of correctly predicting the next article given a user session. The corresponding loss function to be minimized, originally proposed in the \emph{DSSM} \cite{huang2013learning}, follows the pairwise learning-to-rank paradigm \cite{liu2015representation, rahimi2019listwise}, as shown in Equation \ref{eq:acc_loss},

\begin{equation} \label{eq:acc_loss}
\text{accuracy\_loss}(\theta) = \frac{1}{|C|} \sum_{ (s, i^+, \mathbb{D}') \in \mathbb{C}} -\text{log} (P(i^+ \mid s, \mathbb{D}')),
\end{equation}

where $ \mathbb{C} $ is the set of user clicks available for training, whose elements are triples of the form $ (s,i^+, \mathbb{D}')$.

Since $ accuracy\_loss(\theta) $ is differentiable w.r.t.\,to $ \theta $ (the model parameters to be learned), we can use back-propagation on gradient-based numerical optimization algorithms in the \emph{NAR} module.

\subsubsection{Balancing Recommendations \emph{Accuracy} and \emph{Novelty}}
\label{sec:new_loss_function}
In order to incorporate the aspect of novelty of the recommendations directly in the learning process, it is proposed in this work a novelty regularization term in the loss function of the \emph{NAR} module. This regularization term has a hyper-parameter which can be tuned to achieve a balance between novelty and accuracy, according to the desired effect for the given application. Note that this approach is not limited to particular instantiations of the \emph{CHAMELEON} meta-architecture, but can be applied to any other neural architecture which takes article's recent popularity as one of the inputs and uses a \emph{softmax} loss function for training \cite{huang2013learning}.

In this proposed approach, it is considered the novelty definition by \cite{vargas2011rank, vargas2015thesis}, which is based on the inverse popularity of an item. The underlying assumption  of this definition is that less popular (long-tail) items are  more likely to be unknown to users and their recommendation will lead to higher novelty levels \cite{karimi2018news}.

The proposed novelty component therefore aims to bias the recommendations of the neural network toward more novel items. The positive items (actually clicked by the user) are not penalized based on their popularity, only the negative samples. The novelty of the negative items is weighted by their probabilities to be the next item in the sequence (computed according to Equation~\ref{eq:prob_click}) to push those items to the top of the recommendation lists that are both novel and relevant.

Formally, the novelty loss component is defined in Equation~\ref{eq:nov_reg},

\begin{equation} \label{eq:nov_reg}
\begin{split}
\text{nov\_loss}(\theta) =&\\
 & \frac{1}{|C|} \sum_{ (s, i^+, D'^{-}) \in C} \frac{\sum_{i \in D'^{-}} P(i \mid s, D'^{-}) * \text{novelty}(i)}
{\sum_{i \in D'^{-}} P(i \mid s, D'^{-})},
\end{split}
\end{equation}

where $ \mathbb{C} $ is the set of recorded click events for training, $ \mathbb{D}'^{-} $ is a random sample of the negative samples, not including the positive sample as in the accuracy loss function (Equation~\ref{eq:acc_loss}). The novelty values of the items are weighted by their predicted relevance $ P(i \mid s, \mathbb{D}'^{-}) $ in order to push both novel and relevant items towards the top of the recommendations list.

The novelty metric in Equation~\ref{eq:novelty} is defined based on the \emph{recent normalized popularity} of items. The negative logarithm in Equation~\ref{eq:novelty} increases the value of the novelty metric for long-tail items. The computation of the \emph{normalized popularity} sums up to 1.0 for all recommendable items (set $\mathbb{I}$), as shown in Equation~\ref{eq:rec-norm-pop}. Since we are interested in the recent popularity, we only consider the clicks an article has received within a time frame (e.g., in the last hour), as returned by the function $\text{recent\_clicks}(\cdot)$.

\begin{equation} \label{eq:novelty}
\text{novelty}(\text{i}) = -\text{log}_2(\text{rec\_norm\_pop}(i) + 1),
\end{equation}

\begin{equation} \label{eq:rec-norm-pop}
\text{rec\_norm\_pop}(i) = \frac{\text{recent\_clicks}(i)}{\sum_{j \in \mathbb{I}} \text{recent\_clicks}(j)}
\end{equation}

\paragraph{Complete Loss Function}
\label{sec:complete_loss}

The complete loss function proposed in this work combines the objectives of accuracy and novelty, as shown in Equation~\ref{eq:final_loss},

\begin{equation}
\label{eq:final_loss}
L(\theta) = \text{accuracy\_loss}(\theta) - \beta * \text{nov\_loss}(\theta) ,
\end{equation}

where $\beta $ is the tunable hyper-parameter for novelty. Notice that the novelty loss term is subtracted from the accuracy loss, as this term is higher when more novel items are recommended. The values for $\beta $ can either be set based on domain expertise or be tuned to achieve the desired effects.

\section{Comparison with Related Works}
\label{sec:comparison-works}

In the face of Deep Learning rapid advances, an specific architecture may become obsolete some months after its proposal, for using neural network components or methods that are not state-of-the-art anymore.

For this reason, the \emph{CHAMELEON} was structured as a high-level meta-architecture, to be modularized and support changes (RR7). It is composed of inputs, outputs, modules, sub-modules, data repositories, and their interactions. Modules and sub-modules can be instantiated by newly proposed neural network architectures or by existing architectures that prove themselves to work better for specific types of news portals.

The inspiration for the \emph{CHAMELEON} Meta-Architecture came from a survey on tens of works on News Recommender Systems and on Deep Learning methods applied to RS, as described in Chapters 2 and 3. It was designed specifically to address the challenges on news recommendation.

To the best of the knowledge from this research so far, the only works presenting a deep learning architecture for news recommendation were \cite{song2016multi}, \cite{kumar2017, kumar2017word}, \cite{park2017deep}, \cite{okura2017embedding}, \cite{zhang2018deep}, and \cite{wang2018dkn}. Those neural architectures for news recommendation were described in more detail on Chapter~3. In Table~\ref{tab:comparison_related_works} it is consolidated a feature matrix with the main differences among those architectures and the \emph{CHAMELEON}. Those differences are discussed throughout the rest of this section.

\begin{table}[h!t]
\centering
\caption{Comparison among Deep Neural Architectures for News RS}
\vspace{10pt}
\footnotesize
\begin{tabular}{p{2cm}|p{1.4cm}p{1.4cm}p{1.4cm}p{1.4cm}p{1.4cm}p{1.4cm}p{1.8cm}}

\hline
  \emph{Features}
  & \emph{\cite{song2016multi}}
  & \emph{\cite{kumar2017}}
  & \emph{\cite{park2017deep}}
  & \emph{\cite{okura2017embedding}}
  & \emph{\cite{wang2018dkn}}
  & \emph{\cite{zhang2018deep}}
  & \emph{CHAMELEON}
   \\

\hline

Session-based recommendation & No & No & YES & YES & No & YES & YES \\
\hline
Multiple quality objectives  & No & No & No & No & No & No & YES \\
\hline
Temporal articles relevance decay   & No & No & No & No & No & YES (exponential function) & YES (decay learned by the model) \\
\hline
Time-aware negative sampling  & No & No & YES & No & N/A & No & YES  \\
\hline
Modeling of article context
(recency, recent popularity)  & No & No & No & No & No & No & YES \\
\hline
User Context & No & No & No & No & No & No & YES \\
\hline
Textual Content Features & Letter trigrams processed by MLP & doc2vec & PV-DBoW & Denoising auto-encoder & Enrichment with a Knowledge Graph
 & Character-level CNN
 & Sup. and Unsup. 
ACR instantiations (CNN, GRU)
 \\
\hline
Architecture for sequence modeling & RNN & RNN & RNN & RNN & DNN with attention & RNN & RNN \\
\hline
Loss function & Log likelihood from \emph{DSSM} (pairwise ranking) & Log likelihood from \emph{DSSM} (pairwise ranking) & BPR, TOP1 (pairwise ranking) & Pairwise ranking with position bias & Log loss (pointwise ranking) & BPR, TOP1 (pairwise ranking) & Log likelihood from \emph{DSSM} (pairwise ranking) \\

\hline
\end{tabular}

\label{tab:comparison_related_works}
\end{table}

In general, some of those works addressed sequence-based recommendation, ignoring the temporal intervals between user sessions. Only \cite{park2017deep}, \cite{okura2017embedding}, and \cite{zhang2018deep} address session-based news recommendations, like \emph{CHAMELEON}.

\subsection{The \emph{GRU4Rec} and Derived Work on Session-based RS Using RNN}

Since the \emph{GRU4Rec}, proposed in the seminal work of \cite{hidasi2016} (described in Section \ref{sec:gru4rec}), and the subsequent work on \cite{hidasi2016parallel, hidasi2018recurrent}, a research line has emerged on the usage of RNN on session-based recommendations.  \cite{wu2016recurrent} have proposed an RNN to predict the probability that the user will access an item given heterogeneous feedback of this user and \cite{liu2016context} and \cite{smirnova2017contextual} have proposed some adaptations to feed RNNs with user contextual information. 


One limitation of \emph{GRU4Rec} in the news domain is that the method can only recommend items that appeared in the training set, because it is trained to predict scores for a fixed number of items. Another potential limitation is that RNN-based approaches, that only use item IDs for learning with no side information, might not be much better or even worse in terms of prediction accuracy than simpler approaches. Detailed analyses of this phenomenon can be found in \cite{jannach2017recurrent,jugovac2018streamingrec,ludewig2018evaluation}. The \emph{CHAMELEON} uses articles content and context embeddings to represent items and also user context to represent users, dealing smoothly with the incoming stream of new articles and users interactions.

In \cite{park2017deep}, it was proposed a deep neural architecture for news recommendation, which was described in Section~\ref{sec:news_rec_dnn_park2017}. There are some similarities to \emph{CHAMELEON}, as their architecture uses RNN for session-based recommendation. They also leverage the articles content -- in their case a bag-of-words representation -- to provide hybrid recommendation to combat the item cold-start problem.

It is reported in their study that they had tried many paired ranking losses functions, such as \emph{BPR} and \emph{TOP1} \cite{hidasi2016}. During the design of the \emph{CHAMELEON}, those losses functions were also explored, but it was observed that the \emph{DSSM}-based loss function resulted in a much higher accuracy.

Differently from their approach, which takes the articles categories in a post-processing reranking heuristic, the \emph{CHAMELEON} includes the article metadata (e.g., category, author) as an input feature for the RNN, deeply integrated with other features, such as the article context (e.g., popularity, recency) and the user context (e.g., time, device, location). 

Another key difference is that their model projects linearly the output of the RNN to the same space of their article content embeddings to be able to compute similarities, whereas in \emph{CHAMELEON} the similarity is computed based on \emph{User-Personalized Contextual Article Embeddings}, which are generated as a non-linear combination of the content of the articles with user-personalized and contextual information.

\subsection{The \emph{DSSM} and Derived Works on Multi-view Learning}

The \emph{Deep Semantic Structured Model (DSSM)} was proposed in \cite{huang2013learning} for the query ranking purpose, as described in Section \ref{sec:DSSM}. Essentially, the \emph{DSSM} can be seen as a multi-view learning model that often is composed of two or more neural networks for each individual view. In the original two-view \emph{DSSM} model, the left network represents the query view and the right network represents the document view.

\cite{elkahky2015multi} have adapted the \emph{DSSM} for the recommendation task, as described in Section \ref{sec:MV-DNN}. They map users and items to a latent space, where the similarity between users and their preferred items is maximized. This architecture was named the \emph{Multi-View Deep Neural Network (MV-DNN)}.

The \emph{DSSM} and the \emph{MV-DNN} multi-learning view models were important inspirations to the \emph{CHAMELEON} Meta-Architecture, in special, for the approach of item recommendation, by ranking similar embeddings. Usual works on neural networks for RS use a fixed set of items (output neurons), which would be unsuitable to recommend fresh articles, i.e., not seen during training.

\cite{song2016multi} have pointed out that, in \emph{MV-DNN} and \emph{DSSM} architectures, the user view and item view were static, representing a holistic view of users' interests, during a certain period of time. They proposed an extension to those models named the \emph{TDSSM} \cite{song2016multi}, which replaces the left network with item static features, and the right network with two subnetworks to model user static features (with MLP) and user
temporal features (with RNNs). Their case study was on news RS, making that work particularly interesting for this research, as presented in Section \ref{sec:TDSSM}.

To model users interests at varying time spots, they have used two LSTMs in different rates -- the \emph{fast-rate} LSTM used daily signals and the \emph{slow-rate} LSTM leveraged weekly signals. Although, they did not model user sessions explicitly. Users and items representation were pre-trained by a \emph{DSSM}, for which authors do not describe the used features and textual representation. In the proposed \emph{CHAMELEON}, items and users representations are learned directly from news content and users behaviours across sessions.

As introduced in Section~\ref{sec:embeddings_based}, the neural architecture proposed by \cite{okura2017embedding} for session-based news recommendation have some similarities to \emph{CHAMELEON}. They created a modified denoising autoencoder to learn textual content representation from articles. RNNs are used to learn and infer users representation. In this sense, it could be considered as an instantiation of the \emph{ACR} and \emph{NAR} modules of \emph{CHAMELEON}.

Among the differences, they use a pairwise ranking loss not based in \emph{softmax} and they do not include neither article context (e.g., popularity and recency) and user context (e.g., time, device) when providing recommendation, which may limit their accuracy. Finally, their negative sampling approach considers only non-clicked articles listed vertically, which may reinforce a selection bias created by the current ranking system.

\cite{kumar2017} have proposed the \emph{RA-DSSM} architecture, as described in Section \ref{sec:kumar2017}. It was also an extension of the \emph{DSSM}, now applied to news recommendations domain. The \emph{RA-DSSM} could be seen as a partial instantiation of \emph{CHAMELEON's} \emph{NAR} module. They leveraged \emph{bidirectional LSTM} with an attention mechanism to provide session-based recommendations. 

The \emph{RA-DSSM} provide sequence-based (not session-based) recommendation, ignoring the temporal aspects of news readership. Their evaluation protocol is not realistic to emulate news portals behaviour, as previously discussed in Section~\ref{sec:kumar2017}.

Like this proposed \emph{CHAMELEON}, the \emph{RA-DSSM} do not represent the articles only by their IDs. Instead, they represent articles content as \emph{doc2Vec} embeddings \cite{le2014distributed}, which are unsupervisedly pre-trained on the text. On the other hand, the \emph{Article Content Representation (ACR) module} of \emph{CHAMELEON} trains news content embeddings from articles' text in a \emph{supervised} or \emph{unsupervised} approach. 

Finally, the \emph{RA-DSSM} does not use any contextual information about the user and articles, which may profoundly limit its accuracy in a extreme cold-start scenario like news recommendation, as previously discussed. 

\subsection{Deep Feature Extraction from Textual Data}

Deep learning has been shown to be able to automatically extract powerful features from unstructured data. Recent research on hybrid RS has leveraged deep learning to extract features from textual data, like \emph{ConvMF} \cite{kim2016convolutional}, \emph{D-Attn} \cite{seo2017interpretable}, \emph{DeepCoNN} \cite{zheng2017joint}, and \emph{TransNets} \cite{catherine2017transnets}. Therefore, those hybrid RS studies leverage textual data from side information, like movie or product reviews.

\cite{bansal2016ask} proposed a neural architecture for scientific articles recommendation. They used an end-to-end neural network based on GRUs, to jointly model papers' textual content and item ID in multi-task objectives: predicting papers tags and users preferences on items. According to the authors, such multi-task learning lead to better content representations. It would hardly scale in the news domain, since the number of user interactions on news portals is massive and, on their architecture, full processing of text is required to obtain articles content representation.

In \cite{wang2018dkn} it was introduced a neural architecture for news recommendation named \emph{DKN}, described in more details in Section~\ref{sec:dkn}. It models the sequence of past users clicks (sequence-aware, not session-based RS) for CTR prediction. Differently than \emph{CHAMELEON}, they do not use the actual content of the article, but only its title. Then, they search in a knowledge graph the entities mentioned in the title, and then expand the content representation with the found entities and their immediate neighbours in the graph. They also use CNNs to process the textual content, but their architecture have an additional channel to process the found entities representations.

Furthermore, in \emph{DKN} the sequence of user clicks is processed by a DNN with attention mechanism, while the \emph{CHAMELEON} processes the sequence of clicks by using an RNN. 

The \emph{DKN} is a sequence-aware RS and not a session-based RS such as \emph{CHAMELEON}, ignoring the important temporal aspects of news readership. It models the sequence of all user clicks, regardless the elapsed time between the user sessions and clicks, which may vary from minutes to months.

Finally, the evaluation protocol proposed by \cite{wang2018dkn} is not realistic for news recommendation, as it samples candidate news regardless of their publishing date, and also do not evaluate aspects other than accuracy, such as novelty and diversity.

Other related work is \cite{zhang2018deep}, described in more detail in Section~\ref{sec:deep_jonn}. They have proposed a neural architecture -- the \emph{DeepJoNN} -- for session-based news recommendation, that uses a character-level CNN to process articles' input features and processes the sequence of session clicks with an RNN.

The \emph{DeepJoNN} does not use any contextual features from the user or articles (e.g., recent popularity). They also do not model the temporal aspects of news readership, randomly sampling articles published in different periods together for training.

Furthermore, the authors of \emph{DeepJoNN} have proposed a fixed exponential function to model the temporal decay of news relevance. On the other hand, \emph{CHAMELEON} uses the recency as an input feature, letting the network to learn an arbitrary function to model the news relevance decay, depending on the content of the article and on the user and article contexts.


\subsection{Balancing Accuracy and Novelty in Recommender Systems}
\label{sec:balance_acc_x_novelty}

The novelty of a recommended item can be defined in different ways, e.g., as the non-obviousness of the item suggestions \cite{Herlocker2004} or in terms of how different an item is with respect to what has already been experienced by a user or the community \cite{vargas2011rank}. Recommending solely novel or unpopular items can, however, be of limited value when they do not match the users' interests well. Therefore, the goal of a recommender is often to balance these competing factors, i.e., make somewhat more novel and thus risky recommendations, while at the same time ensuring high accuracy.

In the literature, a number of ways have been proposed to \emph{quantify} the degree of novelty, including alternative ways of considering popularity information \cite{zhou2010solving} or the distance of a candidate item to user's profile \cite{karimi2018news, nakatsuji2010classical, rao2013taxonomy}. In \cite{vargas2011rank}, the authors propose to measure novelty as the opposite of popularity of an item, under the assumption that less popular (long-tail) items are more likely to be unknown to users and their recommendation will, hopefully, lead to higher novelty levels. In our work, we will also consider the popularity-based novelty of the recommendations and adapt existing novelty metrics from the literature.

Regarding the treatment of trade-off situations, different technical approaches are possible. One can, for example, try to re-rank an accuracy-optimized recommendation list, either to meet globally defined quality levels \cite{adomavicius2011improving} or to achieve recommendation lists that match the preferences of individual users \cite{JugovacJannachLerche2017eswa}. Another approach is to vary the weights of the different factors to find a configuration that leads to both high accuracy and novelty \cite{garcin2013personalized}.

Finally, one can try to embed the consideration of trade-offs within the learning phase, e.g., by using a corresponding regularization term. In \cite{coba2018novelty}, the authors propose a method called \emph{Novelty-aware Matrix Factorization (NMF)}, which tries to simultaneously recommend accurate and novel items. Their proposed regularization approach is pointwise, meaning that the novelty of each candidate item is considered individually.

In our recommendation approach, we consider trade-offs in the regularization term as well. Differently, from \cite{coba2018novelty}, however, our approach is not focused on matrix factorization, but rather on neural models that are derived from the \emph{DSSM}. Furthermore, the objective function in our work uses a pairwise ranking approach to learn how to enhance the novelty level of the top-n recommendations.

\chapter{CHAMELEON Instantiations and Experiments Design}
\label{sec:chapter_5}

In this chapter, it is described: some instantiations of the \emph{CHAMELEON} meta-architecture; their implementation strategies; the evaluation methodology, including metrics and datasets; and the experimental design.


The results of the evaluation will be discussed later in Chapter~\ref{sec:chapter_6}.

\section{The \emph{CHAMELEON} Instantiations}

In this section, it is described the \emph{CHAMELEON} instantiations for experiments. The \emph{The ACR} module is instantiated as three concrete architectures -- two \emph{supervised} and other \emph{unsupervised}. It is also described an instantiation of the \emph{NAR} module.

\subsection{The \emph{ACR} Module Instantiations}

In this section, it is described the \emph{ACR} module instantiations for the experiments -- \emph{supervised} and \emph{unsupervised}. Their  training methods are detailed in Annex~\ref{sec:acr-training}.

The input features and target attributes (only for \emph{supervised} instantiations) for the \emph{ACR} module are presented in Table~\ref{tab:acr-features}.

\begin{table}[!htbp]
\begin{threeparttable}
\centering
\caption{Features used by the \emph{Article Content Representation (ACR)} module.}

\label{tab:acr-features}
\vspace{10pt}
\footnotesize
\begin{tabular}{p{3cm}lp{8cm}}
\hline
 \emph{Features} &
 \emph{Type} &
 \emph{Description} \\
\hline
\multicolumn{3}{p{6cm}}{\textbf{Input features}}\\ \hline
Textual Content & Emb. & Article text represented as a sequence of word embeddings, pre-trained for the language of the dataset.\footnotemark \\
\hline
\multicolumn{3}{p{6cm}}{\textbf{Target features (\emph{supervised} only}}\\ \hline
Category & Categ. & The category of the article, defined by editors. \\
Keywords* & Categ. & Human-labeled keywords for the Adressa dataset. \\
\hline
\end{tabular}

\end{threeparttable}
\end{table}
\footnotetext{Portuguese: Pre-trained \emph{word2Vec} \emph{skip-gram} model (300 dimensions) available at \url{http://nilc.icmc.usp.br/embeddings}; Norwegian: a \emph{skip-gram} model (100 dimensions) available at \url{http://vectors.nlpl.eu/repository} (model \#100).}

The input for the \emph{ACR} module is the sequence of word embeddings of the news article content. More formally, let $x_i \in \mathbb{R}^k $ be the $k$-dimensional word embedding vector corresponding to the $i$-th word in the document. A document of length $n$ (padded when necessary) is represented, as shown in Equation \ref{eq:cnn-concat}

\begin{equation} \label{eq:cnn-concat}
x_{1:n} = x_1 \oplus x_2 \oplus ... \oplus x_n,
\end{equation}

where $ \oplus $ is the concatenation operator.

\subsubsection{The \emph{ACR} Module - Supervised Instantiation}
\label{sec:acr_supervised_instantiation}

\begin{figure}[h!t]
	\centering\includegraphics[width=10cm]{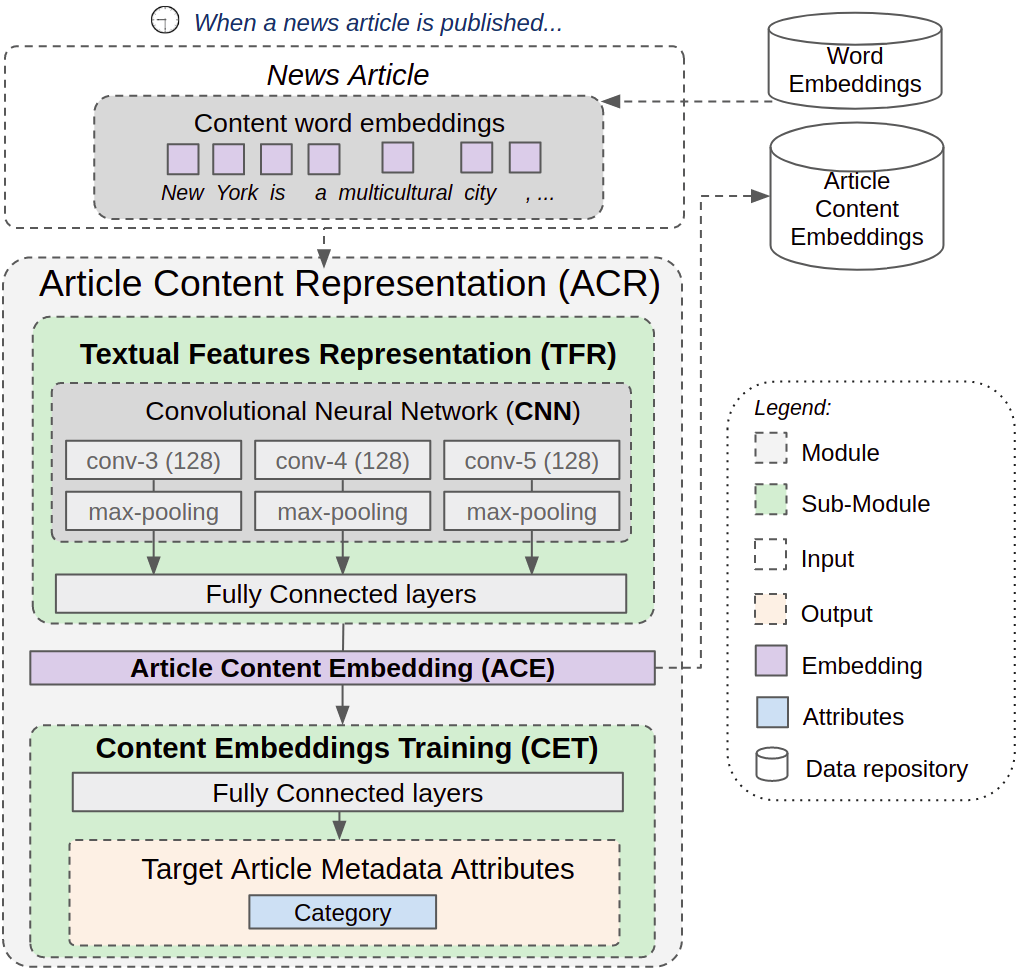}	\caption{A \emph{supervised} CNN-based instantiation of the \emph{ACR} module}
	\label{figure:acr_instantiation_supervised_cnn}
\end{figure}

\begin{figure}[h!t]
	\centering\includegraphics[width=10cm]{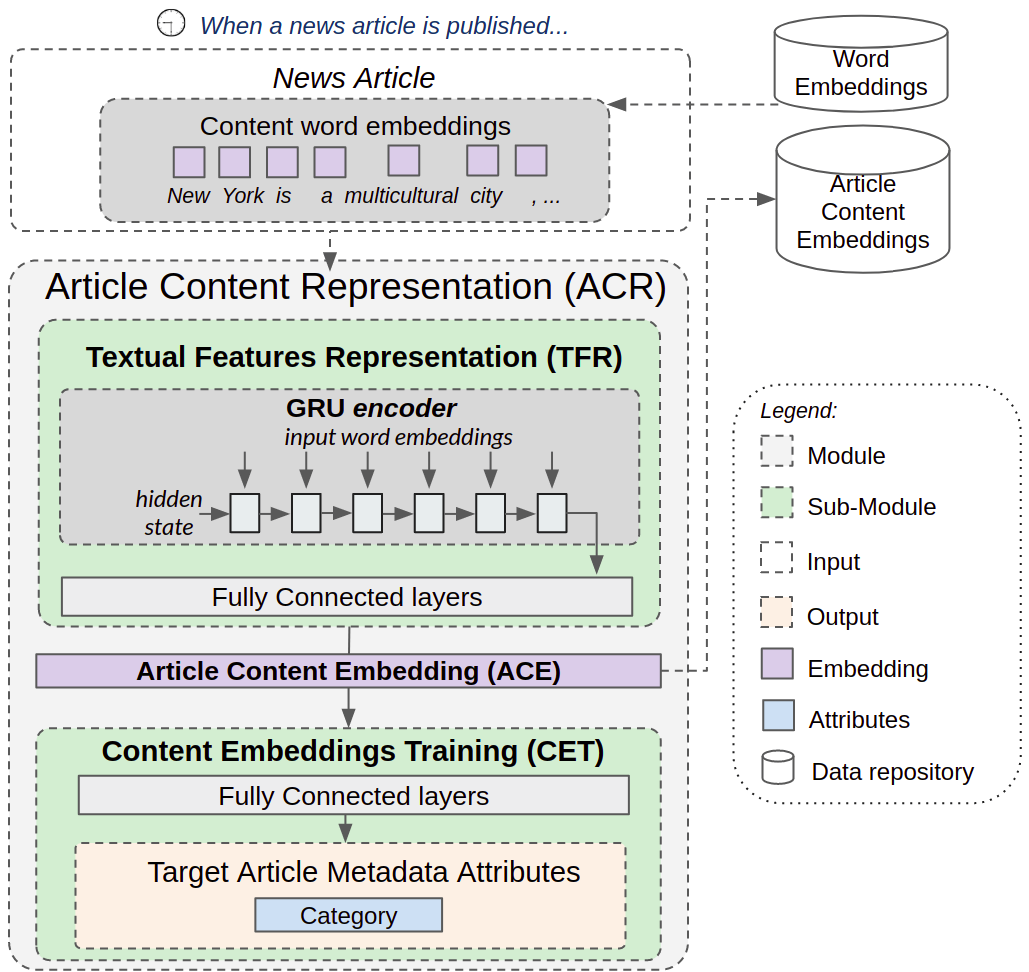}	\caption{A \emph{supervised} RNN-based instantiation of the \emph{ACR} module}
	\label{figure:acr_instantiation_supervised_rnn}
\end{figure}

It is proposed two \emph{supervised} instantiations for the \emph{ACR} module -- one CNN-based and the other RNN-based -- presented in Figures~\ref{figure:acr_instantiation_supervised_cnn}~and~\ref{figure:acr_instantiation_supervised_rnn}, respectively. Those different instantiations are further implemented in the experiments to analyze the effect of different textual representation in recommendation quality.

Their \emph{sub-modules} instantiations for \emph{supervised} scenario are described next.

\paragraph{A CNN-based Instantiation of the \emph{Text Feature Representation (TFR)} Sub-module}
\label{sec:tfr_supervised_cnn_instantiation}

In the CNN-based instantiation of the \emph{Text Feature Representation (TFR)} sub-module, it was instantiated as an architecture based on Convolutional Neural Networks (CNN), inspired on \cite{kim2014convolutional}. CNNs have being widely used in Deep NLP research as a simple, fast, and powerful method for feature extraction from textual data. 

It has a three independent convolutional layers, each one with filter region sizes of 3, 4, and 5. Each filter performs the convolution operation in a sliding window of word embeddings -- in this case, sliding over 3, 4, or 5 words at a time. More formally, let $ x_{i:i+j} $ refer to the concatenation of words $ x_i, x_{i+1},...,x_{i+j}$. A convolution operation involves a filter $w \in \mathbb{R}^{sk} $, which is applied to a window of $ s $ word embeddings of $ k $  dimensions to produce a new feature. For example, as shown in Equation \ref{eq:cnn-feature}, a feature $ c_i $ is generated from a window of words $ x_{i:i+s-1} $.

\begin{equation} \label{eq:cnn-feature}
c_i = f(w \cdot x_{i:i+s-1} + b)
\end{equation}

Here, $ b \in \mathbb{R} $ is a bias term, $ \cdot $ is the dot product (a sum over element-wise multiplications), and $ f $ is a non-linear function: the \emph{Rectified Linear Units (ReLU)}. A filter is applied to each possible window of words in the document $ {x_{1:s},x_{2:s+1},...,x_{n-s+1:n}} $ to produce a feature map  $ c \in \mathbb{R}^{n-s+1} $, as shown in Equation \ref{eq:cnn-feature-map}.

\begin{equation} \label{eq:cnn-feature-map}
c = [c_1,c_2,...,c_{n-s+1}]
\end{equation}

A number of filters $ g $ is defined for each specific filter size $ s $. Each of these filters will produce a feature map for each sliding window over the input text. 

News articles' length may vary, but it is necessary to have a fixed dimension for the \emph{Article Content Embedding}. Thus, for each filter, a max-pooling operation is performed over its produced feature maps, returning its largest value, which are concatenated in a unified feature map $ m \in \mathbb{R}^{g} $ for each filter size $ s $. Finally, the unified feature maps for all filter sizes are concatenated in a vector $ z \in \mathbb{R}^{gs} $ -- the Article Content Embedding of the \emph{ACR} module.

\paragraph{An RNN-based Instantiation of the \emph{Text Feature Representation (TFR)} Sub-module}
\label{sec:tfr_supervised_rnn_instantiation}

In the RNN-based instantiation of the \emph{Text Feature Representation (TFR)} sub-module, it was instantiated based on a type of RNN architecture -- the Gated Recurrent Unit (GRU) \cite{chung2014empirical}, which is less computationally expensive than Long Short-Term Memory (LSTM) \cite{gers2001long, hochreiter1997long}. 

A GRU network updates its hidden state as follows:

\begin{equation}\label{eq:gru-1}
z_t = \sigma_g(W_z x_t + V_z h_{t-1} + b_z)
\end{equation}

\begin{equation}\label{eq:gru-2}
r_t = \sigma_g(W_r x_t + V_r h_{t-1} + b_r)
\end{equation}

\begin{equation}\label{eq:gru-3}
\hat{h}_t = tanh(W_h x_t + V_h (r_t \odot h_{t-1}) + b_h )
\end{equation}

\begin{equation}\label{eq:gru-4}
h_t = (1 - z_t) \cdot h_{t-1} + z_t \cdot \hat{h}_t
\end{equation}

where $ \odot $ denotes element-wise multiplication, $ \sigma_g $ and $ tanh $ are sigmoid and tangent functions respectively, $ W $, $ V $, and $ b $ are parameter matrices and vectors. The reset gate $ r_t $ and update gate $ z_t $ control how information is updated to the new hidden state.

In this \emph{TFR} sub-module instantiation, the GRU layer processes the input text word-by-word.

The hidden states at each step are \emph{max-pooled} to generate a representation of the article content. This representation is processed by an additional feed forward fully connected layer, which outputs the \emph{Article Content Embedding (ACE)}. The \emph{ACE} is trained by the \emph{CET} sub-module, described next.

\paragraph{\emph{Content Embedding Training (CET)} Sub-module Instantiation}

The \emph{Content Embedding Training (CET)} sub-module was instantiated by simple neural components -- a \emph{dropout} layer followed by a feed-forward layer with non-linear activation function. The network is trained to minimize the \emph{softmax cross-entropy loss} of the predicted categories.








More details on the training of the supervised instantiation of the \emph{ACR} module are found in Annex~\ref{sec:acr-training}.

\subsubsection{The \emph{ACR} Module - Unsupervised Instantiation}
\label{sec:acr_unsupervised}

\begin{figure}[h!t]
	\centering\includegraphics[width=10cm]{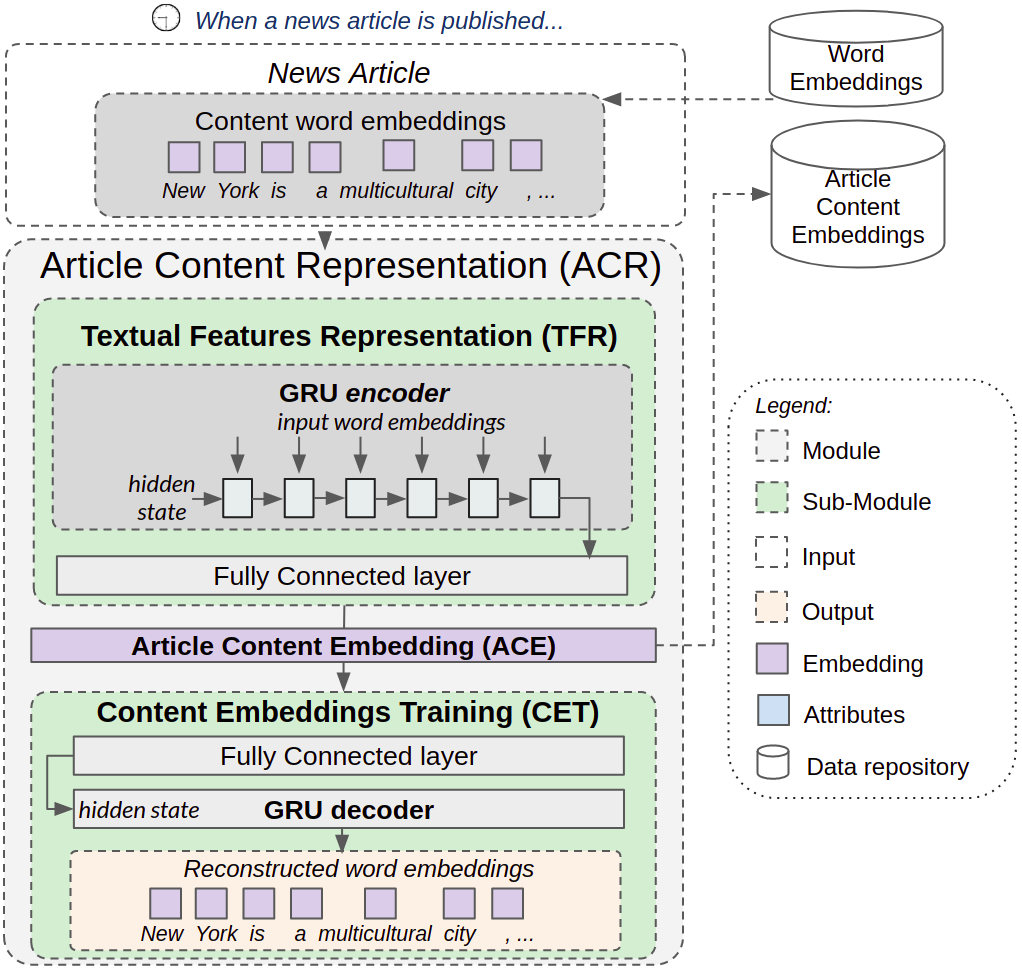}	\caption{An \emph{unsupervised} instantiation of the \emph{ACR} module}
	\label{figure:acr_instantiation_unsupervised}
\end{figure}

An \emph{unsupervised} instantiation of the \emph{ACR} module is presented in Figure~\ref{figure:acr_instantiation_unsupervised}.

As described in Annex~\ref{sec:acr-training}, the \emph{unsupervised} instantiation is based in a \emph{Sequence Autoencoder} \cite{dai2015semi, srivastava2015unsupervised} - a type of RNN-based Encoder-Decoder  architecture. The article text is encoded to generate a compressed representation and decoded to reconstruct the original input text, represented as a sequence of word embeddings.

\paragraph{The \emph{Text Feature Representation (TFR)} Sub-module Instantiation}

The \emph{TFR} sub-module acts as an encoder of the input text. It is instantiated with a \emph{GRU} \cite{chung2014empirical} \emph{encoder layer}, which process the input word embeddings one-by-one. 

The last hidden state of the \emph{GRU encoder layer} is used as input of a feed-forward layer with \emph{tanh} activation function, which projects the content representation to the specified dimension for the \emph{Article Content Embedding (ACE)}.

In the experiments implementation, the first 30 words from the concatenation of article's title and content was considered as input. It was added dome random noise to the input word embeddings, an approach known as \emph{Denoising Autoencoder} \cite{alain2014regularized}), to work as a regularizer.

As suggested by \cite{sutskever2014sequence} and \cite{srivastava2015unsupervised}, the sequence of input words is reversed (while keeping as target words the original words ordering), because doing so has been shown to improve reconstruction accuracy and to introduce short term dependencies in data that make the optimization problem easier.

\paragraph{The \emph{Content Embedding Training (CET)} Sub-module Instantiation}

The \emph{CET} sub-module acts as a decoder, trying to reconstruct the original text from the encoded \emph{ACE}.

It is composed by a feed-forward layer, which project the \emph{ACE} back to the dimension (number of units) of the \emph{GRU decoder layer}.

The \emph{GRU decoder layer} have its hidden state initialized by the previous feed-forward layer. 

The first input for the \emph{GRU decoder layer} is a padding embedding and expected output is the embedding of the first word of article's text. Next, the first word is used as input and the second word as output, and so forth.

The network is trained to minimize the \emph{Mean Squared Error (MSE)} between the correct and predicted word embeddings, as detailed in Annex~\ref{sec:acr-training}.

\subsection{The \emph{NAR} Module Instantiations}

To feed the \emph{Next Article Recommendation (NAR)} module, rich features were extracted from the user interactions logs, as detailed in Table~\ref{tab:nar-features}. The input features were prepared as follows. 

Categorical features with low cardinality (i.e., with less than 10 distinct values) were one-hot encoded and features with high cardinality were represented as trainable embeddings. Numerical features were standardized with \emph{z-normalization}. 

The \emph{Article Content Embeddings} are L2-normalized, so that each embedding has zero mean and unit size.

The dynamic features for \emph{novelty} and \emph{recency} were normalized based on a sliding window of the recent clicks (within the last hour), so that they can accommodate both repeating changes in their distributions over time, e.g., within different periods of the day, and abrupt changes in global interest, e.g., due to breaking news.

\begin{table}[!htbp]
\begin{threeparttable}
\centering
\caption{Features used by the \emph{Next-Article Recommendation (NAR) } module}
\label{tab:nar-features}
\vspace{10pt}
\footnotesize
\begin{tabular}{p{2.5cm}llp{6cm}}
\hline
 \emph{Group} &
 \emph{Features} &
 \emph{Type} &
 \emph{Description} \\
\hline
\multicolumn{4}{p{8cm}}{\textbf{Dynamic article features}}\\ \hline
Article Context & Novelty & Num. & The novelty of an article, computed based on its normalized recent popularity, as described in Equation~\ref{eq:novelty}.  \\
 & Recency & Num. & Computed as the logarithm of the elapsed days (with hours represented as the decimal part) since an article was published: $ \text{log}_2 $((current\_date - published\_date)+1).  \\ \hline
\multicolumn{4}{p{8cm}}{\textbf{Static article features}}\\ \hline
Id & Id & Emb. & Trainable embeddings for article IDs. \\
Content & ACE & Emb. & The \emph{Article Content Embedding} representation learned by the \emph{ACR} module. \\
Metadata & Category & Cat. & Article category \\
 & Author * & Cat. & Article author \\ \hline
\multicolumn{4}{p{8cm}}{\textbf{User context features}}\\ \hline
Location & Country, Region, City* & Categ. & Estimated location of the user \\
\multirow{3}{2.5cm}{Device} & Device type & Categ.& Desktop, Mobile, Tablet, TV** \\
 & OS & Categ. & Device operating system\\
 & Platform** & Categ. & Web, mobile app\\
\multirow{2}{2.5cm}{Time} & Hour of the day & Num. & Hour encoded as cyclic continuous feature (using sine and cosine)\\
 & Day of the week & Num. & Day of the week \\
Referrer & Referrer type & Categ. & Type of referrer: e.g., direct access, internal traffic, search engines, social platforms, news aggregators\\
\hline
\end{tabular}

\begin{tablenotes}\footnotesize
\item * Only available for the Adressa dataset.
\item ** Only available for the G1 dataset.
\end{tablenotes}

\end{threeparttable}
\end{table}

In the following sections, the sub-modules instantiations of the \emph{NAR} module are described.

\subsubsection{The \emph{Contextual Article Representation (CAR)} Sub-module Instantiation}

The \emph{CAR} sub-module is instantiated as follows.

First, all input features $f$ are concatenated and normalized by a layer with trainable scalar parameters to center ($ \gamma $) and scale ($ \beta $) input features, as shown in Equation~\ref{eq:center_scale_features}.

\begin{equation} \label{eq:center_scale_features}
\text{norm}(f) = (f * \gamma) + \beta,
\end{equation}

This is followed by two feed-forward fully connected layers, with \emph{Leaky ReLU} \cite{maas2013rectifier} and \emph{tanh} activation functions, respectively, whose output is the \emph{User-Personalized Contextual Article Embedding}.

\subsubsection{The SEssion Representation (SER) Sub-module Instantiation}

The \emph{SER} sub-module uses an RNN to model the sequence of user interactions. We empirically tested different RNN cells, like variations of LSTM \cite{hochreiter1997long} and GRU \cite{chung2014empirical}, whose results were very similar. At the end, we selected the \emph{\emph{Update Gate RNN} (UGRNN)} cell \cite{collins2016capacity}, as it led to slightly higher accuracy. The \emph{UGRNN} architecture is a compromise between \emph{LSTM}/\emph{GRU} and a vanilla RNN. In the \emph{UGRNN} architecture, there is only one additional gate, which determines whether the hidden state should be updated or carried over \cite{collins2016capacity}. Adding a new (non bi-directional) RNN
layer on top of the previous one also led to some accuracy improvement.

The RNN layer is followed by a sequence of two feed-forward layers, with \emph{Leaky ReLU} and \emph{tanh} activation functions, respectively, whose output is the \emph{Predicted Next-Article Embedding}.

\subsubsection{The Recommendations Ranking (RR) Sub-module Instantiation}

The \emph{RR} sub-module is responsible to rank candidate articles based on the relevance of an article for a given session, based on embeddings similarity.

As described in Section~\ref{sec:acc_loss} and reported in \cite{moreira2018news}, it is possible to use the \emph{cosine similarity}.

For these experiments, also reported in \cite{moreira2019contextual}, the architecture was instantiated so that it was flexible to learn an arbitrary matching function $\text{sim}(\cdot) $. It was composed by the element-wise product of the embeddings, followed by a number of feed-forward layers with non-linear activations (\emph{Leaky ReLU} \cite{maas2013rectifier}), as shown in Equation~\ref{eq:sim},

\begin{equation} \label{eq:sim}
\text{sim}(nae, cae) = \phi(nae \odot cae),
\end{equation}

where $ \phi(\cdot) $ represents a sequence of fully-connected layers with non-linear activation functions and the last layer outputs a single scalar representing the relevance of an article as the predicted next article. In our study, $ \phi(\cdot) $ consisted of a sequence of 4 feed-forward layers with a \emph{Leaky ReLU} activation function \cite{maas2013rectifier}, with 128, 64, 32, and 1 output units.

Finally, the model is trained to minimize the \emph{NAR} module loss function, described in Section~\ref{sec:complete_loss}.

\section{Baseline Recommendation Methods}
\label{sec:baseline_algs}
In our experiments, we consider (a) different variants of our instantiation of the \emph{CHAMELEON} meta-architecture to assess the value of considering additional types of information and (b) a number of session-based recommender algorithms, described in Table~\ref{tab:baselines}. While some of the chosen baselines appear conceptually simple, recent work has shown that some of them are able to outperform very recent neural approaches for session-based recommendation tasks \cite{jannach2017recurrent,ludewig2018evaluation,jugovac2018streamingrec,ludewig2019performance}. Furthermore, the simple methods, unlike neural-based approaches, can be continuously updated over time and take newly published articles into account.

\begin{table}[!htbp]
\centering
\caption{Baseline session-based recommender algorithms used in the experiments.}
\label{tab:baselines}
\vspace{10pt}
\footnotesize
\begin{tabular}{p{3cm}p{9cm}}
\hline

\multicolumn{2}{p{8cm}}{Neural Methods}\\ \hline

\textit{GRU4Rec} & A landmark neural architecture using RNNs for session-based recommendation \cite{hidasi2016}, described in Section~\ref{sec:gru4rec}. For this experiment, we used the \textit{GRU4Rec} \emph{v2} implementation, which includes the improvements reported in \cite{hidasi2018recurrent}.\tnote{1} We furthermore improved the algorithm's negative sampling strategy for the scenario of news recommendation.\tnote{2} \\ \hline

\textit{SR-GNN} & A recently published state-of-the-art architecture for session-based recommendation based on Graph Neural Networks \cite{wu2019session}, described in Section~\ref{sec:sr-gnn}. Their authors reported superior performance over other neural architectures such as \textit{GRU4Rec} \cite{hidasi2016}, \textit{NARM} \cite{Li2017narm} and \textit{STAMP} \cite{Liu2018stamp}. \\
\hline

\multicolumn{2}{p{8cm}}{Association Rules-based Methods}\\ \hline
\emph{Co-Occurrence (CO)} &  Recommends articles commonly viewed together with the last read article in previous user sessions. This algorithm is a simplified version of the association rules technique, having two as the maximum rule size (pairwise item co-occurrences) \cite{ludewig2018evaluation,jugovac2018streamingrec}.\\
\emph{Sequential Rules (SR)} &  The method uses association rules of size two. It however considers the sequence of the items within a session. A rule is created when an item \emph{q} appeared after an item \emph{p} in a session, even when other items were viewed between \emph{p} and \emph{q}. The rules are weighted by the distance \emph{x} (number of steps) between \emph{p} and \emph{q} in the session with a linear weighting function $ w_{\text{\textsc{SR}}} = 1/x $ \cite{ludewig2018evaluation};\\
\hline
\multicolumn{2}{l}{Neighborhood-based Methods}\\ \hline
\emph{Item-kNN} & Returns the most similar items to the last read article using the cosine similarity between their vectors of co-occurrence with other items within sessions. 
This method has been commonly used as a baseline when neural approaches for session-based recommendation were proposed, e.g., in  \cite{hidasi2016}.\\
\emph{Vector Multiplication Session-Based kNN (V-SkNN)} & This method compares the entire active session with past (neighboring) sessions to determine items to be recommended. The similarity function emphasizes items that appear later within the session. The method proved to be highly competitive in the evaluations in \cite{jannach2017recurrent,ludewig2018evaluation,jugovac2018streamingrec}. \\ \hline
\multicolumn{2}{l}{Other Methods}\\ \hline
\emph{Recently Popular (RP)} & This method recommends the most viewed articles within a defined set of recently observed user interactions on the news portal (e.g., clicks during the last  hour). Such a strategy proved to be very effective in the \emph{2017 CLEF NewsREEL Challenge} \cite{ludmann2017recommending}. \\
\emph{Content-Based (CB)} &  For each article read by the user, this method suggests recommendable articles with similar content to the last clicked article, based on the cosine similarity of their \emph{Article Content Embeddings}. \\ \hline
\end{tabular}
\end{table}

\section{Evaluation Methodology}
\label{sec:eval_methodology}
One main goal of our experimental analyses is to make our evaluations as realistic as possible. We therefore did not use the common evaluation approach of random train-test splits and cross-validation. 

Instead, we use a temporal offline evaluation method \cite{moreira2018news}, which simulates a streaming flow of user interactions (clicks) and new articles being published, whose value quickly decays over time. Since in practical environments it is highly important to very quickly react to incoming events \cite{ludmann2017recommending, kille2017clef}, the baseline recommender methods were constantly updated over time.

\emph{CHAMELEON}'s \emph{NAR} module supports online learning, as it is trained on mini-batches. In our evaluation protocol, we decided to emulate a streaming scenario, in which each user session is used for training only once. Such a scalable approach is different from many model-based recommender systems, like \emph{GRU4Rec}, which require training for some epochs on a large set of recent user interactions to reach competitive accuracy results.

\subsection{Evaluation Protocol} 

\begin{figure}[h!t]
	\centering\includegraphics[width=0.99\linewidth]{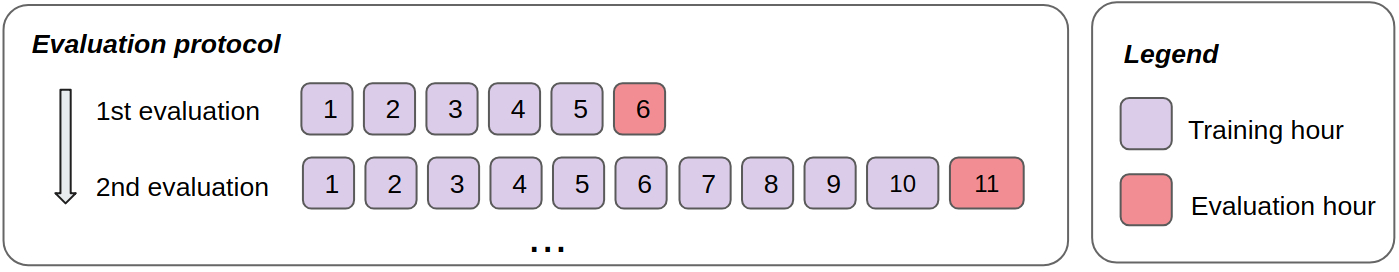}	\caption{Illustration of the evaluation protocol. After training for 5 hours, the sessions of the next hour are evaluated.}
	\label{figure:eval_protocol}
\end{figure}

The evaluation process works as follows:

\begin{itemize}
\item The recommenders are continuously trained on users' sessions ordered by time and grouped by hours. Each five hours, the recommenders are evaluated on sessions from the next hour, as exemplified in Figure~\ref{figure:eval_protocol}. With this interval of five hours (not a divisor of 24 hours), it was possible to sample different hours of the day across the dataset for  evaluation. After the evaluation of the next hour was done, this hour is also  considered for training, until the entire dataset is  covered.\footnote{Our dataset consists of 16 days. We used the first two days to learn an initial model for the session-based algorithms and report the averaged measures after that warm-up period.} It is important to notice that, while the baseline methods are continuously updated during the evaluation hour, \emph{CHAMELEON}'s model is not. This allows us to emulate a realistic scenario in production where the neural network is trained and deployed once an hour to serve recommendations for the next hour;
\item For each session in the evaluation set, we incrementally ``revealed'' one click after the other to the recommender, as done, e.g., in \cite{hidasi2016} and \cite{quadrana2017personalizing}; 
\item For each click to be predicted, we created a random set containing 50 recommendable articles that were \emph{not} viewed by the user in the session (negative samples) plus the true next article (positive sample), as done in \cite{koren2009} and \cite{Cremonesi2010}. We then evaluate the algorithms in the task of ranking those 51 items;
\item Given these rankings, standard information retrieval metrics can be computed.
\end{itemize}

For a realistic evaluation, it is important that the chosen negative samples consist of articles which would be of some interest to readers and which were also available for recommendation in the news portal at a given point of time. For the purpose of this study, we therefore selected as recommendable articles the ones that received at least one click by any user in the preceding hour. To finally select the negative samples, we implemented a popularity-based sampling strategy similar to the one from \cite{hidasi2016}.

\subsection{Metrics}
\label{sec:metrics}
To measure quality factors such as accuracy, item coverage, novelty, and diversity, we have selected a set of top-N metrics from the literature. We chose the cut-off threshold at N=10, representing about 20\% of the list containing the 51 sampled articles (1 positive sample and 50 negative samples). 

The accuracy metrics used in our study -- Hit Rate and Mean Reciprocal Rank -- originate from Information Retrieval field and are very popular in Recommender Systems.

The \emph{Hit Rate (HR@n)} is the percentage of times in which \emph{relevant} items (the next clicked ones) are retrieved among the top-N ranked items, as shown in Equation~\ref{eq:hitrate}, 


\begin{equation} \label{eq:hitrate}
HR@n = \frac{1}{Q}\sum_{i=1}^Q 
\begin{cases} 1 & \mbox{if } \mbox{rank}_i <= n \\ 0, & \mbox{otherwise}  \end{cases}
,
\end{equation}

where $ Q $ is a sample of recommendation lists, $\text{rank}_i$ refers to the rank position of the \emph{relevant} item for the $i$-th recommendation list. If $\text{rank}_i$ is greater than $ n $, this recommendation scores as $0$.

The \emph{Mean Reciprocal Rank (MRR@n)} is a ranking metric that is sensitive to the position of the true next item (\emph{relevant} item) in the list. As shown in Equation~\ref{eq:mrr}, it is the average of reciprocal ranks of results for a sample of recommendation lists $ Q $,

\begin{equation} \label{eq:mrr}
MRR@n = \frac{1}{Q}\sum_{i=1}^Q 
\begin{cases} \frac{1}{\mbox{rank}_i}, & \mbox{if } \mbox{rank}_i <= n \\ 0, & \mbox{otherwise}  \end{cases}
,
\end{equation}

where $\text{rank}_i$ refers to the rank position of the \emph{relevant} item for the $i$-th recommendation list. If $\text{rank}_i$ is greater than $ n $, this recommendation scores as $0$.

Both metrics are common when evaluating session-based recommendation algorithms \cite{hidasi2016,ludewig2018evaluation,jugovac2018streamingrec}.

As an additional metric, we considered \emph{Item Coverage (COV@n)}, which is sometimes also called ``aggregate diversity'' \cite{adomavicius2011improving}. The idea here is to measure to what extent an algorithm is able to diversify the recommendations and to make a larger fraction of the item catalog visible to the users. 
The \emph{COV@n} metric is defined in Equation~\ref{eq:cov},

\begin{equation} \label{eq:cov}
COV@n = \frac{\mid \mathbb{S} \mid}{\mid \mathbb{T} \mid}
,
\end{equation}

where $ \mathbb{S} $ is the set with distinct articles that appeared in any top-N recommendation list and $ \mathbb{T} $ is the set with all distinct \emph{recommendable} items \cite{jannach2015recommenders}. In our case, an item is considered \emph{recommendable} if it has been clicked by any user in the last hour.

To measure novelty and diversity, we adapted the evaluation metrics that were proposed in \cite{vargas2011rank, castells2015novelty, vargas2015thesis}. We provide details of their implementation in Annex~ \ref{sec:novelty_diversity_metrics}.

The novelty metrics \emph{ESI-R@n} and \emph{ESI-RR@n} are based on item popularity, returning higher values for long-tail items. The \emph{ESI-R@n} (Expected Self-Information with  Rank-sensitivity) metric includes a rank discount, so that items in the top positions of the recommendation list have a higher effect on the metric. The \emph{ESI-RR@n} (Expected Self-Information with Rank- and Relevance-sensitivity) metric not only considers a rank discount, but also combines novelty with accuracy, as the relevant (clicked) item will have a higher impact on the metric if it is among the top-n recommended items.
Our diversity metrics are based on the \emph{Expected Intra-List Diversity (EILD)} metric. Analogously to the novelty metrics, there are variations to account for rank-sensitivity (\emph{EILD-R@n}) and for both rank- and relevance-sensitivity (\emph{EILD-RR@n}).

For our experiments, all recommender algorithms were tuned towards higher accuracy (\emph{MRR@10}) for each dataset using random search on a hold-out validation set. The resulting best hyper-parameters are documented in Annex~\ref{sec:annex_hyperparams}.

\subsection{Datasets}
\label{sec:datasets}
Two large news portals datasets are used for the experiments. The datasets contain recorded user interactions and information about the published articles:

\begin{itemize}
    \item \emph{Globo.com} (\emph{G1}) dataset -  Globo.com is the most popular media company in Brazil. This dataset was originally shared by us in \cite{moreira2018news}, with a second version shared in \cite{moreira2019contextual} \footnote{\url{https://www.kaggle.com/gspmoreira/news-portal-user-interactions-by-globocom}}, which also includes contextual information. The dataset was collected from the G1 news portal, which has more than 80 million unique users and publishes over 100,000 new articles per month; and
    \item \emph{SmartMedia Adressa dataset} - This dataset contains approximately 20 million page visits from a Norwegian news portal \cite{gulla2017adressa}. In our experiments we used the full-version dataset, which is available upon request\footnote{\url{http://reclab.idi.ntnu.no/dataset}}, and includes article text and click events of about 2 million users and 13,000 articles.
\end{itemize}

Both datasets include the textual content of the news articles, article metadata (such as publishing date, category, and author), and logged user interactions (page views) with contextual information. Since we are focusing on session-based news recommendations and short-term users preferences, it is not necessary to train algorithms for long periods. Therefore, and because articles become outdated very quickly, we have selected for the experiments all available user sessions from the first 16 days for both datasets.

In a pre-processing step, like in \cite{ludewig2018evaluation, epure2017recommending, twardowski2016modelling}, we organized the data into sessions using a 30 minute threshold of inactivity as an indicator of a new session. Sessions were then sorted by timestamp of their first click. From each session, we removed repeated clicks on the same article, as we are not focusing on the capability of algorithms to act as reminders as in \cite{LercheJannachEtAl2016}. Sessions with only one interaction are not suitable for next-click prediction and were discarded. Sessions with more than 20 interactions (stemming from \emph{outlier} users with an unusual behavior or from bots) were truncated.

The characteristics of the resulting pre-processed datasets are shown in Table~\ref{tab:datasets}. Coincidentally, the datasets are similar in many statistics, except for the number of articles. For the \emph{G1} dataset, the number of recommendable articles (clicked by at least one user) is much higher than for the Adressa dataset. The higher \emph{Gini index} of the articles' popularity distribution also indicates that the clicks in the Adressa dataset are more biased to popular articles, leading to a higher inequality in clicks distribution than for the G1.

\begin{table}[h!t]
\centering
\caption{Statistics of the datasets used for the experiments.}
\label{tab:datasets}
\vspace{10pt}
\begin{tabular}{p{8cm}rr}
\hline
 & \emph{Globo.com (G1)}
 & \emph{Adressa} \\ \hline
Language  & Portuguese & Norwegian  \\ 
Period (days)  & 16 & 16 \\  
\# users   & 322,897 & 314,661  \\ 
\# sessions & 1,048,594 & 982,210 \\  
\# clicks  & 2,988,181 & 2,648,999 \\  
\# articles   & 46,033 & 13,820  \\ 

Avg. sessions length (\# clicks / \# sessions)  & 2.84  & 2.70  \\
\emph{Gini index} (of the article pop. distribution) &  0.952 &  0.969  \\
\hline
\end{tabular}
\end{table}

\section{Implementation}

The architecture instantiations of the \emph{CHAMELEON} were implemented using \emph{TensorFlow} \cite{abadi2016tensorflow}, a popular and efficient numerical computation framework for deep learning. \emph{TensorFlow} computations are described in a graph model, which may be executed onto a wide variety of different hardware platforms, including Graphics Processing Unit (GPU) cards \cite{abadi2016tensorflow}.


The source code with the implementations of \emph{CHAMELEON} and the baseline methods were published\footnote{\url{https://github.com/gabrielspmoreira/chameleon\_recsys}} to make the experiments results reproducible.

\chapter{Main Results and Discussion}
\label{sec:chapter_6}

In this chapter, we present the analysis of the main results of the experiments and discuss our findings under the perspective of our research questions.

As previously stated in Chapter~\ref{sec:chapter_1}, the experiments were designed to address the following Research Questions (RQ):
\begin{itemize}
\item \emph{RQ1} - How does a contextual and hybrid RS based on the proposed neural meta-architecture perform in the news domain, in terms of recommendation quality factors (accuracy, item coverage, novelty, and diversity), compared to other traditional and state-of-the-art approaches for session-based recommendation?
\item \emph{RQ2} - What is the effect on news recommendation quality factors of leveraging different types of information in a neural-based contextual hybrid RS?
\item \emph{RQ3} - What is the effect on news recommendation quality of using different textual representations, produced by statistical NLP and Deep NLP techniques?
\item \emph{RQ4} - Is a hybrid RS based in the proposed meta-architecture able to reduce the problem of \emph{item cold-start} in the news domain, compared to other existing approaches for session-based recommendation?
\item \emph{RQ5} - Is it possible for a neural-based RS to effectively balance the trade-off between the recommendation quality factors of \emph{accuracy} and \emph{novelty}?
\end{itemize}

Those questions are addressed by a number of experiments on two news portals datasets (Section~\ref{sec:datasets}), comparing \emph{CHAMELEON} instantiations with baseline algorithms (Section~\ref{sec:baseline_algs}) using the evaluation methodology proposed in Section~\ref{sec:eval_methodology}.

For all tables presented in this Chapter, the best results for a metric are printed in bold face.
If the best results are significantly different\footnote{As errors around the reported averages were normally distributed, we used paired Student's t-tests with Bonferroni correction for significance tests.}
from measures of all other algorithms, they are marked with *** when $p<0.001$, with ** when $p<0.01$, and with * symbol when $p<0.05$.

\section{Evaluation of Recommendation Quality (RQ1)}
\label{sec:results_session_based}
In this section, it is investigated how  session-based recommendation methods, including the proposed  hybrid  RNN-based  architecture (\emph{CHAMELEON}), perform in the news domain in terms of recommendation quality factors -- accuracy, item coverage, novelty, and diversity. The accuracy analysis was initially explored in \cite{moreira2018news} and the final experiments, including other quality factors and more baselines, were reported in \cite{moreira2019contextual}.

At first, it is analyzed the obtained accuracy results and then it is discussed the other quality factors.

For these experiments, it was used an instantiation of \emph{CHAMELEON}'s \emph{NAR} module, which leverages all available information: the article Id, the article context, the article content (embeddings), the article metadata and the user context features, which are described in detail in Table~\ref{tab:nar-features}. 

\subsection{Accuracy Analysis}
\label{sec:results_accuracy}

Table~\ref{tab:accuracy_benchmarks} shows the accuracy results obtained from different algorithms in terms of the \emph{HR@10} and \emph{MRR@10} accuracy metrics. The reported values correspond to the average of measures obtained from each evaluation hour, according to the evaluation protocol described in Section~\ref{sec:eval_methodology}.

\begin{table}[h!t]
\centering
\caption{The Accuracy results from \emph{G1} and \emph{Adressa}}
\vspace{10pt}
\footnotesize
\begin{tabular}{lll|ll}

\hline
&
\multicolumn{2}{c}{\textbf{G1 dataset}} &
\multicolumn{2}{c}{\textbf{Adressa dataset}}
\\
\hline
  \emph{Algorithm}
  & \emph{HR@10}
  & \emph{MRR@10}
  & \emph{HR@10}
  & \emph{MRR@10}
 \\

\hline

\emph{CHAMELEON} & \textbf{0.6738}*** & \textbf{0.3458}*** & \textbf{0.7018}*** & \textbf{0.3421}*** \\
\emph{SR} & 0.5900 & 0.2889 & 0.6288 & 0.3022 \\
\emph{Item-kNN} & 0.5707 & 0.2801 & 0.6179 & 0.2819 \\
\emph{CO} & 0.5689 & 0.2626 & 0.6131 & 0.2768 \\
\emph{V-SkNN} & 0.5467 & 0.2494 & 0.6140 & 0.2723 \\
\textit{SR-GNN} & 0.5144 & 0.2467 & 0.6122 & 0.2991 \\
\textit{GRU4Rec} & 0.4669 & 0.2092 & 0.4958 & 0.2200 \\
\emph{RP} & 0.4577 & 0.1993 & 0.5648 & 0.2481 \\
\emph{CB} & 0.3643 & 0.1676 & 0.3307 & 0.1253 \\

\hline
\end{tabular}

\label{tab:accuracy_benchmarks}
\end{table}

In this comparison, \emph{CHAMELEON} outperforms the other baseline algorithms on both datasets and on both accuracy metrics by a large margin. The \emph{SR} method performs second-best. The \emph{CB} is the least accurate recommendation method, as it is not influenced by articles popularity like the other methods, but solely by content similarity.

Generally, the accuracy improvement obtained by \emph{CHAMELEON} over the best baseline method (\emph{SR}) is higher for the \textit{G1} dataset. This can be explained by the facts that (a) the number of articles in the \emph{G1} dataset is more than 3 times higher than in the other dataset and (b) the \emph{G1} dataset has a lower popularity bias, see the \emph{Gini index} in Table~\ref{tab:datasets}. As a result, algorithms that have a higher tendency to recommend popular items are less effective for datasets with a more balanced click distribution (i.e., a lower \emph{Gini index}). Looking, for example, at the algorithm that simply recommends recently-popular articles (\emph{RP}), it is possible to see that its performance is much higher for the \emph{Adressa} dataset, even though the best obtained measures (by \emph{CHAMELEON}) are almost similar for both datasets.

It is also possible to observe that the other neural approaches (i.e., \textit{SR-GNN} and \textit{GRU4Rec}) were not able to provide better accuracy than non-neural baselines for session-based news recommendation.
One of the reasons might be that in a real-world scenario---as emulated in our evaluation protocol---neural-based models (including \emph{CHAMELEON}) cannot be updated as often as the baseline methods, due to challenges of asynchronous model training and frequent deployment. 

Additionally, \textit{CHAMELEON}'s architecture was designed to be able to recommend fresh articles not seen during training. \textit{SR-GNN} and \textit{GRU4Rec} in contrast, cannot make recommendations for items that were not encountered during training, which limits their accuracy in a realistic evaluation. In our datasets, for example, we found that about 3\% (\emph{Adressa} dataset) and 4\% (\emph{G1} dataset) of the item clicks in each evaluation hour were on fresh articles, i.e., on articles that were not seen in the preceding training hours.

From the two neural methods, the newer graph-based \textit{SR-GNN} method was performing much better than \textit{GRU4Rec} in our problem setting. 
However, the \textit{SR-GNN} does not achieve the accuracy of \textit{CHAMELEON}, even when \textit{CHAMELEON} is not leveraging any additional side information other than the article ID (configuration \textit{IC1} in Table \ref{tab:input_types_results}), as it will be presented in an \emph{ablation study} in Section~\ref{sec:features_results}.

\begin{figure}[h!t]
    \includegraphics[width=\textwidth]{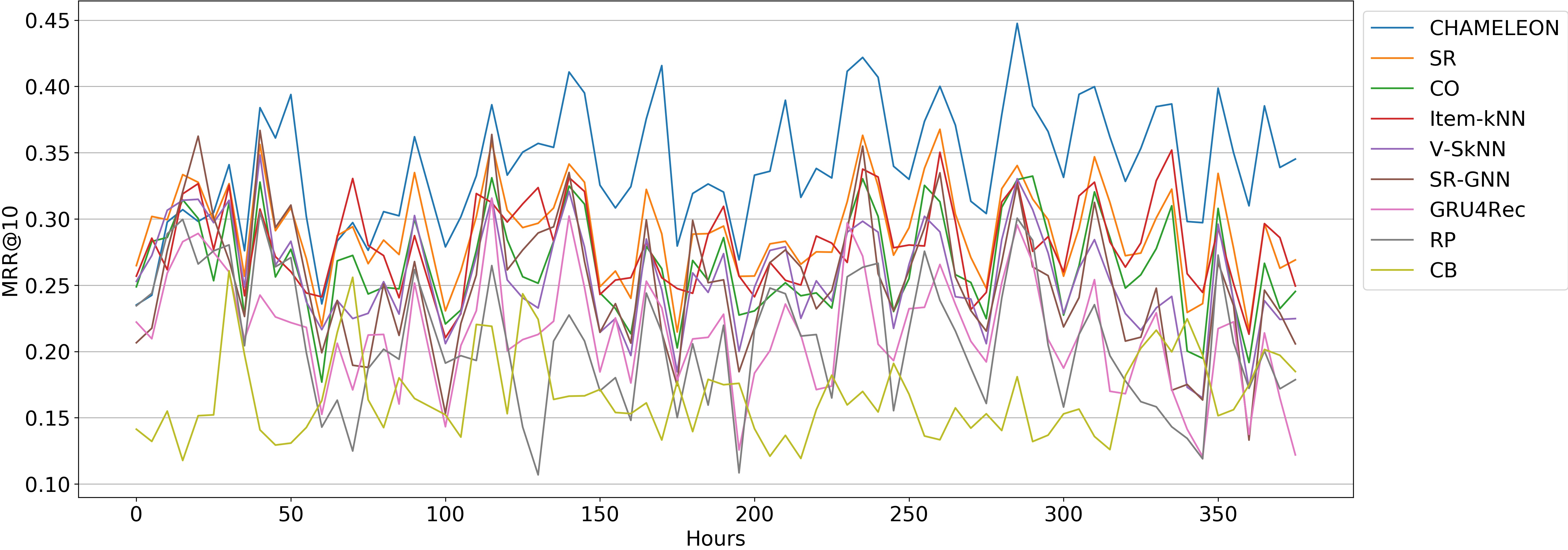}
    \caption{\emph{G1} dataset (16 days) - Accuracy (\emph{MRR@10}) after every 5 hours}
    \label{fig:accuracy_over_time_g1}
\end{figure}

\begin{figure}[h!t]

    \includegraphics[width=\textwidth]{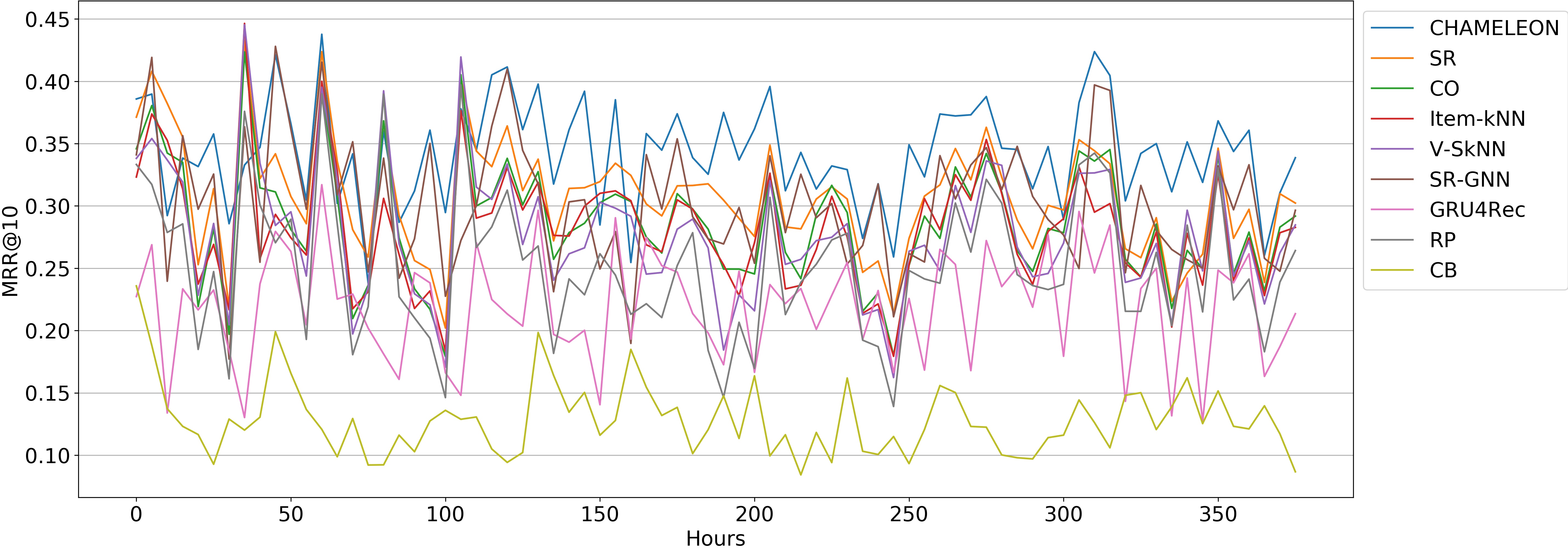}
    \caption{\emph{Adressa} dataset (16 days) - Accuracy (\emph{MRR@10}) after every 5 hours}
    \label{fig:accuracy_over_time_adressa}
\end{figure}

In Figures~\ref{fig:accuracy_over_time_g1}~and~\ref{fig:accuracy_over_time_adressa}, it is plotted the obtained accuracy values (\emph{MRR@10}) of the different algorithms along the 16 days, with an evaluation after every 5 hours. It is possible to see that, after training for some hours, \emph{CHAMELEON} clearly recommends with higher accuracy than all other algorithms.

\subsubsection{Analysis of Recommendation Accuracy by Session Size}

\begin{figure}[h!t]

  \subfigure[a][\emph{G1} dataset]{
    \includegraphics[width=10cm]{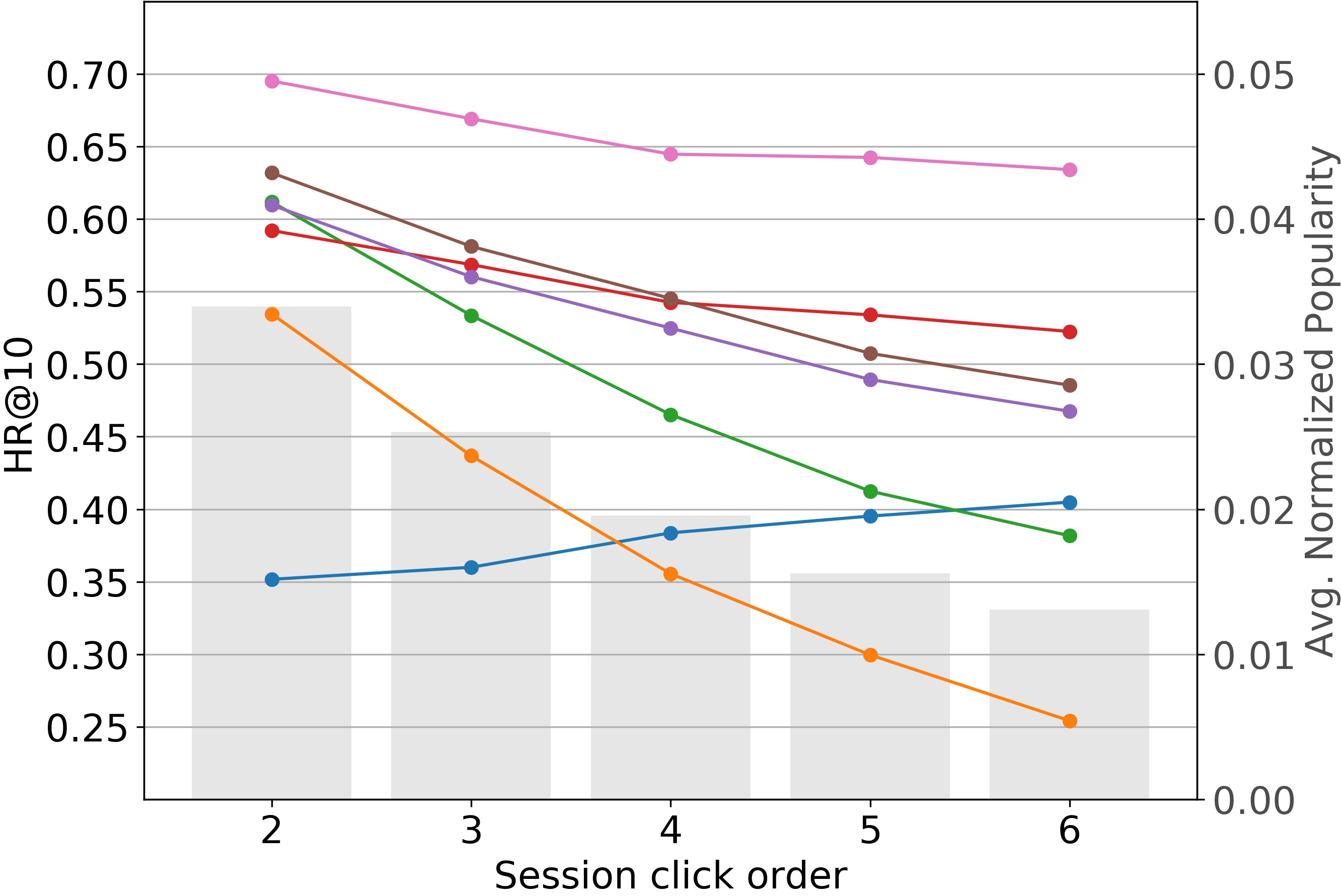}}
    
  \subfigure[b][\emph{Adressa} dataset]{
    \includegraphics[width=13cm]{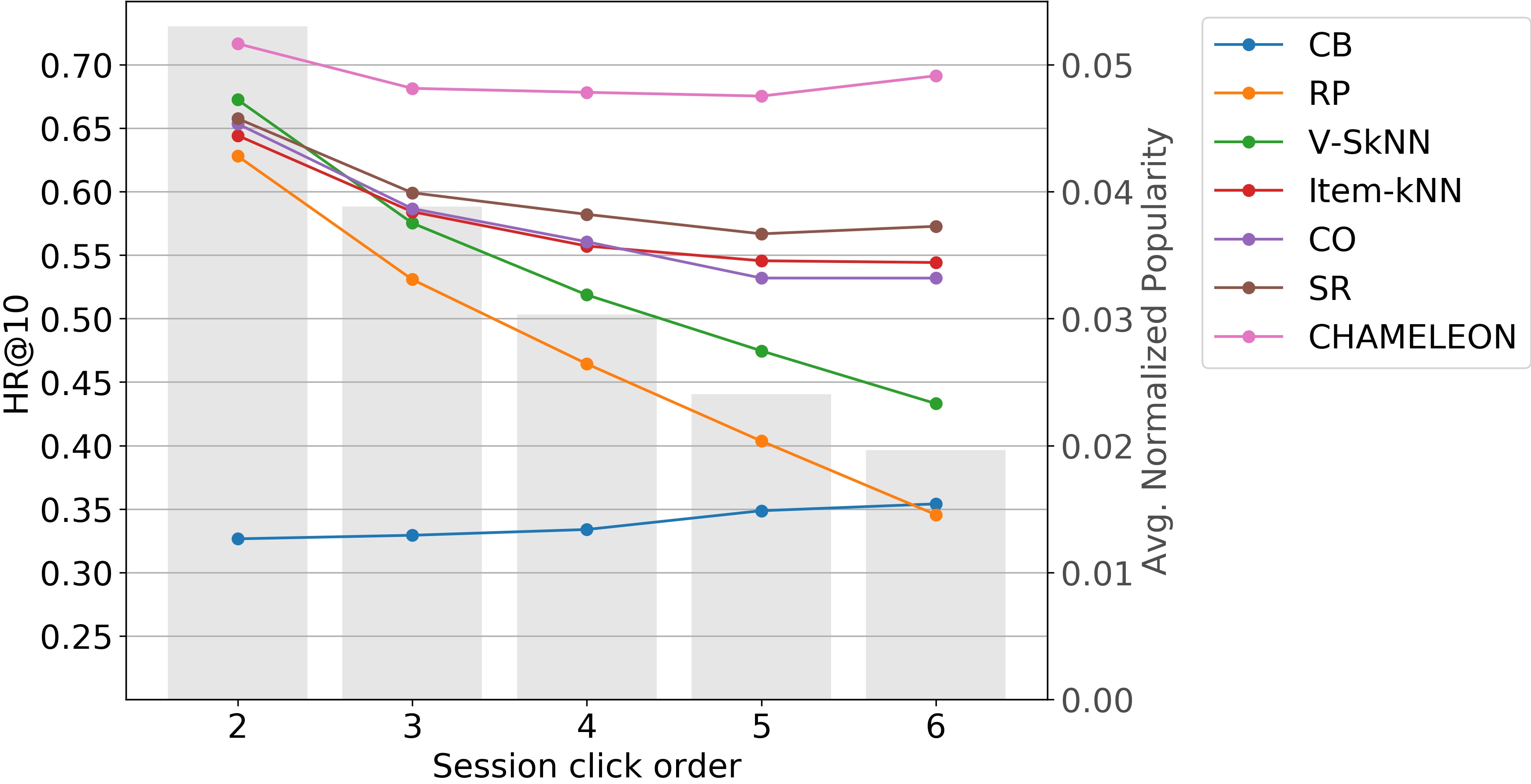}}
  
  \caption{The Recommendation Accuracy (\emph{HR@10}) of algorithms (lines) x Avg. Normalized Popularity (bars) by session click order}
  \label{fig:accuracy_by_session_pos}
\end{figure}

Additionally, it was conducted an analysis on how the accuracy of next-article recommendation varies for latter clicks within sessions, i.e., the accuracy to predict the 2nd, 3rd, 4th, 5th, and 6th clicks. In Figure~\ref{fig:accuracy_by_session_pos}, the \emph{HR@10} for each algorithm is segmented by session click order, and gray bars represent the average \emph{normalized popularity} (Equation~\ref{eq:rec-norm-pop}) at that session position.

Interestingly, it is possible to observe for both datasets, a higher popularity-bias in first session clicks, which reduces for latter clicks. A possible explanation for that may reside on the fact that generally news portals highlight very popular articles in their homepages to attract users attention, influencing more the first clicks of users sessions. 

The accuracy of the \emph{RP} method decays almost linearly with the decrease of normalized popularity, as expected. As the \emph{CB} method is agnostic to articles popularity, it was able to perform even slightly better for latter clicks, as users may start to browse for related content. After the \emph{CB} method, the \emph{CHAMELEON} ranks second in terms of lower relative decrease in accuracy, showing robustness and flexibility to leverage side-information other than popularity to provide accurate recommendations for latter session clicks. 

\subsubsection{Analysis of Accuracy Correlation}
It was also analyzed the \emph{Pearson} correlation of recommendation accuracy results across the evaluation hours of the evaluated algorithms. A high correlation between the accuracy of two methods means that they vary in the same direction, depending on the changes across time of users' global behaviour (e.g., users interests focused in more popular or more niche topics).

\begin{figure}[h!t]
\begin{center}
  \subfigure[a][\emph{G1} dataset]{
    \includegraphics[width=10cm]{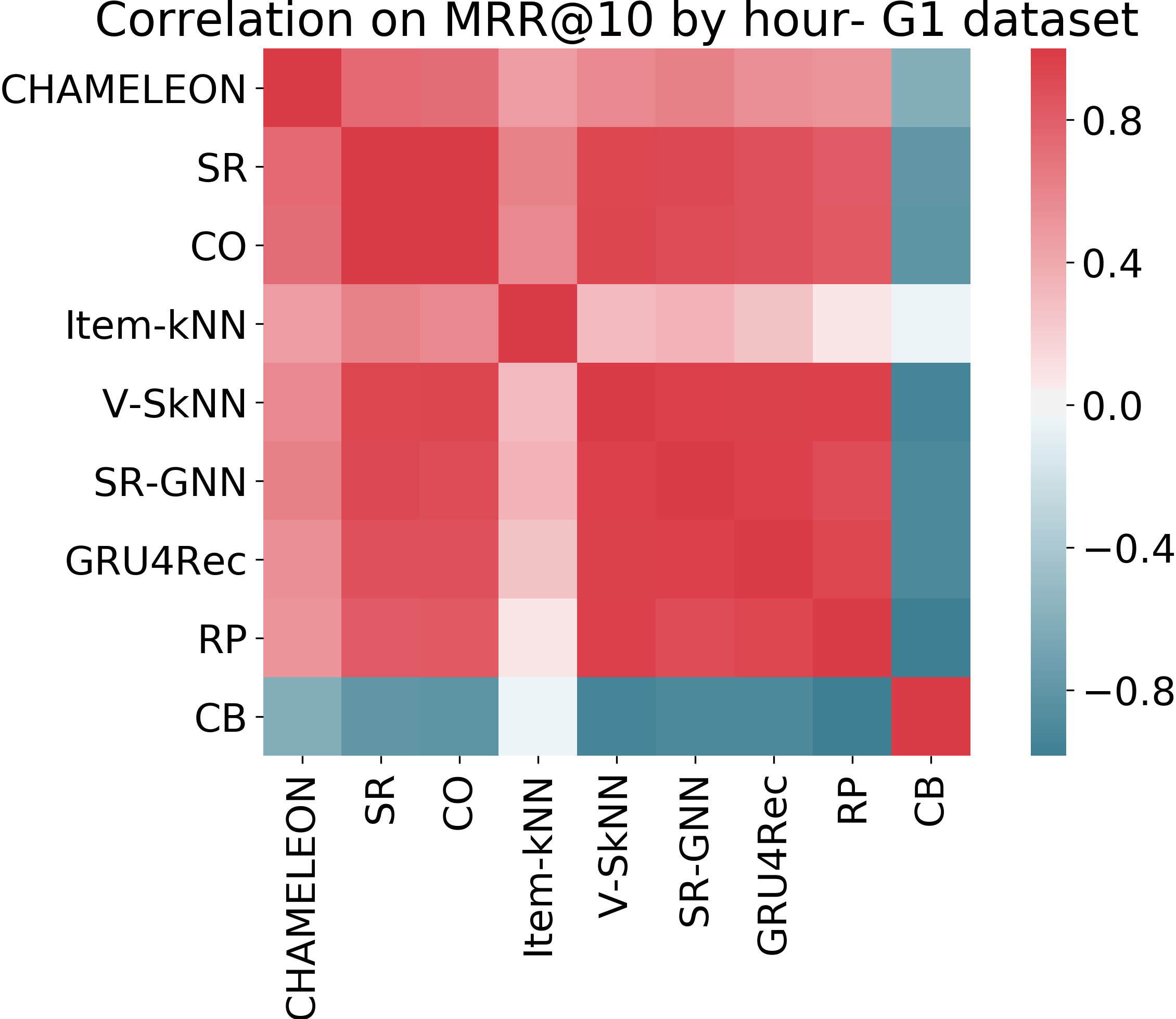}}
    
  \subfigure[b][\emph{Adressa} dataset]{
    \includegraphics[width=10cm]{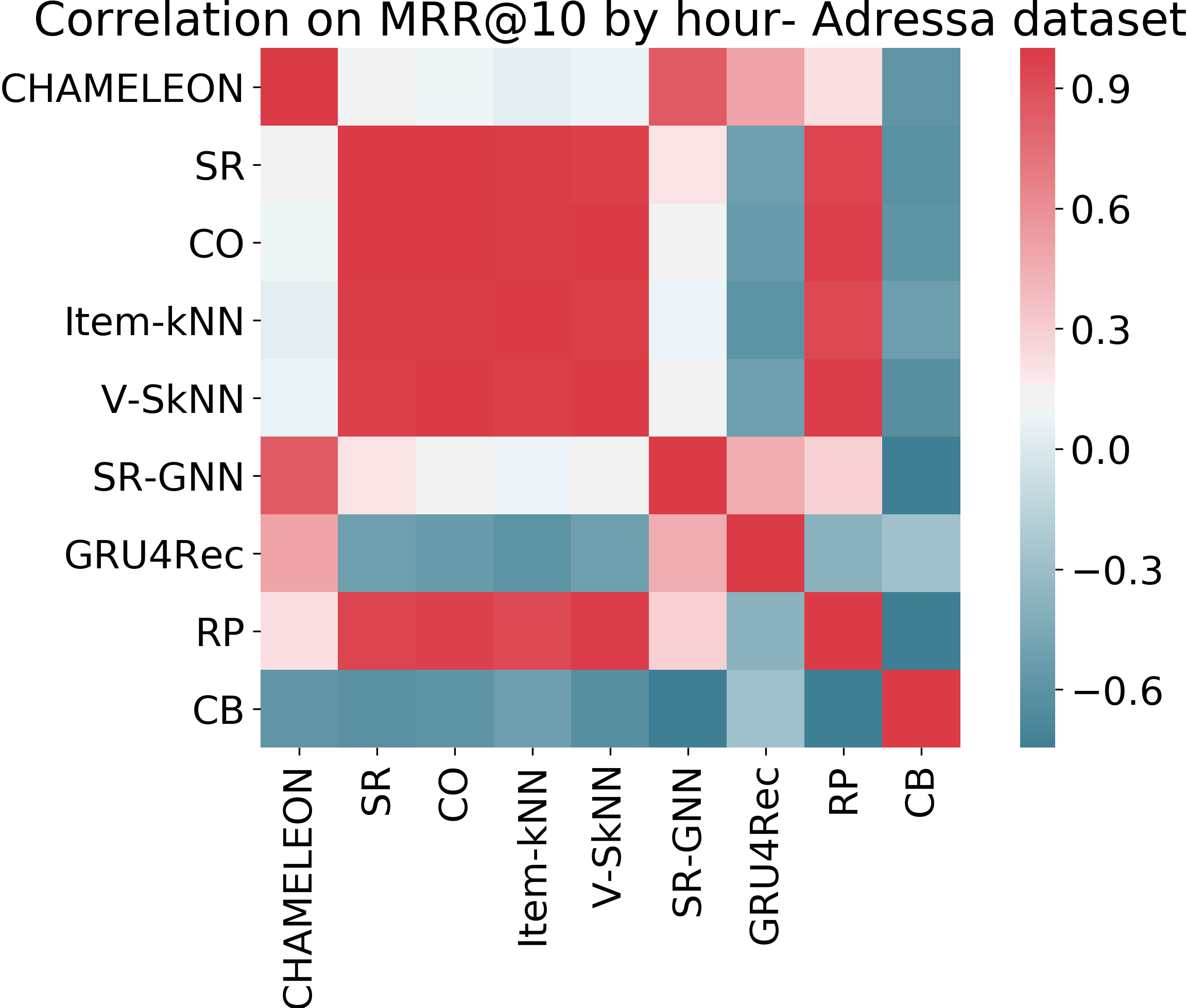}}
  
  \caption{The Correlation matrices on \emph{MRR@10} by hour}
  \label{fig:corr_matrix}
\end{center}
\end{figure}

Figure~\ref{fig:corr_matrix} presents a correlation matrix of algorithms' accuracy (\emph{MRR@10}) across evaluation hours.

As expected, the \emph{Content-Based algorithm (CB)} is the less correlated one compared to all other algorithms. It solely uses content similarity for recommendation, ignoring previous users interactions. The other baselines are well correlated, for using solely item IDs and being popularity biased. 

For the \emph{G1} dataset, the baselines with the highest correlation with \emph{CHAMELEON} were \emph{SR} (0.68), \emph{CO} (0.65), and \emph{SR-GNN} (0.55), and for the \emph{Adressa} dataset, they were \emph{SR-GNN} (0.74), \emph{GRU4Rec} (0.58), and \emph{RP} (0.57).

It is also possible to see that for the \emph{Adressa} dataset, popularity-based algorithms have their recommendation accuracy  more correlated to \emph{RP} than in \emph{G1} dataset. Even \emph{CHAMELEON}'s correlation with \emph{RP} algorithm is higher for \emph{Adressa} (0.57) than for \emph{G1} (0.41) dataset. This is another evidence that the \emph{Adressa} dataset has a stronger popularity-bias compared to the \emph{G1} dataset.

\subsection{Analysis of Additional Quality Factors}
\label{sec:additional_quality_factors}
It was also investigated recommendation quality factors other than accuracy in our research---item coverage, novelty, and diversity. The results are shown in Table~\ref{tab:other_factors_benchmarks}, and can be summarized as follows:

\begin{table}[h!t]
\centering
\caption{The evaluation of other quality factors for the \emph{G1} and \emph{Adressa} datasets}
\label{tab:other_factors_benchmarks}
\resizebox{\textwidth}{!}{%
\begin{tabular}{llllll}
\hline
 & \emph{Item Coverage}
 & \multicolumn{2}{c}{\emph{Novelty}}
 & \multicolumn{2}{c}{\emph{Diversity}} \\
 \emph{Recommender}
 & \emph{COV@10}
 & \emph{ESI-R@10}
 & \emph{ESI-RR@10}
 & \emph{EILD-R@10}
 & \emph{EILD-RR@10}
 \\
\hline
\multicolumn{6}{p{8cm}}{\textbf{G1 dataset}}\\ \hline
\emph{CHAMELEON} & 0.6373 & 6.4177 & \textbf{0.7302}*** & 0.3620 & \textbf{0.0419}*** \\
\emph{SR} & 0.2763 & 5.9747 & 0.5747 & 0.3526 & 0.0374 \\
\emph{Item-kNN} & 0.3913 & 6.5909 & 0.6301 & 0.3552 & 0.0361 \\
\emph{CO} & 0.2499 & 5.5728 & 0.5126 & 0.3570 & 0.0352 \\
\emph{V-SkNN} & 0.1355 & 5.1760 & 0.4411 & 0.3558 & 0.0339 \\
\textit{SR-GNN} & 0.3196 & 5.4280 & 0.5093 & 0.3668 & 0.0350  \\
\textit{GRU4Rec} & 0.6333 & 5.2332 & 0.3925 & 0.3662 & 0.0310  \\
\emph{RP} & 0.0218 & 4.4904 & 0.3259 & \textbf{0.3750}*** & 0.0296 \\
\emph{CB} & \textbf{0.6774} & \textbf{8.1531}*** & 0.5488 & 0.2789 & 0.0193 \\
\hline
\multicolumn{6}{p{8cm}}{\textbf{Adressa dataset}}\\ \hline
\emph{CHAMELEON} & 0.7926 & 5.3410 & \textbf{0.6083}*** & 0.2123 & \textbf{0.0250}*** \\
\emph{SR} & 0.4604 & 5.4443 & 0.5277 & 0.2188 & 0.0235 \\
\emph{CO} & 0.4220 & 5.0789 & 0.4748 & 0.2138 & 0.0222 \\
\emph{Item-kNN} & 0.5314 & 5.4675 & 0.5091 & \textbf{0.2246} & 0.0228 \\
\emph{V-SkNN} & 0.1997 & 4.6018 & 0.4112 & 0.2112 & 0.0217 \\
\textit{SR-GNN} & 0.5197 & 5.1013 & 0.5125 & 0.2214 & 0.0241 \\
\textit{GRU4Rec} & 0.5143 & 5.0571 & 0.3782 & 0.2131 & 0.0184  \\
\emph{RP} & 0.0542 & 4.1465 & 0.3486 & 0.2139 & 0.0200 \\
\emph{CB} & \textbf{0.8875}*** & \textbf{7.6715}*** & 0.4104 & 0.0960 & 0.0060 \\ \hline
\end{tabular}
}
\end{table}

\begin{itemize}
\item \textbf{Item Coverage} - The \emph{COV} metric is computed considering as \emph{recommendable} articles the ones that have been clicked by any user in the last hour. For this metric, \emph{CHAMELEON} has a much richer spectrum of articles that are included in its top-10 recommendations compared to other algorithms, suggesting a higher level of personalization. The only method with a higher coverage was the \emph{CB} method, which however is not very accurate. This is expected for a method that is agnostic of article's popularity. The \emph{RP} method has the lowest item coverage, as expected, for recommending only the most popular articles;
\item \textbf{Novelty} - The \emph{CB} method also recommends the least popular, and thus more novel articles, according to the \emph{ESI-R} metric. This effect has been observed in other works such as \cite{castells2015novelty, celma2008new}, which is expected as this is the only method that does not take item popularity into account in any form. \emph{CHAMELEON} ranks third on this metric for the \emph{G1} dataset and is comparable to the other algorithms for \emph{Adressa}\footnote{it will be shown later, in Section~\ref{sec:tradeoff_results}, how the novelty of \emph{CHAMELEON} can be increased based on the novelty regularization method proposed in Section~\ref{sec:new_loss_function}.}. Looking at novelty in isolation is, however, not sufficient, which is why it is included the relevance-weighted novelty  metric (\emph{ESI-RR}) as well. When novelty and relevance (accuracy) are combined in one metric, it turns out that \emph{CHAMELEON} leads to the best values on both datasets; and
\item \textbf{Diversity} - Most algorithms are quite similar in terms of the \emph{EILD-R@10} metric. The \emph{CB} method has the lowest diversity by design, as it always recommends articles with similar content. When article relevance is taken into account along with diversity with the \emph{EILD-RR@10} metric, it is possible to see again that \emph{CHAMELEON} is more successful than others in providing recommendations with both diversity and accuracy.
\end{itemize}

In general, it can be seen that \emph{CHAMELEON}, besides being the most accurate recommendation method, was also able to provide the best combination of accuracy with novelty and diversity, and the second best in terms of item coverage.

\section{Analyzing the Importance of Input Features for the \emph{NAR} Module (RQ2)}

\label{sec:features_results}
\emph{CHAMELEON} leverages a number of input features to provide more accurate recommendations. To understand the effects of including those features in our model, it was performed a number of experiments with feature sets (described previously in Table~\ref{tab:nar-features}) combined in different Input Configurations (IC)\footnote{This process is sometimes referred to as \emph{ablation study}.}.
Table~\ref{tab:input_config} shows five different configurations where we start only with the article IDs (\emph{IC1}) and incrementally add more features until we have the  model with all input features (\emph{IC5}).


\begin{table}[!htbp]
\centering
\caption{\emph{The Input Configurations (IC)} for the \emph{NAR} module}
\label{tab:input_config}
\vspace{10pt}
\footnotesize
\begin{tabular}{p{2.5cm}p{11cm}}
\hline
 \textbf{Input config.}
 & \textbf{Feature Sets}
 \\ \hline
\emph{IC1} & Article Id \\
\emph{IC2} & \emph{IC1} + Article Context (Novelty and Recency) \\
\emph{IC3} & \emph{IC2} + the \emph{Article Content Embeddings (ACE)} learned by the \emph{supervised} instantiation of the \emph{ACR} module, described in Section~\ref{sec:acr_supervised_instantiation} \\
\emph{IC4} & \emph{IC3} + Article Metadata \\
\emph{IC5} & \emph{IC4} + User Context \\
\hline
\end{tabular}
\end{table}

Table~\ref{tab:input_types_results} shows the results of this analysis, also reported in \cite{moreira2019contextual}. 
In general, it can be seen that both accuracy (\emph{HR@10} and \emph{MRR@10}) and item coverage (\emph{COV@10}) improve when more input features are considered in the \emph{NAR} module.
The largest improvements in terms of accuracy for both datasets can be observed when the \emph{Article Content Embeddings} (\emph{IC3}) are included. The feature sets of \emph{User Context} (\emph{IC5}) and \emph{Article Context} (\emph{IC2}) also played an important role when generating the recommendations.

\begin{table}[h!t]
\centering
\caption{The Effects of different input feature configurations on recommendation quality.}
\label{tab:input_types_results}
\resizebox{0.9\textwidth}{!}{%
\begin{tabular}{llllll}
\hline

 \emph{Recommender}
 & \emph{HR@10}
 & \emph{MRR@10}
 & \emph{COV@10}
 & \emph{ESI-R@10}
 & \emph{EILD-R@10} \\
\hline  \hline
\multicolumn{6}{p{8cm}}{\textbf{G1 dataset}}\\ \hline

\emph{IC1} & 0.5708 & 0.2674 & 0.6084 & 6.2597 & \textbf{0.4515} \\
\emph{IC2} & 0.6073 & 0.2941 & 0.6095 & 6.1841 & 0.3736 \\
\emph{IC3} & 0.6472 & 0.3366 & 0.6296 & 6.1507 & 0.3625 \\
\emph{IC4} & 0.6483 & 0.3397 & 0.6316 & 6.1573 & 0.3621 \\
\emph{IC5} & \textbf{0.6738}*** & \textbf{0.3458}* & \textbf{0.6373} & \textbf{6.4177}** & 0.3620 \\

\hline  \hline
\multicolumn{6}{p{8cm}}{\textbf{Adressa dataset}}\\ \hline

\emph{IC1} & 0.6779 & 0.3260 & 0.7716 & 5.3296 & \textbf{0.2190} \\
\emph{IC2} & 0.6799 & 0.3273 & \textbf{0.8034} & 5.2636 & 0.2187 \\
\emph{IC3} & 0.6906 & 0.3348 & 0.7820 & 5.2771 & 0.2103 \\
\emph{IC4} & 0.6906 & 0.3362 & 0.7882 & 5.2900 & 0.2123 \\
\emph{IC5} & \textbf{0.7018}*** & \textbf{0.3421}**  & 0.7926 & \textbf{5.3410} & 0.2123 \\
\hline
\end{tabular}
}
\end{table}

It is also possible to observe cases where measures become lower with the addition of new features. For both datasets, for example, the diversity of \emph{CHAMELEON}'s recommendations in terms of the \emph{EILD-R} metric decreases with additional features, in particular when the \emph{Article Content} features is included at \emph{IC3}. This is expected, as recommendations become generally more similar when content features are used in a hybrid RS.

Looking at the \emph{IC3} configuration, it  can be observed that for the \emph{G1} dataset the positive effect of textual content in recommendation accuracy is much more than for the \emph{Adressa} dataset. A possible explanation for this difference can lie in the nature of the available metadata of the articles, which are used as target attributes during the \emph{supervised} training of the \emph{ACR} module. In the \emph{G1} dataset, for example, we have 461 article categories, which is much more than for the \emph{Adressa} dataset, with 41 categories. Furthermore, the distribution of articles by category is more unbalanced for \emph{Adressa} (\emph{Gini index} = 0.883) than for \emph{G1} (\emph{Gini index} = 0.820). In theory, fine-grained metadata can lead to content embeddings clustered around distinctive topics, which may be useful to recommend related content.
 
In the next section, it is explored in more detail the effect of different types of textual representation.

\section{Analyzing the Effect of Textual Representation in Recommendations (RQ3)}
\label{sec:content_representation}

In the particular domain of news recommendation, the use of hybrid techniques which consider the textual content of a news item, have often shown to be preferable to deal with item cold-start, see e.g., \cite{chu2009personalized,liu2010personalized,li2011scene,rao2013personalized,lin2014personalized,li2014modeling,trevisiol2014cold,epure2017recommending}.

In this section, as reported in \cite{moreira2019inra}, it is investigated to what extent the choice of the mechanism for encoding articles' textual content can affect the recommendation quality. For that, it was considered different types of textual representation, based on statistical and Deep NLP techniques, trained using \emph{supervised} and \emph{unsupervised} approaches.

For these experiments, it was used as basic inputs for the \emph{NAR} module the following \emph{feature sets} (described previously in Table~\ref{tab:nar-features}) -- \emph{Article Context}, \emph{Article Metadata}, and \emph{User Context}. The \emph{No-ACE} configuration uses only the basic inputs described above\footnote{The \emph{Article Id} trainable embedding was not used as input for these experiments, for a better sensitivity of the effect of different content representations on recommendation quality.}. The other configurations add to the basic features the different textual representations presented in Table~\ref{tab:feature-extraction}.

\begin{table}[h!]
\begin{threeparttable}
\centering
\caption{The alternative content processing techniques to generate \emph{ACEs}.}
\vspace{-5pt}

\label{tab:feature-extraction}
\footnotesize
\begin{tabular}{p{3cm}lp{10cm}}
\hline
 \emph{Technique} &
 \emph{Input} &
 \emph{Description} \\
\hline
\multicolumn{3}{p{6cm}}{\textbf{Supervised}}\\ \hline
\emph{CNN} & \emph{word2vec} & The \emph{supervised} CNN-based instantiation of the \emph{CHAMELEON's} \emph{ACR} module described in Section~\ref{sec:tfr_supervised_cnn_instantiation}. \\
\emph{GRU} & word2vec & The \emph{supervised} RNN-based instantiation of the \emph{CHAMELEON's} \emph{ACR} module described in Section~\ref{sec:tfr_supervised_rnn_instantiation}.\\

\hline
\multicolumn{3}{p{6cm}}{\textbf{Unsupervised}}\\ \hline
\emph{SDA-GRU} & Raw text & The \emph{unsupervised} instantiation of the \emph{ACR} module, based on Sequence Denoising \emph{GRU} Autoencoders, as described in Section~\ref{sec:acr_unsupervised} \\

\emph{LSA} & Raw text & Traditional Latent Semantic Analysis (LSA) \cite{deerwester1990lsa}. We used a variation based on \emph{TF-IDF} vectors \cite{ramos2003using} and \emph{Truncated SVD} \cite{halko2011finding}. \\
\emph{W2V*TF-IDF} & \emph{word2vec} &  \emph{TF-IDF} weighted \emph{word2vec} \cite{lilleberg2015support}, a technique to represent a piece of text as the average of its word embeddings weighted by \emph{TF-IDF} \cite{ramos2003using}. \\ 
\emph{doc2vec} & Raw text & Paragraph Vector (a.k.a doc2vec) \cite{le2014distributed} learns fixed-length feature representations from variable-length pieces of texts, which are trained via the distributed memory and distributed bag of words models. \\

\hline
\end{tabular}
\end{threeparttable}
\vspace{-10pt}
\end{table}

Regarding to the content preprocessing, as there were some very long articles, the text was truncated after the first 12 sentences, and concatenated with the title. The \emph{Article Content Embeddings (ACE)} produced by the selected techniques were \emph{L2}-normalized to make the feature scale similar, but also to preserve high similarity scores for embeddings from similar articles.

The experiments consisted of training \emph{NAR} module of \emph{CHAMELEON} using the \emph{ACE} produced by those different techniques for content representations and evaluating the provided recommendations. The results for the \emph{G1} and \emph{Adressa} datasets are presented in Tables~\ref{tab:g1_content_results} and~\ref{tab:adressa_content_results} \footnote{Results for the diversity metrics are not reported for been meaningless for this comparison of different \emph{ACE}, because the diversity metrics are computed based on \emph{ACE} similarities.}. 

\begin{table}[h!t]
\centering
\caption{The Results for the \emph{G1} dataset, with \emph{CHAMELEON} using different \emph{ACEs}.}
\label{tab:g1_content_results}
\begin{tabular}{lllll}
\hline
 \emph{Recommender}  & \emph{HR@10}  & \emph{MRR@10}  & \emph{COV@10}  & \emph{ESI-R@10} \\
\hline  \hline

\emph{No-ACE} & 0.6281 & 0.3066 & 0.6429 & 6.3169 \\ \hline
\multicolumn{5}{p{6cm}}{\emph{Supervised}} \\ \hline
\emph{CNN} & 0.6585 & 0.3395 & \textbf{0.6493} & 6.2874 \\
\emph{GRU} & 0.6585 & 0.3388 & 0.6484 & 6.2674 \\ \hline
\multicolumn{5}{p{6cm}}{\emph{Unsupervised}} \\ \hline
\emph{SDA-GRU} & 0.6418 & 0.3160 & 0.6481 & 6.4145 \\
\emph{W2V*TF-IDF} & 0.6575 & 0.3291 & 0.6500 & 6.4187 \\
\emph{LSA} & \textbf{0.6686}*** & \textbf{0.3423} & 0.6452 & 6.3833 \\
doc2vec & 0.6368 & 0.3119 & 0.6431 & \textbf{6.4345} \\
\hline

\hline  \hline
\end{tabular}
\vspace{0pt}
\end{table}

\vspace{-5pt}
\begin{table}[h!t]
\centering
\caption{The Results for the \emph{Adressa} dataset, with \emph{CHAMELEON} using different \emph{ACEs}.}
\vspace{-5pt}
\label{tab:adressa_content_results}
\begin{tabular}{lllll}
\hline
 \emph{Recommender}  & \emph{HR@10}  & \emph{MRR@10}  & \emph{COV@10}  & \emph{ESI-R@10} \\
\hline  \hline

\emph{No-ACE} & 0.6816 & 0.3252 & \textbf{0.8185} & 5.2453  \\ \hline
\multicolumn{5}{p{6cm}}{\emph{Supervised}} \\ \hline
\emph{CNN} & 0.6860 & 0.3333 & 0.8103 & 5.2924  \\
\emph{GRU} & 0.6856 & 0.3327 & 0.8096 & 5.2861 \\ \hline

\multicolumn{5}{p{6cm}}{\emph{Unsupervised}} \\ \hline
\emph{SDA-GRU} & 0.6905 & 0.3360 & 0.8049 & 5.3170 \\
\emph{W2V*TF-IDF} & 0.6913 & 0.3402 & 0.7976 & 5.3273  \\
\emph{LSA} & \textbf{0.6935} & \textbf{0.3403} & 0.8013 & 5.3347  \\
\emph{doc2vec} & 0.6898 & 0.3402 & 0.7968 & \textbf{5.3417}  \\
\hline  \hline
\end{tabular}
\vspace{-10pt}
\end{table}



In general, we can observe that considering content information is in fact highly beneficial in terms of recommendation accuracy. It is also possible to see that the choice of the article representation matters.

Surprisingly, the long-established \emph{LSA} method was the best performing technique to represent the content for both datasets in terms of accuracy, even when compared to more recent techniques using pre-trained word embeddings, such as the \emph{CNN} and \emph{GRU}.

For the \emph{G1} dataset, the \emph{Hit Rates} (\emph{HR}) were improved by around 7\% and the \emph{MRR} by almost 12\% when using the \emph{LSA} representation instead of the \emph{No-ACE setting}. For the \emph{Adressa} dataset, the difference between the \emph{No-ACE} settings and the hybrid methods leveraging text are less pronounced. The improvement using \emph{LSA} compared to the \emph{No-ACE setting} was around 2\% for \emph{HR} and 5\% for \emph{MRR}.

Furthermore, for the \emph{Adressa} dataset, it is possible to observe that all the \emph{unsupervised} methods (\emph{SDA-GRU}, \emph{LSA}, \emph{W2V*TF-IDF}, and \emph{doc2vec}) for generating \emph{ACEs} performed better than the \emph{supervised} ones, differently from the \emph{G1} dataset.
A possible explanation can be that the \emph{supervised} methods depend more on the \emph{quality} and \emph{depth} of the available article metadata information  used as the target for the training.

In terms of coverage (\emph{COV@10}) and novelty(\emph{ESI-R@10}), it turns out that for the \emph{G1} dataset, using some content  representations can lead to slightly higher values compared to the \emph{No-ACE} settings and to the unsupervised approaches.

\section{Reducing the Item Cold-Start Problem for News  Recommendation (RQ4)}
\label{sec:item_cold_start_results}

The news domain is one of the recommendation scenarios where the item cold-start problem is more intense, as thousands of articles are published every day in a large news portal \cite{spangher2015}, and they become obsolete in just a few hours \cite{das2007}. If an algorithm takes to long to learn which users might be interested in a fresh article, such content might not be relevant for the majority of users anymore when it finally starts to get recommended.

In this section, it is investigated whether the proposed hybrid RNN-based architecture (\emph{CHAMELEON}) is able to reduce the problem of item cold-start in the news domain, compared to existing approaches for session-based recommendation.

For that, it was designed and performed experiments to measure how long does it take since the article became available (first user click) until an algorithm recommends that article among the top-N items for the first time.

In our training and evaluation protocol, the recommender algorithms are processed by \emph{mini-batches} of user sessions. In this sense, it is highly desirable that an algorithm starts recommending a fresh item after observing just a few user sessions (i.e., a low number of \emph{batches}).

For these experiments, the item cold-start problem is measured here as the number of \emph{training batches} between the first user click in the article and its first top-N recommendation by a given algorithm. For that, it is proposed the metric \emph{\# Batches before First Recommendation (BFR@n)}:

\begin{equation} \label{eq:coldstart}
BFR@n(a) = \text{frb}@n(a) - \text{fcb}(a),
\end{equation}

where $ \text{fcb}(a) $ is the sequential number of the batch where the article $ a $ was clicked by any user for the first time (a \emph{proxy} for article's publishing time), and $ \text{frb}@n(a) $ is the  sequential number  of the batch with the first top-$n$ recommendation containing the article $a$, provided by a recommender algorithm.

For these experiments, it was used a subset of the \emph{G1} and \emph{Adressa} datasets, considering the first 9 days of user interactions. The metric \emph{BFR@n} was computed for all investigated session-based recommenders, using a \emph{ batch size} of \textbf{64 samples} for both datasets. 

In Tables~\ref{tab:g1_item_cold_start}~and~\ref{tab:adressa_item_cold_start}, it is possible to observe statistics of the distribution of the \emph{BFR@n} measures for each dataset and algorithm. For example, the 75\% percentile column means that  75\% of the first articles' recommendation required at most that number of \emph{training batches} to be recommended, since the article was published. 

\begin{table}[h!t]
\centering
\caption{The analysis of the item cold-start problem by algorithm for the \emph{G1} dataset}
\label{tab:g1_item_cold_start}
\resizebox{\textwidth}{!}{%
\begin{tabular}{l|rrrrrrrrrrr|p{3cm}}
\hline
\textbf{Algorithm} & \multicolumn{11}{|c|}{\textbf{\textit{Statistics of the BFR@n measures}}} & \multirow{2}{3cm}{\textbf{Global Item Coverage (\%)}} \\

 & \textbf{Min.}
 & \textbf{1\%}
 & \textbf{10\%}
 & \textbf{25\%}
 & \textbf{50\%}
 & \textbf{75\%}
 & \textbf{90\%}
 & \textbf{99\%}
 & \textbf{Max.}
 & \textbf{Mean.}
 & \textbf{Std.}
 & 
 \\
\hline
\emph{CHAMELEON} & 0 & 0 & 0 & 0 & 0 & 19 & 57 & 3146 & 9704 & 102 & 629 & 0.927 \\
\emph{CB} & 0 & 0 & 0 & 0 & 0 & 11 & 36 & 409 & 8872 & 35 & 315 & 0.962 \\
\emph{Item-kNN} & 1 & 1 & 8 & 27 & 95 & 455 & 1859 & 7053 & 9869 & 627 & 1353 & 0.354 \\
\emph{V-SkNN} & 1 & 1 & 18 & 68 & 289 & 828 & 2205 & 6643 & 9180 & 777 & 1291 & 0.175 \\
\emph{SR} & 1 & 1 & 22 & 79 & 286 & 842 & 2455 & 7510 & 9869 & 856 & 1471 & 0.247 \\
\emph{CO} & 1 & 1 & 33 & 116 & 379 & 1019 & 2864 & 7467 & 9869 & 977 & 1527 & 0.239 \\
\emph{RP} & 1 & 1 & 142 & 646 & 1722 & 3197 & 4921 & 7523 & 8980 & 2125 & 1847 & 0.023 \\
\hline
\end{tabular}
}
\end{table}

\begin{table}[h!t]
\centering
\caption{The analysis of the item cold-start problem by algorithm for \emph{Adressa} dataset}
\label{tab:adressa_item_cold_start}
\resizebox{\textwidth}{!}{%
\begin{tabular}{l|rrrrrrrrrrr|p{3cm}}
\hline
\textbf{Algorithm} & \multicolumn{11}{|c|}{\textbf{\textit{Statistics of the BFR@n measures}}} & \multirow{2}{3cm}{\textbf{Global Item Coverage (\%)}} \\

 & \textbf{Min.}
 & \textbf{1\%}
 & \textbf{10\%}
 & \textbf{25\%}
 & \textbf{50\%}
 & \textbf{75\%}
 & \textbf{90\%}
 & \textbf{99\%}
 & \textbf{Max.}
 & \textbf{Mean.}
 & \textbf{Std.}
 & 
 \\
\hline
\emph{CHAMELEON} & 0 & 0 & 0 & 0 & 0 & 14 & 41 & 2177 & 8087 & 69 & 448 & 0.950 \\
\emph{CB} & 0 & 0 & 0 & 0 & 0 & 7 & 19 & 64 & 7355 & 14 & 169 & 0.987 \\
\emph{Item-kNN} & 1 & 1 & 3 & 10 & 62 & 1408 & 4068 & 7318 & 8600 & 1087 & 1812 & 0.324 \\
\emph{V-SkNN} & 1 & 1 & 6 & 16 & 48 & 887 & 4057 & 7869 & 8804 & 955 & 1832 & 0.148 \\
\emph{SR} & 1 & 1 & 6 & 21 & 231 & 1913 & 4632 & 7591 & 8625 & 1331 & 1961 & 0.232 \\
\emph{CO} & 1 & 1 & 8 & 29 & 340 & 2366 & 4664 & 7482 & 8803 & 1451 & 1994 & 0.229 \\
\emph{RP} & 1 & 1 & 13 & 30 & 53 & 96 & 219 & 2483 & 8204 & 151 & 540 & 0.040 \\
\hline
\end{tabular}
}
\end{table}

As an exercise to simulate how the number of batches translates to time, it is possible to use the average number of sessions being clicked every minute for the datasets: \textbf{48 sessions/minute} for \emph{G1} dataset and \textbf{44 sessions/minute} for \emph{Adressa} dataset \footnote{In practice, the number of user sessions is much lower during dawn (12pm to 6am).}.
For example, if \emph{CHAMELEON} has shown a $ BFR@n = 19 $ at the 75\% percentile, that means that for 75\% of the articles it would require less than \textbf{25 minutes} ($ (19 \text{ batches} \times 64 \text{ sessions (batch\_size)}) / 48 \text{ (sessions/minute)} $) to provide their first top-N recommendation after articles get published. 

The column \emph{Global Item Coverage (\%)} for these tables is computed by the \emph{COV@n} metric (Equation~\ref{eq:cov}), but considering as \emph{recommendable} all previously clicked items and not only the articles clicked in the last hour, like in the experiments reported in Section~\ref{sec:additional_quality_factors}.

For both datasets, the \emph{CB} is the most resilient algorithm against the item cold-start problem. This algorithm, by design, can recommend an article immediately after being published, because it does not consider user behaviour for recommending, but only the articles content similarity. It has shown $ BFR@n = 11 $ for \emph{G1} and $ BFR@n = 7 $ for \emph{Adressa} dataset, at the 75\% percentile.

The \emph{CHAMELEON} recommender follows \emph{CB} closely in terms of effectively combating the item cold-start problem -- $ BFR@n = 19 $ for \emph{G1} and $ BFR@n = 14 $ for \emph{Adressa}, at the 75\% percentile) -- but with a much higher accuracy than \emph{CB}, as seen in Section~\ref{sec:results_accuracy}. 

Additionally, at the 50\% percentile, it is noticeable that both \emph{CHAMELEON} and \emph{CB} are able to start immediately recommending a fresh article when processing the batch where it was clicked for the first time by any user ($ BFR@n = 0 $).

The other algorithms are more popularity-biased and suffer with the item cold-start problem. Their \emph{BFR@n} measures are much higher and their \emph{Global Item Coverage} much lower than \emph{CHAMELEON} and \emph{CB}. 

It is also interesting to observe that, for the \emph{G1} dataset, the \emph{RP} is the worse algorithm in terms of the item cold-start problem. Although, for the \emph{Adressa} dataset, the \emph{RP} is the third best one (after \emph{CB} and \emph{CHAMELEON}), another indicative of that users clicks in \emph{Adressa} news portal are more popularity-biased.

\begin{figure}[h!t]
    \centering
    \includegraphics[width=0.85
    \textwidth]{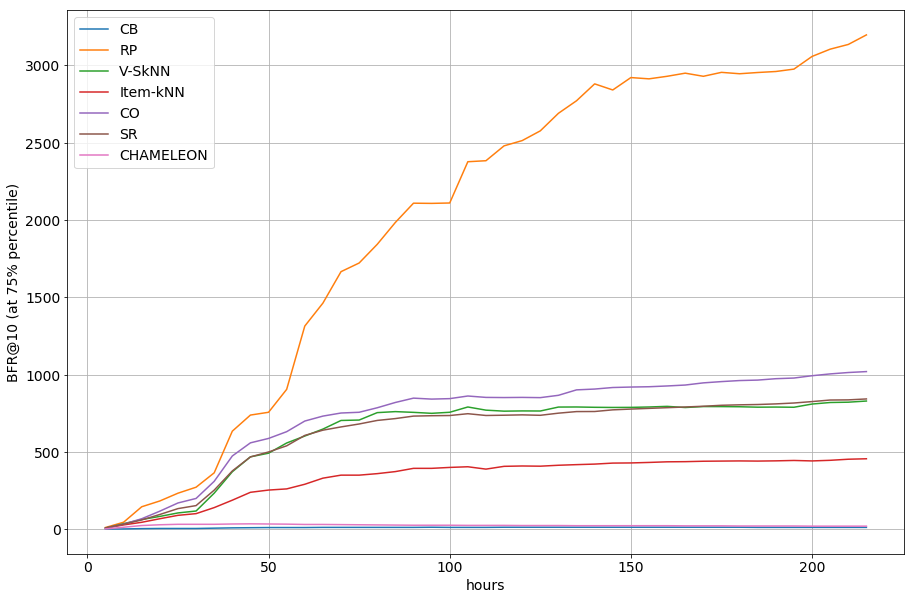}
    \caption{\emph{G1} dataset (9 days) - Evolution over time of the \emph{BFR@10 at 75\% percentile} of the recommendation algorithms}
    \label{fig:g1_item_cold_start_over_time}
\end{figure}

\begin{figure}[h!t]
    \centering
    \includegraphics[width=0.85\textwidth]{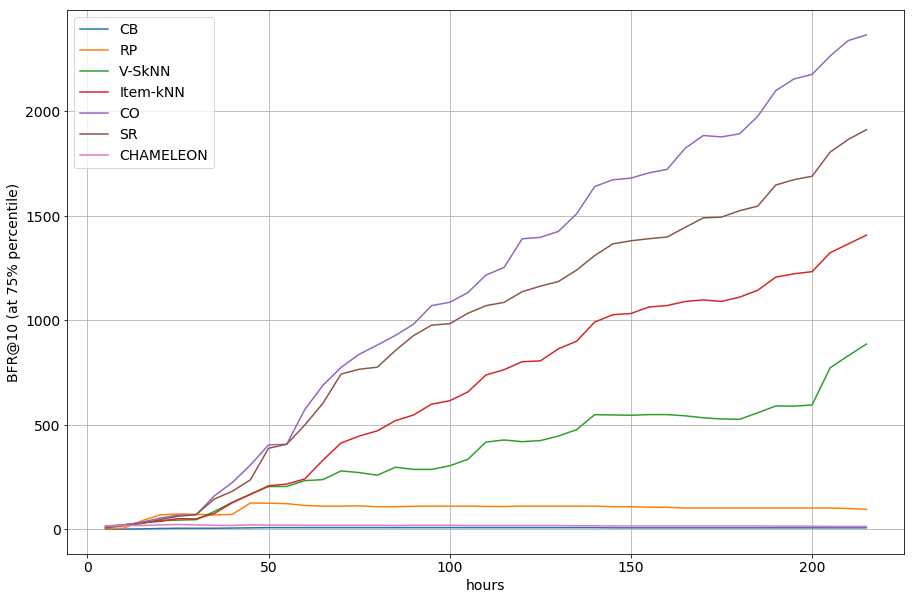}
    \caption{\emph {Adressa} dataset (9 days) - Evolution over time of the \emph{BFR@10 at 75\% percentile} of the recommendation algorithms}
    \label{fig:adressa_item_cold_start_over_time}
\end{figure}

In Figures~\ref{fig:g1_item_cold_start_over_time}~and~\ref{fig:adressa_item_cold_start_over_time}, it is possible to observe the temporal evolution of the item cold-start problem for the session-based algorithms measured as the $ BFR@n $ metric at the 75\% percentile. In general, for most algorithms, the item cold-start problem worsens as time goes and the item catalog increases. The only methods that are robust against that problem over time are the \emph{CB} and \emph{CHAMELEON}. For the \emph{Adressa} dataset, the item cold-start problem also stabilizes at some point for the \emph{RP} algorithm, but a much higher level.

With these experiments, it is possible to conclude that the proposed hybrid RNN-based architecture -- \emph{CHAMELEON} (\emph{NAR} module) -- is able to effectively combat the item cold-start problem much better than the other algorithms, and similarly to the \emph{CB} algorithm, but with a much higher accuracy.

\section{Balancing Accuracy and Novelty with \emph{CHAMELEON} (RQ5)}
\label{sec:tradeoff_results}

As previously discussed in Section~\ref{sec:balance_acc_x_novelty}, it is very important to be able to provide recommendations of long-tail items and not only the most popular ones, in special in the news domain. Therefore, there is a trade-off between accuracy and novelty, as recommending popular items have usually a higher chance to match the interests of most users.

In this section, as reported in \cite{moreira2019contextual}, it is analyzed the effectiveness of our proposed approach to balance accuracy and novelty within \emph{CHAMELEON}, as described in Section~\ref{sec:new_loss_function}. Specifically, it was conducted a sensitivity analysis for the novelty regularization factor ($\beta$) in the proposed loss function, presented in Table~\ref{tab:nov_div_sensitivity}.

\begin{table}[h!t]
\centering
\caption{The evaluation of \emph{CHAMELEON}'s loss regularization factor for novelty ($ \beta $)}
\label{tab:nov_div_sensitivity}
\vspace{10pt}
\resizebox{0.55\textwidth}{!}{%
\begin{tabular}{rlll}
\hline
 \emph{Reg. factors}
 & \emph{ESI-R@10}
 & \emph{MRR@10}
 & \emph{COV@10}
 \\
\hline
\multicolumn{4}{p{6cm}}{\textbf{G1 dataset}}\\ \hline
$ \beta=0.0 $  & 6.4177 & \textbf{0.3458} & 0.6373 \\
$ \beta=0.1 $  & 6.9499 & 0.3401 & 0.6785 \\
$ \beta=0.2 $  & 7.7012 & 0.3222 & 0.6962 \\
$ \beta=0.3 $  & 8.5763 & 0.2933 & 0.7083 \\
$ \beta=0.4 $  & 9.3054 & 0.2507 & 0.7105 \\
$ \beta=0.5 $  & \textbf{9.8012} & 0.2170 & \textbf{0.7123}* \\

\hline
\multicolumn{4}{p{6cm}}{\textbf{Adressa dataset}}\\ \hline
$ \beta=0.0 $  & 5.3410 & \textbf{0.3421} & 0.7926 \\
$ \beta=0.1 $  & 5.8279 & 0.3350 & 0.8635 \\
$ \beta=0.2 $  & 7.5561 & 0.2948 & 0.9237 \\
$ \beta=0.3 $  & 9.4709 & 0.2082 & 0.9353 \\
$ \beta=0.4 $  & 10.2500 & 0.1560 & \textbf{0.9376} \\
$ \beta=0.5 $  & \textbf{10.5184} & 0.1348 & 0.9365 \\
\hline
\end{tabular}
}
\end{table}

\begin{figure}[h!t]
\centering
  \subfigure[a][\emph{G1} dataset]{
    \includegraphics[width=10cm]{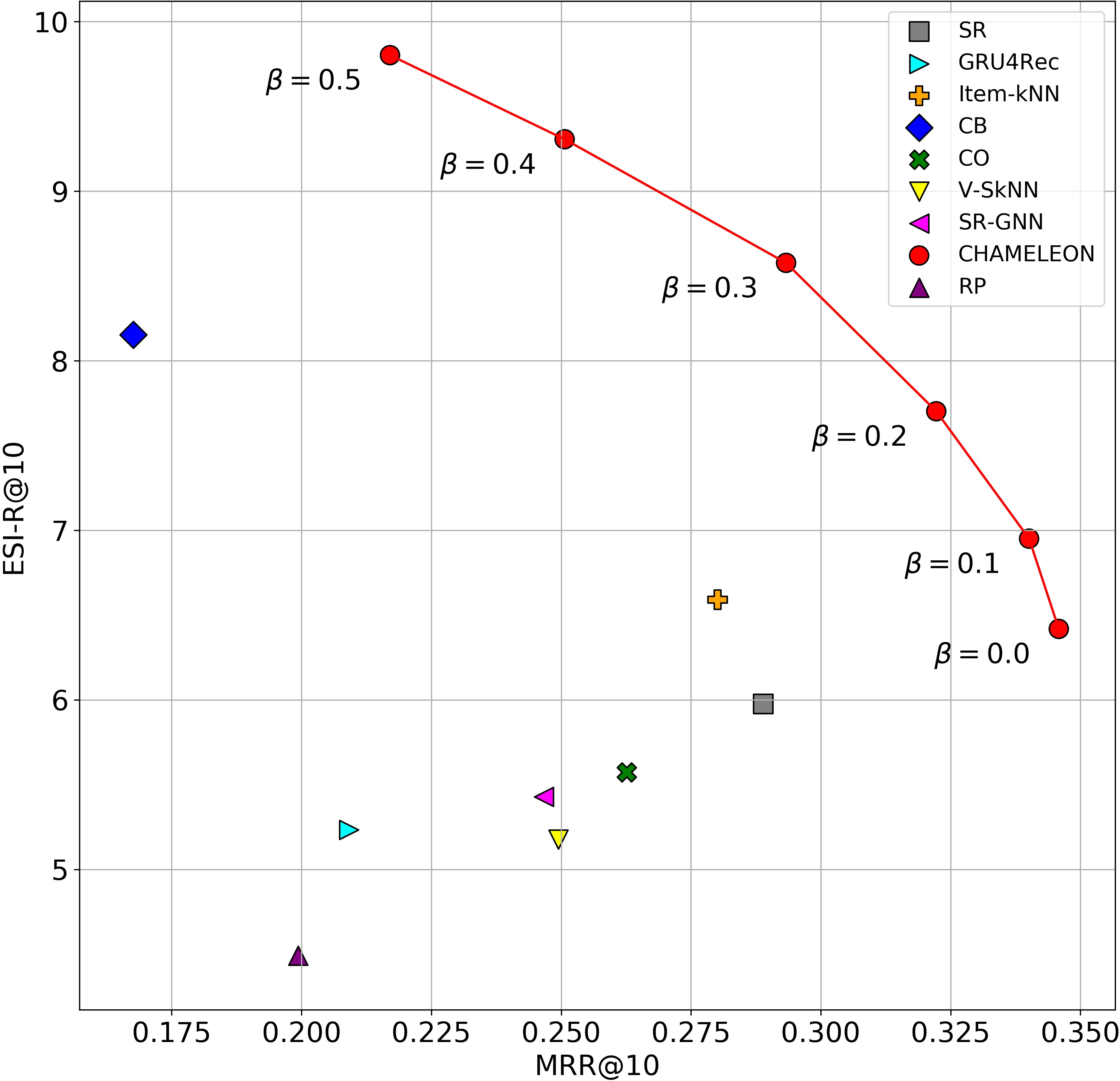}}
    
  \subfigure[b][\emph{Adressa} dataset]{
    \includegraphics[width=10cm]{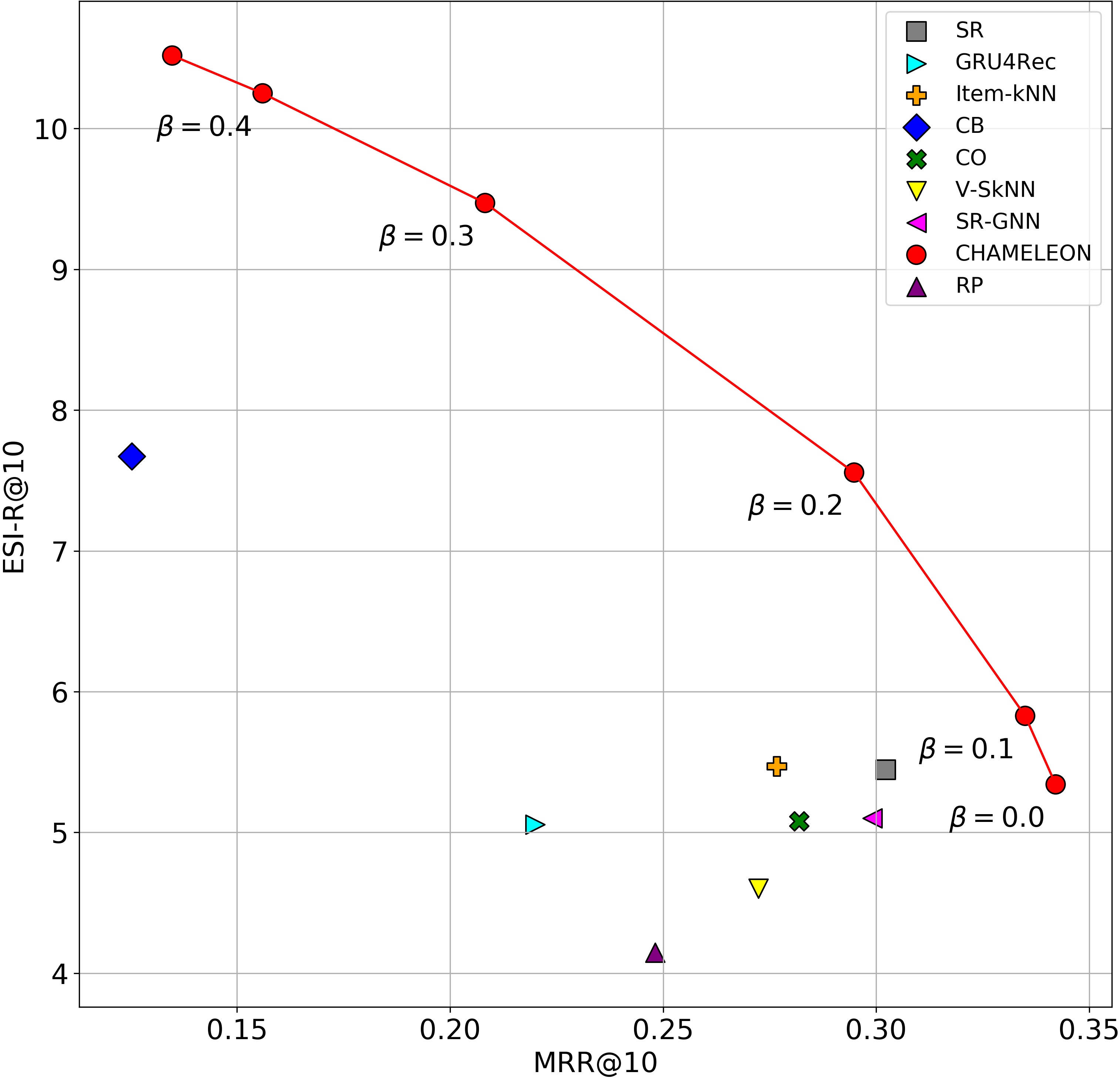}}
  
  \caption{The trade-off between Accuracy (MRR@10) and Novelty (ESI-R) for different values of $\beta$.}
  \label{fig:tradeoff_novelty}
\end{figure}

As expected, increasing the value of $\beta$  increases the novelty of the recommendations and also leads to higher item coverage. Correspondingly, the accuracy values decrease with higher levels of novelty.  Figure~\ref{fig:tradeoff_novelty} shows a scatter plot that illustrates some effects and contrasts of the obtained results in our evaluation. The trade-off between accuracy (\emph{MRR@10}) and novelty (\emph{ESI-R@10}) for \emph{CHAMELEON} can be clearly identified. We also plot the results for the baseline methods there for reference. This comparison reveals that tuning  $\beta$ helps us to end up with recommendations that are both more accurate and more novel than the ones by the baselines. 

With the proposed approach to balance accuracy and novelty within \emph{CHAMELEON}, an administrator of a news portal have the possibility to tune the desired level of novelty ($\beta$) and be assured that the negative effect on recommendation accuracy will be as minimum as possible, provided that the model is trained to optimize for both quality factors.


\chapter{Conclusion}
\label{chap:conclusion}

In this chapter, this research is briefly summarized and it is provided recommendations and suggestions for future works in this area.

First of all, a survey on news recommendation reported in Chapter~\ref{sec:chapter_2} was conducted to identify the key challenges of news recommendation such as the fast flow of incoming articles and the very short lifetime of the recommended items. In general, most users are anonymous, so that the recommendations cannot be based on long-term preference profiles as in other domains.

Based in this survey, it was proposed a conceptual model of the factors that affect the news relevance for user, involving: the article context (popularity, recency); the user context (e.g. time, location, device, referrer); users' short- and long-term interests; and global factors (popular topics, seasonality, and breaking events).

In Chapter~\ref{sec:chapter_3}, it was investigated the state-of-art on Deep Learning methods and techniques applied to Recommender Systems. In a special, the main inspirations for this work were: (1) session-based recommendation with RNNs \cite{hidasi2016}; (2) scalable recommendations based on \emph{multi-view} learning models \cite{huang2013learning, elkahky2015multi}; (3) textual feature extraction using Deep Learning \cite{bansal2016ask, tuan20173d}; and (4) deep learning architectures for news recommendation \cite{song2016multi, kumar2017}.


The main objective of this research was to propose a Deep Learning meta-architecture -- the \emph{CHAMELEON} --  designed to tackle the specific challenges of news recommendation and improve the recommendation quality of news portals.

The \emph{CHAMELEON} meta-architecture \cite{moreira2018chameleon}, described in Chapter~\ref{sec:chapter_4}, consists of a modular reference neural architecture,  designed to tackle specific challenges of the news recommendation domain, in special: a) the short lifetime of the recommendable items; and b) the lack of longer-term preference profiles of users. 

The main technical contribution of \emph{CHAMELEON} lies in an effective combination of \emph{content} and \emph{contextual} features and a sequence modeling technique based on Recurrent Neural Networks for higher-quality session-based news recommendation. 

In Chapter~\ref{sec:chapter_5}, it was proposed different instantiations of \emph{CHAMELEON}'s \emph{ACR} and \emph{NAR} modules as concrete neural architectures, inspired on the state-of-art research on Deep Learning and Recommender Systems.

A temporal offline evaluation method was also proposed to address the dynamics of news readership, where articles context (recent popularity and recency) is constantly changing. Baseline recommender methods were continuously trained on streaming user clicks and the \emph{CHAMELEON} was trained once an hour to emulate a scheduled model deployment in a production environment. 

It was also proposed some adaptations on rank- and relevance-sensitive \emph{novelty} and \emph{diversity} metrics from \cite{vargas2011rank,vargas2015thesis}, which are described in Annex~\ref{sec:novelty_diversity_metrics}.

The proposed \emph{CHAMELEON} architecture instantiations and the selected baseline session-based recommenders were implemented for this research. A comprehensive number of experiments were performed in datasets from two real-words news portals -- \emph{Globo.com (G1)} and \emph{Adressa} -- to answer the stated \emph{research questions}.

The experiments results, presented in Chapter~\ref{sec:chapter_6}, have demonstrated the effectiveness of the proposed approach for news recommendation (\emph{CHAMELEON}) in the quality of recommendation. 

The accuracy, item coverage, and novelty obtained by \emph{CHAMELEON} were much higher than by the baseline algorithms (\emph{RQ1}) \cite{moreira2018news,moreira2019contextual}. It was also analyzed the effect of session sizes in the recommendation accuracy and the correlation of accuracy results among the algorithms.


It was also investigated the effect on recommendation quality of different sets of information for contextual and hybrid recommendations (\emph{RQ2}) \cite{moreira2019contextual}. The results showed that each of the proposed features (e.g., articles popularity and recency, article content embeddings, user context) adds predictive power and helps a hybrid news recommender system to provide better recommendations.

The effect of different textual representations on news recommendation quality was also investigated (\emph{RQ3}) \cite{moreira2019inra}. It was possible to observe that article's textual content really improves the accuracy in a hybrid recommendation approach, in special for news portals with less popularity-bias, i.e., \emph{G1}. Therefore, there is a small difference on the results obtained with a number of different statistical and deep learning techniques to represent textual content, which was a somewhat surprising result.


It was also analyzed the effects of the intense item cold-start problem in the news domain (\emph{RQ4}). It was shown that popularity-biased algorithms suffer more from that problem, whereas the algorithms leveraging articles' content -- \emph{CB} and \emph{CHAMELEON} -- showed to be much robust against the item cold-start problem.

Finally, it was evaluated our proposal to balance the competing \emph{accuracy} and \emph{novelty} recommendation quality factors through a parameterizable multi-task loss function, in the context of news session-based recommendations using neural networks (\emph{RQ5}) \cite{moreira2019contextual}. A sensitivity analysis revealed that our approach is effective and can lead to recommendations that are both more novel and more accurate than the recommendations of the other algorithms in our comparison.

After the design, implementation, execution, and analysis of experiments with real world news portals datasets, it was possible to conclude that the proposed Deep Learning Meta-Architecture for News Recommendation -- the \emph{CHAMELEON} -- was able to effectively surpass the recommendation quality of the other session-based algorithms in many aspects: accuracy, item coverage, novelty, and robustness against the item cold-start problem.

\section{Contributions}

In summary, the main contribution of this research work was the proposal of a Deep Learning meta-architecture -- the \emph{CHAMELEON} -- designed to tackle the specific challenges of news recommendation.




Some complementary contributions of this research work were:

\begin{enumerate}  
    
    \item The elaboration of a conceptual model of factors that affect the relevance of news articles, based on an extensive survey on news recommender systems;
    \item The instantiation of \emph{CHAMELEON}'s \emph{ACR} and \emph{NAR} modules as four concrete architectures, which were used in the experiments to answer the stated research questions;
    \item The implementation of the architecture instantiations of \emph{CHAMELEON} and of the baseline methods, which were open-sourced for the reproducibility of the experiments and for supporting advances in this research line \footnote{\url{https://github.com/gabrielspmoreira/chameleon_recsys}};
    \item A temporal offline evaluation of news recommender algorithms, to emulate a real-world scenario of continuously training a neural model with streaming user clicks and deploying a new trained model once an hour, to provide recommendations for upcoming user sessions; and
    \item The adaptation of rank- and relevance-sensitive \emph{novelty} and \emph{diversity} metrics, originally proposed by \cite{vargas2011rank,vargas2015thesis}.
    
\end{enumerate}

Furthermore, one additional contribution of this research work was the preparation and sharing of two novel datasets for evaluation of hybrid recommender systems for news (\emph{Globo.com (G1)} \footnote{\url{https://www.kaggle.com/gspmoreira/news-portal-user-interactions-by-globocom}}) and articles  (\emph{CI\&T Deskdrop} \footnote{\url{https://www.kaggle.com/gspmoreira/articles-sharing-reading-from-cit-deskdrop}}).

\section{Recommendations}

It is recommended some extensions of this research to investigate the following aspects of news recommendation:

\begin{itemize}
    \item To perform online evaluation (A/B tests) of \emph{CHAMELEON} in a live news portal, to assess its performance and recommendation quality in a production scenario;
    \item To explore mechanisms to balance more than two quality factors, with a particular look at enhancing the diversity of the recommendations while preserving accuracy;
    \item To support session-aware recommendations, in which information from past user sessions is used to model his long-term interests. Some interesting starting points are the research from \cite{quadrana2017personalizing} and \cite{ruocco2017inter}, which uses hierarchical RNNs to model both the sequences of user sessions and sequences of clicks within sessions;
    \item To explore different approaches to sample negative items for recommendation training, as stronger negative samples tends to force the network to capture subtle patterns for more accurate recommendations. One idea would be to sample recent and popular articles at the current geographic region of the user. Other inspirations on negative sampling for session-based recommendation can be found in \cite{hidasi2018recurrent}; and
    \item To explore other recent Deep NLP techniques to produce better article content embeddings (e.g., \emph{Transformers}, \emph{BERT}) for hybrid recommendation.
\end{itemize}

\section{Suggestions for Future Works}

It is suggested the instantiation of \emph{CHAMELEON}'s \emph{NAR} module with neural network architectures different than RNNs to model the sequence of user interactions, such as CNNs,  \emph{Quasi-Recurrent Neural Network (QRNN)} \cite{bradbury2016quasi}, attention mechanisms \cite{Li2017narm} \cite{Liu2018stamp}, and graph neural networks \cite{wu2019session}.

Regarding attention mechanism, it is suggested their investigation as an interesting alternatives to provide recommendations explanation, as pointed out also in \cite{zhang2019deep}. Explainable deep learning is important both for users, allowing them to
understand the factors behind the network's recommendations \cite{seo2017interpretable, xiao2017attentional}, and also for practitioners, probing weights and activations to understand more about the model \cite{seo2017interpretable, xiao2017attentional, tay2018multi}.

An specific challenge not particularly addressed in this research, which was not investigated to a large extent in the literature as well is that of ``outliers'' in the user profiles. Specifically, there might be a certain level of noise in the user profiles. In the case of news recommendation, this could be random clicks by the user or user actions that result from a click-bait rather than from genuine user interest. As proposed in previous works \cite{Saiaetal2016asemantic,Said:2018:CIR:3231208.3231212,DBLP:conf/ic3k/SaiaBC14}, it is suggested an investigation to identify such outliers and noise in the context of session-based news recommendation to end up with a better estimate of the true user intent within a session.

It is also suggested the adaptation of \emph{CHAMELEON} to provide contextual and hybrid session-based recommendations in other domains like e-commerce, entertainment, and media.

\annex
\chapter{The ACR Module - Training of the Supervised and Unsupervised instantiations} 
\label{sec:acr-training}

In this Annex, it is presented the methods for training two instantiations of the \emph{ACR} module architecture: (1) \emph{supervised} and (2) \emph{unsupervised}. 

The \emph{supervised} architecture trains the \emph{Article Content Embeddings (ACE)} to classify article's metadata attributes (e.g., category) from its textual content. This architecture depends on human-labeled articles metadata to train the \emph{ACE}, such as the categories that are usually defined by editors of news portals.

The \emph{unsupervised} architecture trains the \emph{ACEs} as a Sequence Autoencoder, which encodes the input text as the \emph{ACE} and reconstructs the original text from that representation. This method depends solely on articles' text and no other metadata.

In both cases, the \emph{ACR} module is trained to optimize a loss function on a mini-batch of news articles. After training, the learned \emph{ACEs} are persisted in a repository, for further usage by the \emph{NAR} module.

\section{The Supervised ACR Module Training Method}

The \emph{supervised} variation of the \emph{ACR} module is trained according to the following steps:

\begin{enumerate}
    \item The \emph{TFR} sub-module extracts features from its input text (using a \emph{CNN} or \emph{RNN}-based architecture) and generates an Article Content Embedding (\emph{ACE}) based on its inputs: the article's textual content, represented as a sequence of word embeddings; and
    \item The \emph{CET} sub-module trains the \emph{ACE} to classify articles metadata attributes.
\end{enumerate}

For inference of the \emph{ACE} of newly published articles, only step 1 is necessary.

Articles metadata attributes may have a single label for each instance (e.g., categories) or multiple labels, when there is more than one value for an attribute (e.g., tags, entities). 

For single-label classification, it is used the \emph{softmax} function to normalize the output layer as a probability distribution that sums up to one, as follows:

\begin{equation} \label{eq:softmax}
\hat{y_i} = \sigma(x_j) = \frac{e^{x_j}}{\sum_i e^{x_i}}.
\end{equation}

Cross-entropy log loss is used for optimization, as follows

\begin{equation} \label{eq:crossentropyloss}
l(\theta) = -\frac{1}{N}(\sum_{i=1}^N y_i \cdot \text{log}(\hat{y_i})) + \lambda \lVert \theta \rVert,
\end{equation}

where  $ y $ is a vector with the one-hot encoded label for each instance, $ \hat{y} $ is the vector with the output probabilities for each class, previously normalized by \emph{softmax}, $ \theta $ represents model parameters to be learned, and $ \lambda $  controls the importance of the regularization term, to avoid overfitting.

For multi-label classification, the \emph{sigmoid} function (Equation \ref{eq:sigmoid}) is used instead of softmax, because in that case classes probabilities should be independent from each other:

\begin{equation} \label{eq:sigmoid}
s(x) = \frac{1}{1 + e^{-x}}
\end{equation}

When more than one article metadata attribute is available for classification (e.g. categories, tags, and entities), it is possible to use \emph{Multi-Task Learning (MTL)}. 

\begin{figure}[h]
	\centering
	\includegraphics[width=7cm]{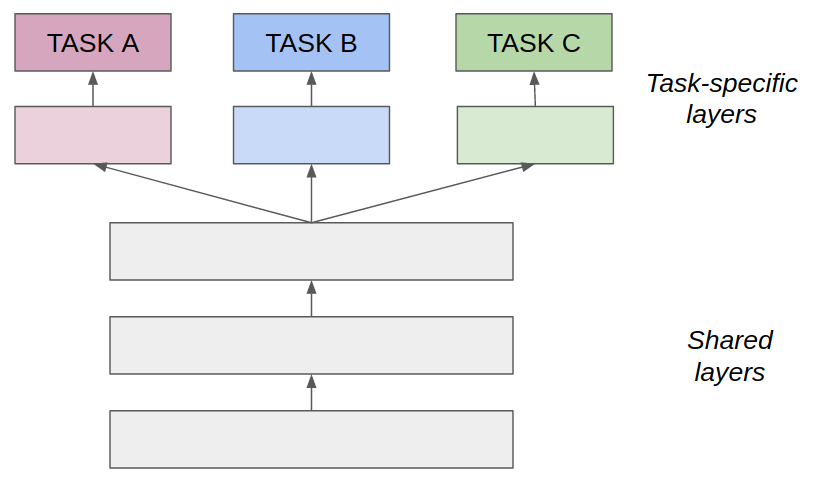}
	\caption{An illustration of a Multi-Task Learning architecture}
	\label{figure:multi-task}
\end{figure}

The \emph{MTL} is typically done with either hard or soft parameter sharing of hidden layers. Hard parameter sharing is the most commonly used approach to \emph{MTL} in neural networks and goes back to \cite{caruna1993multitask}. It is generally applied by sharing hidden layers for all tasks, while keeping task-specific output layers \cite{ruder2017overview}, like represented in Figure \ref{figure:multi-task}.

By sharing representations between related tasks, the model may generalize better, as can be seen in successful MTL applications in NLP \cite{collobert2008unified}, speech recognition \cite{deng2013new}, computer vision \cite{girshick2015fast}, and recommender systems \cite{bansal2016ask}.

The \emph{MTL} is usually trained by following one of these strategies: (1) alternate optimization step of each task's parameters, or (2) summing up losses for each classification task in a single optimization step. 

In the \emph{ACR} module, strategy (2) was chosen, as it is usually recommended when you have a single dataset with multiple labels, like in a news articles domain.


\section{The Unsupervised ACR Module Training Method}

It is also proposed an \emph{unsupervised} instantiation of the \emph{ACR} module as an adaptation of the Sequence Autoencoder \cite{dai2015semi, srivastava2015unsupervised}. In such Encoder-Decoder RNN-based architecture, the article text is encoded to generate a compressed representation and decoded to reconstruct the original input text.

\begin{figure}[h]
	\centering
	\includegraphics[width=12cm]{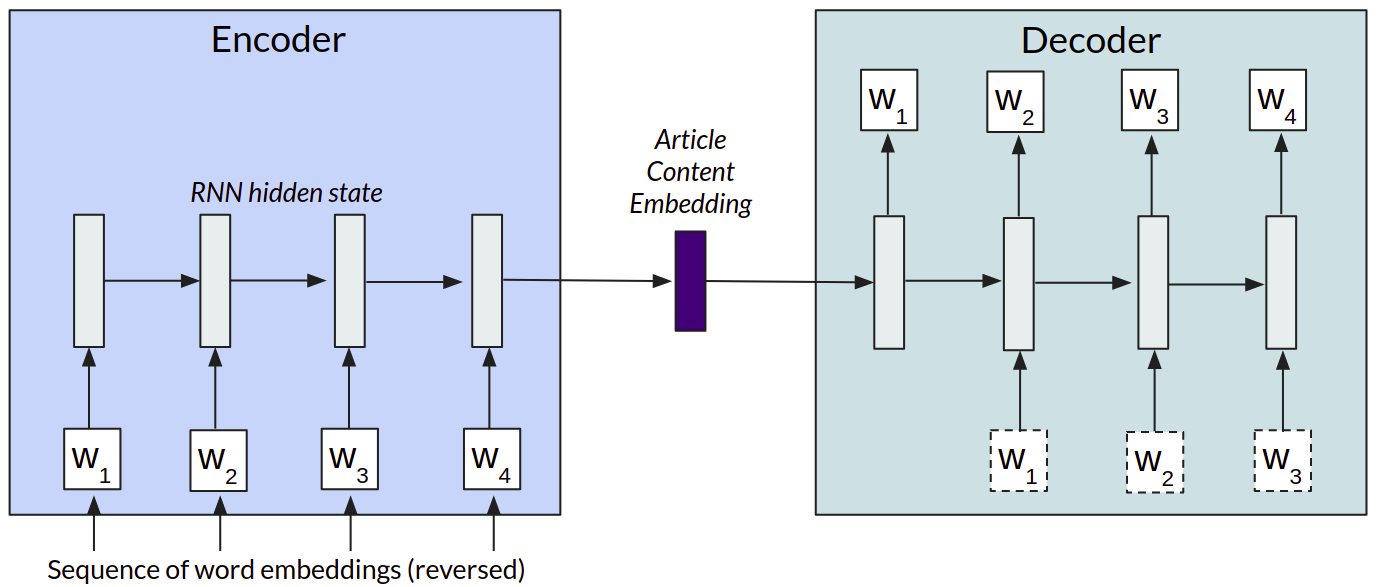}
	\caption{An illustration of a \emph{Sequence Autoencoder}, adapted from \cite{srivastava2015unsupervised} }
	\label{figure:seq-autoencoder}
\end{figure}

For the \emph{ACR} module, the \emph{TFR} and \emph{CET} sub-modules are the \emph{encoder} and \emph{decoder}, respectively, as illustrated in Figure~\ref{figure:seq-autoencoder}. They can be instantiated as an RNN-based architecture, such as the Gated Recurrent Units (GRU) \cite{cho2014properties, chung2014empirical}.

The training method for the \emph{unsupervised} instantiation of the \emph{ACR} module works as follows:

\begin{enumerate}
    \item The \emph{TFR} sub-module (\emph{encoder}) compressed the input text, represented as a sequence of word embeddings, into an Article Content Embedding (\emph{ACE}); and
    \item The \emph{CET} sub-module (\emph{decoder}) have its RNNs hidden state initialized with a non-linear projection of the \emph{ACE} produced by the \emph{encoder} and tries to reconstruct the original input text word-by-word, represented as word embeddings.
\end{enumerate}

This \emph{unsupervised} instantiation of the \emph{ACR} module is trained to minimize the \emph{Mean Squared Error (MSE)} between the sequence of predicted words $\hat{w}_i$ and correct words $w_i$, represented by their embeddings, as shown in Equation~\ref{eq:mse}.

\begin{equation} \label{eq:mse}
MSE = \frac{1}{n}\sum_{i=1}^{n} {(w_i - \hat{w}_i)^2}
\end{equation}

\chapter{Novelty and Diversity Metrics} 
\label{sec:novelty_diversity_metrics}
This Annex describes our adaptation of the novelty and diversity metrics, originally proposed by \cite{vargas2011rank,vargas2015thesis}. They were tailored to fit the specific problem of session-based news recommendation. Generally, for the purpose of this investigation, novelty is evaluated in terms of \emph{Long-Tail Novelty}. Items with high novelty correspond to long-tail items, i.e., items that were clicked on by few users, whilst low novelty items correspond to more popular items.

\subsection{The ESI-R@n}

The \emph{Expected Self-Information with Rank-sensitivity} metric, presented in Equation~\ref{eq:esi-r}, was adapted from the \emph{Mean Self-Information (MSI)} metric (Equation~\ref{eq:msi}) proposed by \cite{castells2015novelty}, with the addition of a rank discount. 

\begin{equation} \label{eq:msi}
\text{MSI} = \frac{1}{L}
\sum_{i \in L}  -\text{log}_2{p(i|\text{known},\theta)} 
\end{equation}

The term  $ -\text{log}_2{p(i)} $ represents the core of this metric, which comes from the \emph{self-information} (also known as \emph{surprisal}) metric of Information Theory, which quantifies the amount of information conveyed by the observation of an event \cite{castells2015novelty}. Applying the $ \text{log}(\cdot) $ function emphasizes the effect of highly novel items. It is defined $ L = {i_1, ... , i_N} $ as a recommendation list of size $ N = | L | $.

\begin{equation} \label{eq:esi-r}
\text{ESI-R}(L) = \frac{1}{\sum_{j=1}^N \text{disc}(j)}
\sum_{k=1}^N  -\text{log}_2{p(i_k)} \times \text{disc}(k)
\end{equation}

In this setting, the probability $ p(i) $ of an item being part of a random user interaction under free discovery is the normalized recent popularity, i.e., $ p(i) = \text{rec\_norm\_pop}(i) $, previously presented in Equation~\ref{eq:rec-norm-pop}.
In Equation~\ref{eq:esi-r}, $ \text{disc}(\cdot) $ is a logarithmic rank discount, defined in Equation~\ref{eq:disc}, that maximizes the impact of novelty for top ranked items, under the assumption that their characteristics will be more visible to users compared to the rest of the top-n recommendation list.

\begin{equation} \label{eq:disc}
\text{disc}(k) = \frac{1}{\text{log}_2(k+1)}
\end{equation}

\subsection{The ESI-RR@n}
Analyzing quality factors like accuracy, novelty, and diversity in isolation can be misleading. Some Information Retrieval (IR) metrics, such as $ \alpha\-\text{nDCG} $, therefore  consider novelty contributions only for relevant items for a given query \cite{castells2015novelty}.
As proposed by \cite{vargas2011rank}, a relevance-sensitive novelty metric should likewise assess the novelty level based on the recommended items that are actually relevant to the user.

Thus, it was used a variation of a novelty metric to account for relevance---\emph{Expected Self-Information with Rank- and Relevance-sensitivity (ESI-RR@n)}. It weights the novelty contribution by the relevance of an item for a user $ p(rel|i,u)$ \cite{castells2015novelty}. It was adapted the proposal from \cite{vargas2015thesis}, as shown in Equation~\ref{eq:p_rel}.

\begin{equation} \label{eq:p_rel}
p(rel|i,u) = \text{relevance}(i,u) =
\begin{cases}
      1.0, & \text{if}\ i \in \mathbb{I}_u \\
      b, & \text{otherwise}
    \end{cases},
\end{equation}

where $ \mathbb{I}_u $ is the set of items the user interacted within the ongoing session, and $ b $ is a background probability of an unobserved interaction (negative sample) being also somewhat relevant for a user. The lower the value of $ b $ (e.g., $ b=0 $) the higher the influence of relevant items (accuracy) in this metric. The author of \cite{vargas2015thesis} used an empirically determined value of $ b = 0.02 $, based on his experiments on balancing diversity and novelty. In this study, it was arbitrarily set $ b = 0.02 $, so that all the 50 negative samples would sum up to the same relevance (1.0) of a positive (clicked) item.

Equation~\ref{eq:esi-rr} shows how the \emph{ESI-RR@n} metric is computed. 
\begin{equation} \label{eq:esi-rr}
\text{ESI-RR}(L) =
 C_k \sum_{k=1}^N  -\text{log}_2{p(i_k)} \times \text{disc}(k) \times \text{relevance}(i_k,u),
\end{equation}

Equation~\ref{eq:rr-normalization-term} defines the term $ C_k $, which computes the weighted average based on ranking discount.

\begin{equation} \label{eq:rr-normalization-term}
C_k = \frac{1}{\sum_{k' = 1}^N \text{disc}(k')}
\end{equation}

Like in \cite{vargas2015thesis}, the relevance is not normalized, so that more relevant items among the top-n recommendations lead to a global higher novelty.

\subsection{The EILD-R@n}
Diversity was measured based on the \emph{Expected Intra-List Diversity} metric proposed by \cite{vargas2011rank}, with variations to account for rank-sensitivity (\emph{EILD-R@n}) and for both rank- and relevance-sensitivity (\emph{EILD-RR@n}).

Intra-List Diversity measures the dissimilarity of the recommended items with respect to the other items in the recommended list. In this case, the distance metric $ d(\cdot) $ defined in Equation~\ref{eq:cos_dist} is the cosine distance.
\begin{equation} \label{eq:cos_dist}
d(a,b) = (1 - \text{sim}(a,b))  / 2,
\end{equation}
Here, $a$ and $b$ are the \emph{Article Content Embeddings} of two articles and $ \text{sim}(a,b) $ is their cosine similarity. As the cosine similarity ranges from -1 to +1, the cosine distance is scaled to the range [0,1].

The \emph{Expected Intra-List Diversity with Rank-sensitivity (EILD-R@n)} metric, defined in Equation~\ref{eq:eild-r}, is the average intra-distance between items pairs weighted by a logarithmic rank discount $ \text{disc}(\cdot) $, defined in Equation~\ref{eq:disc}. Given a recommendation list $ L = {i_1, ... , i_N} $ of size $ N = | L | $, the \emph{EILD-R@n} metric is computed as follows.

\begin{equation} \label{eq:eild-r}
\begin{split}
\text{EILD-R}(L) = &\\
\frac{1}{\sum_{k'=1}^N \text{disc}(k')}
\sum_{k=1}^N \text{disc}(k) \frac{1}{ \sum_{l'=1:l' \neq k}^N \text{rdisc}(l', k) } \sum_{l=1:l \neq k}^N d(i_k, i_l) \times  \text{rdisc}(l, k)
\end{split}
\end{equation}

The term $ \text{rdisc}(l, k) $, defined in Equation~\ref{eq:rdisc}, represents a relative ranking discount, considering that an item $ l $ that is ranked before the target item $ k $ has already been discovered. In this case, items ranked after $ k $ are assumed to lead to a decreased diversity perception as the relative rank between $ k $ and $ l $ increases.

\begin{equation} \label{eq:rdisc}
\text{rdisc}(l, k) = \text{disc}(\text{max}(0, l-k))
\end{equation}

\subsection{The EILD-RR@n}
The \emph{Expected Intra-List Diversity with Rank- and Relevance-sensitivity} finally measures
the average diversity between item pairs, weighting items by rank discount and relevance, analogously to the \emph{ESI-RR@n} metric, as shown in Equation~\ref{eq:eild-rr}.

\begin{equation} \label{eq:eild-rr}
\begin{split}
\text{EILD-RR}(L) = &\\
C_k \sum_{k=1}^N \text{disc}(k) \times \text{relevance}(i_k,u) C_l \sum_{l=1:l \neq k}^N d(i_k, i_l) \text{rdisc}(k, l) \times  \text{relevance}(i_l,u)
\end{split}
\end{equation}

Here, $ C_k $ (Equation~\ref{eq:rr-normalization-term}) and $ C_l $ (Equation~\ref{eq:eild-internal-normalization-term}) are normalization terms representing a weighted average based on rank discounts.

\begin{equation} \label{eq:eild-internal-normalization-term}
C_l = \frac{1}{ \sum_{l'=1:l \neq k}^N \text{rdisc}(k, l') }
\end{equation}

\chapter{Algorithms' Hyper-parameters for Experiments} 
\label{sec:annex_hyperparams}
In Table~\ref{tab:hyperparams}, it is presented the best hyper-parameters found for each algorithm and dataset. They were tuned for accuracy (\emph{MRR@10}) on a hold-out validation set, by running random search within defined ranges for each hyper-parameter\footnote{The methods \emph{CO}, \emph{RP}, and \emph{CB} do not have hyper-parameters.}. More information about the hyper-parameters can be found in our shared code and in the papers where the baseline methods were proposed.

\begin{table}[!htbp]
\centering
\caption{The best hyper-parameters per algorithm and dataset}
\label{tab:hyperparams}
\vspace{10pt}
\tiny
\begin{tabular}{lp{3.0cm}p{6.5cm}rr}
\hline
 \emph{Method} &
 \emph{Parameter} &
 \emph{Description} &
 \emph{G1} &
 \emph{Adressa}
 \\
\hline

\emph{CHAMELEON} & batch\_size & Number of sessions considered for each mini-batch & 256 & 64 \\
 & learning\_rate & Learning rate for each training step (mini-batch) & 1e-4 & 3e-4 \\
 & reg\_l2 & L2 regularization of the network's parameters & 1e-5 & 1e-4 \\
 & softmax\_temperature & Used to control the ``temperature'' of the softmax function & 0.1 & 0.2 \\
 & CAR\_embedding\_size & Size of the User-Personalized Contextual Article Embedding & 1024 & 1024 \\
 & rnn\_units & Number of units in an RNN layer & 255 & 255 \\
 & rnn\_num\_layers & Number of stacked RNN layers & 2 & 2 \\

\hline

\emph{SR} & max\_clicks\_dist & Maximum number of clicks to walk back in the session from the currently viewed item. & 10 & 10 \\
 & dist\_between\_clicks
 \_decay & Decay function for the distance between two items clicks within a session (e.g., linear, same, div, log, quadratic) & div & div \\
\hline

\emph{Item-kNN} & reg\_lambda & Smoothing factor for the popularity distribution to normalize item vectors for co-occurrence similarity & 20 & 20 \\
 & alpha & Balance between normalizing with the support counts of the two items. 0.5 gives cosine similarity, 1.0 gives confidence. & 0.75 & 0.50 \\
\hline

\emph{V-SkNN} & sessions\_buffer\_size & Buffer size of last processed sessions & 3000 & 3000 \\
 & candidate\_sessions
 \_sample\_size & Number of candidates near the sessions to sample & 1000 & 2000 \\
  & nearest\_neighbor
  \_session\_for\_scoring & Nearest neighbors to compute item scores & 500 & 500 \\
  & similarity & Similarity function (e.g., Jaccard, cosine) & cosine & cosine \\
  & sampling\_strategy & Strategy for sampling (e.g., recent, random) & recent & recent \\
  & first\_session\_clicks
  \_decay & Decays the weight of first user clicks in active session when finding neighbor sessions (e.g. same, div, linear, log, quadratic) & div & div \\
\hline

\textit{SR-GNN} & batch\_size & Batch size & 128 & 128 \\
 & n\_epochs & Number of training epochs & 10 & 10 \\
 & hidden\_size & Number of units on hidden state & 200 & 200 \\
 & l2\_lambda & Coefficient of the $ L_2 $ regularization & 1e-5 & 2e-5 \\
 & propagation\_steps & GNN propagation steps  & 1 & 1 \\
 & learning\_rate & Learning rate & 1e-3 & 1e-3 \\
 & learning\_rate\_decay & Learning rate decay factor & 0.15 & 0.1 \\
 & learning\_rate\_decay\_steps & number of steps after which the learning rate decays & 3 & 3 \\
 & nonhybrid & Enables/disables the Hybrid mode & True & True \\

\hline
\textit{GRU4Rec} & batch\_size & Batch size & 128 & 128 \\
 & n\_epochs & Number of training epochs & 3 & 3 \\
 & optimizer & Training optimizer & Adam & Adam \\
 & loss & The loss type & bpr-max-0.5 & bpr-max-0.5 \\
 & layers & Number of GRU units in the layers & [300] & [300] \\
 & dropout\_p\_hidden & Dropout rate & 0.0 & 0.0 \\
 & learning\_rate & Learning rate & 1e-4 & 1e-4 \\
 & l2\_lambda & Coefficient of the $ L_2 $ regularization  & 1e-5 & 2e-5 \\
 & momentum & if not zero, Nesterov momentum will be applied during training with the given strength & 0 & 0 \\
 & embedding & Size of the embedding used, 0 means not to use embedding & 0 & 0 \\

\hline
\hline
\end{tabular}

\end{table}

\appendix
\chapter{Deep Learning Background} 
\label{ape:dl-background}

This Appendix provides a background on Deep Learning techniques, methods, and neural architectures.

Machine learning is able to extract patterns from raw data and map input representations to outputs. Representation learning extends machine learning, in the sense that it is also able to learn the representation itself \cite{Goodfellow-et-al-2016}.

Deep Learning solves a central problem in representation learning, by introducing representations or concepts that are expressed in terms of other simpler representations \cite{Goodfellow-et-al-2016}. All deep models tries to learn good representations from observed data, in such a way that they are modeled as being generated by interactions of many hidden factors \cite{betru2017}. 


Deep models can be trained by either supervised or unsupervised approaches in neural network architectures with several layers, forming a hierarchy. Each subsequent layer extracts a progressively more abstract representation of the input data and builds upon the representation from the previous layer, typically by computing a nonlinear transformation of its input. The parameters of these transformations are optimized by training the model on a dataset \cite{zheng2016} \cite{deng2014book}. 

Typical shallow neural networks consists of one or two layers. Although these models achieved good performance and were dominant in the 90's, due to its limited representation learning capacity, they have difficulties in modeling unstructured data such as text, images, and audios \cite{zheng2016}. 

In 2006, \cite{hinton2006} have shown that with a layer-wise training strategy, a Deep Belief Network (DBN) could be successfully trained to predict hand written digits. This was the first successful attempt to train a deep model. Before that, researchers had not seriously exploited deep models due to lack of data and computational power \cite{zheng2016}.

Distributed representations or embeddings form the basis of deep learning \cite{betru2017}. Embeddings are learned real value vectors representing entities. They are also known as latent feature vectors, or latent representations. Embeddings vectors of similar entities are also similar.

Deep learning models were initially applied to the field of: Computer Vision, Audio, Speech Recognition, and Language Processing. They outperformed many state-of-the-art models, like reported in \cite{lecun1998} \cite{hinton2006}  \cite{bengio2007} \cite{lee2009}. 

\section{Architectures of Deep Learning}
Deep learning architectures can be composed by the combination of many different architectures. In this section, some types of deep architectures are briefly described.


\subsection{The Restricted Boltzmann Machine (RBM)}

The Boltzmann Machine is a network of symmetrically connected, neuron-like units, that make stochastic decisions about whether to be on or off. 

The Restricted Boltzmann Machine (RBM) consists of a special BM, composed by a layer of visible units and a layer of hidden units with no visible-visible or hidden-hidden connections \cite{resnick1994} \cite{deng2014book} \cite{zheng2016} \cite{betru2017}.

The Deep Boltzmann Machine (DBM) is a special Boltzmann Machine where the hidden units are organized in a deep layered manner, where only adjacent layers are connected, and there are no visible-visible or hidden-hidden connections within the same layer \cite{zheng2016} \cite{deng2014book} \cite{betru2017}.

\subsection{The Deep Belief Network (DBN)}

The Deep Belief Network (DBN) consists of probabilistic generative models composed of multiple layers of stochastic hidden variables. The top two layers have indirectly symmetric connections between them. The lower layers receive top-down, directed connections from the layer above \cite{zheng2016} \cite{deng2014book} \cite{betru2017}.

A DBN is formed with a stack of RBM. In its first two layers, it should be trained a two layer RBM with one visible layer and one hidden layer. Then, activation probabilities of the hidden layer forms a visible layer to learn another hidden layer. In this manner, the RBM can be stacked to learn a multi-layer DBN  \cite{zheng2016}.

\subsection{The Deep Feed-Forward Neural Network (FFNN)}

The Deep Feed-Forward Neural Network (FFNN) is a multilayer perceptron network with many hidden fully connected layers. The parameters if its layers are trained with the Back-Propagation (BP) method, using gradient-based numerical optimization \cite{chauvin1995}. 

Usually a FFNN is composed of a sequence of Fully Connected (FC) layers that perform affine transformations on the output of the previous layer. Generally, a non-linear activation function is applied to model complex relationships among its input features , like shown as follows

\begin{equation} \label{eq:hidden-layer}
h_i = f(w_i \cdot h_{i-1} + b_i),
\end{equation}

where $ h_i $ is the output of a hidden layer, $ w_i $ and $ b_i $ are the layer's weight vector and bias vector, respectively, $ h_{i-1} $ is the previous layer, and $ f(\cdot) $ is a non-linear activation function.

The Rectified Linear Units (ReLU) activation function, shown in Equation \ref{eq:relu}, is usually preferred over other saturating non-linear functions, like hyperbolic tangent (tanh) or sigmoid, because neural networks with ReLU layers may be trained several times faster \cite{krizhevsky2012imagenet},  without a significant penalty to generalization accuracy.

\begin{equation} \label{eq:relu}
f(x) = max(0,x)
\end{equation}

Other common ingredient of FFNN is a Dropout layer. Dropout is a popular regularization technique for Deep Learning, which stochastically disables a configured fraction of its neurons. This prevent neurons from co-adapting and forces them to learn individually useful features, providing better generalization. 

Part of the success of FFNN is that it can accommodate a larger number of hidden units and perform better parameter initialization methods. Even when learned parameters are at local optimal, FFNN can perform much better than those with less hidden units \cite{zheng2016} \cite{deng2014book} \cite{betru2017}.

\subsection{The Deep Auto Encoder (DAE)}

The Deep Auto-Encoder (DAE) is a special type of DNN, which output target is the data input itself, often pre-trained with DBN or by using distorted training data to regularize the learning, using denoising auto-encoders \cite{vincent2008extracting}.

By forcing the input and output to be the same, the output of the middle layer can be regarded as dense representations \cite{zheng2016}. 
\cite{hinton2006b} proposed a pre-training technique to learn deep auto encoders with multiple layers. This technique involves treating each neighboring set of two layers as an RBM. In this manner, the pre-training procedure approximates a good parameter initialization. Then, they use a back-propagation technique to fine-tune the pre-trained model \cite{zheng2016} \cite{deng2014book}.

\subsection{The Convolutional Neural Network (CNN)}
The Convolutional Neural Network (CNN) \cite{lecun1998} architecture have achieved state-of-the-art results in computer vision, speech recognition, and have been shown to be competitive in several NLP tasks.

By applying convolution operations (known as kernels or filters) at different levels of granularity, a CNN can extract features that are useful for learning tasks and reduce the needs of manual feature engineering \cite{tuan20173d}.

The CNN is a variant of FFNN with two components: 1) convolution layer, for generating local features; and 2) pooling (or sub-sampling) layer, for representing data as more concisely by selecting only representative local features (i.e., features having the highest score, via the activation functions) from the previous layer, which is usually a convolution layer \cite{kim2016convolutional}.

\begin{figure}[h]
	\centering
	\includegraphics[width=15cm]{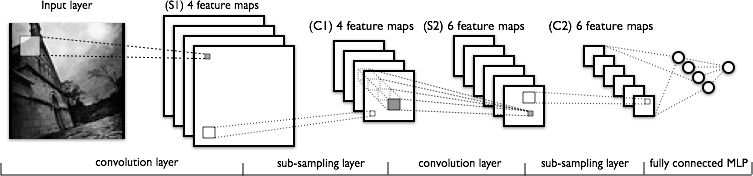}
	\caption{A typical CNN architecture}
	\label{figure:cnn}
\end{figure}

In its convolutional layers, the outputs of the previous layer are fed into a set of convolutional filters and generate a set of filtered results. Then, these results are sub-sampled based on their activations in a following sub-sampling layer. Convolutional layers and sub-sampling layers can be alternatively added to build a deep CNN model, as shown in Figure \ref{figure:cnn}. 

As a class of deep models for learning features, the CNN learns a hierarchy of increasingly complex features. Without building hand-crafted features, these methods utilize layers with convolving filters that are applied on top of pre-trained word embeddings.

The weight sharing in the convolutional layer, together with properly chosen pooling schemes, endows the CNN with some invariance properties \cite{huang2007}. Moreover, in benefiting from the shared weights, CNNs have fewer parameters to be learned than traditional feed-forward neural networks \cite{zheng2016} \cite{deng2014book}. Such weight sharing imposes a general assumption that the input can be decomposed into a set of local regions with then same nature , thus, could be processed with the same set of transformations.

The Meta-Architecture proposed in this research has CNNs as one of the feature extractors mechanism from news textual content.

\subsection{The Recurrent Neural Network (RNN)}

Several methods have been proposed to incorporate temporal information into RS, at some stage \cite{campos2014time}. In \cite{vinagre2015overview} they are differentiated as follows:

\begin{itemize}
	\item \textit{Time-aware RS} - considers time as a contextual feature during the training phase. Timestamps serve as an additional source of information by which the model is enriched. The rationale behind is that user behavior underlies certain habits and regularities that repeat in regular time intervals, consequently allowing a more accurate prediction of similar patterns in the future; and
	\item \textit{Time-dependent RS} - consider user preference data as chronologically ordered sequences, assuming that the most intrinsic property is that time establishes an order for events. Input is required in chronological form, while exact time spans do not need to be taken into account. The algorithms consequently do not aim at modeling time as being cyclic, but rather at adapting to changes.
\end{itemize}

The RNN has been devised to model variable-length sequence data. The main difference between the RNN and the traditional DFFNN is the existence of an internal hidden state, or memory, in the units that compose the network \cite{hidasi2016}.

An RNN has a number of particular characteristics: (1) It is sensitive to sequences' order, (2) It does not require hand-engineered features to model sequences, and (3) it is easy to leverage large unlabeled datasets, by pretraining the RNN parameters with unsupervised language modeling objectives \cite{bansal2016ask}.

The RNN has provided substantial performance gains in a variety of natural language processing applications such as language modeling \cite{mikolov2010recurrent} and machine translation \cite{cho2014learning}.

The RNN uses gradient based methods, like Back-Propagation Though Time (BPTT) \cite{williams1995gradient}, to learn its parameters. Although errors signals flowing back in time exponentially depends on the magnitude of the weights, this implies that the back-propagated error quickly either vanishes or blows up \cite{hochreiter1997long}. Thus, a standard RNN has a limitation in learning from longer time lags between relevant input and target events \cite{hochreiter1997long} \cite{gers2001long}.

Gated RNN architectures \cite{chung2014empirical} are designed to overcome this limitation by including gating units trained to control information flow through the network, thereby learning to keep information over a long period of time. Gates essentially learn when and by how much to update the hidden state of the unit. Both Long Short-Term Memory (LSTM) \cite{gers2001long} \cite{hochreiter1997long} and Gated Recurrent Unit (GRU) \cite{cho2014properties, chung2014empirical} networks have shown advantages in real-world applications \cite{donkers2017sequential}.

Gated RNNs architectures have not been designed with the recommendation domain in mind. In particular, they are not optimized for taking interaction between user and system into account \cite{donkers2017sequential}. 

\begin{figure}[h]
	\centering
	\includegraphics[height=5cm]{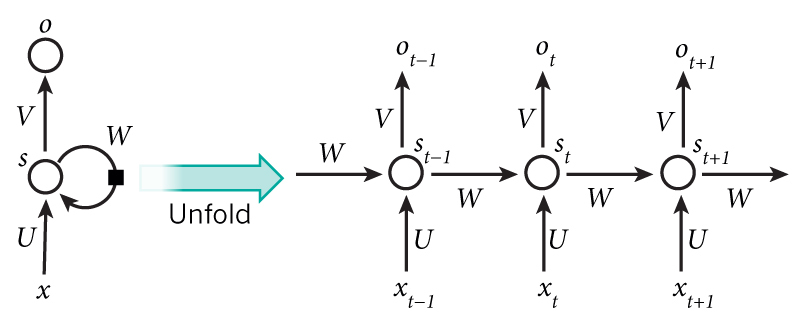}
	\caption{An RNN and its unfolding in time for forward computation}
	\label{figure:rnn}
\end{figure}

Figure \ref{figure:rnn} illustrates an RNN being unfolded (or unrolled) in time for forward computation, where $ x_t $ is the input at time $ t $, $ s_t $ is the hidden state at time step $ t $, and $ o_t $ is the output at step $ t $. For example, if the sequence we care about is a sentence of 5 words, the network would be unrolled into a 5-layer neural network, one layer for each word.

When the RNN is unrolled, it becomes a type of Deep Feed-Forward Neural Network (FFNN), but instead of using different parameters in each layer like in FFNN, it shares the same weights across every time step. 

A recent research line, starting from the seminal work of \cite{hidasi2016}, have adapted RNNs for session-based recommendations, where the sequence of user interactions in a session is leveraged to predict his next interaction in that session.

The Meta-Architecture proposed in this research uses RNNs to model users sessions in news portals.

\subsection{The \emph{Deep Structured Semantic Model (DSSM)}}
\label{sec:DSSM}
\begin{figure}[ht]
	\centering
	\includegraphics[height=6cm]{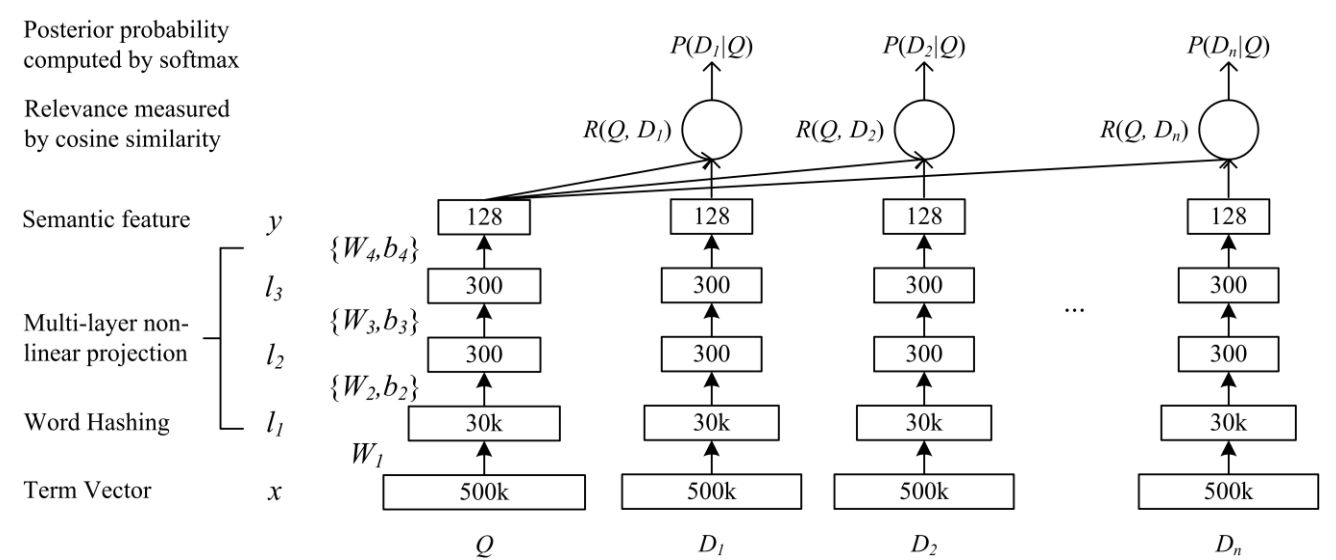}
	\caption{The \emph{Deep Semantic Structured Model (DSSM)}, from \cite{huang2013learning}}
	\label{figure:dssm}
\end{figure}

The \emph{Deep Semantic Structured Model (DSSM)} was proposed in \cite{huang2013learning} for ranking purpose and was later on extended to the recommendation scenarios in \cite{elkahky2015multi}, and is shown in Figure \ref{figure:dssm}. 

The \emph{DSSM} uses a FFNN to map the raw term vector (i.e., a bag-of-words textual representation) of a query or a document to its latent semantic vector, where the first layer, also known as the word hashing layer, converts the term vector to a letter-\emph{trigram} vector to scale up the training. The final layer's neural activities form the embedding vector representation in the semantic space. In document retrieval, the relevance score between a document and a query is the cosine similarity of their corresponding semantic concept vectors. The \emph{DSSM} is reported to give superior IR performance to other semantic models \cite{shen2014latent}. 

Essentially, the \emph{DSSM} can be seen as a multi-view learning model often composed of two or more neural networks for each individual view. In the original two-view \emph{DSSM} model, the left network represents the query view and the right network represents the document view.
The input of each neural network can be arbitrary types of features, e.g., letter-trigram used in the original paper \cite{huang2013learning}, or bag of unigram features used in \cite{elkahky2015multi}. 

Each input feature vector goes through non-linear transformations in the feed forward network to output an embedding vector, which is often much smaller than the original input space. The learning objective of the \emph{DSSM} is to maximize the cosine similarity between the two output vectors. During training, a set of positive examples and randomly sampled negative examples are generated in each mini-batch to minimize the cosine similarity with the positive examples. The log-likelihood
loss function used to learn the model parameters follows the pair-wise learning-to-rank paradigm  \cite{liu2015representation, rahimi2019listwise}.

\cite{gao2014modeling} demonstrate the effectiveness of the \emph{DSSM} by using two interesting tasks: automatic highlighting and contextual entity search. These tasks outperform not only the classic document models not using semantics but also the state-of-the-art of topic models. They trained the \emph{DSSM}, by using browsing transitions between Wikipedia documents.

\begin{figure}[h]
	\centering
	\includegraphics[height=7cm]{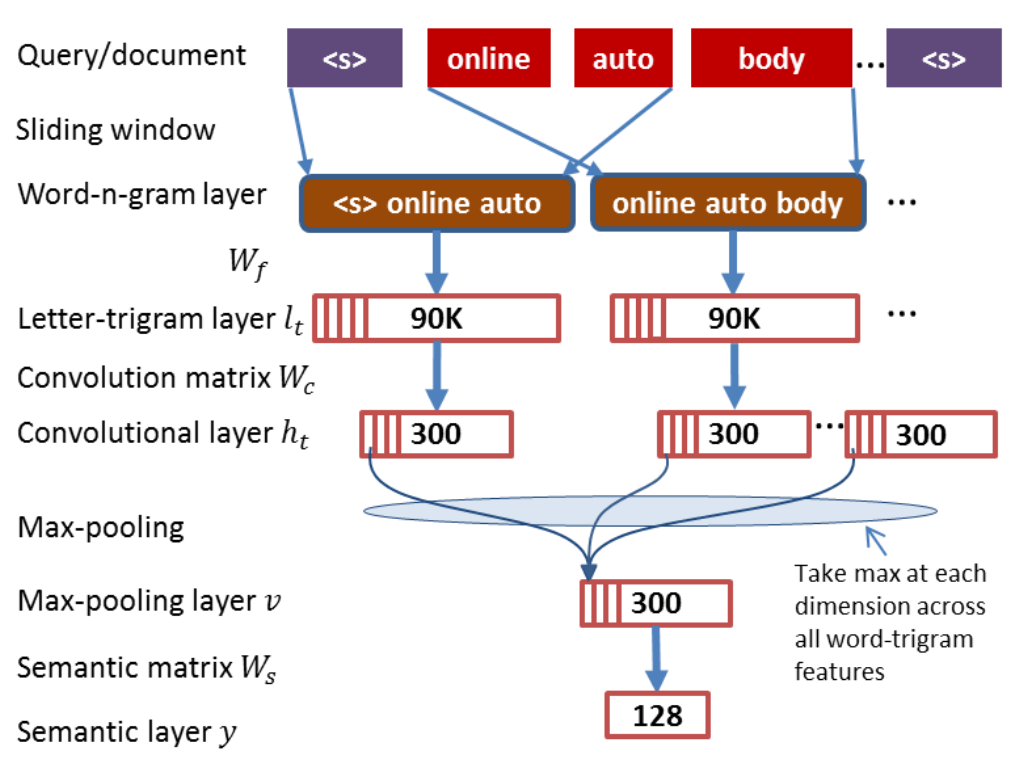}
	\caption{The \emph{Convolutional Latent Semantic Model (CLSM)}, from \cite{shen2014latent}}
	\label{figure:clsm}
\end{figure}

\cite{shen2014latent} have proposed the \emph{Convolutional Latent Semantic Model (CLSM)}, shown in Figure \ref{figure:clsm}.

This was done, based on the \emph{DSSM} and, according to authors, it was the first successful attempt in applying CNN-like methods to Information Retrieval (IR). One main difference from the conventional CNN is that the convolution operation in the \emph{CLSM} is implicitly applied on the letter-trigram representation space. Such convolutional architecture was more effective for IR to capture salient local features than simply summing over contributions from all words uniformly. 

The \emph{DSSM} and derived work were some of the main inspirations for the Meta-Architecture proposed in this research.

\section{Deep Learning for NLP}
\label{sec:deep-nlp}
The Natural Language Processing (NLP) is the use of human language by a computer. It includes applications like machine translation, entity recognition, disambiguation, parsing, part-of-speech tagging, among others. Many NLP applications are based on language models that define a probability distribution over sequences of words or characters, in a natural language \cite{Goodfellow-et-al-2016}.

Deep models based on CNN and RNN have shown their effectiveness for various NLP tasks \cite{zheng2016}, like sentence modeling \cite{kalchbrenner2014}, machine translation \cite{meng2015}, semantic parsing \cite{yih2014}, and many other NLP tasks \cite{santos2015}. Many inspirations for the application of Deep Learning in Recommender Systems came from successful Deep NLP techniques.

The \emph{Quasi-Recurrent Neural Network (QRNN)} is a NLP architecture that provides a sequence modeling that alternates convolutional layer in parallel across time-steps, and a minimalist recurrent pooling function that applies in parallel across channels \cite{bradbury2016quasi}. According to its authors, the \emph{QRNN} has advantages over the RNN, in terms of increased parallelism and even better predictive accuracy.

Neural language models are designed to overcome the curse of dimensionality problem for modeling NLP sequences. Words representation, originally sparse word vectors -- one-hot vectors whose dimensionality is the vocabulary size, are projected to a lower dimensional vector space, a.k.a distributed representations. They are able to recognize that two words are similar without loosing the ability to encode each word representation (embedding) as distinct from the other \cite{bengio2003neural}. 

Such word representations may explicitly encode many linguistic regularities and patterns. For example, the result of a vector calculation $vec("Madrid") - vec("Spain") + vec("France")$ is closer to $vec("Paris")$ than to any other word vector \cite{mikolov2013distributed}. These word embeddings have shown very good results in many NLP tasks \cite{mikolov2013, turian2010, collobert2011, bengio2003, zheng2016}.

\emph{Word2Vec} \cite{mikolov2013distributed} is a popular method for learning high-quality vector representations of words from large amounts of unstructured text data. It defines the context of a word as the surrounding words, based in the famous statement of the English linguist John Rupert Firth: "You shall know a word by the company it keeps" \cite{firth1957synopsis}.

\begin{figure}[h]
	\centering
	\includegraphics[height=8cm]{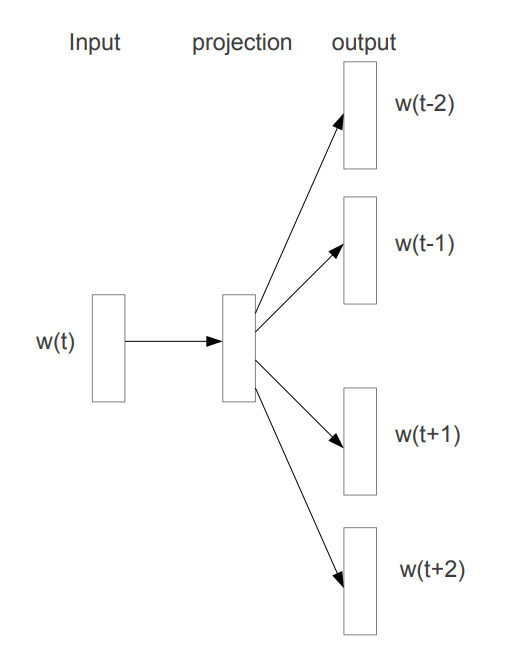}
	\caption{The Word2Vec Skip-gram model, from \cite{mikolov2013distributed}}
	\label{figure:skipgram}
\end{figure}

\emph{Word2Vec} embeddings can be trained by using two strategies: \emph{CBOW} and  \emph{Skip-gram}. The \emph{CBOW} is trained to predict a target word based on their surrounding words (context). The Skip-gram employs an inverse strategy, trying to predict the surrounding words based on the target word, as shown in Figure \ref{figure:skipgram}. Word embeddings trained by using the Skip-gram strategy have shown to be more robust and representative, specially when trained on large text corpus.

\emph{GloVe} is another popular method to obtain vector representations for words by using unsupervised learning. Training is performed on aggregated global word-word co-occurrence statistics from a corpus, and resulting representations showcase interesting linear substructures of the word vector space \cite{pennington2014glove}.

The \emph{Text CNN} \cite{kim2014convolutional} offers a simple solution for sentence-level embeddings by using convolutions. The convolutions act as a form of n-gram filters, allowing the network to embed sentence-level information and specializing word embeddings to higher-order tasks such as text classification or sentiment analysis. However, the kernel size in a CNN is fixed during training. To achieve good enough performance, sometimes an ensemble of multiple CNNs with different kernel sizes has to be used. A more natural and adaptive way of modeling text sequences would be to use gated RNN models \cite{hochreiter1997long, cho2014learning, sutskever2014sequence}.

In this research, articles' textual contents are represented as a sequence of pre-trained word embeddings, used as input for feature extraction by Deep NLP architectures.


\Glossary{Session-Based Recommender System}{Is a type of a Context-Aware RS (CARS), which leverages only information available in the current session for recommendations, ignoring the behaviour of past sessions of the user \cite{epure2017recommending}.}

\Glossary{Session-Aware Recommender System}{A type of Context-Aware RS (CARS) which leverages information about both the current session and past user sessions (differently than Session-Based RS) when available, to model user's preferences \cite{twardowski2016modelling}.}

\Glossary{Matrix Factorization}{Linear algebra methods that factorizes a matrix into a product of matrices. Traditional collaborative filtering techniques are based on Matrix Factorization of the User-Item matrix.}

\Glossary{Cold-start problem}{A well-known challenge in recommender systems research, specially for Collaborative Filtering methods. The \emph{User cold-start problem} occurs when most users have not provided enough interactions to receive accurate recommendations. Similarly, the \emph{Item cold-start problem} takes place when most active items have not received enough interactions to be effectively recommended.}

\Glossary{Content-Based Filtering (CBF)}{Family of recommendation methods that matches users and items by means of their attributes, ignoring the behaviour of other users \cite{Burke2007}.}

\Glossary{Collaborative Filtering (CF)}{Family of recommendation methods based on similar behavior among users in the system. They ignore users and item attributes. The main methods for CF are memory-based (eg. Nearest Neighbors) and model-based (e.g., Clustering, Association Rules, Matrix Factorization, Neural Networks) \cite{jannach2010}.}

\Glossary{Hybrid Filtering}{Family of recommendation methods that combines CF and CBF to provide recommendations with higher quality and to deal with the Cold-start problem, which usually takes place when using only CF methods \cite{Burke2002, Burke2007}.}

\Glossary{Context-Aware Recommender System (CARS)}{Family of recommendation methods that try to improve the quality of recommendations by adapting them to the specific contextual situation of the user \cite{Adomavicius2011}.}

\Glossary{Graph Neural Networks (GNN)}{Type of neural network architectures designed to generate representations for graphs \cite{scarselli2008graph, li2015gated}.}.

\Glossary{Convolutional Neural Network (CNN)}{Type of neural network architecture that apply convolution operations (known as kernels or filters) at different levels of granularity, under the assumption that neighbouring values are correlated in the input structure \cite{lecun1998}. They are able to reduce the number of parameters of the network, compared to MLP-only networks. CNNs have been successfully applied in Computer Vision and NLP.}

\Glossary{Recurrent Neural Network (RNN)}{Type of neural network architecture that naturally deals with sequence data. The RNNs have a hidden state, or memory, which represents past information previously processed. The RNN uses Back-Propagation Though Time (BPTT) \cite{williams1995gradient}, to learn its parameters. Gated RNN architectures \cite{chung2014empirical}, such as LSTM and GRU, are designed to overcome the vanishing gradient problem from the original RNN, by including gating units trained to control information flow through the network, thereby learning to keep information over a longer period of time.}

\Glossary{Deep Learning (DL)}{A family of Representation Learning methods that express representations or concepts in terms of composition of other simpler representations \cite{Goodfellow-et-al-2016}. Deep models usually have many stacked layers, and try to learn good representations from observed data, in such a way that they are modeled as being generated by interactions of many hidden factors \cite{betru2017}. Deep learning models have achieved super-human performance in tasks from Computer Vision, Audio, Speech Recognition, and Language Processing}.

\Glossary{Multi-Task Learning (MTL)}{The approach of training a model for more than one task. By sharing representations between related tasks, the model may generalize better, as can be seen in successful MTL applications in NLP \cite{collobert2008unified}, speech recognition \cite{deng2013new}, computer vision \cite{girshick2015fast}, and recommender systems \cite{bansal2016ask}.}

\Glossary{Natural Language Processing (NLP)}{Use of human language by a computer. It includes applications like machine translation, entity recognition, disambiguation, parsing, part-of-speech tagging, among others. Many NLP applications are based on language models that define a probability distribution over sequences of words or characters, in a natural language \cite{Goodfellow-et-al-2016}. Deep models based on CNN and RNN have shown their effectiveness for various NLP tasks \cite{zheng2016}, like sentence modeling \cite{kalchbrenner2014}, machine translation \cite{meng2015}, semantic parsing \cite{yih2014}, and many other NLP tasks \cite{santos2015}.}

\Glossary{Recommendation System (RS)}{A subclass of information filtering system that seeks to predict the relevance of an item for a given user. Recommender systems are utilized in a variety of domains, such as e-commerce, entertainment, media, and financial services companies.}

\Glossary{Stochastic Gradient Descent (SGD)}{An iterative method for optimizing an objective function with suitable smoothness properties (e.g.m differentiable). Methods derived from SGD are commonly used for training linear and deep neural networks. They are specially useful in big data applications, because SGD allows the network to be incrementally trained in steps composed by a single sample or by a mini-batch of samples (Mini-Batch Gradient Descent).}

\Glossary{Term Frequency-Inverse Document Frequency (TF-IDF)}{A numerical statistic that is intended to reflect how important a word is to a document in a collection or corpus. A relevant word for a document is one that is frequent in that document and rare in other documents of the corpus}

\Glossary{Multi-View Learning}{Multi-view or multi-modal learning aims to learn how to model latent factors for each view and jointly optimizes all the functions to improve the generalization performance \cite{zhao2017multi}}

\Glossary{Representation Learning}{Set of techniques that allows a system to automatically discover the representations needed for feature detection or classification from raw data. This reduces the need of manual feature engineering and allows a machine to both learn the features and use them to perform a specific task.}

\Glossary{Loss function}{In mathematical optimization and decision theory, a loss function, cost function, or objective function is a function that maps an event or values of one or more variables onto a real number intuitively representing some "cost" associated with the event. An optimization problem seeks to minimize a loss function.}

\Glossary{Session}{A sequence of user interactions that takes place within a given time frame \cite{quadrana2017personalizing}}.

\Glossary{G1 dataset}{Dataset provided for this research by \emph{Globo.com}, the most popular media company in Brazil. This dataset was originally shared by us in \cite{moreira2018news}, with a second version shared in \cite{moreira2019contextual}. More details in Section~\ref{sec:datasets}.}

\Glossary{Adressa dataset}{Dataset from the Adressa \cite{gulla2017adressa}, a Norwegian news portal which have been commonly used in news recommendation research. More details in Section~\ref{sec:datasets}.}


\Glossary{GRU4Rec}{A landmark neural architecture using RNNs for session-based recommendation \cite{hidasi2016}, described in Section~\ref{sec:gru4rec}.}

\Glossary{SR-GNN}{A recently published state-of-the-art architecture for session-based recommendation based on Graph Neural Networks \cite{wu2019session}, described in Section~\ref{sec:sr-gnn}.}

\Glossary{Item-kNN}{Baseline algorithm that returns the most similar items to the last read article using the cosine similarity between their vectors of co-occurrence with other items within sessions. This method has been commonly used as a baseline when neural approaches for session-based recommendation were proposed, e.g., in  \cite{hidasi2016}.}

\Glossary{Vector Multiplication Session-Based kNN (V-SkNN)}{Baseline algorithm that compares the entire active session with past (neighboring) sessions to determine items to be recommended. The similarity function emphasizes items that appear later within the session. The method proved to be highly competitive in the evaluations in \cite{jannach2017recurrent,ludewig2018evaluation,jugovac2018streamingrec}.}

\Glossary{Co-Occurence (CO)}{Baseline algorithm that recommends articles commonly viewed together with the last read article in previous user sessions. This algorithm is a simplified version of the association rules technique, having two as the maximum rule size (pairwise item co-occurrences) \cite{ludewig2018evaluation,jugovac2018streamingrec}.}

\Glossary{Sequential Rules (SR)}{Baseline algorithm that uses association rules of size two. It however considers the sequence of the items within a session. A rule is created when an item \emph{q} appeared after an item \emph{p} in a session, even when other items were viewed between \emph{p} and \emph{q}. The rules are weighted by the distance \emph{x} (number of steps) between \emph{p} and \emph{q} in the session with a linear weighting function $ w_{\text{\textsc{SR}}} = 1/x $ \cite{ludewig2018evaluation}.}

\Glossary{Recently Popular (RP)}{Baseline algorithm that recommends the most viewed articles within a defined set of recently observed user interactions on the news portal (e.g., clicks during the last  hour). Such a strategy proved to be very effective in the \emph{2017 CLEF NewsREEL Challenge} \cite{ludmann2017recommending}.}

\Glossary{Content-Based (CB)}{Baseline algorithm that recommends articles with similar content to the last clicked article, based on the cosine similarity of their \emph{Article Content Embeddings}.}


\Glossary{Hit Rate (HR@n) metric}{The percentage of times in which \emph{relevant} items (the next clicked ones) are retrieved among the top-N ranked items. More details in Section~\ref{sec:metrics}.}
\Glossary{Mean Reciprocal Ranking (MRR@n) metric}{A ranking metric that is sensitive to the position of the true next item (\emph{relevant} item) in the list. More details in Section~\ref{sec:metrics}.}

\Glossary{Item Coverage metric (COV@n)}{Sometimes also called ``aggregate diversity'' \cite{adomavicius2011improving}. The idea behind this metric is to measure to what extent an algorithm is able to diversify the recommendations and to make a larger fraction of the item catalog visible to the users. More details in Section~\ref{sec:metrics}.}

\Glossary{Expected Self-Information with Rank-sensitivity (ESI-R@n) metric}{Measures novelty as the negative log of items normalized popularity, with a rank discount. More details in Annex~\ref{sec:novelty_diversity_metrics}.}

\Glossary{Expected Self-Information with Rank- and Relevance-sensitivity (ESI-RR@n) metric}{Measures novelty as the negative log of items normalized popularity, with a rank and a relevance discount. More details in Annex~\ref{sec:novelty_diversity_metrics}.}

\Glossary{Expected Intra-List Diversity with Rank-sensitivity (EILD-R@n) metric}{Measures diversity as the average intra-distance between items pairs in a recommendation list, weighted by a logarithmic rank discount. More details in Annex~\ref{sec:novelty_diversity_metrics}.}

\Glossary{Expected Intra-List Diversity with Rank- and Relevance-sensitivity (EILD-RR@n) metric}{Measures diversity as the average intra-distance between items pairs in a recommendation list, weighted by a logarithmic rank and a relevance discount. More details in Annex~\ref{sec:novelty_diversity_metrics}.}

\Glossary{Batches before First Recommendation (BFR@n) metric}{A metric that measures the number of \emph{training batches} between the first user click in the article and the its first top-N recommendation by a given algorithm. More details in Section~\ref{sec:item_cold_start_results}.}

\Glossary{Accuracy}{A recommendation quality factor that measures the ability of an RS to provide recommendations that matches users interest, i.e., items actually interacted by users}

\Glossary{Novelty}{A recommendation quality factor that measures the ability of an RS to unknown or novel items for the users. For the purpose of this investigation, novelty is evaluated in terms of \emph{Long-Tail Novelty}, i.e., the ability of an RS to recommend long-tail (non-popular) items \cite{vargas2011rank,vargas2015thesis}.}

\Glossary{Diversity}{A recommendation quality factor that measures the ability of an RS to recommend a diversified list of items. The most common metrics for diversity are based on dissimilarity between recommended items \cite{zhang2002novelty, ziegler2005improving, rodriguez2012multiple, li2014modeling}. Lack of diversity may lead to poor user experiences \cite{li2014modeling}}

\Glossary{Item Coverage}{A recommendation quality factor that measures the percentage of active items that are ever recommended in top-N lists \cite{maksai2015predicting}.}

\Glossary{Gini Index}{The \emph{Gini Index} measures the inequality among values of a frequency distribution (for example, levels of income). A Gini coefficient of zero expresses perfect equality, where all values are the same. A Gini coefficient of one (or 100\%) expresses maximal inequality among values.}

\Glossary{Recency}{The recency of an item is the elapsed time since the article was initially made available for users in the system. In the news domain, for example, it is the time since an article was published.}

\Glossary{Recent Popularity}{For the purpose of this research, the recent popularity of an item is the number of user interactions an item have received in the last hour.}

\Glossary{Recent Normalized Popularity}{For the purpose of this research, the Normalized Popularity is the ratio between the number of interactions an item have received divided by the sum of all user interactions in a website, during the last hour}

\Glossary{Cosine similarity}{A measure of similarity between two non-zero vectors of an inner product space that measures the cosine of the angle between them. The cosine similarity is particularly used in positive space, where the outcome is neatly bounded in [0,1].}

\Glossary{Mini-batch}{A set of data samples used for training a model using Mini-Batch Gradient Descent.}

\Glossary{Multi-label classification}{A classification problem in which there is no constraint on how many of the classes the instance can be assigned to}.

\Glossary{Unsupervised learning}{A class of machine learning tasks that is able to find patterns in data set, without pre-existing labels.}

\Glossary{Supervised learning}{A class of machine learning tasks of learning a function that maps an input to an output (label) based on example input-output pairs.}

\Glossary{Pearson correlation coefficient}{A measure of the linear correlation between two variables.}

\Glossary{Deep Feed-Forward Neural Network (FFNN)}{A Multilayer Perceptron Network with many hidden fully connected layers. The parameters of its layers are trained with the Back-Propagation (BP) method, using SGD optimization.}


\Glossary{Article Content Representation (ACR)}{The \emph{CHAMELEON}'s module responsible to learn a distributed representation (an embedding) of articles' textual content.}

\Glossary{Next-Article Recommendation (NAR)}{The \emph{CHAMELEON}'s module responsible to generate next-article recommendations for ongoing user sessions.}

\Glossary{Contextual Article Representation (CAR)}{The sub-module of \emph{NAR} module responsible to combine the inputs for the \emph{NAR} module: (1) the pre-trained Article Content Embedding; (2) the Article Context features; and (3) the User Context features.}

\Glossary{Recommendations Ranking (RR)}{The sub-module of \emph{NAR} module responsible to recommend articles for a user session.}

\Glossary{SEssion Representation (SER)}{The sub-module of \emph{NAR} module responsible to model users' short-term  preferences, based on the sequence of interactions (reading news) in user's active session.}

\Glossary{Textual Features Representation (TFR)}{The sub-module of \emph{ACR} module responsible to learn relevant features directly from the article textual content.}

\Glossary{Content Embeddings Training (CET)}{The sub-module of \emph{ACR} module responsible to train the \emph{Article Content Embeddings (ACE)}, in a supervised or unsupervised approach.}

\Glossary{CHAMELEON}{The Deep Learning Meta-Architecture for News Recommender Systems proposed in this research.}

\Glossary{Article Content Embedding (ACE)}{A distributed representation (embedding) of the textual content of an article.}

\Glossary{Predicted Next-Article Embedding}{The output of the \emph{SER} sub-module, which is the predicted representation of a news content the user would like to read next in the active session.}

\Glossary{User-Personalized Contextual Article Embedding}{The output of the \emph{CAR} sub-module, which is a a non-linear combination of the content of the articles with user-personalized and contextual information.}

\Glossary{Article Metadata}{For the purpose of this work, this is a feature set composed by the following attributes of an article: Category and Author.}

\Glossary{Article Context}{For the purpose of this work, this is a feature set composed by the dynamic features of \emph{Novelty} and the \emph{Recency} of an article.}

\Glossary{Neural Architecture Search (NAS)}{A family of techniques for automating the design of artificial neural networks.}

\Glossary{User Context}{For the purpose of this work, this is a feature set composed by the user \emph{Location}, \emph{Device}, \emph{Time}, and \emph{Referrer}.}

\Glossary{No-ACE}{The \emph{No-ACE} Input Configuration used in the experiments to answer (\emph{RQ3}), that uses only features from the following feature sets: \emph{Article Context}, \emph{Article Metadata}, and the \emph{User Context}, without using a textual content representation. More details in Section~\ref{sec:content_representation}}


\Glossary{Deep Structured Semantic Model (DSSM)}{A deep neural network for learning semantic representations of entities in a common continuous semantic space and measuring their semantic similarities \cite{huang2013learning}. More details in Appendix~\ref{sec:DSSM}.}

\Glossary{W2V*TF-IDF}{A baseline representation of textual content, created by averaging the \emph{word2vec} embeddings of the title and first words of an article and weigthing by their \emph{TF-IDF} scores.}

\Glossary{Latent Semantic Analysis (LSA)}{A classical method of computing high-dimensional semantic vectors for words from their co-occurrence statistics \cite{deerwester1990lsa}.}

\Glossary{Sequence Denoising GRU Autoencoder (SDA-GRU)}{For the purpose of this work, it is the \emph{unsupervised} instantiation of the \emph{ACR} module, based on Sequence Denoising \emph{GRU} Autoencoders, as described in Section~\ref{sec:acr_unsupervised}.}

\Glossary{Long Short-Term Memory (LSTM)}{A Gated RNN architecture designed to deal with the vanishing gradient problem of original RNNs \cite{gers2001long, hochreiter1997long}.}

\Glossary{Gated Recurrent Unit (GRU)}{A Gated RNN similar to the LSTM with forget gate, but with fewer parameters than LSTM, as it lacks an output gate \cite{cho2014properties, chung2014empirical}.}

\Glossary{Update Gate RNN (UGRNN)}{The \emph{UGRNN} architecture is a compromise between \emph{LSTM}/\emph{GRU} and a vanilla RNN. In the \emph{UGRNN} architecture, there is only one additional gate, which determines whether the hidden state should be updated or carried over \cite{collins2016capacity}.}

\Glossary{Multi-View Deep Neural Network (MV-DNN)}{A Deep Learning architecture to map users and items to a latent space, where the similarity between users and their preferred items is maximized \cite{elkahky2015multi}. More details in Section~\ref{sec:MV-DNN}.}

\Glossary{word2vec}{The most popular method for learning high-quality vector representations of words from large amounts of unstructured text data. It defines the context of a word as the surrounding words, based in the famous statement of the English linguist John Rupert Firth: "You shall know a word by the company it keeps" \cite{firth1957synopsis}. More details in Section~\ref{sec:deep-nlp}.}

\Glossary{doc2vec}{The Paragraph Vector (a.k.a doc2vec) \cite{le2014distributed} learns fixed-length feature representations from variable-length pieces of texts, which are trained via the distributed memory and distributed bag of words models.}

\Glossary{GloVe}{A popular method to obtain vector representations for words by using unsupervised learning. Training is performed on aggregated global word-word co-occurrence statistics from a corpus, and resulting representations showcase interesting linear substructures of the word vector space \cite{pennington2014glove}.}

\Glossary{TOP1}{A pairwise ranking loss function proposed for \emph{GRU4Rec} in \cite{hidasi2018recurrent}.}

\Glossary{Bayesian Personalized Ranking (BPR)}{A classical pairwise ranking loss proposed in \cite{rendle2009bpr}.}

\itaglossary
\printglossary

\renewcommand\bibname{\itareferencesnamebabel} 
\bibliography{Referencias/referencias}

\FRDitadata{17 de dezembro de 2019}
\FRDitadocnro{DCTA/ITA/TD-035/2019} 
\FRDitaorgaointerno{Instituto Tecnol\'ogico de Aeron\'autica -- ITA}
\FRDitapalavrasautor{Recommender Systems; Deep Learning; News Recommender Systems; Session-based Recommender Systems}
\FRDitapalavrasresult{Sistemas de recomenda\c{c}\~{a}o; Not\'{i}cias; Aprendizagem (intelig\^{e}ncia artificial); Redes neurais; Computa\c{c}\~{a}o.}

\FRDitapalavraapresentacao{ITA, S\~ao Jos\'e dos Campos. Curso de Doutorado. Programa de P\'os-Gradua\c{c}\~ao em Engenharia Eletr\^onica e Computa\c{c}\~ao. \'Area de Inform\'atica. Orientador: Prof.~Dr. Adilson Marques da Cunha. Defesa em 09/12/2019. Publicada em 2019.} 
\FRDitaresumo{}
\FRDitaOpcoes{N}{O}
\itaFRD

\end{document}